 \documentclass[preprint,11pt,a4paper]{elsarticle}

\usepackage{graphicx}

\usepackage{amssymb}
\usepackage{amsmath}

\usepackage{lineno}

\biboptions{compress}

\usepackage[cm]{fullpage}
\usepackage[title]{appendix}
\usepackage{cases}
\usepackage{siunitx}
\usepackage{doi}
\usepackage{hyperref}
\hypersetup{
    colorlinks,
    linkcolor={red!50!black},
    citecolor={blue!50!black},
    urlcolor={blue!80!black}
}
\usepackage[font=small,labelfont=bf]{caption}
\usepackage[font=footnotesize,labelfont=bf,subrefformat=parens]{subcaption}
\usepackage{xcolor}
\usepackage{booktabs}  
\usepackage{multirow}
\usepackage{mathtools}

\journal{Journal of Non-Newtonian Fluid Mechanics}

\begin{document}

\newcommand{\vf}[1]{\underline{#1}}
\newcommand{\tf}[1]{\underline{\underline{#1}}}
\newcommand{\pd}[2]{\frac{\partial #1}{\partial #2}}
\newcommand{\ucd}[1]{\overset{\scriptscriptstyle \triangledown}{\tf{#1}}}
\newcommand\Rey{\operatorname{\mathit{Re}}}
\newcommand\Str{\operatorname{\mathit{Sr}}}
\newcommand\Wei{\operatorname{\mathit{Wi}}}
\newcommand\Deb{\operatorname{\mathit{De}}}
\newcommand\Bin{\operatorname{\mathit{Bn}}}
\allowdisplaybreaks

\begin{frontmatter}



\title{A Finite Volume method for the simulation of elastoviscoplastic flows and its application to 
the lid-driven cavity case}


\author[up]{Alexandros Syrakos\corref{cor1}}
\ead{syrakos@upatras.gr, alexandros.syrakos@gmail.com}

\author[up]{Yannis Dimakopoulos}
\ead{dimako@chemeng.upatras.gr}

\author[up]{John Tsamopoulos}
\ead{tsamo@chemeng.upatras.gr}

\cortext[cor1]{Corresponding author}

\address[up]{Laboratory of Fluid Mechanics and Rheology, Department of Chemical Engineering, 
University of Patras, Patras, Greece}

\begin{abstract}
We propose a Finite Volume Method for the simulation of elastoviscoplastic flows, modelled after 
the extension to the Herschel-Bulkley model by Saramito [J.\ Non-Newton.\ Fluid Mech.\ 158 (2009) 
154--161]. The method is akin to methods for viscoelastic flows. It is applicable to cell-centred 
grids, both structured and unstructured, and includes a novel pressure stabilisation technique of 
the ``momentum interpolation'' type. Stabilisation of the velocity and stresses is achieved through 
a ``both sides diffusion'' technique and the CUBISTA convection scheme, respectively. A second-order 
accurate temporal discretisation scheme with adaptive time step is employed. The method is used to 
obtain benchmark results of lid-driven cavity flow, with the model parameters chosen so as to 
represent Carbopol. The results are compared against those obtained with the classic 
Herschel-Bulkley model. Simulations are performed for various lid velocities, with slip and no-slip 
boundary conditions, and with different initial conditions for stress. Furthermore, we investigate 
the cessation of the flow, once the lid is suddenly halted. 
\end{abstract}


\end{frontmatter}



\section{Introduction}
\label{sec: introduction}

Viscoplastic (VP) fluids are a class of materials whose distinctive property is that they flow as  
fluids if subjected to large enough stresses but behave as solids if the applied stress is below a 
critical value, termed the yield stress. Although solids can also undergo plastic deformation, 
viscoplastic fluids are characterised by reversibility of the structural changes caused during 
plastic flow once the flow ceases \cite{Coussot_2017, Coussot_2018}. The class of VP fluids includes 
a variety of materials such as foams, emulsions, colloids and physical gels, with possibly different 
microscopic mechanisms being responsible for the emergence of a yield stress in each case 
\cite{Bonn_2017}. Viscoplastic flows are of major relevance in many industries (oil, construction, 
cosmetics, foodstuffs, etc.) and many natural flows can be classified as such (mud and lava flows, 
landslides, avalanches etc.).

Viscoplastic fluids were first studied in depth by Eugene Bingham \cite{Coussot_2017} who proposed 
the renowned constitutive equation named after him to describe their behaviour \cite{Bingham_1922}. 
A short time later, Herschel and Bulkley \cite{Herschel_1926} extended the Bingham model to describe 
also the shear-thinning (or shear-thickening) post-yield behaviour that most of these materials 
exhibit, by assuming a power-law dependence of the viscosity on the shear rate. These models were 
originally proposed in scalar form, but it was not long until a full tensorial form was proposed, 
employing the von Mises yield criterion \cite{Hohenemser_1932}. The empirical Herschel-Bulkley (HB) 
constitutive equation has been found to represent well the behaviour of yield stress fluids under 
steady shear flow, and is arguably the most popular model for VP fluids. It is commonly given in the 
following form:
\begin{equation} \label{eq: HB stress}
 \left\{
   \begin{array}{ll}
     \tau \leq \tau_y \quad & \Rightarrow \quad \tf{\dot{\gamma}} \;=\; 0
     \\
     \tau  >   \tau_y \quad & \Rightarrow \quad \tf{\tau} \;=\; 
                              \left( \dfrac{\tau_y}{\dot{\gamma}}  \;+\;
                                     k \dot{\gamma}^{n-1} \right) \tf{\dot{\gamma}} 
   \end{array}
 \right.
\end{equation}
where $\tf{\tau}$ is the deviatoric stress tensor; $\tf{\dot{\gamma}} = \nabla \vf{u} + (\nabla 
\vf{u})^{\mathrm{T}}$ is the rate-of-strain tensor, $\vf{u}$ being the fluid velocity vector and T 
denoting the tensor transpose; and $\tau \equiv (\frac{1}{2} \tf{\tau}:\tf{\tau})^{\frac{1}{2}}$ and 
$\dot{\gamma} \equiv (\frac{1}{2} \tf{\dot{\gamma}}:\tf{\dot{\gamma}})^{\frac{1}{2}}$ are their 
magnitudes. The HB model includes three parameters, the yield stress $\tau_y$, the consistency index 
$k$, and the exponent $n$; the latter is usually in the range $0.2-0.8$ \cite{Bonn_2017, 
Coussot_2018}, which represents a shear-thinning behaviour.

Equation \eqref{eq: HB stress} predicts two possible phases of the material, either rigid solid 
($\tau < \tau_y$) or fluid ($\tau > \tau_y$), separated by the yield surface ($\tau = \tau_y$). In 
order to express the von Mises yield criterion, $\tf{\tau}$ in Eq.\ \eqref{eq: HB stress} is defined 
to be the \textit{deviatoric} component of the total stress tensor $\tf{\sigma}$; such a 
decomposition of $\tf{\sigma}$ into deviatoric (traceless), $\tf{\sigma}{}_d$, and isotropic, 
$\tf{\sigma}{}_i$, components is useful for solids:
\begin{equation} \label{eq: total stress into isotropic and deviatoric}
 \tf{\sigma} \;=\; \underbrace{
 \tf{\sigma} \;-\; \frac{1}{3} \, \mathrm{tr}(\tf{\sigma}) \, \tf{I}}_{\tf{\sigma}{}_d}
 \;+\;
 \underbrace{\frac{1}{3} \, \mathrm{tr}(\tf{\sigma}) \, \tf{I}}_{\tf{\sigma}{}_i}
\end{equation}
where $\mathrm{tr}(\tf{\sigma}) \equiv \sum_i \sigma_{ii}$ is the trace of $\tf{\sigma}$. 
Therefore, in the common form \eqref{eq: HB stress} of the HB equation, $\tf{\tau}$ identifies 
with $\tf{\sigma}{}_d$ of Eq.\ \eqref{eq: total stress into isotropic and deviatoric}. However, from 
a constitutive equation for a fluid one expects a decomposition of $\tf{\sigma}$ into the 
\textit{extra-stress} tensor (for which the symbol $\tf{\tau}$ is usually reserved) which expresses 
the forces that arise due to deformation of the fluid, and an isotropic pressure component which is 
responsible for enforcing continuity in incompressible fluids:
\begin{equation} \label{eq: total stress into tau and p}
 \tf{\sigma} \;=\; \tf{\tau} \;-\; p\tf{I}
\end{equation}
where $\tf{I}$ is the identity tensor. Equation \eqref{eq: HB stress} does not suffice to define 
the HB extra-stress tensor because it allows the possibility that this tensor has an isotropic 
part, for which no information is given. This ambiguity is removed by the tacit assumption that the 
HB extra-stress tensor is traceless, so that decompositions \eqref{eq: total stress into isotropic 
and deviatoric} and \eqref{eq: total stress into tau and p} become equivalent and $\tf{\tau}$ of 
Eq.\ \eqref{eq: HB stress} also happens to equal the extra-stress tensor (hence the symbol 
$\tf{\tau}$ instead of $\tf{\sigma}{}_d$ in \eqref{eq: HB stress}). This makes the HB fluid a 
generalised Newtonial one, for which the extra and deviatoric stresses are identical since the 
former is traceless due to the incompressible continuity equation. When elasticity is introduced 
into the HB equation in Sec.\ \ref{sec: equations}, the extra-stress $\tf{\tau}$ will no longer be 
traceless and thus can no longer identify with $\tf{\sigma}{}_d$.

In the solid region, the concepts of extra-stress and pressure are not meaningful, since the HB 
solid does not deform, and the continuity equation is already enforced by the rigidity condition 
without a need for pressure. The symbols $\tf{\tau}$ and $p$ are still used there, but it must be 
kept in mind that they simply refer to the deviatoric and isotropic components of $\tf{\sigma}$. In 
other words, in the HB solid the stress decomposition is written as in Eq.\ \eqref{eq: total stress 
into tau and p}, but what is really meant is the decomposition \eqref{eq: total stress into 
isotropic and deviatoric}.

With some manipulations, noticing that in the fluid branch the stress tensor and the rate-of-strain 
tensor are parallel and thus the unit tensors $\tf{\dot{\gamma}} / \dot{\gamma}$ and $\tf{\tau} / 
\tau$ are equal, the solid and fluid branches of Eq.\ \eqref{eq: HB stress} can be combined into a 
single expression for $\tf{\dot{\gamma}}$ \cite{Saramito_2016}:
\begin{equation} \label{eq: HB rate-of-strain}
 \tf{\dot{\gamma}} \;=\; \left( \frac{\max(0, \tau-\tau_y)}{k} \right)^{\frac{1}{n}}
                         \frac{1}{\tau} \; \tf{\tau}
\end{equation}

Viscoplastic constitutive equations such as the HB have been studied extensively during the past 
decades. Their discontinuity at the yield surfaces and their inherent indeterminacy of the stress 
tensor within the unyielded regions require specialised numerical techniques for performing flow
simulations \cite{Mitsoulis_2017, Saramito_2017, Dimakopoulos_2018}. Furthermore, there is the 
question of whether their assumptions of a completely rigid (inelastic) solid phase and a purely 
viscous fluid phase are physically realistic. Allowing for some deformation of the solid phase under 
stress seems more natural and indeed experimental studies on Carbopol, a prototypical material often 
used in experimental studies on viscoplasticity, have shown that prior to yielding it exhibits 
elastic deformation under stress \cite{Piau_2007, Dinkgreve_2017}. Elastic effects can be observed 
also in the fluid phase. For example, bubbles rising in Carbopol solutions usually acquire the shape 
of an inverted teardrop, with a cusp at their leeward side \cite{Dubash_2007, Lopez_2018}, but the 
classical Bingham and HB viscoplastic models fail to predict such behaviour \cite{Tsamopoulos_2008, 
Dimakopoulos_2013}; on the other hand, such shapes are observed also in viscoelastic fluids, and are 
correctly captured by viscoelastic constitutive equations \cite{Fraggedakis_2016_JFM}. Similarly, 
for the settling of spherical particles in Carbopol, classical VP models cannot predict phenomena 
such as the loss of fore-aft symmetry under creeping flow conditions and the formation of a negative 
wake behind the sphere, but these phenomena are predicted if elasticity is incorporated into the 
constitutive modelling \cite{Fraggedakis_2016_SM}.

Therefore, recently the focus has been shifting towards constitutive equations that incorporate 
both plasticity and elasticity, usually called \textit{elastoviscoplastic} (EVP) constitutive 
equations. Actually, EVP constitutive modelling dates back to the beginning of the previous century 
-- a nice historical overview can be found in \cite{Saramito_2007}. Although several EVP models 
have been proposed (see e.g.\ the literature reviews in \cite{Saramito_2007, SouzaMendes_2012}), 
many of them appear only in scalar form. To be applicable in complex two- and three-dimensional 
flow simulations (2D/3D), a full tensorial form is required; some such models are compared in 
\cite{Fraggedakis_2016_JNNFM}.

Complex simulations also require that these models be accompanied by appropriate numerical solvers 
characterised by accuracy, robustness and efficiency. Usually, FEM solvers are employed in EVP flow 
simulations (e.g.\ \cite{Cheddadi_2013, Frey_2015, Fraggedakis_2016_JNNFM}). An alternative, very 
popular discretisation / solution method in Computational Fluid Dynamics is the Finite Volume 
Method (FVM); it has been successfully applied to viscoplastic (e.g.\ \cite{Syrakos_2013, 
Syrakos_2016b}) and viscoelastic (e.g.\ \cite{Oliveira_1998, Favero_2010, Afonso_2012, 
Syrakos_2018}) flows individually, but not to EVP flows, to the best of our knowledge (a hybrid 
FE/FV method was used in \cite{Belblidia_2011}). In this paper a FVM for the simulation of EVP flows 
is described. The EVP constitutive equation chosen is that of Saramito \cite{Saramito_2009} which 
introduces elasticity into the classic HB model, to which it reduces in the limit of inelastic 
behaviour. This model shall be referred to as the Saramito-Herschel-Bulkley (SHB) model. We chose 
this model because of its simplicity, its potential as revealed in a number of recent studies on 
materials such as foams \cite{Cheddadi_2012} and Carbopol \cite{Lacaze_2015}, and because of the 
popularity of the classic HB model. Nevertheless, the general framework of the presented FVM should 
be applicable to a range of other EVP models, particularly those that can be regarded as 
modifications / extensions of viscoelastic constitutive equations.

The presentation of the method in Sections \ref{sec: method: discretisation} and \ref{sec: method: 
solution} and its validation in Sec.\ \ref{sec: method: validation} are followed by application of 
the method to simulate EVP flow in a lid-driven cavity. The lid-driven cavity test case is arguably 
the most popular benchmark test case for new numerical methods for flow simulations. As such, it 
has been used also as a benchmark problem for viscoplastic \cite{Syrakos_2013, Syrakos_2014} and 
viscoelastic \cite{Sousa_2016} flows; there exist also the EVP flow studies of \cite{Martins_2013, 
Frey_2015}, but with a different EVP model that incorporates a kind of regularisation. In the 
present study the parameters of the SHB model are chosen so as to represent Carbopol, which is 
regarded as a simple VP fluid (more complex behaviour such as thixotropy \cite{Syrakos_2015} and 
kinematic hardening \cite{Dimitriou_2014} are not considered, but may be incorporated into the 
model in the future). The lid-driven cavity problem constitutes a convenient ``playground'' for 
testing the numerical method, testing the behaviour of the SHB model under conditions of complex 2D 
flow, comparing its predictions against those of the classic HB model, and providing benchmark 
results. The tests include varying the lid velocity to vary the flow character as quantified by 
dimensionless numbers, flow cessation (which occurs in finite time for VP flows), and varying the 
initial conditions to investigate the issue of multiplicity of solutions of the SHB model.

\section{Governing equations}
\label{sec: equations}

The flow is governed by the continuity, momentum, and constitutive equations. The first two are:
\begin{equation} \label{eq: continuity}
\nabla \cdot \left( \rho \vf{u} \right) \;=\; 0
\end{equation}
\begin{equation} \label{eq: momentum}
 \pd{(\rho\vf{u})}{t} \;+\; \nabla \cdot \left( \rho \vf{u} \vf{u} \right) \;=\; -\nabla p \;+\;
 \nabla\cdot\tf{\tau}
\end{equation}
%
where $t$ is time, and $\rho$ is the density of the material, assumed to be constant; the rest 
of the variables have been defined in Sec.\ \ref{sec: introduction}. The right-hand side of Eq.\ 
\eqref{eq: momentum} can be written collectively as $\nabla \cdot \tf{\sigma}$.

Closure of the system of governing equations requires that the extra stress tensor be related to 
the flow kinematics through a constitutive equation. In the present work we use the EVP equation 
of Saramito \cite{Saramito_2009}, which assumes that the total deformation of the material is equal 
to the sum of an elastic strain $\gamma_e$ (applicable in both the solid and fluid phases) and the 
provisional (if the yield stress is exceeded) viscous deformation $\gamma_v$ predicted by the HB 
model \eqref{eq: HB rate-of-strain}. This behaviour is depicted schematically in Fig.\ \ref{fig: 
model schematic} by a mechanical analogue. In terms of the rate-of-strain this is written as:
\begin{equation} \label{eq: constitutive}
 \underbrace{
 \vphantom{
   \left( \frac{\max(0, \tau_d-\tau_y)}{k} \right)^{\frac{1}{n}} \frac{1}{\tau} \; \tf{\tau}
 }
 \frac{1}{G} \, \ucd{\tau} 
 }_{\dot{\tf{\gamma}}{}_e}
 \;+\;
 \underbrace{
 \left( \frac{\max(0, \tau_d-\tau_y)}{k} \right)^{\frac{1}{n}} \frac{1}{\tau_d} \; \tf{\tau}
 }_{\dot{\tf{\gamma}}{}_v}
 \;=\; 
 \dot{\tf{\gamma}}
\end{equation}
where $G$ is the elastic modulus, the triangle denotes the upper-convected time derivative
\begin{equation} \label{eq: upper convected derivative}
 \ucd{\tau} \;\equiv\; \pd{\tf{\tau}}{t} \;+\; \vf{u} \cdot \nabla \tf{\tau} 
            \;-\; (\nabla \vf{u})^{\mathrm{T}} \!\cdot \tf{\tau} \;-\; \tf{\tau} \cdot \nabla 
\vf{u} 
\end{equation}
and $\tau_d \equiv (\frac{1}{2} \tf{\tau_d}:\tf{\tau_d})^{\frac{1}{2}}$ is the magnitude of the 
deviatoric part of the EVP stress tensor,
\begin{equation} \label{eq: deviatoric stress}
 \tf{\tau}{}_d \;\equiv\; \tf{\tau} \;-\; \frac{1}{3} \mathrm{tr}(\tf{\tau}) \tf{I}
\end{equation}
Note that because $\tf{\tau}$ and $\tf{\sigma}$ differ only by an isotropic quantity ($-p\tf{I}$, 
Eq.\ \eqref{eq: total stress into tau and p}) it holds that $\tf{\tau}{}_d = \tf{\sigma}{}_d$. An 
important difference with the classic HB model \eqref{eq: HB rate-of-strain} is that, due to the 
last two terms of the upper convected derivative \eqref{eq: upper convected derivative}, the trace 
of $\tf{\tau}$ is now not necessarily zero and thus, in general, $\tf{\tau}{}_d \neq \tf{\tau}$, 
necessitating explicit use of $\tau_d$ inside the ``max'' term of Eq.\ \eqref{eq: constitutive} in 
order to express the von Mises yield criterion. The full 3D formula \eqref{eq: deviatoric stress} 
must be used for the deviatoric stress even in 1D or 2D simulations: a 2D stress state where all 
$\tau_{ij}$ are zero except $\tau_{11} = \tau_{22} \neq 0$ is not isotropic ($\tau_{33} \neq 
\tau_{11}, \tau_{22}$) and $\tau_d$ is not zero.

\begin{figure}[tb]
  \centering
  \includegraphics[scale=1.00]{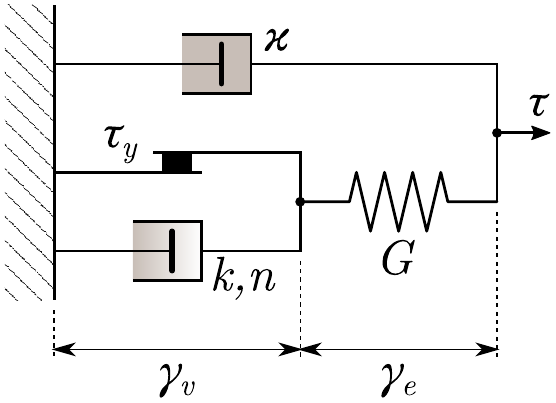}
  \caption{A mechanical analogue of the Saramito-Herschel-Bulkley (SHB) model \cite{Saramito_2009}.}
  \label{fig: model schematic}
\end{figure}

Other qualitative differences with the HB model exist. For example, the SHB model allows non-zero 
rate of strain in the unyielded regions, arising from elastic deformations of the solid phase. 
Conversely, it is theoretically possible that the rate of strain is zero in yielded material because 
the stresses there have not had enough time to relax. In particular, in the SHB model, contrary to 
the HB model, the stresses do not respond instantaneously to changes in the rate of strain; rather, 
the EVP stress tensor is also a function of all past states of $\dot{\tf{\gamma}}$, as in 
viscoelastic fluids.

Furthermore, as noted in \cite{Cheddadi_2008, Cheddadi_2012}, the combination of plastic and elastic 
terms in the SHB Eq.\ \eqref{eq: constitutive} can produce complex behaviour not observed in either 
purely viscoplastic or purely viscoelastic flows. One such feature is that the extra stress tensor 
and the velocity gradient can vary discontinuously across yield surfaces \cite{Cheddadi_2008}. 
Another feature is that flows are inherently transient, in the sense that there exist an infinitude 
of steady states which depend in a continuous manner on the initial conditions. Even if it is just 
the steady state that is sought, one cannot simply discard the time derivatives in Eqs.\ \eqref{eq: 
momentum} and \eqref{eq: constitutive}; the steady state must be obtained by a transient simulation 
with appropriate initial conditions. It is not only the steady state stresses that depend on the 
initial conditions, but also the steady state velocity. This issue was studied in 
\cite{Cheddadi_2012} in the context of cylindrical Couette flow. In fact, actual EVP materials do 
behave this way in experiments, with the steady state depending on the residual stresses present in 
the unyielded, stationary material at the start of the experiment \cite{Cheddadi_2012}. The residual 
stresses are stresses that are ``trapped'' inside unyielded material because there is no relaxation 
mechanism there. E.g.\ for stationary ($\vf{u} = 0$) unyielded material, Eq.\ \eqref{eq: 
constitutive} predicts $\partial\tf{\tau} / \partial t = 0$. In experiments, residual stresses do 
develop during the preparation of the material and are difficult to eliminate. 

This contrasts the behaviour of classic VP models such as the HB, where there is a stress 
indeterminacy in the unyielded regions: there exist infinite $\tf{\sigma}$ fields within these 
regions that have the same divergence $\nabla \cdot \tf{\sigma}$ which yields $\dot{\tf{\gamma}} = 
0$ when substituted in the momentum equation \eqref{eq: momentum}. Each of these fields is a valid 
solution. The HB steady-state does not depend on the initial conditions, and can be obtained 
directly, by dropping the time derivative from Eq.\ \eqref{eq: momentum}. The HB model makes no 
connection between this indeterminacy and the initial conditions, and in fact it even allows that 
stresses in the unyielded regions vary discontinuously in time. No such indeterminacy is exhibited 
by the SHB model, the stresses in both the yielded and unyielded regions being precisely determined 
for given initial conditions.

In the limit of very high elastic modulus, $G \rightarrow \infty$, the SHB material becomes so 
stiff that it behaves as an inelastic fluid or solid; the importance of the first term on the 
left-hand side of the SHB constitutive equation \eqref{eq: constitutive} diminishes and that 
equation tends to become identical to the classic HB equation \eqref{eq: HB rate-of-strain}. 
However, it must be kept in mind that the difference in character between the SHB and HB models is 
in some respects retained no matter how large the value of $G$ is. This concerns mostly the stress 
state in the unyielded regions, which for the SHB model remains uniquely determined by the initial 
conditions whereas that of the HB model is indeterminate. Also, as discussed in Appendix 
\ref{appendix: SHB tau in limit of large G}, in the HB model $\tf{\tau}$ was defined to equal 
$\tf{\sigma}{}_d$ whereas for the SHB model the identification of $\tf{\tau}$ with $\tf{\sigma}{}_d$ 
as $G \rightarrow \infty$ is not guaranteed. 

Figure \ref{fig: model schematic} shows that the complete SHB model contains an additional viscous 
component labelled $\kappa$ not discussed thus far. This is a Newtonian component, of viscosity 
$\kappa$, which makes the unyielded phase behave as a Kelvin-Voigt solid. The entire extra stress 
tensor is then
\begin{equation} \label{eq: entire extra stress}
  \tf{\tau_e} \;=\; \tf{\tau} \;+\; \kappa \dot{\tf{\gamma}}
\end{equation}
The SHB model then reduces to the Oldroyd-B viscoelastic model when $\tau_y = 0$ and $n = 1$. 
However, in the main results of Sec.\ \ref{sec: results} we will use $\kappa = 0$.

Finally, concerning the boundary conditions, we consider solid wall boundaries. We will mostly 
employ the no-slip condition, but since (elasto)viscoplastic materials are usually slippery we will 
also employ a Navier slip condition, according to which the relative velocity between the fluid and 
the wall, in the tangential direction, is proportional to the tangential stress. For two-dimensional 
flows this is expressed as follows: Let $\vf{n}$ be the unit vector normal to the wall, and $\vf{s}$ 
be the unit vector tangential to the wall within the plane in which the equations are solved. Let 
also $\vf{u}$ and $\vf{u}_w$ be the fluid and wall velocities, respectively. Then,
\begin{equation} \label{eq: navier slip}
 \left( \vf{u} - \vf{u}_w \right) \cdot \vf{s} \;=\; \beta \left( \vf{n} \cdot \tf{\tau} \right) 
\cdot \vf{s}
\end{equation}
where the parameter $\beta$ is called the slip coefficient.

\subsection{Nondimensionalisation of the governing equations}

In the present work we will solve the dimensional governing equations. Nevertheless, 
nondimensionalisation of the equations reveals a number of dimensionless parameters that 
characterise the flow. So, let $L$ and $U$ be characteristic length and velocity scales of the 
problem, with $T = L/U$ the associated time scale, and choose the following characteristic value 
of stress, $S$ (the Newtonian viscosity $\kappa$, Eq.\ \eqref{eq: entire extra stress}, is 
omitted): 
\begin{equation} \label{eq: characteristic stress}
S \;\equiv\; \tau_y \;+\; k \left( \frac{U}{L} \right)^n
\end{equation}
Then, we define the dimensionless variables, denoted with a tilde (\textasciitilde), as $x_i = 
\tilde{x}_i L$, $t = \tilde{t} T$, $\vf{u} = \tilde{\vf{u}} U$, $p = \tilde{p} S$, and $\tf{\tau} = 
\tilde{\tf{\tau}} S$. Substituting these into the governing equations \eqref{eq: continuity}, 
\eqref{eq: momentum}, and \eqref{eq: constitutive}-\eqref{eq: upper convected derivative}, we obtain 
their dimensionless forms:
\begin{equation} \label{eq: continuity ND}
 \tilde{\nabla} \cdot \tilde{u} \;=\; 0
\end{equation}
\begin{equation} \label{eq: momentum ND}
 \Rey \left( \pd{\tilde{\vf{u}}}{\tilde{t}} \;+\;
             \tilde{\nabla} \cdot (\tilde{u} \tilde{u}) \right)
 \;=\;
 - \tilde{\nabla} \tilde{p} \;+\;
   \tilde{\nabla} \cdot \tilde{\tf{\tau}}
\end{equation}
\begin{equation} \label{eq: constitutive ND}
 \Wei \left( 
             \pd{\tilde{\tf{\tau}}}{\tilde{t}} \;+\;
             \tilde{\vf{u}} \cdot \tilde{\nabla} \tilde{\tf{\tau}} \;-\;
             ( \tilde{\nabla} \tilde{\vf{u}} )^{\mathrm{T}} \cdot \tilde{\tf{\tau}} \;-\;
             \tilde{\tf{\tau}} \cdot \tilde{\nabla} \tilde{\vf{u}}
      \right)
 \;+\;
 \left( \frac{\max \left( 0, \tilde{\tau}_d - \Bin \right)}{1 - \Bin} \right)^{\frac{1}{n}}
   \frac{1}{\tilde{\tau}_d} \; \tilde{\tf{\tau}}
 \;=\;
 \tilde{\dot{\tf{\gamma}}}
\end{equation}
where $\tilde{\nabla} = (1/L) \nabla$ and $\tilde{\dot{\tf{\gamma}}}  = \dot{\tf{\gamma}} / (U/L)$. 
The following dimensionless numbers appear:
\begin{align}
 \label{eq: Reynolds number}
 \text{Reynolds number} \qquad & \Rey \;\equiv\; \frac{\rho U^2}{S}
 \\[0.17cm]
 \label{eq: Weissenberg number}
 \text{Weissenberg number} \qquad & \Wei \;\equiv\; \frac{S}{G}
 \\[0.17cm]
 \label{eq: Bingham number}
 \text{Bingham number} \qquad & \Bin \;\equiv\; \frac{\tau_y}{S}
\end{align}

A more standard choice for $S$ would account only for viscous effects, omitting plasticity: $S = 
k(U/L)^n$. This would lead to more standard definitions of the Reynolds and Bingham numbers:
\begin{alignat}{2}
 \label{eq: Reynolds number standard}
 \Rey' \;&\equiv\; \frac{\rho U^{2-n} L^n}{k} \;&&=\; \frac{\Rey}{1 - \Bin}
 \\[0.17cm]
 \label{eq: Bingham number standard}
 \Bin' \;&\equiv\; \frac{\tau_y L^n}{k U^n} \;&&=\; \frac{\Bin}{1 - \Bin}
\end{alignat}
However, it was shown in \cite{Syrakos_2016a} (see also the discussion in \cite{Thompson_2016}) that 
in viscoplastic flows $\Rey$ suffices as a standalone indicator of inertial effects in the flow, 
whereas $\Rey'$ does not (the inertial character must be inferred from the values of $\Rey'$ and 
$\Bin'$ combined). Hence the definition \eqref{eq: characteristic stress} was preferred. Concerning 
the Bingham number, $\Bin \in [0,1]$ and $\Bin' \in [0,\infty)$ are simply related through Eq.\ 
\eqref{eq: Bingham number standard} and carry exactly the same information. Because of familiarity 
with $\Bin' $ that is almost universally used in the literature, we will use both $\Bin$ and $\Bin'$ 
in this study.

The Weissenberg number $\Wei$ is an indicator of elastic effects in the flow. Unlike the $\Rey$ and 
$\Bin$ numbers, it is not a ratio between stresses of different nature or momentum fluxes. In fact, 
as seen in the mechanical analogue of Fig.\ \ref{fig: model schematic}, omitting the Newtonian 
component $\kappa$, the viscoplastic and elastic components are connected in series and thus carry 
the same load. Therefore, the representative stress $S$, although defined by considerations 
pertaining to the viscoplastic component of the material behaviour (Eq.\ \eqref{eq: characteristic 
stress}), is borne also by the elastic component of the material structure and thus $\Wei \equiv 
S/G$ is a typical elastic deformation. In the literature, the standard form for the Weissenberg 
number is $\Wei = \lambda U / L$, where $\lambda$ is a relaxation time, which becomes equivalent to 
our definition if we define
\begin{equation} \label{eq: lamda and eta}
 \lambda \;\equiv\; \frac{\eta}{G} \qquad \text{where} \qquad \eta \equiv \frac{S}{U/L}
\end{equation}
The relaxation time $\lambda$ is proportional to the apparent viscosity $\eta$ (the more viscous 
the flow is, the slower the recovery from elastic deformations) and inversely proportional to the 
elastic modulus $G$ (the stiffer the material is, the faster it recovers). Note that $\lambda$ and 
$\eta$ as defined by \eqref{eq: lamda and eta} are not material constants but reflect also the 
influence of the flow, as they depend on $U$ and $L$ in addition to the material parameters 
$\tau_y$, $G$, $k$ and $n$ (Eq.\ \eqref{eq: characteristic stress}). The definition $\Wei = \lambda 
U / L$ is also interpreted as a typical elastic strain: it is the product of the strain rate $U/L$ 
by the time 
period $\lambda$ during which the material has not yet significantly relaxed.

Finally, an important dimensionless number is the yield strain,
\begin{equation} \label{eq: yield strain}
 \gamma_y \;=\; \frac{\tau_y}{G} \;=\; \Bin \cdot \Wei
\end{equation}
which depends only on the material parameters and not on kinematic or geometric parameters of the 
flow such as $U$ and $L$. The fact that it is equal to the product of $\Bin$ and $\Wei$ means that 
these two numbers are not independent but their product is constant for a given material. Since 
$\Bin \in [0,1]$, it follows that $\Wei \in [\gamma_y, \infty)$. That $\Wei$ is bounded from below 
by $\gamma_y$ follows also directly from the definitions \eqref{eq: Weissenberg number} and 
\eqref{eq: characteristic stress}, and derives from the characteristic stress definition \eqref{eq: 
characteristic stress} which assumes the existence of unyielded regions; in such regions the strain, 
of which $\Wei$ is a measure, is higher than the yield strain $\gamma_y$.

\section{Discretisation of the governing equations}
\label{sec: method: discretisation}

\subsection{Preliminary considerations}
\label{ssec: method: discretisation: preliminary}

\begin{figure}[tb]
  \centering
  \includegraphics[scale=0.75]{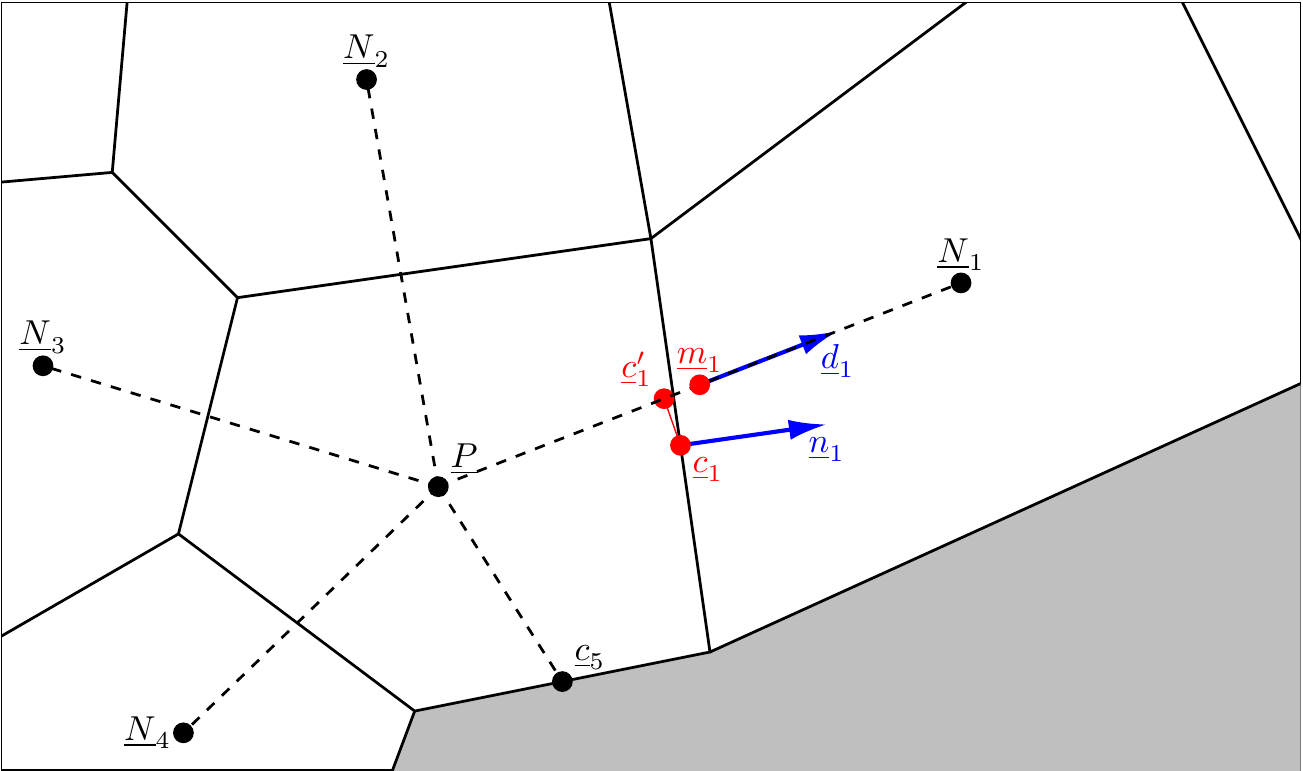}
  \caption{Cell $P$ and its neighbouring cells, each having a single common face with $P$. Its faces 
and neighbours are numbered in anticlockwise order, with face $f$ separating $P$ from its neighbour 
$N_f$. The shaded area lies outside the domain, so face 5 is a boundary face.  The geometric 
characteristics of face 1 are displayed. The position vectors of the centroids of cells $P$ and 
$N_f$ are denoted as $\vf{P}$ and $\vf{N}_f$; $\vf{c}_f$ is the centroid of face $f$ and $\vf{c}'_f$ 
is its closest point on the line connecting $\vf{P}$ and $\vf{N}_f$; $\vf{m}_f$ is the midpoint 
between $\vf{P}$ and $\vf{N}_f$; $\vf{n}_f$ is the unit vector normal to face $f$, pointing outwards 
of $P$, and $\vf{d}_f$ is the unit vector pointing from $\vf{P}$ towards $\vf{N}_f$. The volume of 
cell $P$ is denoted as $\Omega_P$.}
  \label{fig: grid nomenclature}
\end{figure}

In this section we propose a Finite Volume Method (FVM) for the discretisation of the governing 
equations. The method will be described for two-dimensional problems, but extension to three 
dimensions is straightforward. The first step is the tessellation of the domain $\Omega$ into a 
number of non-overlapping volumes, or \textit{cells}. Each cell is bounded by straight faces, each 
of which separates it from another single cell or from the exterior of the domain (the latter are 
called boundary faces). Figure \ref{fig: grid nomenclature} shows such a cell $P$ along with its 
faces and neighbours, and the associated nomenclature. Our FVM is applicable to grids of arbitrary 
polygonal cells, although for the results of Sec.\ \ref{sec: results} we will employ Cartesian grids 
due to the regularity of the problem geometry. Grids will be labelled after a characteristic cell 
length $h$.

\begin{figure}[tb]
  \centering
  \includegraphics[scale=0.60]{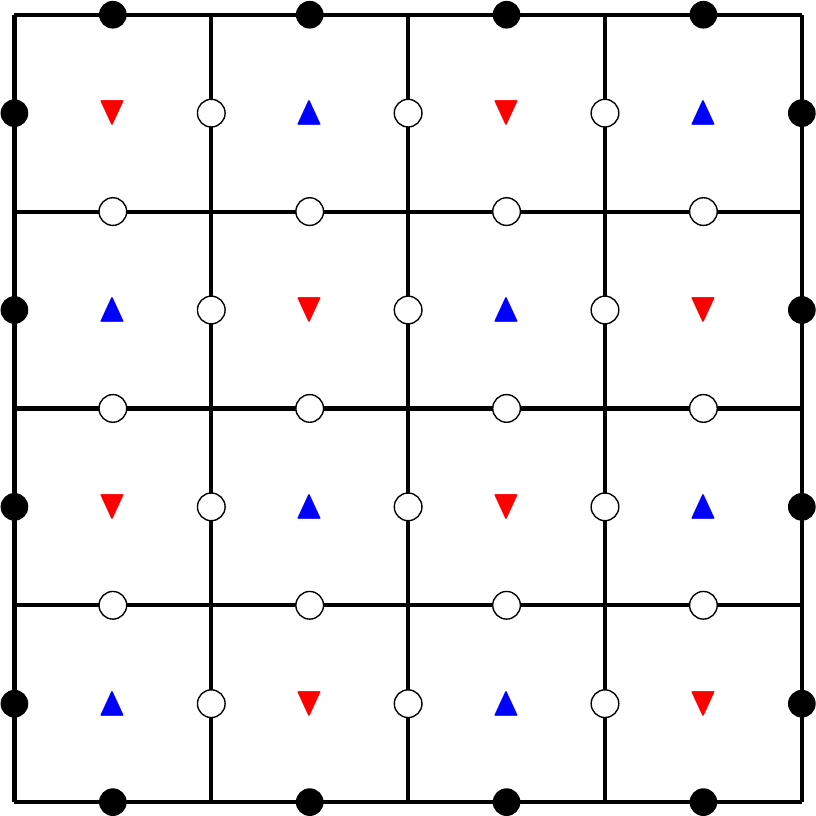}
  \caption{A checkerboard distribution of values on a Cartesian grid. Different values are stored 
at points \textcolor{blue}{$\blacktriangle$} and \textcolor{red}{$\blacktriangledown$}.}
  \label{fig: oscillations schematic}
\end{figure}

The discretisation procedure will convert the governing equations into a large set of algebraic 
equations involving only (approximate) values of the dependent variables at the cell centroids. 
These values will be denoted using the cell index as a subscript, i.e.\ $\phi_P$ is the value of the 
variable $\phi$ at the centroid of cell $P$; if the variable name already includes a subscript then 
the cell index will be separated by a comma (e.g.\ $\tau_{d,P}$ is the deviatoric stress at cell 
$P$). We aim for a second-order accurate method, i.e.\ one whose discretisation error scales as 
$h^2$. With the present governing equations, a difficulty arises from the fact that they all contain 
only first spatial derivatives because this allows the development of spurious oscillations in the 
solution. For example, assume that on the grid of Fig.\ \ref{fig: oscillations schematic}, a 
quantity $\phi$ has the value zero at the boundary points ($\bullet$), the value $+1$ at points 
(\textcolor{blue}{$\blacktriangle$}), and the value $-1$ at points 
(\textcolor{red}{$\blacktriangledown$}). FVM integration of first derivatives of $\phi$ over each 
cell eventually comes down to a calculation involving values of $\phi$ at face centres ($\circ$) 
and ($\bullet$), due to application of the Gauss theorem. If we use linear interpolation to 
approximate the values of $\phi$ at the face centroids ($\circ$) from the values at the cell centres 
on either side of each face, then we obtain $\phi = 0$ at every face centre ($\circ$). This 
ultimately results in obtaining a zero integral of the first derivative of $\phi$ over any cell. 
Thus, interpolation filters out the oscillating cell-centre field and leaves a smooth field over 
the face centres, which in turn produces an image of the discrete operator that varies smoothly from 
one cell to the next.

To see how this creates a problem, consider the FVM discretisation on grid $h$ of the following PDE
\begin{equation} \label{eq: PDE}
f(\phi) \;=\; b
\end{equation}
where $b$ is a known function. The FVM produces a system of algebraic equations that can be written 
as
\begin{equation} \label{eq: FVM system}
 F_h(\phi_h) \;=\; B_h
\end{equation}
where $\phi_h$ is the vector of unknown values of $\phi$ at the cell centroids, $F_h$ is the 
discrete operator representing the integral of the differential operator $f$ over each cell, and 
$B_h$ is the vector of integrals of the known function $b$ over each cell. The aforementioned 
filtering action of the face interpolation scheme means that $F_h$ will filter out any oscillations 
in $\phi_h$, producing a smooth image $F_h(\phi_h)$; therefore, the solution $\phi_h$ to the system 
\eqref{eq: FVM system} can be oscillatory even if the right-hand side $B_h$ is smooth. This is not 
just a remote possibility, but it does occur in practice, a notorious example being the spurious 
pressure oscillations produced by the FVM solution of the Navier-Stokes equations (see 
\cite{Syrakos_06a} for a demonstration). But, if the discretisation schemes incorporated in the 
operator $F_h$ are designed so as to reflect any oscillations in the vector $\phi_h$ onto the image 
$F_h(\phi_h)$, then for a smooth right-hand side $B_h$ the system \eqref{eq: FVM system} can only 
have a smooth solution $\phi_h$. For the Navier-Stokes equations, the main concern is the pressure 
oscillations because any velocity oscillations will be reflected onto the second derivatives of 
velocity present in the momentum equations and are therefore inhibited. But in our case, all three 
of the governing equations, \eqref{eq: continuity}, \eqref{eq: momentum} and \eqref{eq: 
constitutive}--\eqref{eq: upper convected derivative}, only contain first derivatives and thus we 
have to be concerned about the possibility of spurious oscillations in all three variables $\vf{u}$, 
$p$, and $\tf{\tau}$. In the following subsections the adopted measures will be described.

In what follows we will need a ``characteristic viscosity'', a quantity with units of viscosity 
somewhat characteristic of the flow's viscous character. Our first choice is the following:
\begin{equation} \label{eq: apparent viscosity}
 \eta_a \;\equiv\; \kappa \;+\; \frac{S}{U/L}
\end{equation}
where $S$ is given by \eqref{eq: characteristic stress}. This value is constant throughout the 
domain. A second option we tried is similar to the coefficient of the DAVSS-G technique proposed in 
\cite{Sun_1999} and varies throughout the domain, depending on the ratio between the magnitudes of 
the stress and rate-of-strain tensors:
\begin{equation} \label{eq: apparent viscosity DAVSS}
 \eta_a \;\equiv\; \kappa \;+\;
                   \frac{1}{2} \, \frac{S}{U/L} \;+\;
                   \frac{1}{2} \, \frac{S \,+\, \tau_P \,+\, \tau_{N_f}}
                                       {U/L \,+\, \dot{\gamma}_P \,+\, \dot{\gamma}_{N_f}}
\end{equation}
The above formula gives $\eta_a$ at face $f$ of cell $P$. With definition \eqref{eq: apparent 
viscosity DAVSS}, $\eta_a$ never falls below $\kappa + 0.5 S / (U/L)$, tends to $\kappa + S/(U/L)$ 
when both $\tau$ and $\dot{\gamma}$ are small, and tends to $\kappa + 0.5 S / (U/L) + \tau / 
\dot{\gamma}$ when both $\tau$ and $\dot{\gamma}$ are large. Also, it does not tend to infinity when 
$\dot{\gamma} \rightarrow 0$, although it has no upper bound.

The discretisation of several terms (e.g.\ the calculation of $\tf{\dot{\gamma}} \equiv \nabla 
\vf{u} + (\nabla \vf{u})^{\mathrm{T}}$ and its magnitude) requires the use of a discrete gradient 
operator which approximates $\nabla \phi$ at the cell centres. We employ the least-squares operator 
described in detail in \cite{Syrakos_2017}, denoted here as $\nabla_h^q$, the superscript $q$ 
being the exponent employed; in \cite{Syrakos_2017} it was shown that on smooth structured grids the 
choice $q=1.5$ engenders second-order accuracy ($\nabla_h^{1.5} \phi = \nabla \phi + O(h^2)$) while 
with other $q$ values the accuracy degrades to first-order at boundary cells. On irregular 
(unstructured) grids all choices of $q$ result in first order accuracy everywhere. Nevertheless, 
first-order accurate gradients suffice for second-order accuracy of the differentiated variables, 
i.e.\ are compatible with second-order accurate FVMs \cite{Syrakos_2017}.

Finally, linear interpolation for calculating face centre values on unstructured grids is performed 
as
\begin{equation} \label{eq: CDS}
 \bar{\phi}_{c_f} \;=\; \bar{\phi}_{c'_f} \;+\; \bar{\nabla}^q_h \phi_{c'_f} \!\cdot\! (\vf{c}_f - 
\vf{c}'_f) 
\end{equation}
where the overbar denotes an interpolated value, and in the right-hand side both $\phi$ and its 
gradient are interpolated linearly to point $\vf{c}'_f$ (the closest point to the face centre lying 
on the line joining $\vf{P}$ to $\vf{N}_f$, Fig.\ \ref{fig: grid nomenclature}) according to the 
formula
\begin{equation} \label{eq: linear interpolation}
 \bar{\phi}_{c'_f} \;\equiv\; \frac{\| \vf{c}'_f - \vf{N}_f \|}{\| \vf{N}_f - \vf{P} \|} \, \phi_P
                   \;+\;      \frac{\| \vf{c}'_f - \vf{P} \|}{\| \vf{N}_f - \vf{P} \|} \, \phi_{N_f}
\end{equation}

\subsection{Discretisation of the continuity equation}
\label{ssec: method: discretisation: continuity}

Integrating Eq.\ \eqref{eq: continuity} over cell $P$ and applying the divergence theorem, we get 
the mass flux balance for that cell:
\begin{equation} \label{eq: continuity integral}
 \sum_f \int_{s_f} \vf{n} \cdot \left( \rho \vf{u} \right) \,\mathrm{d}s
 \;=\; 0
\end{equation}
where $s_f$ is the surface of face $f$, $\vf{n}$ is the outward (of $P$) unit vector normal to that 
face, and $\mathrm{d}s$ is an infinitesimal surface element. The integrals summed are the outward 
mass fluxes through the respective faces of cell $P$. They are approximated by midpoint rule 
integration with additional stabilising terms:
\begin{equation} \label{eq: mass flux}
 \int_{s_f} \vf{n} \cdot \left( \rho \vf{u} \right) \,\mathrm{d}s
 \;\approx\;
 \rho \, s_f \, \left[ 
 \bar{\vf{u}}{}_{c_f} \!\cdot \vf{n}{}_f
 \;+\; u^{p+}_f \;-\; u^{p-}_f
 \right]
 \;\equiv\;
 \dot{M}_f
\end{equation}
where
\begin{align} 
\label{eq: up+}
 u^{p+}_f \;&\equiv\; a_f^{mi} \left( p_P - p_{N_f} \right)
\\[0.2cm]
\label{eq: up-}
 u^{p-}_f \;&\equiv\; a_f^{mi} \: \bar{\nabla}^q_h p_{m_f} \cdot (\vf{P} - \vf{N}{}_f) 
 ,\qquad
 \bar{\nabla}^q_h p_{m_f} \;\equiv\; \frac{1}{2} \left( \nabla^q_h p_P + \nabla^q_h p_{N_f} \right)
\\[0.2cm]
\label{eq: ami}
 a_f^{mi} \;&\equiv\; \frac{1}{\rho \left( \|\vf{u}_{N_f} - \vf{u}_P \| \:+\: \dfrac{h_f}{\Delta t} 
                      \right) \;+\; \dfrac{2\eta_a}{h_f}} 
 , \qquad
 h_f \;\equiv\; \left[ \frac{1}{2} \left( \Omega_P + \Omega_{N_f} \right) \right]^{\frac{1}{D}}
\end{align}
where $\Delta t$ is the current time step (see Sec.\ \ref{ssec: method: discretisation: temporal}), 
$\Omega_P$ and $\Omega_{N_f}$ are the cell volumes, and $D$ equals either 2 or 3, for 2D and 3D 
problems, respectively. With the definition \eqref{eq: mass flux}, the discrete version of the 
continuity equation for a cell $P$ is
\begin{equation} \label{eq: continuity integral discrete}
 \sum_f \dot{M}_f \;=\; 0
\end{equation}
To expound the above scheme, we make the following remarks:

\textbf{Remark 1:} If $u_f^{p+}$ and $u_f^{p-}$ are omitted from Eq.\ \eqref{eq: mass flux} then we 
are left with simple midpoint rule integration ($\bar{\vf{u}}{}_{c_f}$ is obtained using the scheme 
\eqref{eq: CDS}). The part of $\dot{M}_f$ due to $u_f^{p+}$ and $u_f^{p-}$ can be viewed as 
belonging to the truncation error of the continuity equation of cell $P$:
\begin{equation} \label{eq: momentum inteprolation truncation error}
 \frac{1}{\Omega_P} \rho \, s_f \left( u_f^{p+} - u_f^{p-} \right) \;=\;
 \frac{\rho \, s_f}{\Omega_P} a_f^{mi} \left[ 
   \left( p_P - p_{N_f} \right) \:-\: \bar{\nabla}^q_h p_{m_f} \cdot (\vf{P} - \vf{N}{}_f) 
 \right] 
\end{equation}
(the truncation error is defined per unit volume, hence we divide by the cell volume $\Omega_P$). 
It is easy to show, by expanding the pressure in a Taylor series about point $\vf{m}_f$ (Fig.\ 
\ref{fig: grid nomenclature}), that $(p_P - p_{N_f}) - \nabla p(\vf{m}_f) \cdot (\vf{P} - \vf{N}_f) 
= O(h^3)$. However, since we use only an approximation to the pressure gradient, $\nabla_h^q p = 
\nabla p + O(h)$, the term in square brackets in Eq.\ \eqref{eq: momentum inteprolation truncation 
error} is $O(h^2)$ (because $O(h) \cdot (\vf{P} - \vf{N}_f) = O(h^2)$). Given also that $\Omega_P = 
O(h^2)$, $s_f = O(h)$ and $a_f^{mi} = O(h)$, the whole term \eqref{eq: momentum inteprolation 
truncation error} is $O(h^2)$, which is compatible with a second-order accurate method such as the 
present one.

\textbf{Remark 2:} The term \eqref{eq: momentum inteprolation truncation error} inhibits spurious 
pressure oscillations by reflecting them on the image of the discrete continuity operator (see Eq.\ 
\eqref{eq: FVM system} and associated discussion). Pressure oscillations cause the term $u_f^{p+}$ 
to oscillate from face to face, and are thus reflected on the continuity image. On the other hand, 
such oscillations are filtered out by the gradient operator $\nabla_h^q$, so that the term 
$u_f^{p-}$ varies smoothly from face to face (it does not pass the oscillations on to the image). 
The $u_f^{p-}$ term is used simply for counterbalancing the truncation error introduced by 
$u_f^{p+}$.

\begin{figure}[tb]
  \centering
  \includegraphics[scale=1.]{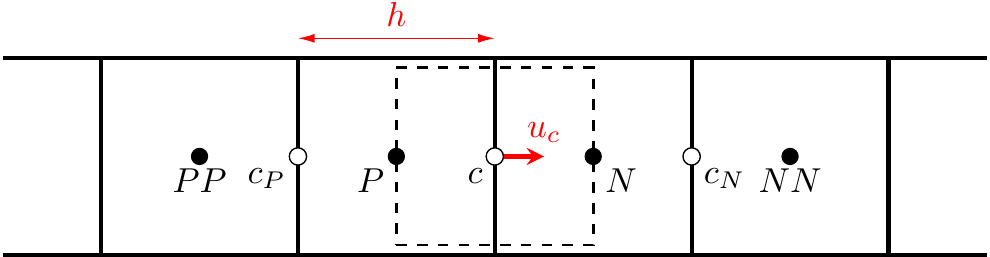}
  \caption{A row of grid cells.}
  \label{fig: 1D grid}
\end{figure}

\textbf{Remark 3:} The coefficient $a_f^{mi}$ is chosen so that the terms $u_f^{p+}$ and $u_f^{p-}$ 
of the mass flux expression \eqref{eq: mass flux}, which have units of velocity, are neither too 
small to have a stabilising effect nor so large that they dominate the mass flux. The ``velocities'' 
$u_f^{p+}$ and $u_f^{p-}$ are functions of local pressure variations, and attempt to quantify the 
contributions of these pressure variations to the velocity field. Pressure and velocity are 
connected through the momentum equation, so consider this equation in the following non-conservative 
form, where we assume that the stress tensor can be approximated through the use of a characteristic 
viscosity such as \eqref{eq: apparent viscosity} or \eqref{eq: apparent viscosity DAVSS} as 
$\tf{\tau} \approx \eta (\nabla \vf{u} + (\nabla \vf{u})^{\mathrm{T}})$:
\begin{equation} \label{eq: ami momentum equation}
 \rho \left( \pd{\vf{u}}{t} \:+\; \vf{u} \cdot \nabla \vf{u} \right) \;=\;
 -\nabla p \;+\; \nabla \cdot (\eta \nabla \vf{u})
\end{equation}
where we have neglected the term $\nabla \cdot (\eta (\nabla \vf{u})^{\mathrm{T}})$, assuming that 
it is small (it is zero if $\eta$ is constant). We are free to make these sorts of approximations 
because all we want is a rough estimate of the effect that the pressure gradient has on velocity. 
So, consider the simple uniform grid, of spacing $h$, shown in Fig.\ \ref{fig: 1D grid}, where $u$ 
denotes the velocity component normal to the face separating cells $P$ and $N$. We will employ a 
simple FV discretisation of Eq.\ \eqref{eq: ami momentum equation} in order to relate the velocity 
$u_c$ at the face centre $c$ to the pressures at the centres of the adjacent cells $P$ and $N$. The 
momentum conservation, Eq.\ \eqref{eq: ami momentum equation}, in the direction $x$ normal to the 
face, for the imaginary cell drawn in dashed line surrounding that face, can be discretised as
\begin{equation} \label{eq: ami momentum eqn discrete}
 \rho \frac{u_c - u_c^{old}}{\Delta t} h^2
 \;+\;
 \rho u_c \left. \frac{\mathrm{d}u}{\mathrm{d}x} \right|_c h^2
 \;=\;
 -\frac{p_N - p_P}{h} h^2
 \;+\;
 \frac{1}{h} \left( \eta_N \frac{u_{c_N} - u_c}{h} - \eta_P \frac{u_c - u_{c_P}}{h} \right) h^2
\end{equation}
where $u_c^{\mathrm{old}}$ was the velocity at time $\Delta t$ ago. Assuming that $(\eta_N + 
\eta_P) 
u_c \approx 2\eta_c u_c$ we can solve this for $u_c$:
\begin{equation} \label{eq: ami 1D}
 u_c \;=\; \frac{1}{
 \rho \left( \left. \dfrac{\mathrm{d}u}{\mathrm{d}x} \right|_c h \:+\; \dfrac{h}{\Delta t} \right)
 \;+\; \dfrac{2\eta_c}{h}}
 \, \left( p_P - p_N\right) \;+\; \cdots
\end{equation}
where the dots ($\cdots$) denote the terms that are not related to pressure. The above equation 
provides a quantification of the local effect of pressure gradient on velocity. It was derived from 
a simplistic one-dimensional consideration; for more general flows, the coefficient $a_f^{mi}$ 
\eqref{eq: ami} can be seen to be a generalisation of the coefficient multiplying the pressure 
difference in Eq.\ \eqref{eq: ami 1D}. It should be noted that the one-dimensional momentum 
equation \eqref{eq: ami momentum eqn discrete} accounts for the viscous force due to the velocity 
variation in the direction perpendicular to the face, but omits that due to the velocity variation 
in the direction parallel to the face; had the latter also been accounted for, the viscous term in 
the denominator of \eqref{eq: ami 1D}, and of $a_f^{mi}$ \eqref{eq: ami}, would have been $4\eta/h$ 
(or $6\eta/h$ in 3D) instead of $2\eta/h$, which would have reduced the magnitude of the 
stabilisation terms in the mass flux scheme \eqref{eq: mass flux}, but would also likely somewhat 
increase the accuracy, according to the results of Sec.\ \ref{sec: method: validation}. We did not 
investigate this issue further, but the choice is not crucial as it affects neither the 
stabilisation ability of the technique nor the $O(h^2)$ magnitude of the error \eqref{eq: momentum 
inteprolation truncation error} that it introduces.

\textbf{Remark 4:} An explanation of how the scheme retains its oscillation-inhibiting effect when 
$h \rightarrow 0$ despite $u_f^{p+}$, $u_f^{p-}$ tending to zero is given in Appendix \ref{appendix: 
pressure stabilisation}.

The scheme \eqref{eq: mass flux} -- \eqref{eq: ami} is a variant of the popular technique known as 
\textit{momentum interpolation}, which was originally proposed in \cite{Rhie_1983}. Ever since, many 
variants of this technique have been proposed (see \cite{Zhang_2014} and references therein) almost 
all of which are intertwined with the SIMPLE algebraic solver (exceptions include \cite{Deng_1994, 
Syrakos_06a}). Although this connection with SIMPLE can be useful in some respects 
\cite{Klaij_2015}, and SIMPLE is employed in the present work, we prefer an independent method such 
as the presently proposed because it is more general, transparent, and easily adaptable. It can be 
used with other algebraic solvers such as Newton's method, and it does not lead to dependence of 
the solution on underrelaxation factors.

\subsection{Discretisation of the momentum equation}
\label{ssec: method: discretisation: momentum}

As for the continuity equation, the FVM discretisation of the momentum equation \eqref{eq: 
momentum} begins by integrating it over a cell $P$ and applying the divergence theorem, to get
\begin{equation} \label{eq: momentum integral}
 \int_{\Omega_P} \pd{(\rho\vf{u})}{t} \, \mathrm{d}\Omega
 \;+\;
 \sum_f \int_{s_f} \vf{n} \cdot \left( \rho \vf{u} \vf{u} \right) \, \mathrm{d}s
 \;=\;
 -\sum_f \int_{s_f} \vf{n} \, p \, \mathrm{d} s
 \;+\;
 \sum_f \int_{s_f} \vf{n} \cdot \tf{\tau} \, \mathrm{d}s
\end{equation}
Using midpoint rule integration, the above equation is approximated by
\begin{equation} \label{eq: momentum integral discrete}
 \left. \frac{\partial (\rho \vf{u})}{\partial t} \right|^a_P \Omega_P
 \;+\;
 \sum_f \dot{M}_f \, \bar{\vf{u}}{}_{c_f}
 \;=\;
 - \sum_f \bar{p}_{c_f} \, s_f \, \vf{n}{}_f
 \;+\;
 \sum_f \vf{F}^{\tau}_f
\end{equation}
where $\partial / \partial t|^a_P$, the approximate time derivative at $\vf{P}$, will be defined in 
Sec.\ \ref{ssec: method: discretisation: temporal}, $\dot{M}_f$ is the mass outflux through face $f$ 
defined by Eq.\ \eqref{eq: mass flux}, $\bar{\vf{u}}{}_{c_f}$ and $\bar{p}_{c_f}$ are approximated 
from cell-centre values with the interpolation scheme \eqref{eq: CDS}, and $\vf{F}^{\tau}_f$ is the 
approximation of the force on face $f$ due to the EVP stress tensor $\tf{\tau}$. Note that Eq.\ 
\eqref{eq: momentum integral discrete} is a vector equation, with two (in 2D) or three (in 3D) 
scalar components. The force $\vf{F}^{\tau}_f$ is calculated as
\begin{equation} \label{eq: stress force}
 \int_{s_f} \vf{n} \cdot \tf{\tau} \, \mathrm{d}s
 \;\approx\;
 \vf{F}^{\tau}_f
 \;\equiv\;
 s_f \, \left[ \vf{n}{}_f \cdot \bar{\tf{\tau}}{}_{c_f} \;+\;
               \vf{D}^{\tau+}_f \;-\; \vf{D}^{\tau-}_f \right]
\end{equation}
Again, the stress tensor at the face centroid, $\bar{\tf{\tau}}{}_{c_f}$, is calculated via the 
linear interpolation scheme \eqref{eq: CDS} (applied component-wise). The viscous pseudo-stresses 
$\vf{D}_f^{\tau+}$ and $\vf{D}_f^{\tau-}$ are stabilisation terms employed to suppress spurious 
velocity oscillations. They are vectors, whose $i$-th components are
\begin{align}
\label{eq: D+}
 D_{f,i}^{\tau+} \;&\equiv\; \eta_a \frac{u_{i,N_f} - u_{i,P}}{\|\vf{N}_f - \vf{P}\|}
\\[0.2cm]
\label{eq: D-}
 D_{f,i}^{\tau-} \;&\equiv\; \eta_a \bar{\nabla}^q_h u_{i,m_f} \cdot \vf{d}_f
 ,\qquad
 \bar{\nabla}^q_h u_{i,m_f} \;\equiv\; \frac{1}{2} \left( \nabla^q_h u_{i,P} + \nabla^q_h u_{i,N_f} 
\right)
\end{align}
where $u_i$ is the $i$-th velocity component ($\vf{u} = (u_1, u_2)$ in 2D), and $\vf{d}_f = 
(\vf{N}_f - \vf{P}) / \| \vf{N}_f - \vf{P} \|$ is the unit vector pointing from $\vf{P}$ to 
$\vf{N}_f$ (Fig.\ \ref{fig: grid nomenclature}). It can be seen that both $D_{f,i}^{\tau+}$ and 
$D_{f,i}^{\tau-}$ equal a characteristic viscosity, given by \eqref{eq: apparent viscosity} or 
\eqref{eq: apparent viscosity DAVSS}, times the velocity gradient in the direction $\vf{d}_f$, 
albeit calculated differently: the gradient as computed in $D_{f,i}^{\tau+}$ is sensitive to 
velocity oscillations whereas that computed in $D_{f,i}^{\tau-}$ is not. The mechanism of 
oscillation suppression is similar to that for momentum interpolation: in the presence of spurious 
velocity oscillations, the smooth part of $D_{f,i}^{\tau+}$ is cancelled out by $D_{f,i}^{\tau-}$, 
but the oscillatory part produces oscillations in the operator image (in $F_h(\phi_h)$, in the 
terminology of Eq.\ \eqref{eq: FVM system}).

In the context of colocated FVMs for viscoelastic flows, this technique was first proposed in 
\cite{Oliveira_1998}, inspired from the corresponding ``momentum interpolation'' technique for 
pressure. A question concerning the method is the appropriate choice of the parameter $\eta_a$. The 
original method \cite{Oliveira_1998} as well as some subsequent variants (e.g.\ \cite{Matos_2009, 
Niethammer_2017}), similarly to the original momentum interpolation \cite{Rhie_1983}, derived the 
coefficient from the SIMPLE matrix of the linearised discrete constitutive equation, arriving at 
complicated expressions. The present simpler approach is essentially equivalent to that adopted by 
\cite{Pimenta_2017, Fernandes_2017} who, for viscoelastic flows, set the coefficient $\eta_a$ equal 
to the polymeric viscosity. The aim is that $\vf{D}_f^{\tau+}$ and $\vf{D}_f^{\tau-}$ are of the 
same order of magnitude as the EVP stress acting on the cell face, and this can be achieved using a 
characteristic viscosity $\eta_a$ defined as the ratio of a typical stress to a typical rate of 
strain for the given problem.

The present technique can also be interpreted as a ``both-sides diffusion'' technique 
\cite{Fernandes_2017} where the momentum equation discretised by the FVM is not \eqref{eq: 
momentum} but an equivalent equation where the same diffusion term $\nabla \cdot (\eta_a \nabla 
\vf{u})$ has been subtracted from both sides:
\begin{equation} \label{eq: momentum bsd}
 \pd{(\rho\vf{u})}{t} \;+\; \nabla \cdot \left( \rho \vf{u} \vf{u} \right)
 \;-\; \nabla \cdot (\eta_a \nabla \vf{u})
 \;=\;
 -\nabla p \;+\; \nabla\cdot\tf{\tau}
 \;-\; \nabla \cdot (\eta_a \nabla \vf{u})
\end{equation}
The left-hand side such term is not discretised in the same way as the right-hand side term; in 
particular, the component $\vf{D}_f^{\tau+}$ in \eqref{eq: stress force} comes from the 
discretisation of the left-hand side term, whereas the component $\vf{D}_f^{\tau-}$ comes from the 
discretisation of the right-hand side term. The other components of these diffusion terms are 
discretised in exactly the same way for both of them and cancel out, leaving only $\vf{D}_f^{\tau+}$ 
and $\vf{D}_f^{\tau-}$. Since both of these diffusion terms are discretised by central differences 
they contribute $O(1)$ to the truncation error \cite{Syrakos_2017} but $O(h^2)$ to the 
discretisation error. In particular, consider that Eq.\ \eqref{eq: momentum bsd} corresponds to Eq.\ 
\eqref{eq: PDE}, and its FVM discretisation with the present schemes leads to the algebraic system 
\eqref{eq: FVM system}, with the dependent variable $\phi \equiv \vf{u}$ being the velocity. If 
$\phi^e$ is the exact velocity (the solution of Eq.\ \eqref{eq: PDE}) and $\phi$ is the approximate 
velocity obtained by solving the system \eqref{eq: FVM system}, then the discretisation error is 
$\varepsilon = \phi^e - \phi$. The latter is related to the truncation error $\alpha$ by 
$F'_h(\phi_h) \cdot \varepsilon_h \approx - \alpha_h$, where $F'_h(\phi_h)$ is the Jacobian matrix 
of the operator $F_h$ evaluated at the approximate solution $\phi_h$ (see e.g.\ \cite{Syrakos_2012} 
for a derivation). Let us focus on the $\nabla \cdot (\eta_a \nabla \vf{u})$ term of the left-hand 
side of \eqref{eq: momentum bsd}, which is discretised with a compact stencil giving rise to the 
$\vf{D}_f^{\tau+}$ term of the scheme \eqref{eq: stress force} (a similar analysis of the term on 
the right-hand side can be made). Because the diffusion operator is linear, its contribution to the 
truncation error $F'_h(\phi_h) \cdot \varepsilon_h$ is equal to its contribution to 
$F_h(\varepsilon_h)$, i.e.\ it can be calculated by replacing the velocities by their discretisation 
errors in \eqref{eq: D+} (we neglect the rest of the contributions, which cancel out with the 
corresponding ones of the diffusion term of the right side of Eq.\ \eqref{eq: momentum bsd}). So, 
plugging $\varepsilon$ into Eq.\ \eqref{eq: D+} instead of $u_i$ we get a contribution to the 
truncation error of cell $P$ of $(s_f / \Omega_P) \eta_a (\varepsilon_{N_f} - \varepsilon_P) / \| 
\vf{N}_f - \vf{P} \| = O(h^{-2}) (\varepsilon_{N_f} - \varepsilon_P)$. If our method is second-order 
accurate then $\varepsilon = O(h^2)$. On unstructured grids, the variation of discretisation error 
in space has a random component \cite{Syrakos_2012} so that $\varepsilon_{N_f} - \varepsilon_P = 
O(h^2)$ as well, which results in a $O(1)$ contribution to the truncation error. On smooth 
structured grids the discretisation error varies smoothly \cite{Syrakos_2012} so that 
$\varepsilon_{N_f} - \varepsilon_P = O(h^3)$ and each face contributes $O(h)$ to the truncation 
error (on such grids there is also a cancellation between opposite faces for each cell which reduces 
the net contribution to the truncation error to $O(h^2)$).

Another benefit that comes from the incorporation of the stabilisation terms in the scheme 
\eqref{eq: stress force} concerns the algebraic solver, SIMPLE \cite{Patankar_1980}. Within each 
SIMPLE iteration a succession of linearised systems of equations are solved, one for each dependent 
variable. The systems for the velocity components $u_i$ are derived from linearisation of the 
(discretised) momentum equation \eqref{eq: momentum}. The latter, unlike in (generalised) Newtonian 
flows, contains no diffusion terms and the velocity appears directly only in the convection terms of 
the left-hand side. In low Reynolds number flows, these terms play only a minor role and the 
momentum balance involves mainly the pressure and stress. Constructing the linear systems for the 
velocity components only from these inertial terms would lead to bad convergence of the SIMPLE 
algorithm. Nevertheless, velocity plays an important indirect role in the momentum equation through 
the EVP stress tensor, to which it is related via the constitutive equation \eqref{eq: 
constitutive}. Therefore, either the momentum and constitutive equations have to be solved in a 
coupled manner, or the effect of velocity on the stress tensor has to somehow be directly quantified 
in the momentum equation, which is precisely what the stabilisation term $\vf{D}_f^{\tau+}$ 
achieves; in particular, the diffusion term on the left-hand side of Eq.\ \eqref{eq: momentum bsd} 
contributes to the matrix of coefficients of the linear systems for $u_i$, making this matrix very 
similar to those for (generalised) Newtonian flows.

Finally, the momentum balance for cells that lie at the wall boundaries involves the pressure and 
the stress tensor at these boundaries. The pressure is linearly extrapolated from the interior, as 
is a common practice \cite{Ferziger_02}. For stress two possible options are linear extrapolation, 
as for pressure, and the imposition of a zero-gradient condition which is roughly equivalent to 
setting the stress at the boundary face equal to that at the owner cell centre. In 
\cite{Habla_2012}, the former approach was found, as expected, to be more accurate, while the latter 
was found to be more robust. In the present work we employed either linear extrapolation or a 
modified extrapolation which is akin to the scheme \eqref{eq: stress force}. So, if $b$ is a 
boundary face of cell $P$, the linearly extrapolated value of stress at its centre is calculated as
\begin{equation} \label{eq: boundary extrapolation}
 \bar{\tf{\tau}}{}_{c_b} \;=\; \tf{\tau}{}_P \;+\; 
                               \nabla_h^{1} \tf{\tau}{}_P \cdot (\vf{c}_b - \vf{P})
\end{equation}
where the least-squares stress gradient $\nabla_h^{1} \tf{\tau}{}_P$ is calculated with $q=1$ and 
using only the stress values at the centres of $P$ and its neighbour cells. This scheme is used 
also for extrapolation of pressure, but when it comes to stress we can go a step further: the force 
due to the EVP stress tensor on boundary face $b$ can be calculated by a scheme that has precisely 
the form \eqref{eq: stress force} only that the interpolated stress $\bar{\tf{\tau}}{}_{c_f}$ 
\eqref{eq: CDS} is replaced by the extrapolated value $\bar{\tf{\tau}}{}_{c_b}$ \eqref{eq: boundary 
extrapolation} and the $D$ terms are replaced by
\begin{align}
\label{eq: D+ boundary}
 D_{b,i}^{\tau+} \;&\equiv\; \eta_a \frac{u_{i,c_b} - u_{i,P}}{\|\vf{c}_b - \vf{P}\|}
\\[0.2cm]
\label{eq: D- boundary}
 D_{b,i}^{\tau-} \;&\equiv\; \eta_a \bar{\nabla}^q_h u_{i,m_b} \cdot \vf{d}_b
\end{align}
where $\vf{d}_b$ is the unit vector pointing from $\vf{P}$ to $\vf{c}_b$ and the gradient in 
\eqref{eq: D- boundary} is extrapolated to point $\vf{m}_b \equiv (\vf{P} + \vf{c}_b) / 2$ via 
the scheme \eqref{eq: boundary extrapolation} with $\vf{c}_b$ replaced by $\vf{m}_b$. The terms 
\eqref{eq: D+ boundary} -- \eqref{eq: D- boundary} do not have an apparent stabilising effect, and 
in fact our experience has shown that they sometimes can cause convergence difficulties to the 
algebraic solver. Such is the case with wall slip examined in Sec.\ \ref{ssec: results: slip}, 
where we had to set these terms to zero and use simple linear extrapolation of stresses at the 
walls. Nevertheless, we also found that these terms can increase the accuracy of the solution (see 
Sec.\ \ref{sec: method: validation}) and therefore we employed them whenever possible (they were 
used for all the main results except those of Sec.\ \ref{ssec: results: slip}).

\subsection{Discretisation of the constitutive equation}
\label{ssec: method: discretisation: constitutive}

Again, by integrating the SHB equation \eqref{eq: constitutive} -- \eqref{eq: upper convected 
derivative} and applying the divergence theorem we obtain
\begin{align}
\nonumber
 &\int_{\Omega_P} \pd{\tf{\tau}}{t} \, \mathrm{d}\Omega
 \;+\;
 \sum_f \int_{s_f} \vf{n} \cdot \vf{u} \tf{\tau} \, \mathrm{d}s 
 \;=\; 
\\
\label{eq: constitutive integral}
 &G \int_{\Omega_P} \dot{\tf{\gamma}} \, \mathrm{d}\Omega
 \;-\;
 G \int_{\Omega_P}
 \left( \frac{\max(0, \tau_d-\tau_y)}{k} \right)^{\frac{1}{n}} \frac{1}{\tau_d} \; \tf{\tau}
 \, \mathrm{d}\Omega
 \;+\;
 \int_{\Omega_P}
 \left( (\nabla \vf{u})^{\mathrm{T}} \cdot \tf{\tau} \;+\; \tf{\tau} \cdot \nabla \vf{u} \right)
 \mathrm{d} \Omega
\end{align}
The surface integral on the left-hand side derives from application of the divergence theorem to the 
second term on the right-hand side of Eq.\ \eqref{eq: upper convected derivative}, using the 
identity $\vf{u} \cdot \nabla \tf{\tau} = \nabla \cdot (\vf{u}\tf{\tau}) - (\nabla \cdot \vf{u}) 
\tf{\tau}$ and the incompressibility condition $\nabla \cdot \vf{u} = 0$. Equation \eqref{eq: 
constitutive integral} has the form of a conservation equation for stress. In the left-hand side, 
there is the rate of change of stress in the volume plus the outflux of stress, which equal the rate 
of change of stress in a fixed mass of fluid moving with the flow. In the right-hand side, there 
are three source terms which express, respectively, stress generation (the $\dot{\tf{\gamma}}$ 
term), relaxation (the viscous term), and transformation (the upper convective derivative term). 
Equation \eqref{eq: constitutive integral} is then discretised as:
\begin{align}
\nonumber
 &\left. \pd{\tf{\tau}}{t} \right|^a_P \Omega_P
  \;+\;
  \frac{1}{\rho} \sum_f \dot{M}_f \bar{\tf{\tau}}{}^{_C}_{c_f}
  \;=\;
\\
\label{eq: constitutive integral discrete}
 &G \, \Omega_P \, \dot{\tf{\gamma}}{}_P
  \;-\;
  G \, \Omega_P \, \left( \frac{\max(0, \tau_{d,P}-\tau_y)}{k} \right)^{\frac{1}{n}} 
    \frac{1}{\tau_{d,P}} \; \tf{\tau}{}_P
  \;+\;
  \Omega_P \, \left( (\nabla_h^q \vf{u})^{\mathrm{T}}_P \cdot \bar{\tf{\tau}}{}_P
    \;+\; \bar{\tf{\tau}}{}_P \cdot \nabla_h^q \vf{u}_P \right) 
\end{align}
where $\partial / \partial t|^a_P$ is the approximate time derivative at $\vf{P}$ (Sec.\ \ref{ssec: 
method: discretisation: temporal}), $\dot{M}_f$ is the mass flux \eqref{eq: mass flux}, 
$\dot{\tf{\gamma}}{}_P = \nabla_h^q \vf{u}_P + (\nabla_h^q \vf{u})^{\mathrm{T}}_P$ is the discrete 
value of $\dot{\tf{\gamma}}$ at $\vf{P}$, and $\tau_{d,P}$ is the value of $\tau_d$ at $\vf{P}$, 
calculated from $\tf{\tau}{}_P$ via Eq.\ \eqref{eq: deviatoric stress}.

\begin{figure}[tb]
  \centering
  \includegraphics[scale=0.9]{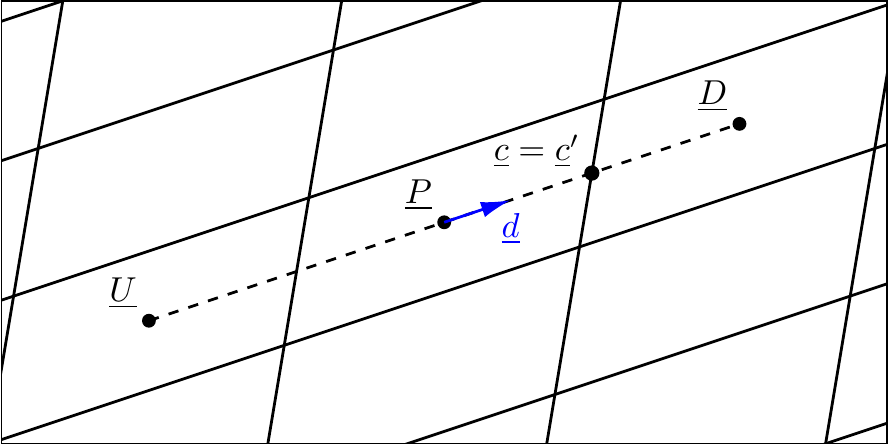}
  \caption{On a uniform structured grid, CUBISTA interpolates the value at face centre $\vf{c}$ from
the values at the downwind ($D$), upwind ($P$), and farther-upwind ($U$) cells (assuming the mass 
flux is directed from cell $P$ to cell $D$).}
  \label{fig: CUBISTA grid}
\end{figure}

In the convection term of Eq.\ \eqref{eq: constitutive integral discrete} (the second term on the 
left-hand side), $\bar{\tf{\tau}}{}^{_C}_{c_f}$ is the value of $\tf{\tau}$ interpolated at the face 
centre $\vf{c}_f$, but not by the linear interpolation \eqref{eq: CDS}. Since the constitutive 
equation lacks diffusion terms, there is a danger of spurious stress oscillations, and a common 
preventive measure is to use so-called \textit{high resolution schemes} \cite{Moukalled_2016} for 
the discretisation of the convection term. In viscoelastic flows, the CUBISTA scheme 
\cite{Alves_2003} has proved to be effective and is widely adopted (e.g.\ \cite{Chen_2013, 
Habla_2014, Comminal_2015, Syrakos_2018}). In the present work, we adapted it for use on 
unstructured grids as follows. To account for skewness, a scheme similar to \eqref{eq: CDS} is used, 
but the value at point $\vf{c}'_f$ is interpolated with CUBISTA:
\begin{equation} \label{eq: CUBISTA skewness}
 \bar{\phi}^{_C}_{c_f}
 \;=\;
 \bar{\phi}^{_C}_{c'_f} \;+\; \bar{\nabla}^q_h \phi_{c'_f} \!\cdot\! (\vf{c}_f - \vf{c}'_f) 
\end{equation}
Note that the second term of the right-hand side is exactly the same as for the scheme \eqref{eq: 
CDS} (the gradient is interpolated linearly). The first term of the right-hand side is the value at 
point $\vf{c}'_f$ interpolated via CUBISTA. CUBISTA is a high-resolution scheme based on the 
Normalised Variable Formulation \cite{Gaskell_1988}, and on structured grids such schemes 
interpolate the value at a face from the values at two upwind cell centres and one downwind cell 
centre; these are the two cells on either side of the face, labelled $P$ and $D$ in Fig.\ \ref{fig: 
CUBISTA grid}, and a farther cell $U$. But on an unstructured grid the farther-upwind cell $U$ is 
not straightforward (or even not possible) to identify. To overcome this hurdle, we follow an 
approach \cite{Jasak_1999, Moukalled_2016} which is based on the observation that, on a uniform 
structured grid such as that shown in Fig.\ \ref{fig: CUBISTA grid}, it holds that
\begin{equation} \label{eq: phi_U}
  \phi_U \;=\; \phi_D \;-\; \nabla_h^q \phi_P \cdot (\vf{D} - \vf{U})
\end{equation}
because the least-squares gradient is such that $\nabla_h^q \phi_P \cdot (\vf{D} - \vf{U}) = \phi_D 
- \phi_U$. On unstructured grids, following \cite{Jasak_1999, Moukalled_2016} we define the location 
$\vf{U} \equiv \vf{P} - (\vf{D} - \vf{P})$ which lies at the diametrically opposite position of 
$\vf{D}$ relative to $\vf{P}$; since the grid is unstructured, it is unlikely that $\vf{U}$ 
coincides with an actual cell centre, but still we can use Eq.\ \eqref{eq: phi_U} to estimate a 
value there, $\phi_U$. Information from the upwind side of cell $P$ is implicitly incorporated into 
$\phi_U$ through the gradient $\nabla_h^q \phi_P$, and if the scheme is used on a uniform structured 
grid then it automatically retrieves the value at the centroid of the actual cell $U$.

So now we have a set of three co-linear and equidistant points, $\vf{U}$, $\vf{P}$ and $\vf{D}$, 
and three corresponding values, $\phi_U$, $\phi_P$ and $\phi_D$, and CUBISTA can be employed to 
calculate the value at point $\vf{c}'_f$, which lies on the same line as these three points. 
Defining the normalised variables as
\begin{equation} \label{eq: def NV xi}
 \xi \;=\; \frac{\| \vf{c}' - \vf{P} \|}{\| \vf{D} - \vf{P} \|}
\end{equation}
\begin{equation} \label{eq: def NV phi_P}
 \tilde{\phi}_P \;=\; \frac{\phi_P - \phi_U}{\phi_D - \phi_U}
\end{equation}
\begin{equation} \label{eq: def NV phi_c}
 \tilde{\phi}_{c'_f}(\xi,\tilde{\phi}_P) \;=\;
 \frac{\bar{\phi}^{_C}_{c'_f} - \phi_U}{\phi_D - \phi_U}
\end{equation}
(note that $\xi$ is just the linear interpolation factor multiplying $\phi_{N_f}$ in Eq.\ 
\eqref{eq: linear interpolation}), CUBISTA blends Quadratic Upwind Interpolation (QUICK) and 
first-order upwinding (UDS) as follows:
\begin{equation} \label{eq: CUBISTA NV formulation}
  \tilde{\phi}_{c'_f}(\xi,\tilde{\phi}_P)
  \;=\;
  \left\{
  \begin{array}{l l l}
    \tilde{\phi}_P
    & \quad \text{if } \tilde{\phi}_P \leq 0
    & \text{(UDS)}
    \\[0.2cm]
    \frac{1}{3} (1 + \xi) (3 + \xi) \tilde{\phi}_P
    & \quad \text{if } 0 < \tilde{\phi}_P < \frac{3}{8}
    & \text{(Transition 1)}
    \\[0.2cm]
    (1-\xi^2) \tilde{\phi}_P \;+\; \frac{1}{2} \xi (1 + \xi)
    & \quad \text{if } \frac{3}{8} < \tilde{\phi}_P \leq \frac{3}{4}
    & \text{(QUICK)}
    \\[0.2cm]
    (1 - \xi)^2 \tilde{\phi}_P \;+\; \xi (2-\xi)
    & \quad \text{if } \frac{3}{4} < \tilde{\phi}_P \leq 1
    & \text{(Transition 2)}
    \\[0.2cm]
    \tilde{\phi}_P
    & \quad \text{if } \tilde{\phi}_P > 1
    & \text{(UDS)}
  \end{array}
  \right.
\end{equation}
From $\tilde{\phi}_{c'_f}$ the un-normalised value $\bar{\phi}^{_C}_{c'_f}$ can be recovered via
\eqref{eq: def NV phi_c} for use in Eq.\ \eqref{eq: CUBISTA skewness}. However, the scheme can be 
conveniently reformulated in terms of un-normalised $\phi$ values using the median 
function\footnote{It can be implemented as $\text{median}(a,b,c) \,=\, \max( \,\min(a,b), \,\min( 
\max(a,b), c ) \,)$.} \cite{Leonard_1996} as:
\begin{align} \label{eq: CUBISTA median}
  \bar{\phi}^{_C*}_{c'_f} \;&=\; \text{median}\left( \quad
  \phi_P, \quad
  \phi_U \,+\, \frac{(1+\xi)(3+\xi)}{3}(\phi_P - \phi_U), \quad
  \phi_D \,-\, (1-\xi)^2 (\phi_D-\phi_P) \quad
  \right)
\\[0.2cm]
  \bar{\phi}^{_C}_{c'_f} \;&=\;  \text{median}\left( \quad
  \phi_P, \quad
  \bar{\phi}^{_C*}_{c'_f}, \quad
  \phi_P \,+\, \frac{\phi_D-\phi_U}{2} \xi \,+\, \frac{1}{2}(\phi_D - 2\phi_P + \phi_U) \xi^2 \quad
  \right)
\end{align}
which also automatically sets $\bar{\phi}^{_C}_{c'_f} = \phi_P$ (UDS) when $\phi_D = \phi_U$. The 
scheme \eqref{eq: CUBISTA NV formulation} is based on QUICK, but switches to UDS when there is a 
local minimum or maximum at $\vf{P}$ ($\tilde{\phi}_P < 0$ or $\tilde{\phi}_P > 1$), which could 
indicate a spurious oscillation. The region $\tilde{\phi}_P \in [3/8, 3/4]$, where the variation 
between the three values $\phi_U$, $\phi_P$ and $\phi_D$ is not far from linear, is considered to be 
``safe'' and QUICK is applied unreservedly there. The upper bound of this region, which for 
non-uniform structured grids is given to be a $\xi$-dependent value in \cite{Alves_2003}, is chosen 
here to be the fixed number $\tilde{\phi}_P = 3/4$ based on the same criterion as in 
\cite{Alves_2003}, namely the condition that the quadratic profile passing through the three points 
$\vf{U}$, $\vf{P}$ and $\vf{D}$ is monotonic (no overshoots / undershoots). In terms of normalised 
variables, this condition requires that $\tilde{\phi}_{c'_f} (\xi, \tilde{\phi}_P)$, as given by the 
QUICK branch of Eq.\ \eqref{eq: CUBISTA NV formulation}, increases monotonically towards the value 1 
as $\xi \rightarrow 1$, which is ensured by the condition $\partial \tilde{\phi}_{c'_f} 
(\xi,\tilde{\phi}_P) / \partial \xi |_{\xi = 1} \geq 0 \Rightarrow \tilde{\phi}_P \leq 3/4$. We also 
note that, if CUBISTA achieves its goal of producing smooth $\phi$ distributions, then grid 
refinement, which brings the points $\vf{U}$, $\vf{P}$ and $\vf{D}$ closer together, will cause 
$\phi$ to vary more linearly in the neighbourhood of the three points, i.e.\ $\tilde{\phi}_P 
\rightarrow 0.5$ as the grid is refined. On fine grids, therefore, CUBISTA reduces to QUICK 
throughout the domain, except at local extrema of $\phi$.

Finally, one can notice an interpolated value $\bar{\tf{\tau}}{}_P$ in the stress transformation
terms in the right-hand side of Eq.\ \eqref{eq: constitutive integral discrete}. An obvious choice 
is to set $\bar{\tf{\tau}}{}_P = \tf{\tau}{}_P$, but it was found in \cite{Zhou_2016} that using 
instead a weighted average of the stress at $P$ and its neighbours can mitigate the high-$\Wei$ 
problem. In the original scheme \cite{Zhou_2016} the weighting was based on the mass fluxes through 
the cell's faces. However, we noticed that this causes a noticeable accuracy degradation and 
instead used the following scheme (written here for 2D problems):
\begin{equation} \label{eq: upper convective terms discretisation}
 (\nabla_h^q \vf{u})^{\mathrm{T}}_P \cdot \bar{\tf{\tau}}{}_P
    \;+\; \bar{\tf{\tau}}{}_P \cdot \nabla_h^q \vf{u}_P
 \;=\;
 \left[ \begin{array}{l}
 2\, \overline{\tau_{11,P}} \, \pd{u_1}{x_1} \:+\: 2\, \tau_{12,P} \, \pd{u_1}{x_2}
 \\
 \tau_{11,P} \pd{u_2}{x_1} \:+\:
 \overline{\tau_{12,P}} (\pd{u_1}{x_1} + \pd{u_2}{x_2}) \:+\:
 \tau_{22,P} \pd{u_1}{x_2} 
 \\
 2\, \overline{\tau_{22,P}} \pd{u_2}{x_2} \;+\; 2\, \tau_{12,P} \pd{u_2}{x_1}
 \end{array} \right]
 \,
 \begin{array}{l}
  \text{\small{(1,1) component}}
  \\
  \text{\small{(1,2) component}}
  \\
  \text{\small{(2,2) component}}
 \end{array}
\end{equation}
where the velocity derivatives are the components of $\nabla_h^q \vf{u}{}_P$ and
\begin{equation} \label{eq: stress reconstruction}
 \overline{\tau_{ij,P}} \;=\; 
 \frac{1}{D \, \Omega_P} \sum_{f=1}^F
 \bar{\tau}_{ij,c_f}^{_C} \, s_f \, (\vf{c}{}_f - \vf{P}) \cdot \vf{n}{}_f
\end{equation}
where $D$ = 2 or 3, for 2D or 3D problems, respectively. The interpolation scheme \eqref{eq: stress 
reconstruction} is 2nd-order accurate (Appendix \ref{appendix: derivation of reconstruction 
scheme}). In the (1,2) component of \eqref{eq: upper convective terms discretisation}, the middle 
term, although equal to zero in the limit of infinite grid fineness due to the continuity equation, 
has been retained for numerical stability. Compared to simply setting $\overline{\tau_{ij,P}} = 
\tau_{ij,P}$, this scheme was found, in the numerical tests of Sec.\ \ref{sec: method: validation}, 
to allow an increase of about 40\% in the maximum solvable $\Wei$ number and to provide a slight 
increase of accuracy. At wall boundaries it was found that using linearly extrapolated stress values 
in \eqref{eq: stress reconstruction} leads to convergence difficulties. Therefore, wall boundary 
values were omitted in the calculation, weighting the values at the remaining faces by their $s_f 
(\vf{c}_f-\vf{P}) \cdot \vf{n}_f$ factors, i.e.\ by replacing $D \, \Omega_P$ in the denominator of 
\eqref{eq: stress reconstruction} by
\begin{equation}
 D \, \Omega_P' \;=\; \sum_{f \notin \{W_P\}} s_f \left( \vf{c}_f -\vf{P} \right) \cdot \vf{n}_f
                \;=\; D \, \Omega_P \;-\; \sum_{f \in \{W_P\}} s_f \left( \vf{c}_f -\vf{P} \right) 
\cdot \vf{n}_f
\end{equation}
where $\{W_P\}$ is the set of wall boundary faces of cell $P$. This is a 1st-order accurate 
interpolation.

In the simulations of Sec.\ \ref{sec: results} we did not encounter the high-$\Wei$ problem, but 
should it arise one can transform the constitutive equation in terms of a more weakly varying 
function of the stress tensor such as the logarithm of the conformation tensor \cite{Fattal_2004, 
Afonso_2009}. This has been applied to the Saramito model in \cite{Vita_2018}.

\subsection{Temporal discretisation scheme}
\label{ssec: method: discretisation: temporal}

The approximate time derivatives in the momentum and constitutive equations are calculated with a 
2nd-order accurate, three-time level implicit scheme with variable time step. Suppose we enter time 
step $i$ and let $\phi_P^i$ be the current, yet unknown, value of $\phi$ at cell $P$. Fitting a 
quadratic Lagrange polynomial between the three points $(t_i, \phi_P^i)$, $(t_{i-1}, \phi_P^{i-1})$ 
and $(t_{i-2}, \phi_P^{i-2})$ and differentiating it at $t = t_i$ gives the following approximate 
time derivative:
\begin{equation} \label{eq: temporal lagrange polynomial derivative}
 \left. \frac{\mathrm{d}\phi}{\mathrm{d}t} \right|_P^a (t_i)
 \;=\;
 l'_{i,i}(t) \phi_P^i
 \;+\;
 l'_{i,i-1}(t) \phi_P^{i-1}
 \;+\;
 l'_{i,i-2}(t) \phi_P^{i-2}
\end{equation}
where
\begin{equation} \label{eq: lagrange l' coefficients}
 l'_{i,i}(t) = 
 \frac{2t_i-t_{i-1}-t_{i-2}}{(t_i-t_{i-1})(t_i-t_{i-2})}
 ,\;\;
 l'_{i,i-1}(t) =
 \frac{2t_i-t_{i}-t_{i-2}}{(t_{i-1}-t_i)(t_{i-1}-t_{i-2})}
 ,\;\;
 l'_{i,i-2}(t) =
 \frac{2t_i-t_{i}-t_{i-1}}{(t_{i-2}-t_i)(t_{i-2}-t_{i-1})}
\end{equation}
All other terms of the governing equations are evaluated at the current time (the scheme is fully 
implicit).

After solving the equations to obtain $\phi_P^{i}$ and deciding (as will be described shortly) the 
next time step size $\Delta t_{i+1} = t_{i+1} - t_i$, to facilitate the solution at the new time 
$t_{i+1}$ we can make a \textit{prediction} by evaluating the aforementioned Lagrange polynomial at 
$t = t_{i+1}$ to obtain a provisional value $\phi_P^{i+1*}$. This value serves as an initial guess 
for solving the equations at the new time step $i+1$, offering significant acceleration. We perform 
such a prediction even for pressure, whose time derivative does not appear in the governing 
equations, as this was found to accelerate the solution of the continuity equation.

The subsequent solution at time $t_{i+1}$ will give us a value $\phi_P^{i+1} \neq \phi_P^{i+1*}$, 
in general. In order to obtain an estimate of the $O(h^2)$ truncation error of scheme \eqref{eq: 
temporal lagrange polynomial derivative} we can augment our set of three points by the addition of 
the point $(t_{i+1},\phi_P^{i+1})$ and fit a \textit{cubic} polynomial through the four points; the 
difference between $\partial \phi / \partial t|^a_P$ and $\partial \phi / \partial t|^c_P$, the 
derivatives predicted at time $t_i$ by the quadratic and cubic polynomials, respectively, is an 
approximation of the truncation error associated with $\partial \phi / \partial t|^a_P$. Omitting 
the details, it turns out that 
\begin{equation} \label{eq: temporal truncation error}
 \left. \pd{\phi}{t} \right|_P^c (t_i) \;-\; \left. \pd{\phi}{t} \right|_P^a (t_i)
 \;=\;
 c \, \frac{\phi_P^{i+1} - \phi_P^{i+1*}}{\Delta t_{i+1}}
\end{equation}
where the factor $c = O(1)$ depends on the relative magnitudes of $\Delta t_{i-1}$, $\Delta t_i$ 
and $\Delta t_{i+1}$ ($c$ = 1/3 for $\Delta t_{i-1} = \Delta t_i = \Delta t_{i+1}$). This allows us 
to keep an approximately constant level of truncation error by adjusting the time step sizes so as 
to keep the right-hand side of Eq.\ \eqref{eq: temporal truncation error} constant. We set the 
following goal, which accounts for all grid cells at once:
\begin{equation} \label{eq: temporal adjustment goal}
 g_t^i 
 \equiv
 \sqrt{ \frac{1}{\Omega} \sum_P 
 \Omega_P \left( \frac{\phi_P^i - \phi_P^{i*}}{\Delta t_i} \right)^2
 }
 \;=\;
 \tilde{g}^0_t \, \frac{\Phi}{T}
\end{equation}
I.e. we want $g_t^i$, the $L_2$ norm of $(\phi_P^i - \phi_P^{i*}) / \Delta t_i$, for any time step 
$i$, to equal a pre-selected non-dimensional target value $\tilde{g}_t^0$ times $\Phi / T$ where 
$\Phi$ is either $U$ (for the momentum equation, where $\phi \equiv u_j$) or $S$ (for the 
constitutive equation, where $\phi \equiv \tau_{jk}$). Suppose then that at some time step $i$ we 
are somewhat off target, i.e.\ Eq.\ \eqref{eq: temporal adjustment goal} is not satisfied; we can 
try to amend this at the new time step $i+1$ by noting that the truncation error, and the 
associated metric $g_t$, are of order $O(\Delta t^2)$, which means that $g_t^{i+1} / g_t^i = (\Delta 
t_{i+1} / \Delta t_i)^2$. This relation allows us to choose $\Delta t_{i+1}$ so that $g_t^{i+1}$ 
will equal $\tilde{g}_t^0 \Phi / T$ (Eq.\ \eqref{eq: temporal adjustment goal}):
\begin{equation} \label{eq: new time step}
 \Delta t_{i+1} \;=\; r_t^i \, \Delta t_i
 \qquad \text{where} \qquad
 r_t^i \;=\; \sqrt{\tilde{g}_t^0 / \tilde{g}_t^i},
 \quad
 \tilde{g}_t^i \;=\; \frac{g_t^i}{\Phi/T}
\end{equation}
In practice, we calculate the adjustment ratios $r_t^i$ for all variables whose time derivatives 
appear in the governing equations (one per velocity component and per stress component) and choose 
the smallest among them. We also limit the allowable values of $r_t^i$ to a range $r_t^i \in 
[r_t^{\min}, r_t^{\max}]$, and also the overall time step size to $\Delta t_i \in [\Delta t_{\min}, 
\Delta t_{\max}]$. Typical values used for these parameters in the simulations of Sec.\ \ref{sec: 
results} are $\tilde{g}_t^0$ = \num{e-4}, $r_t^{\min}$ = \num{0.95}, $r_t^{\max}$ = \num{1.001}, 
$\Delta t_{\min}$ =\num{5e-5} \si{s}, $\Delta t_{\max}$ = \num{2e-3} \si{s} (these may vary slightly 
between simulations). The scheme automatically chooses small time steps when the flow evolves 
rapidly, such as during the early stages of the flow, and large time steps when it evolves slowly, 
such as when the flow is nearing its steady state. A similar scheme is used in 
\cite{Dimakopoulos_2003}.

\section{Solution of the algebraic system}
\label{sec: method: solution}

Discretisation results in one large nonlinear algebraic system per time step. These systems are 
solved using the SIMPLE algorithm, with multigrid acceleration, where the algebraic equations are 
arranged into groups, one for each momentum and constitutive equation component, each of which 
through linearisation produces a linear system for one of the dependent variables. The algorithm is 
applied in the same manner as for viscoelastic flows \cite{Oliveira_1998, Afonso_2012}; in each 
SIMPLE iteration, we solve successively: the linear systems for each stress component, the linear 
systems for the velocity components, and the pressure correction system to enforce continuity. 
The systems are solved with preconditioned conjugate gradient (pressure correction) or GMRES (all 
other variables) solvers.

In the velocity systems, derived from the components of the momentum eq.\ \eqref{eq: momentum 
integral discrete}, the matrix of coefficients is constructed via a contribution from the time 
derivative, a UDS discretisation of the convective term, and the $D_{f,i}^{\tau+}$ (Eq.\ \eqref{eq: 
D+}) part of the EVP force $\vf{F}_f^{\tau}$. The remaining terms are evaluated from their 
currently estimated values and moved to the right-hand side vector, as is the difference between the 
UDS and CDS convection schemes (deferred correction). For the stress systems we follow the 
established practice in viscoelastic flows of constructing the matrix of coefficients with 
contributions only from the terms of the left-hand side of Eq.\ \eqref{eq: constitutive integral 
discrete}, breaking the CUBISTA flux into a UDS component and a remainder which is moved to the 
right-hand side vector (deferred correction). This poor representation of the constitutive equation 
by the matrix of coefficients may be responsible for the observed convergence difficulties of SIMPLE 
at large time steps; for small time steps the role of the time derivative in the constitutive 
equation becomes dominant and convergence is fast.

It was noticed that if the residual reduction within each time step is not at least a couple orders 
of magnitude, then the temporal prediction step (Sec.\ \ref{ssec: method: discretisation: temporal}) 
may not produce good enough initial guesses and the solution may not be smooth in time. To avoid 
this possibility we set a maximum effort per time step of about 20 single-grid SIMPLE iterations 
followed by about 5 W(4,4) multigrid cycles \cite{Syrakos_06b, Syrakos_2013}, and if further 
iterations are needed to accomplish the required residual reduction then the time step is reduced 
by a factor of $r_t^{\min}$ (Sec.\ \ref{ssec: method: discretisation: temporal}). Setting $\Delta 
t_{\max}$ to an appropriate value avoids the need for this.

\section{Validation and testing of the method in Oldroyd-B flow}
\label{sec: method: validation}

For $\tau_y = 0$ and $n = 1$, the SHB model reduces to the Oldroyd-B viscoelastic model, for which 
benchmark results for square lid-driven cavity flow can be found in the literature. We apply our 
method to this problem to validate it and to evaluate alternative options for the FVM components. 
The problem is solved as steady-state, omitting the time derivatives from the governing equations. 
Table \ref{table: validation} lists the computed location and strength of the main vortex, along 
with values reported in the literature, with which there is very good agreement.

\begin{table}[tb]
\caption{$\tilde{x}-$ and $\tilde{y}-$ coordinates $(\tilde{x}_c,\tilde{y}_c)$ of the centre of the 
main vortex for Oldroyd-B flow in a square cavity of side $L$, and associated value of the 
streamfunction there, $\tilde{\psi}_c$. The top wall moves in the positive $x-$direction with 
variable velocity $u(x) = 16 U (x/L)^2(1-x/L)^2$. The coordinates are normalised by the cavity side 
$L$ and the streamfunction by $U L$. The flow is steady, the Reynolds number is zero, and the 
solvent and polymeric viscosities are equal ($\kappa = k$).}
\label{table: validation}
\begin{center}
\begin{small}   
\renewcommand\arraystretch{1.25}   
\begin{tabular}{ l | c c c | c c c }
\toprule
& \multicolumn{3}{c |}{ $\Wei = 0.5$ } & \multicolumn{3}{c}{ $\Wei = 1.0$ }
\\
\midrule
                                &  $\tilde{x}_c$  &  $\tilde{y}_c$  &  $\tilde{\psi}_c$
                                &  $\tilde{x}_c$  &  $\tilde{y}_c$  &  $\tilde{\psi}_c$  \\
\midrule
 Pan et al. \cite{Pan_2009}     & $0.469$ & $0.798$ & $-0.0700$ & $0.439$ & $0.816$ & $-0.0638$ \\
 Saramito  \cite{Saramito_2014} &         &         &           & $0.429$ & $0.818$ & $-0.0619$ \\
 Sousa et al. \cite{Sousa_2016} & $0.467$ & $0.801$ & $-0.0698$ & $0.434$ & $0.814$ & $-0.0619$ \\
 Zhou et al. \cite{Zhou_2016}   & $0.468$ & $0.798$ &           & $0.430$ & $0.818$ &           \\
 Present results                & $0.468$ & $0.799$ & $-0.0698$ & $0.434$ & $0.818$ & $-0.0619$
\\
\bottomrule
\end{tabular}
\end{small}
\end{center}
\end{table}

\begin{figure}[tb]
    \centering
    \begin{subfigure}[b]{0.32\textwidth}
        \centering
        \includegraphics[width=0.84\linewidth]{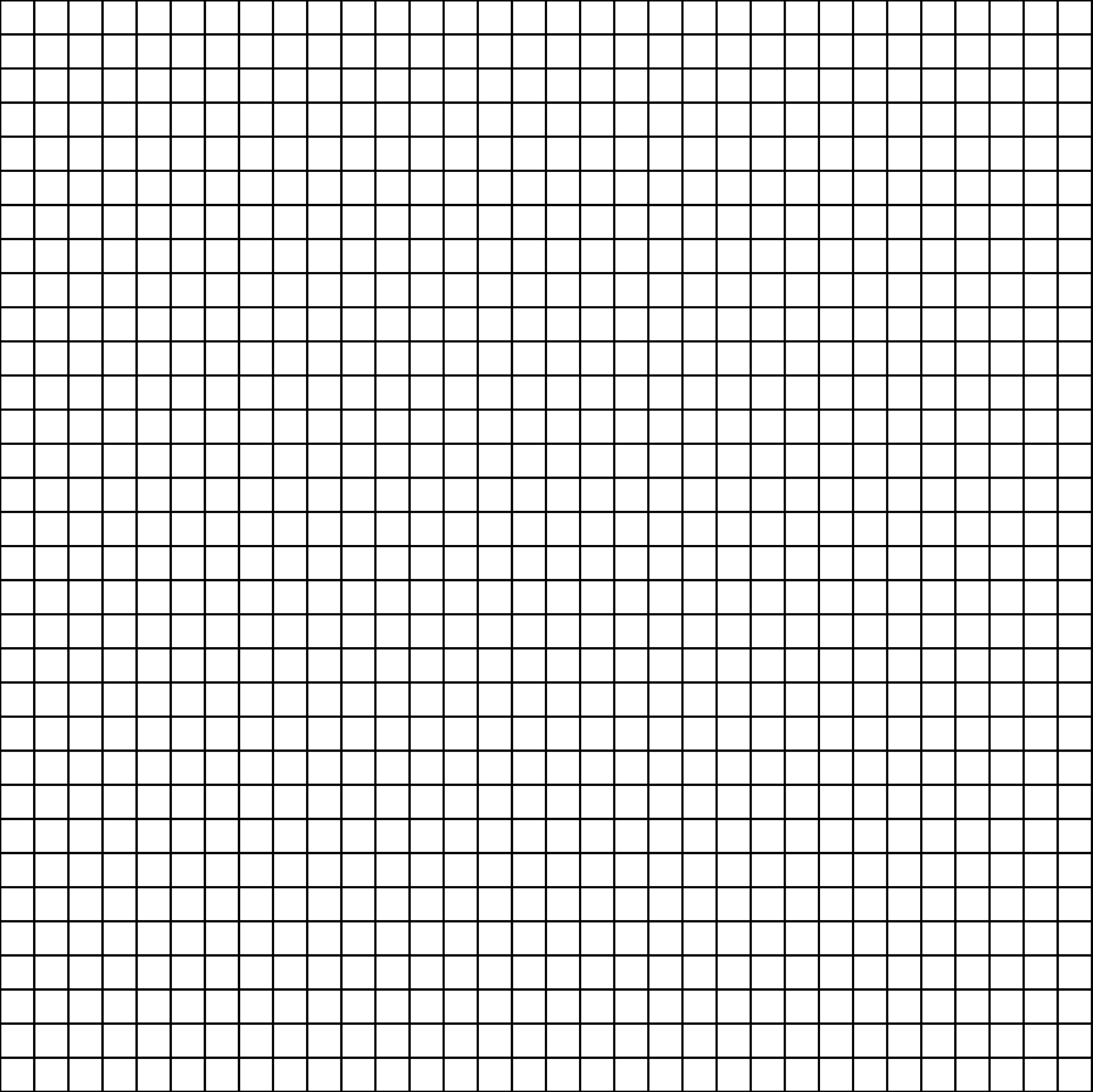}
        \caption{Cartesian uniform (CU)}
        \label{sfig: grid 32 CU}
    \end{subfigure}
    \begin{subfigure}[b]{0.32\textwidth}
        \centering
        \includegraphics[width=0.84\linewidth]{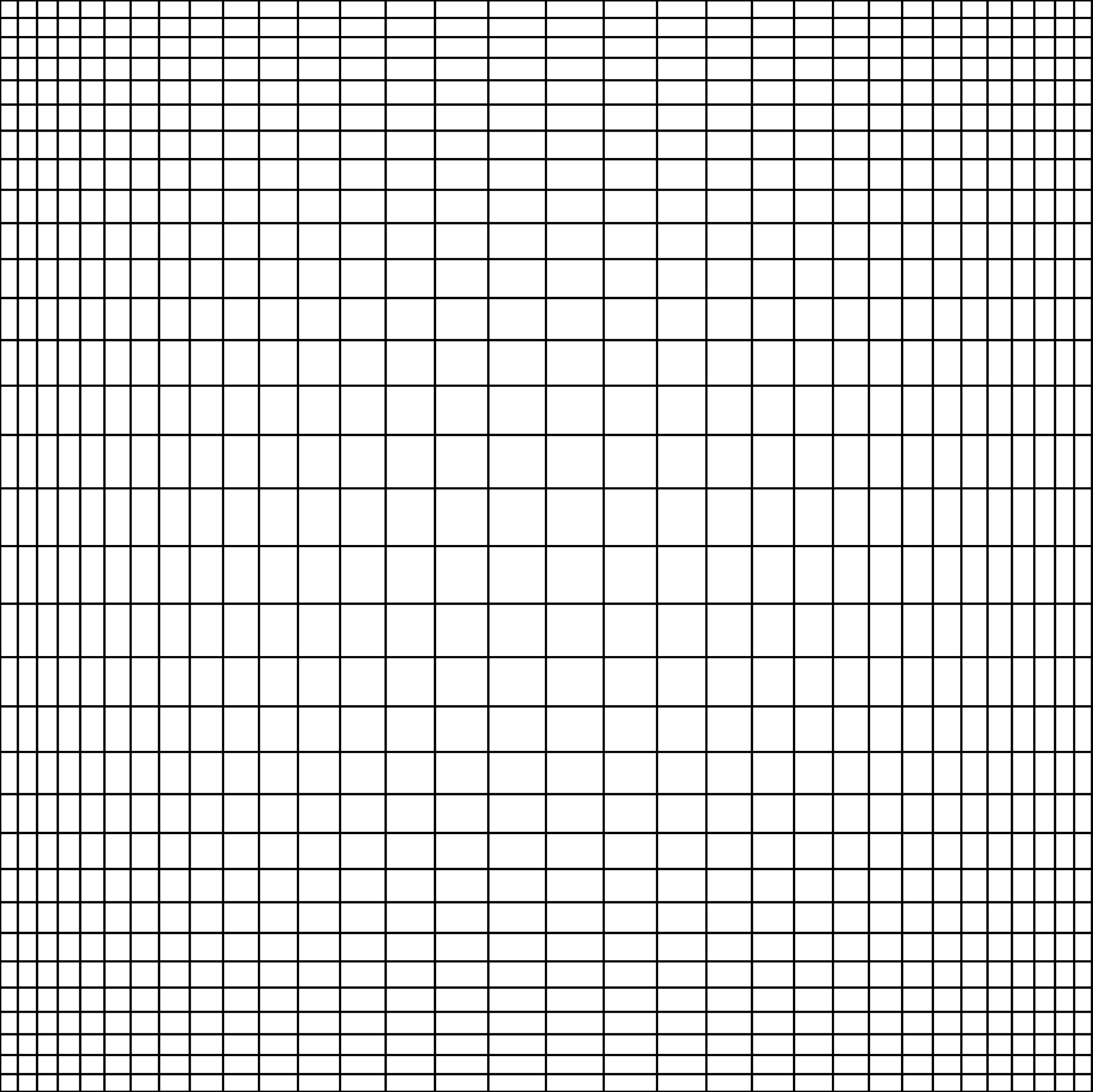}
        \caption{Cartesian non-uniform (CN)}
        \label{sfig: grid 32 CN}
    \end{subfigure}
    \begin{subfigure}[b]{0.32\textwidth}
        \centering
        \includegraphics[width=0.84\linewidth]{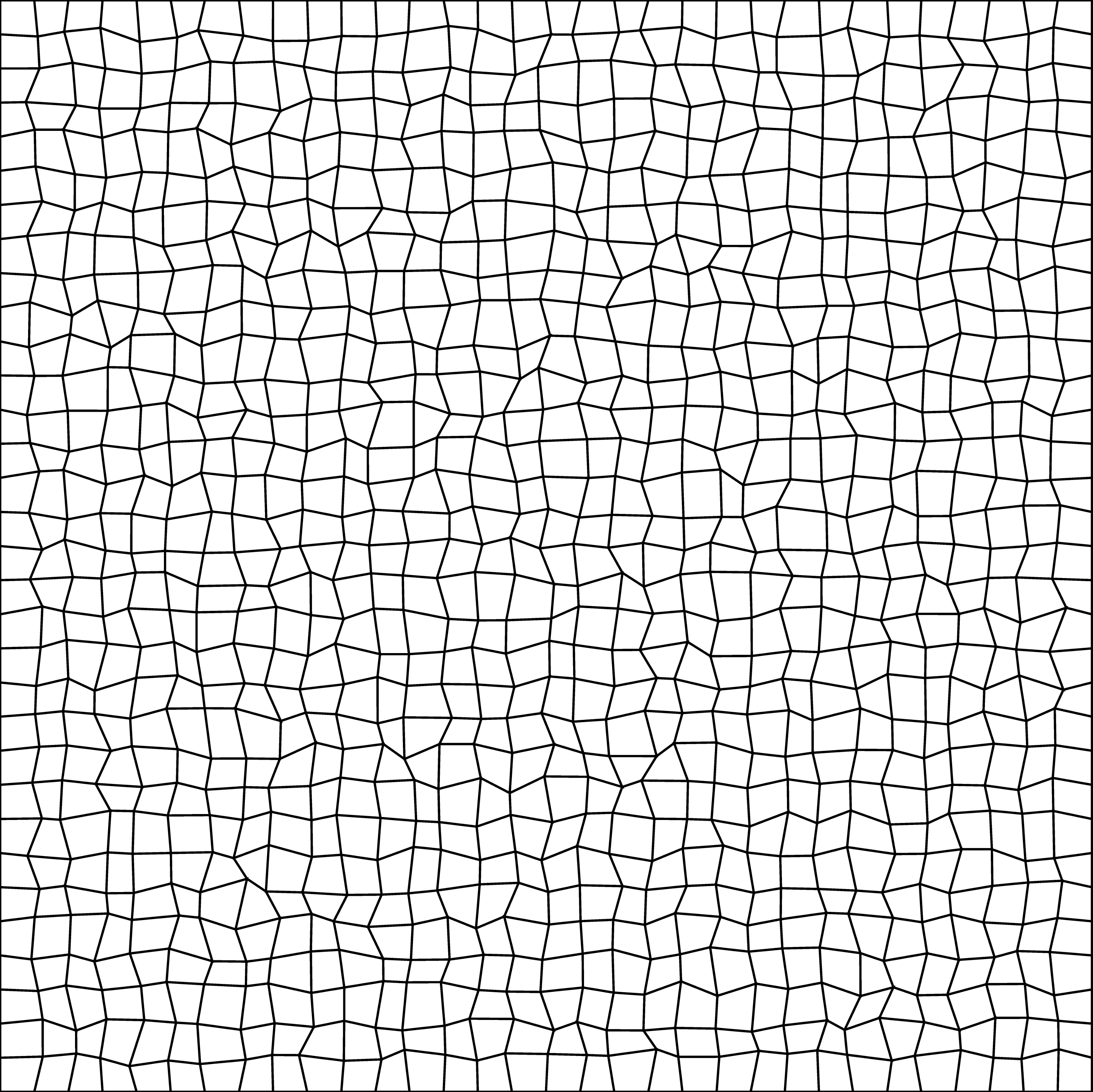}
        \caption{Distorted (D)}
        \label{sfig: grid 32 D}
    \end{subfigure}
    \caption{$32 \times 32$ grids of different kinds, used for the Oldroyd-B benchmark problems.}
  \label{fig: grids Oldroyd-B}
\end{figure}

\begin{figure}[!tb]
    \centering

    \begin{subfigure}[b]{0.32\textwidth}
        \centering
        \includegraphics[width=0.99\linewidth]{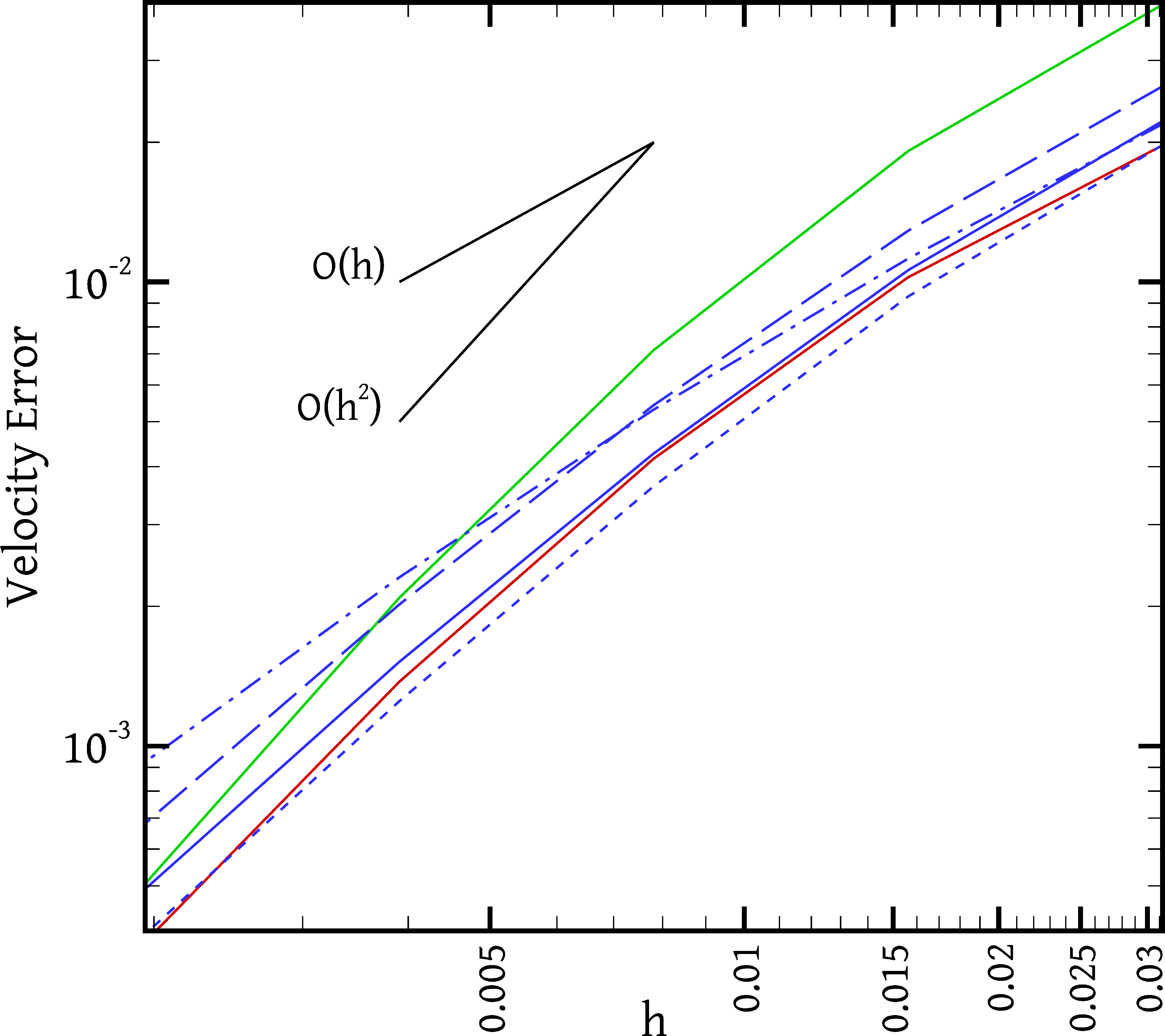}
        \caption{$\Wei = 0.5$, CU grids: $\epsilon_u$}
        \label{sfig: grid convergence 1u}
    \end{subfigure}
    \begin{subfigure}[b]{0.32\textwidth}
        \centering
        \includegraphics[width=0.99\linewidth]{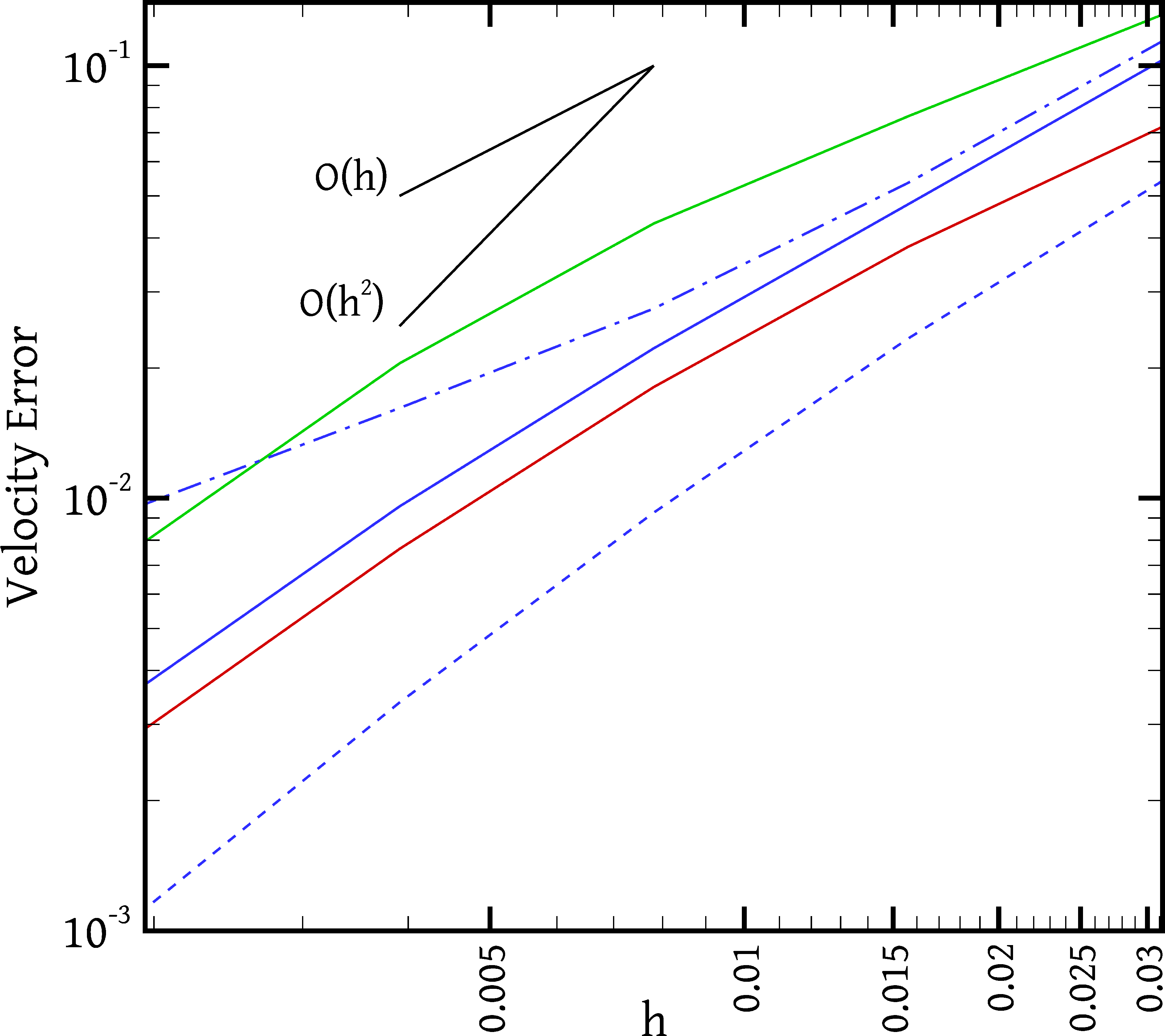}
        \caption{$\Wei = 1.0$: $\epsilon_u$}
        \label{sfig: grid convergence 2u}
    \end{subfigure}
    \begin{subfigure}[b]{0.32\textwidth}
        \centering
        \includegraphics[width=0.99\linewidth]{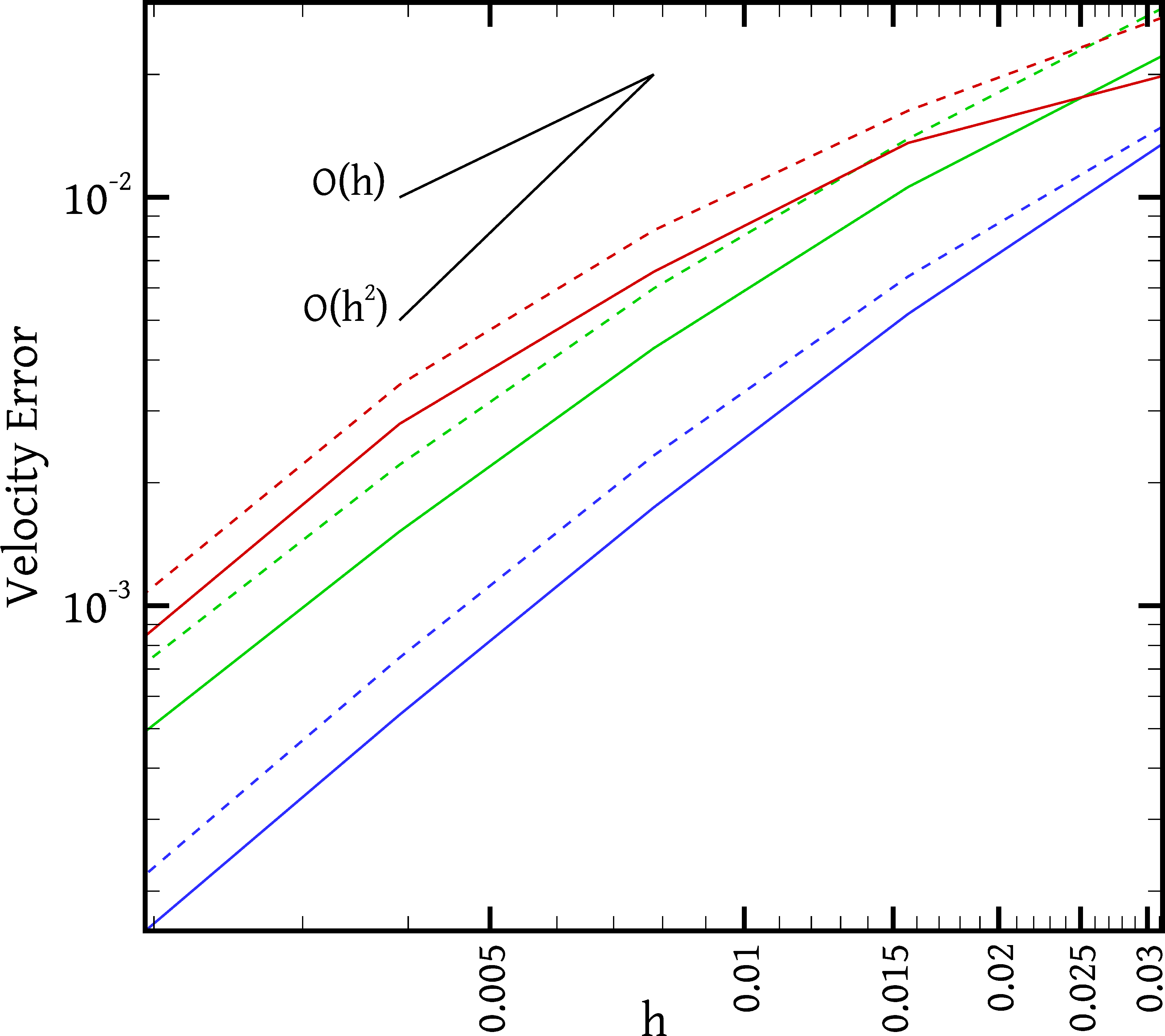}
        \caption{$\Wei = 0.5$; CU,CN,D grids: $\epsilon_u$}
        \label{sfig: grid convergence 3u}
    \end{subfigure}

\vspace{0.5cm}    
    
    \begin{subfigure}[b]{0.32\textwidth}
        \centering
        \includegraphics[width=0.99\linewidth]{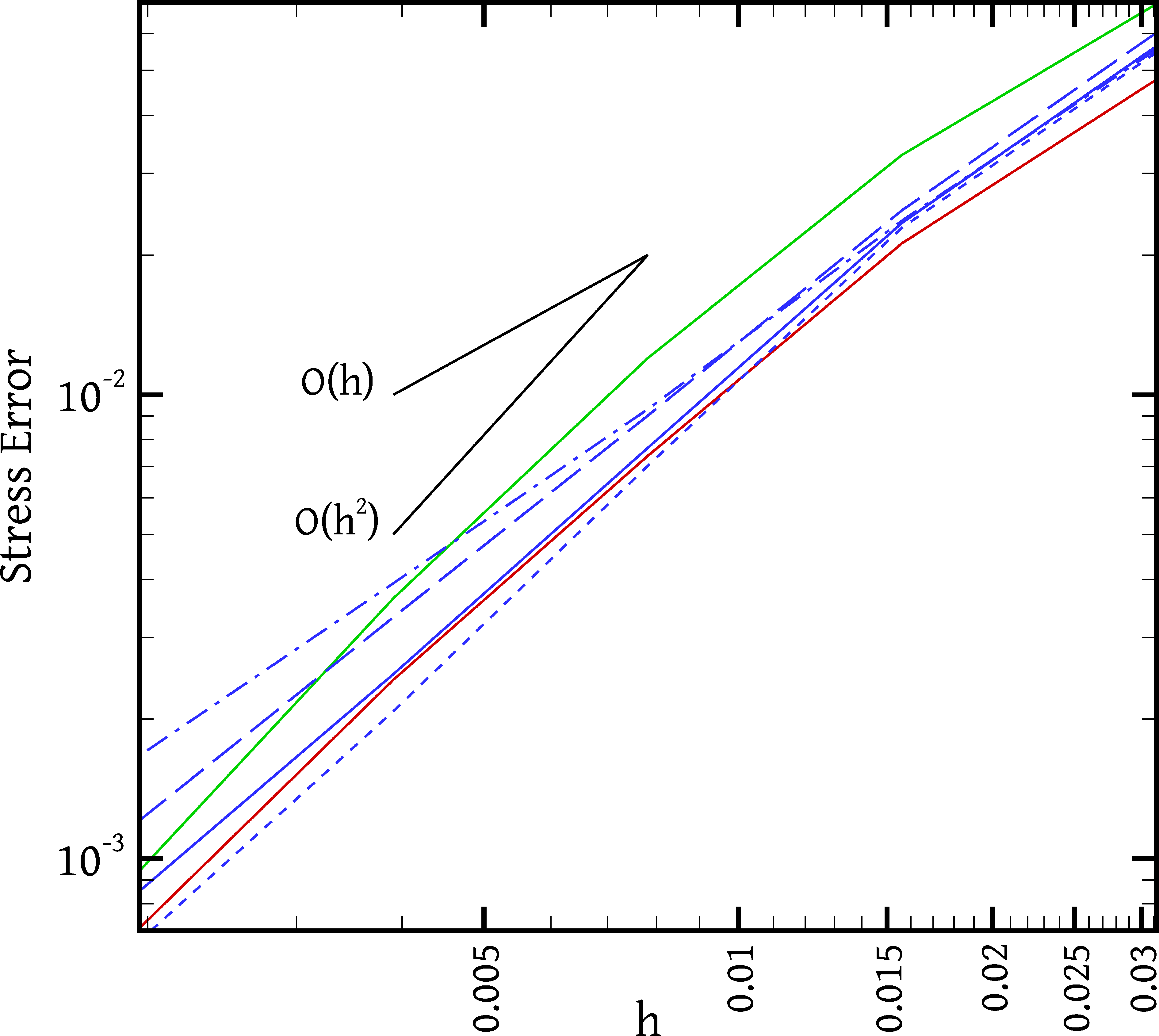}
        \caption{$\Wei = 0.5$, CU grids: $\epsilon_{\tau}$}
        \label{sfig: grid convergence 1s} 
    \end{subfigure}
    \begin{subfigure}[b]{0.32\textwidth}
        \centering
        \includegraphics[width=0.99\linewidth]{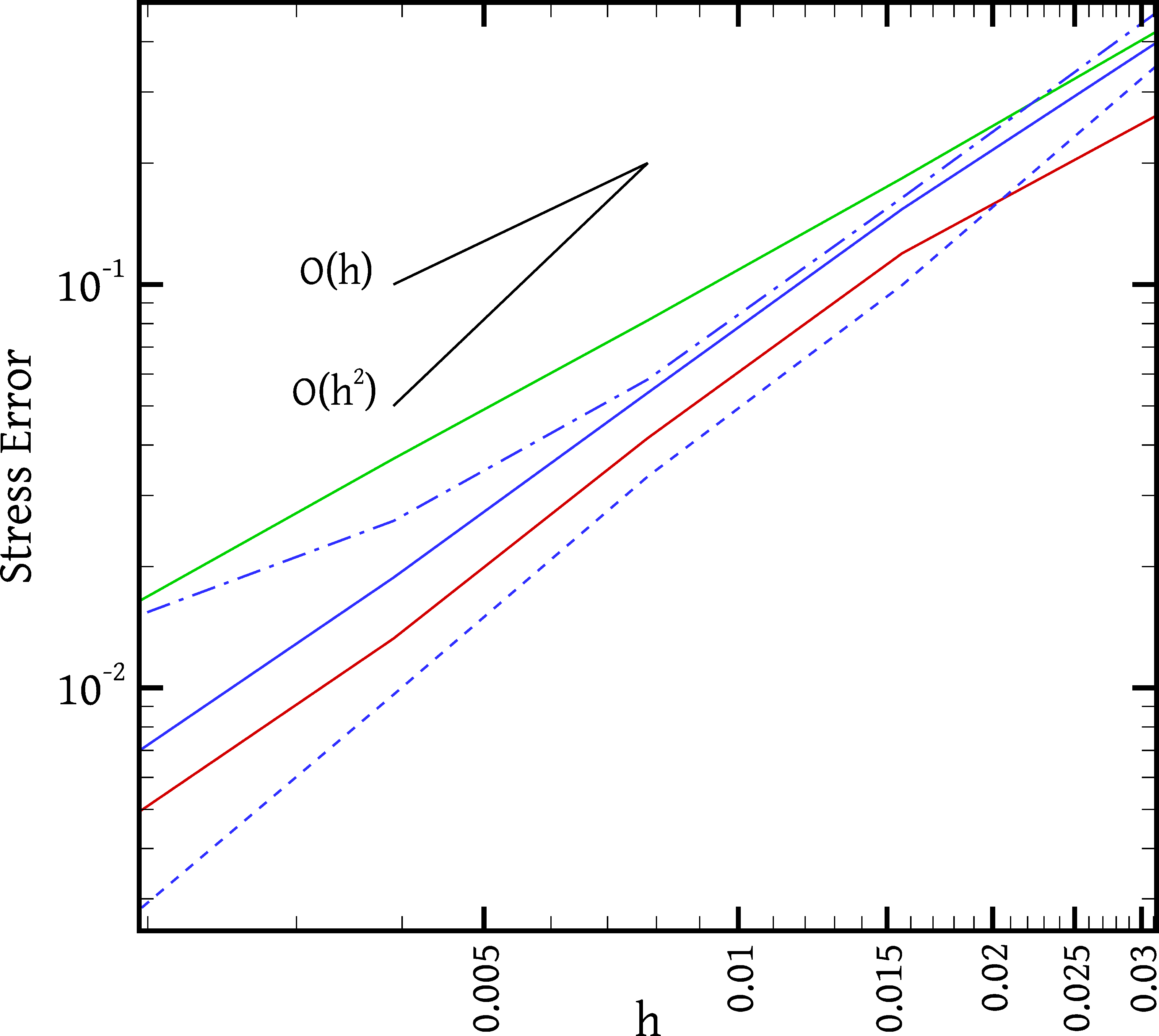}
        \caption{$\Wei = 1.0$: $\epsilon_{\tau}$}
        \label{sfig: grid convergence 2s} 
    \end{subfigure}
    \begin{subfigure}[b]{0.32\textwidth}
        \centering
        \includegraphics[width=0.99\linewidth]{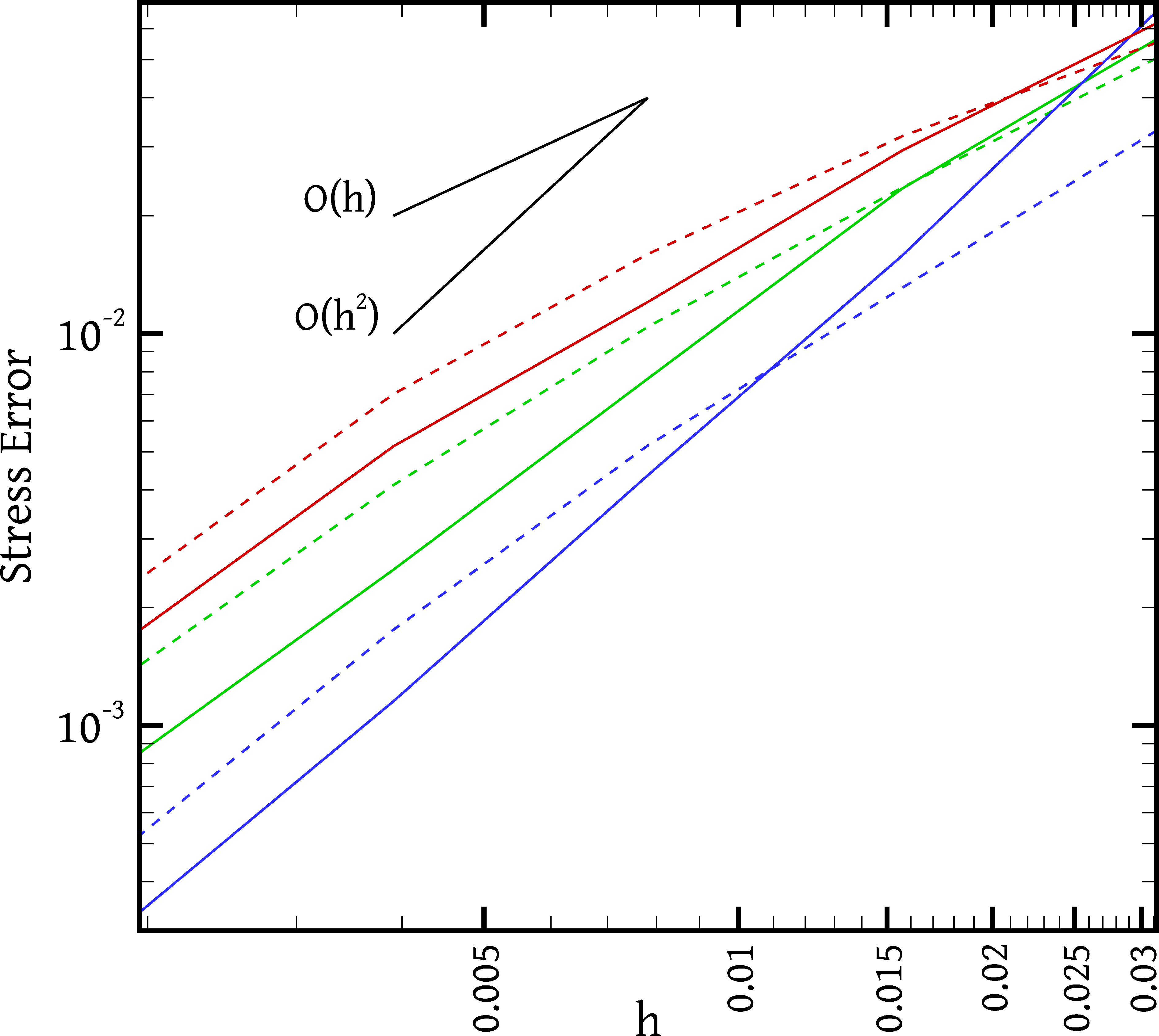}
        \caption{$\Wei = 0.5$; CU,CN,D grids: $\epsilon_{\tau}$}
        \label{sfig: grid convergence 3s} 
    \end{subfigure}

    \caption{Velocity (top row) and stress (bottom row) discretisation errors (Eqs.\ \eqref{eq: 
e_u} and \eqref{eq: e_tau}) of Oldroyd-B flow as a function of the grid spacing $h$, for various 
cases. Unless otherwise stated, the $\nabla_h^{1.5}$ gradient is used and the $D$'s at boundary 
faces are given by \eqref{eq: D+ boundary} -- \eqref{eq: D- boundary}.
\textbf{Figs.\ \subref{sfig: grid convergence 1u}, \subref{sfig: grid convergence 1s}:} $\Wei = 
0.5$, CU grids (Fig.\ \ref{sfig: grid 32 CU}). Green, red and blue lines correspond to setting 
$\overline{\tau_{ij,P}} = \tau_{ij,P}$ in \eqref{eq: upper convective terms discretisation}, using 
the log-conformation technique, and defining $\overline{\tau_{ij,P}}$ by \eqref{eq: stress 
reconstruction} in \eqref{eq: upper convective terms discretisation}, respectively. In the latter 
case, the solid and dash-dot lines correspond to the use of Eqs.\ \eqref{eq: apparent viscosity} and 
\eqref{eq: apparent viscosity DAVSS}, respectively. Dashed lines with short or long dashes 
correspond, respectively, to replacing $S$ by $S/2$ or $2S$ in \eqref{eq: apparent viscosity}.
\textbf{Figs.\ \subref{sfig: grid convergence 2u}, \subref{sfig: grid convergence 2s}:} as for 
Figs.\ \subref{sfig: grid convergence 1u}, \subref{sfig: grid convergence 1s}, but for $\Wei = 
1.0$. The dashed line now corresponds to CN grids (Fig.\ \ref{sfig: grid 32 CN}), using \eqref{eq: 
apparent viscosity} and \eqref{eq: characteristic stress}.
\textbf{Figs.\ \subref{sfig: grid convergence 3u}, \subref{sfig: grid convergence 3s}:} $\Wei = 
0.5$, use of \eqref{eq: apparent viscosity} and \eqref{eq: characteristic stress}. Green, blue, 
and red lines: CU (Fig.\ \ref{sfig: grid 32 CU}), CN (Fig.\ \ref{sfig: grid 32 CN}), and D (Fig.\ 
\ref{sfig: grid 32 D}) grids, respectively. Dashed lines correspond to use of $\nabla^{1.0}_h$ and 
simple linear extrapolation of stresses to the walls (i.e.\ setting $D_{b,i}^{\tau+} = 
D_{b,i}^{\tau-} = 0$ instead of \eqref{eq: D+ boundary} -- \eqref{eq: D- boundary}).}

  \label{fig: grid convergence}
\end{figure}

To estimate discretisation errors, we obtained accurate estimates of the $\Wei = 0.5$ and $\Wei = 
1.0$ solutions by Richardson extrapolation of the solutions obtained on uniform Cartesian grids 
(Fig.\ \ref{sfig: grid 32 CU}) of $512 \times 512$ and $1024 \times 1024$ cells; the procedure is 
described in detail in \cite{Syrakos_2012}. These extrapolated solutions served as ``exact'' 
solutions against which we calculated the discretisation errors of coarser grids. Figure \ref{fig: 
grid convergence} plots average velocity and stress discretisation errors defined as
\begin{equation} \label{eq: e_u}
 \epsilon_u \;\equiv\; \frac{1}{\#\Omega'} \sum_{P \in \Omega'}
                       \| \vf{u}^e_P - \vf{u}_P \| \;/\; \| \vf{u}^e \|_{\mathrm{avg}}
\end{equation}
\begin{equation} \label{eq: e_tau}
 \epsilon_{\tau} \;\equiv\; \frac{1}{\#\Omega'} \sum_{P \in \Omega'}
                            \| \tf{\tau}^e_P - \tf{\tau}_P \| \;/\; \| \tf{\tau}^e \|_{\mathrm{avg}}
\end{equation}
where the superscript $e$ denotes the ``exact'' solution, calculated by Richardson extrapolation. 
The summation is over all cells $P$ whose centres lie inside the area $\Omega' = [0.05L,0.95L] 
\times [0.05L,0.95L]$ ($\#\Omega'$ is the total number of such cells), $L$ being the cavity length, 
i.e.\ a strip of width $0.05L$ touching the boundary is excluded, because the stress magnitude 
($\tau_{xx}$ in particular) appears to grow exponentially over part of the top boundary and this 
inflates the discretisation errors there. The vector and tensor magnitudes in \eqref{eq: e_u} and 
\eqref{eq: e_tau} are $\|\vf{u}\| = (\sum_i u_i^2)^{1/2}$ and $\| \tf{\tau} \| = (\sum_{ij} 
\tau_{ij}^2)^{1/2}$. The definitions \eqref{eq: e_u} and \eqref{eq: e_tau} incorporate a 
normalisation of the errors by the average velocity and stress magnitudes over $\Omega'$: $\| 
\vf{u}^e \|_{\mathrm{avg}} \equiv (1/\Omega') \int_{\Omega'} \| \vf{u}^e \| \mathrm{d}\Omega$ and 
$\| \tf{\tau}^e \|_{\mathrm{avg}} \equiv (1/\Omega') \int_{\Omega'} \| \tf{\tau}^e \| 
\mathrm{d}\Omega$. In Fig.\  \ref{fig: grid convergence} the discretisation errors are plotted with 
respect to the average grid spacing $h = (\Omega / \#\Omega)^{1/2}$ which equals $L / N$ for a $N 
\times N$ grid of any of the kinds shown in Fig.\ \ref{fig: grids Oldroyd-B}.

To ensure that our method is applicable to unstructured grids, although the geometry favours 
discretisation by uniform (Fig.\ \ref{sfig: grid 32 CU}, ``CU'') or non-uniform (Fig.\ \ref{sfig: 
grid 32 CN}, ``CN'') Cartesian grids, we also employed grids obtained from the CU ones by randomly 
perturbing their vertices as described in \cite{Syrakos_2017}. In particular, from a $N \times N$ 
CU grid (Fig.\ \ref{sfig: grid 32 CU}) a $N \times N$ distorted grid is constructed (Fig.\ 
\ref{sfig: grid 32 D}) by moving each interior node $(x,y)$ to a location $(x',y') = (x + \delta x, 
y + \delta y)$ where $\delta x$ and $\delta y$ are randomly selected for each node under the 
restriction that $|\delta x|, |\delta y| < h/4$. This restriction ensures that all resulting grid 
cells are simple convex quadrilaterals. This procedure produces a series of grids whose skewness, 
unevenness and non-orthogonality do not diminish with refinement \cite{Syrakos_2017}, as is typical  
of unstructured grids. Checking the convergence of the method on this sort of grids is important 
because in \cite{Syrakos_2017} it was shown that there are popular FVM discretisations, widely 
regarded as second-order accurate, which actually do not converge to the exact solution with 
refinement on unstructured grids. So, we employed three series of grids, one for each of the grid 
types shown in Fig.\ \ref{fig: grids Oldroyd-B}, having $32 \times 32$, $64 \times 64$, $128 \times 
128$, $256 \times 256$ and $512 \times 512$ cells. The CN grids were constructed as follows: in the 
$512 \times 512$ such grid, the grid spacing at the walls equals $L/1024$ and grows under a constant 
ratio towards the cavity centre. Then, by removing every second grid line we obtain the $256 \times 
256$ grid, and so on for coarser grids.

\begin{figure}[tb]
    \centering
    \begin{subfigure}[b]{0.32\textwidth}
        \centering
        \includegraphics[width=0.95\linewidth]{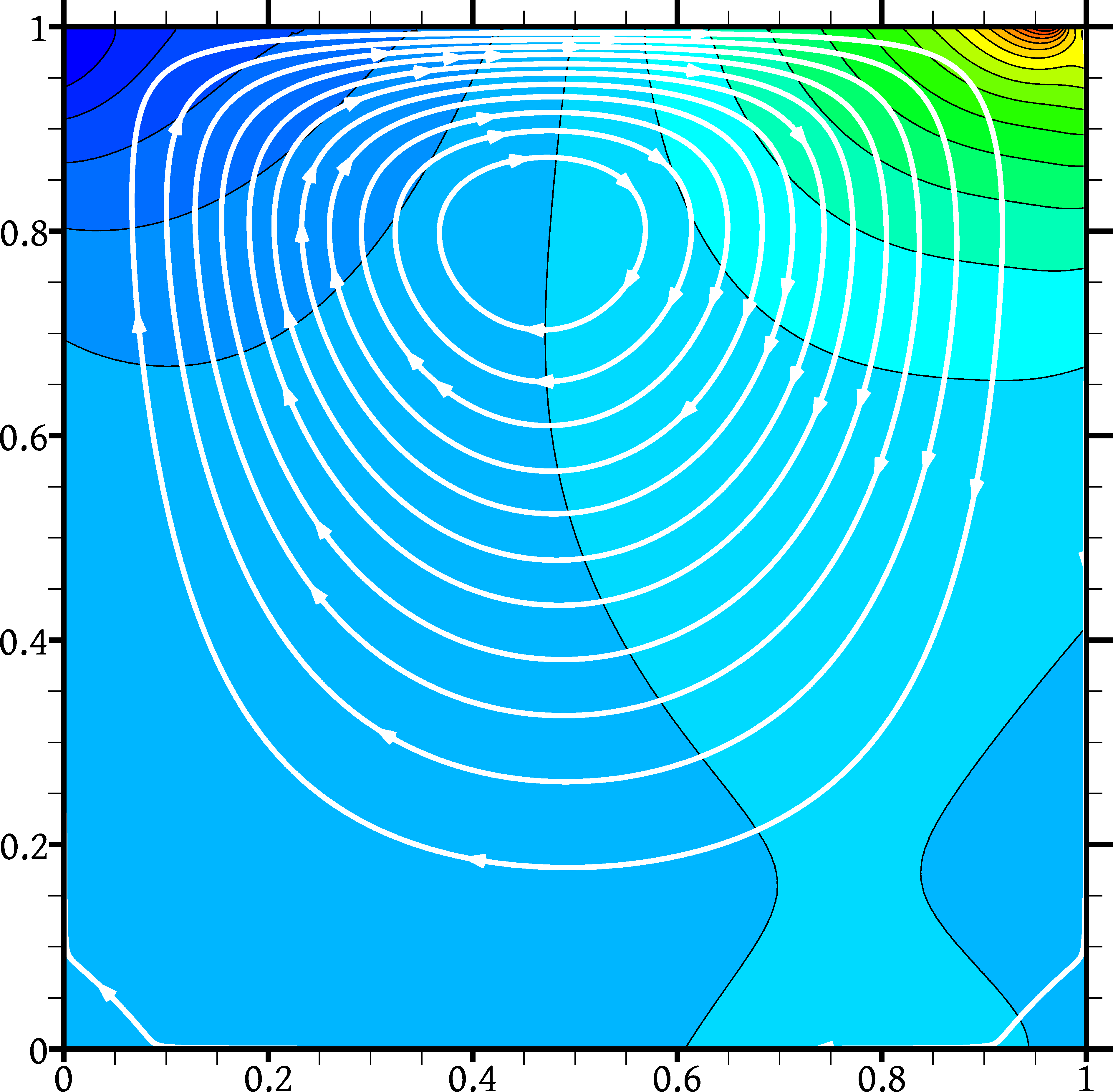}
        \caption{Pressure}
        \label{sfig: Oldroyd-B pressure}
    \end{subfigure}
    \begin{subfigure}[b]{0.32\textwidth}
        \centering
        \includegraphics[width=0.95\linewidth]{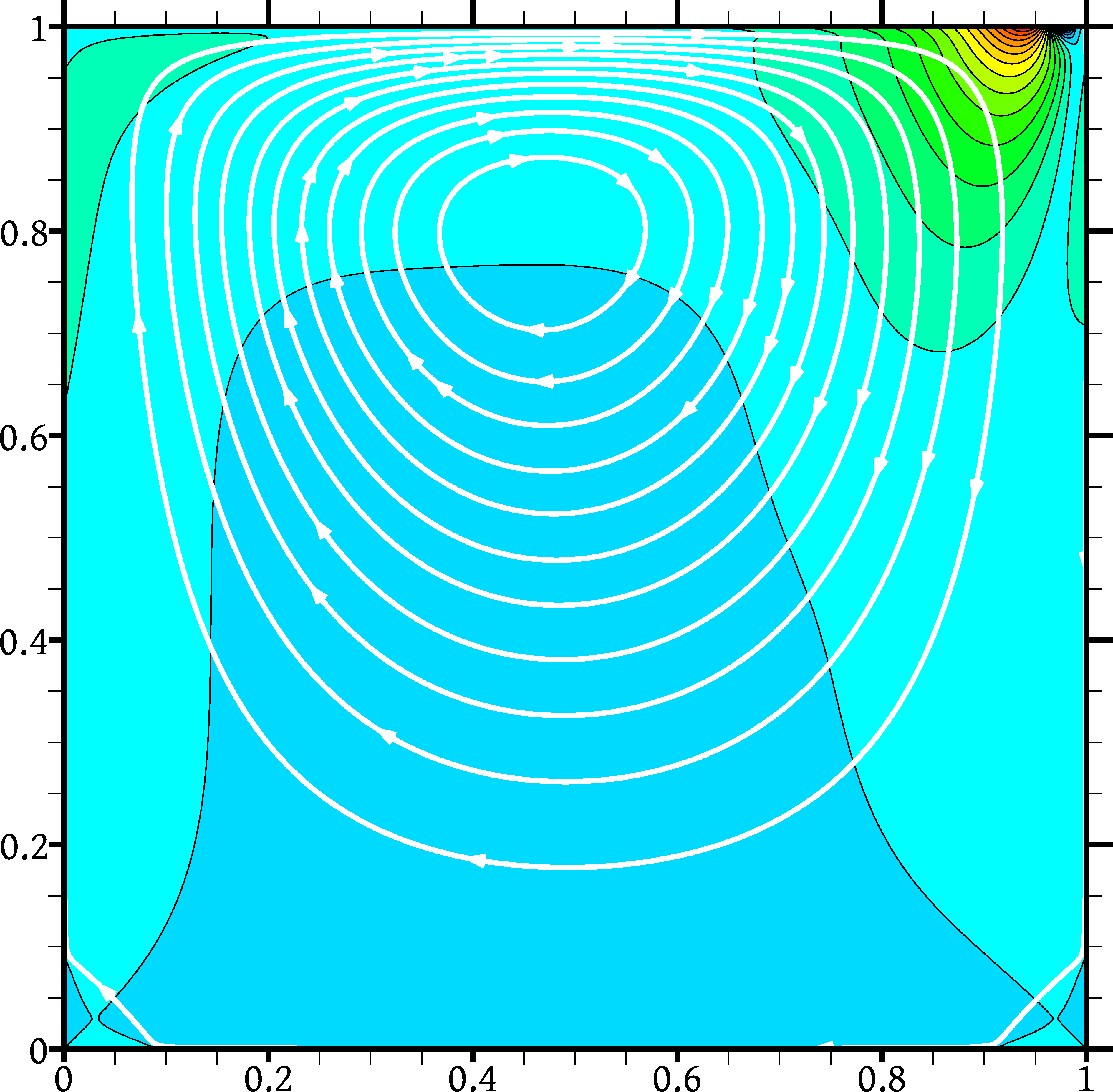}
        \caption{Stress $\tau_{12}$}
        \label{sfig: Oldroyd-B stress xy}
    \end{subfigure}
    \begin{subfigure}[b]{0.32\textwidth}
        \centering
        \includegraphics[width=0.95\linewidth]{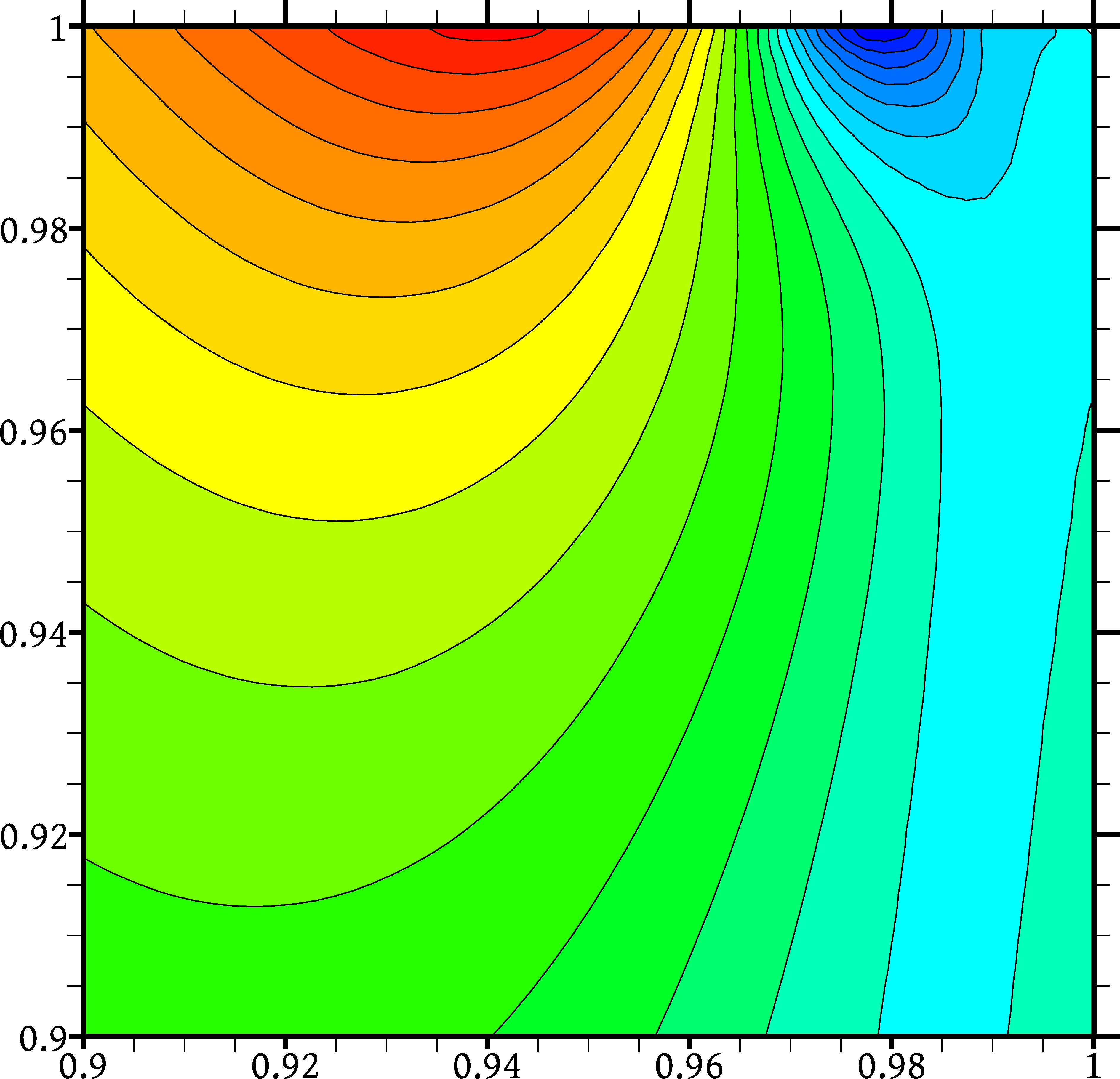}
        \caption{Stress $\tau_{12}$}
        \label{sfig: Oldroyd-B stress xy zoom}
    \end{subfigure}
    \caption{The Oldroyd-B, $\Wei=0.5$ flow field calculated on the $512 \times 512$ distorted 
grid. \subref{sfig: Oldroyd-B pressure} Pressure contours (colour, $\delta p = 4S$ with $S$ given 
by Eq.\ \eqref{eq: characteristic stress}) and streamlines (white lines). \subref{sfig: Oldroyd-B 
stress xy} Contours of $\tau_{12}$ ($\delta \tau_{12} = 2S$) and streamlines. \subref{sfig: 
Oldroyd-B stress xy zoom} Close-up view of \subref{sfig: Oldroyd-B stress xy} near the upper-right 
corner.}
  \label{fig: flow Oldroyd-B}
\end{figure}

The numerical tests led to the following observations. Concerning the oscillations issue, the 
stabilisation strategies achieved their goal as all dependent variables varied smoothly across the 
domain without spurious oscillations; Fig.\ \ref{fig: flow Oldroyd-B} shows smooth pressure and 
$\tau_{12}$ fields and streamlines obtained on the $512 \times 512$ distorted grid. Concerning the 
accuracy of the method, Fig.\ \ref{fig: grid convergence} shows convergence to the exact solution 
with grid refinement on any type of grid, including the randomly distorted type. The order of 
accuracy is close to 1 on coarse grids and increases towards 2 (the design value) as the grid is 
refined, but for the $\Wei = 1$ case (Figs.\ \ref{sfig: grid convergence 2u}, \ref{sfig: grid 
convergence 2s}) it is still quite far from 2  even on the finest ($512 \times 512$) grids. 
Evidently, the convergence rates deteriorate at higher elasticity (compare Figs.\ \ref{sfig: grid 
convergence 1u},\ref{sfig: grid convergence 1s} and \ref{sfig: grid convergence 2u},\ref{sfig: grid 
convergence 2s}). The substandard accuracy performance of the method at high elasticity is most 
likely associated with the exponential stress growth near the lid. The Oldroyd-B predictions can be 
unrealistic, such as predicting unlimited extension of the material at finite extension rates; in 
such cases the accuracy of numerical methods can degrade significantly \cite{Hulsen_2005}. The SHB 
model suffers from these same issues as the Oldroyd-B model for $n \geq 1$, but not for $n < 1$ 
\cite{Saramito_2009}.

With typical EVP materials it is usually the case that $\Wei < 1$, which, combined with the fact 
that their behaviour is described by $n < 1$, avoids the ``high-$\Wei$'' problem. Nevertheless, in 
the tests of the present Section we also tried the log-conformation technique \cite{Fattal_2004, 
Afonso_2009} which is implemented as an option in our code \cite{Syrakos_2018}, and, interestingly, 
Fig.\ \ref{fig: grid convergence} shows that it can be beneficial even at low $\Wei$ as it provides 
enhanced accuracy. Interpolation \eqref{eq: stress reconstruction} also appears to improve the 
accuracy compared to simply setting $\overline{\tau_{ij,P}} = \tau_{ij,P}$, although rate of 
convergence of the latter method appears to be higher for $\Wei = 0.5$ so that it is expected to 
surpass the accuracy of interpolation \eqref{eq: stress reconstruction} on even finer grids. Use of 
\eqref{eq: stress reconstruction} was observed to enable obtaining a solution at slightly higher 
$\Wei$ compared to setting $\overline{\tau_{ij,P}} = \tau_{ij,P}$ (up to $\Wei \approx 1.4$ compared 
to $\Wei = 1$). For the main results of Sec.\ \ref{sec: results} the scheme \eqref{eq: stress 
reconstruction} was selected, given also that it is cheaper than the log-conformation method.

The scheme \eqref{eq: apparent viscosity DAVSS} performs poorly in terms of accuracy (Figs.\ 
\ref{sfig: grid convergence 1u},\ref{sfig: grid convergence 1s} and \ref{sfig: grid convergence 
2u},\ref{sfig: grid convergence 2s}, dash-dot lines) compared to the simpler scheme \eqref{eq: 
apparent viscosity}. Concerning the latter, it is noteworthy that the larger the value of 
$\eta_a$ (adjusted by replacing $S$ with larger / smaller values in \eqref{eq: apparent viscosity}) 
the less accurate the results (Figs.\ \ref{sfig: grid convergence 1u},\ref{sfig: grid convergence 
1s}, lines with long or short dashes). Thus, stabilisation should be exercised with parsimony. 
Henceforth we abandon the scheme \eqref{eq: apparent viscosity DAVSS} and employ scheme \eqref{eq: 
apparent viscosity}.

Fig.\ \ref{fig: grid convergence} shows that packing the grid lines close to the walls -- i.e.\ 
using the CN grids -- achieves a very significant increase of accuracy. This is likely related to 
the aforementioned singular behaviour near the lid. As expected, the discretisation errors are 
largest on the distorted grids (Figs.\ \ref{sfig: grid convergence 3u},\ref{sfig: grid convergence 
3s}, red lines), but not by a great margin compared to the uniform Cartesian grids. The rates of 
error convergence towards zero are similar on all grid types. Another disadvantage of the distorted 
grids is that the high-$\Wei$ problem is intensified: for $\Wei = 1$, we only managed to obtain a 
solution up to grid $64 \times 64$ with the scheme \eqref{eq: apparent viscosity}, and up to grid 
$128 \times 128$ with the scheme \eqref{eq: apparent viscosity DAVSS}. Finally, we note that Figs.\ 
\ref{sfig: grid convergence 3u},\ref{sfig: grid convergence 3s} show 
that the stress extrapolation scheme at the boundaries which uses \eqref{eq: D+ boundary} -- 
\eqref{eq: D- boundary}, in combination with the gradient $\nabla_h^{1.5}$, which retains 2nd-order 
accuracy at the boundaries unlike the more popular $\nabla_h^{1}$ \cite{Syrakos_2017}\footnote{This 
is true only for variables with Dirichlet boundary conditions, i.e.\ the velocity components. For 
pressure and stress, $\nabla_h^{1.5}$ also deteriorates to first-order accuracy. Hence we use the 
more standard gradient $\nabla_h^{1}$ in the extrapolation scheme \eqref{eq: boundary 
extrapolation}.}, leads to a noticeable accuracy improvement compared to simple linear extrapolation 
of stresses.

\section{EVP flow in a lid-driven cavity}
\label{sec: results}

\begin{figure}[tb]
  \centering
  \includegraphics[scale=0.75]{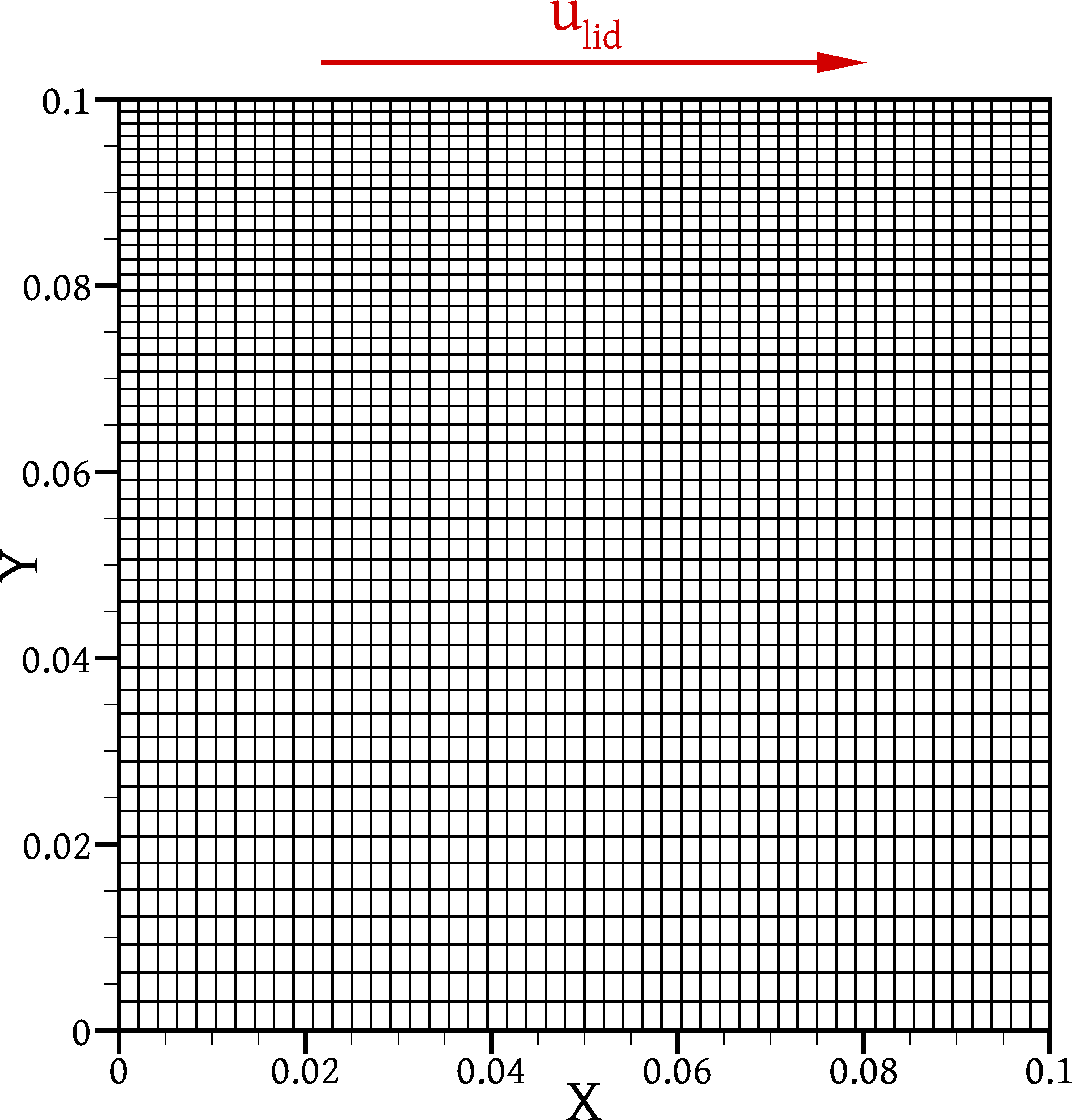}
  \caption{A coarse, $48 \times 48$ cell grid, showing the packing of the cells near the lid (a 
coarse grid is shown for clarity; the EVP lid-driven cavity simulations were performed on a similar 
grid of $384 \times 384$ cells).}
  \label{fig: grid 48x48}
\end{figure}

We now turn our attention to EVP flow in a lid-driven cavity. The SHB model parameters were chosen 
so as to represent the behaviour of Carbopol. Carbopol gels are used very frequently as 
prototypical viscoplastic fluids in experimental studies \cite{Piau_2007, Poumaere_2014}. In 
\cite{Lacaze_2015}, the SHB model was fitted to experimental data for a Carbopol gel of 
concentration 0.2\% in weight, and we rounded those parameters to arrive at our own chosen 
parameters, listed in Table \ref{table: carbopol parameters}. This material has a yield strain (Eq.\ 
\eqref{eq: yield strain}) of $\gamma_y = 0.175$. It is enclosed in a square cavity of side $L$ = 0.1 
\si{m} and the flow is driven by horizontal motion of the top wall (lid) towards the right. We 
employed a grid of $384 \times 384$ cells, of uniform size in the $x$ direction but packed near the 
lid so that the vertical size of the cells touching the lid is $L/625$, and of those touching the 
bottom wall is about $L/252$. A coarse $48 \times 48$ grid of similar packing is shown in Fig.\ 
\ref{fig: grid 48x48}.

\begin{table}[tb]
\caption{Properties of the fluid used in the EVP lid-driven cavity simulations.}
\label{table: carbopol parameters}
\begin{center}
\begin{small}   
\renewcommand\arraystretch{1.25}   
\begin{tabular}{ l | r l }
\toprule
 $\rho$    &   1000  & \si{kg/m^3}
\\
 $\tau_y$  &   70    &  \si{Pa}
\\
 $G$       &  400    &  \si{Pa}
\\
 $k$       &   20    &  \si{Pa.s^n}
\\
 $n$       &   0.40  &
\\
\bottomrule
\end{tabular}
\end{small}
\end{center}
\end{table}

\subsection{The base case}
\label{ssec: results: base case}

We start with a case where the enclosed material is initially at rest and fully relaxed ($\tf{\tau} 
= 0$ everywhere). Starting from rest, the lid accelerates towards the right until it reaches a 
maximum velocity of $U$ = 0.1 \si{m/s} at time $T = L/U$ = 1 \si{s}. The lid velocity remains 
constant thereafter:
\begin{equation} \label{eq: lid velocity}
 u_{\mathrm{lid}} \;=\; U \sin \left( \frac{\min(t,T)}{T} \, \frac{\pi}{2} \right)
\end{equation}
The dimensionless numbers for this case are listed in the ``$U$ = 0.100 \si{m/s}'' column of Table 
\ref{table: dimensionless numbers}.

\begin{table}[tb]
\caption{Values of the dimensionless numbers, and of the relaxation time \eqref{eq: lamda and eta}, 
at different lid velocities.}
\label{table: dimensionless numbers}
\begin{center}
\begin{small}   
\renewcommand\arraystretch{1.25}   
\begin{tabular}{ l | c c c }
\toprule
 $U$ [\si{m/s}]      &  0.025  &  0.100  &  0.400
\\
\midrule
 $\Bin'$             &  6.094  &  3.500  &  2.010
\\
 $\Bin$              &  0.859  &  0.778  &  0.668
\\
 $\Wei$              &  0.204  &  0.225  &  0.262
\\
 $\Rey$              &  0.008  &  0.111  &  1.526
\\
\midrule
 $\lambda$ [\si{s}]  &  0.815  &  0.255  &  0.066
\\
\bottomrule
\end{tabular}
\end{small}
\end{center}
\end{table}

\begin{figure}[tb]
    \centering
    \begin{subfigure}[b]{0.49\textwidth}
        \centering
        \includegraphics[width=0.80\linewidth]{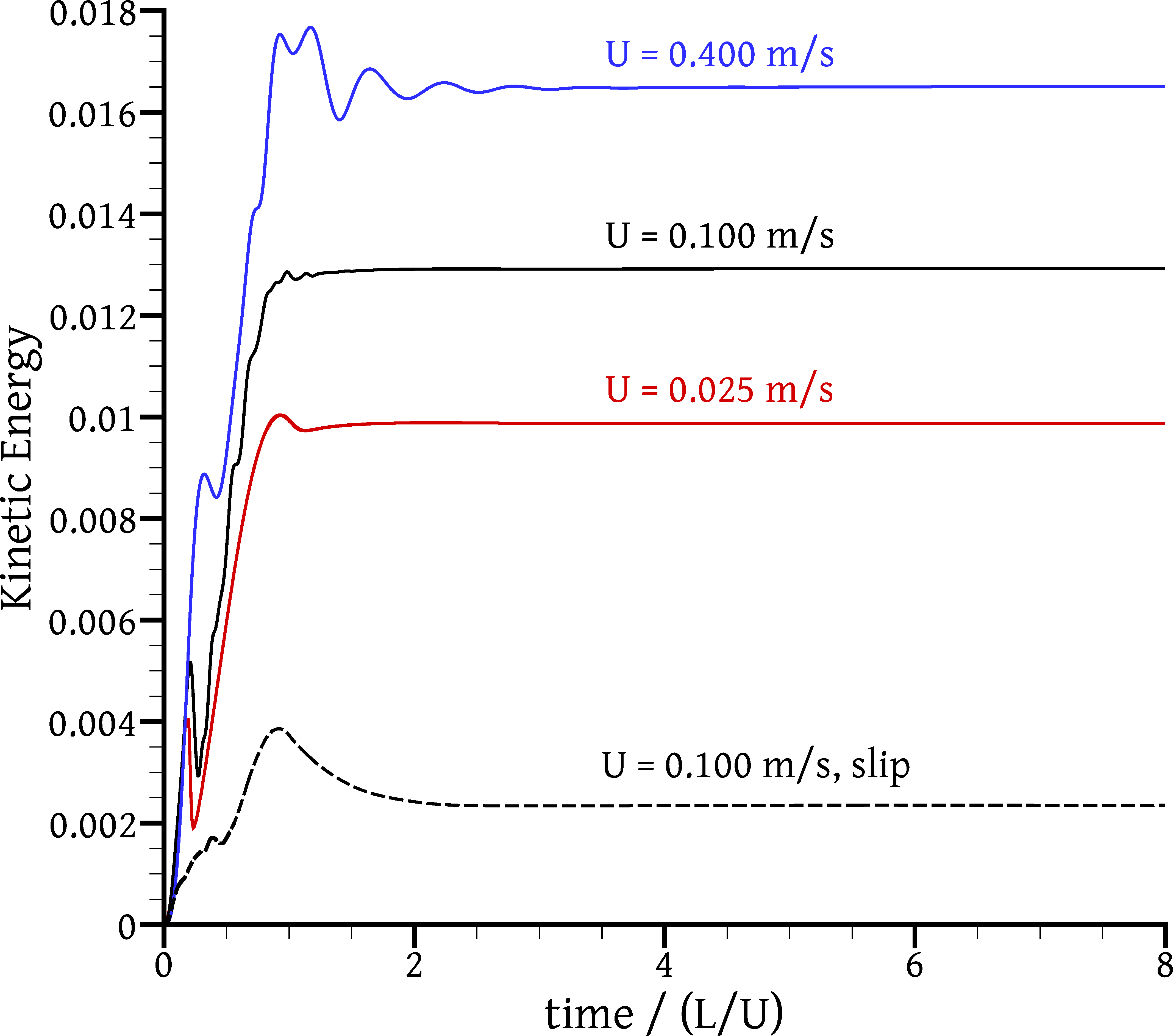}
        \caption{N.K.E.}
        \label{sfig: monitor KE}
    \end{subfigure}
    \begin{subfigure}[b]{0.49\textwidth}
        \centering
        \includegraphics[width=0.80\linewidth]{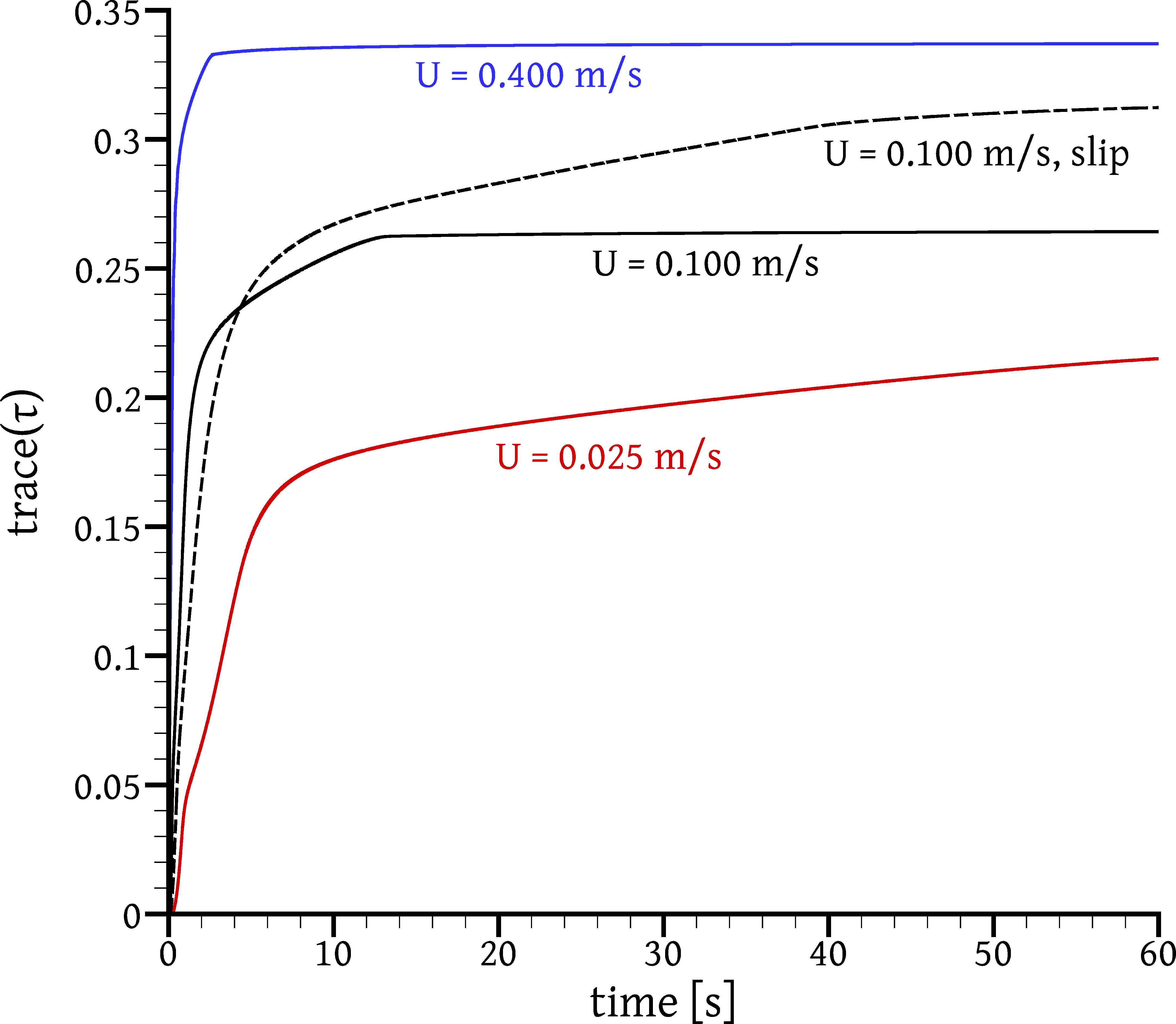}
        \caption{$\mathrm{tr}(\tf{\tilde{\tau}})_{\mathrm{avg}}$}
        \label{sfig: monitor trT} 
    \end{subfigure}
    \caption{\subref{sfig: monitor KE} Normalised kinetic energy (Eq.\ \eqref{eq: kinetic energy}) 
and  \subref{sfig: monitor trT} normalised average trace of $\tf{\tau}$ (Eq. \eqref{eq: average 
trace}), versus time, for different lid velocities. In \subref{sfig: monitor KE} time is normalised 
by $L/U$.}
  \label{fig: monitor}
\end{figure}

Figure \ref{fig: monitor} shows the evolution in time of the normalised kinetic energy of the fluid 
and of the normalised average absolute value of the trace of the stress tensor, calculated as 
\begin{equation} \label{eq: kinetic energy}
 \mathrm{N.K.E.} \;=\; \frac{1}{U^2\,\Omega} \sum_P \Omega_P \| \vf{u}_P \|^2
\end{equation}
\begin{equation} \label{eq: average trace}
 \mathrm{tr}(\tf{\tilde{\tau}})_{\mathrm{avg}} \;=\; 
 \frac{1}{S\,\Omega} \sum_P \Omega_P |\mathrm{tr}(\tf{\tau}{}\!_P)| 
\end{equation}
where $\Omega = L^2$ is the volume of the cavity. It is immediately obvious from Fig.\ \ref{fig: 
monitor} that the kinetic energy assumes a constant value very quickly, but the average stress trace 
evolves much more slowly. To investigate this, the simulation was carried on until a time of $t$ = 
210 \si{s}.

\begin{figure}[tb]
    \centering
    \begin{subfigure}[b]{0.45\textwidth}
        \centering
        \includegraphics[scale=0.90]{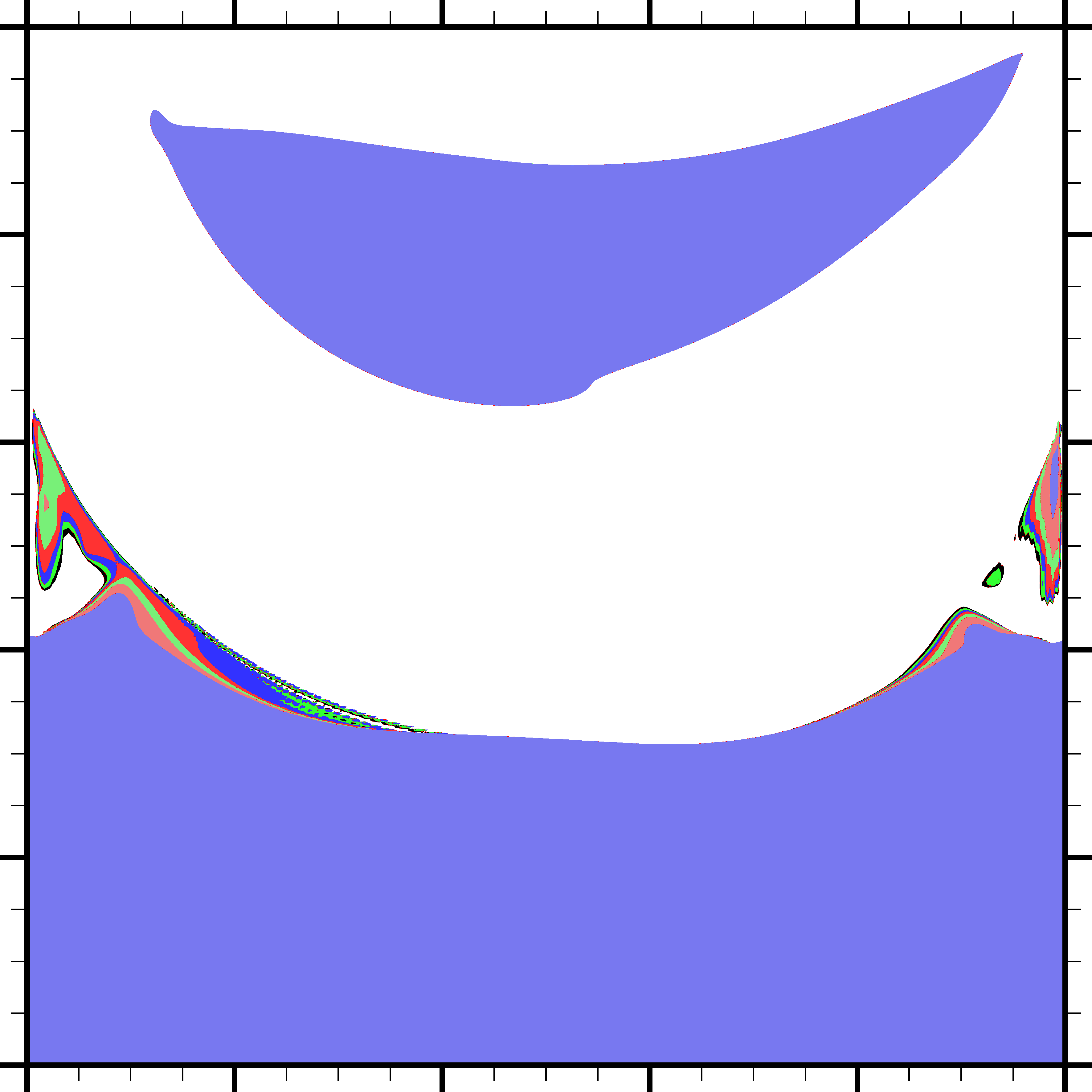}
        \caption{Unyielded areas}
        \label{sfig: base case, yielded vs time}
    \end{subfigure}
    \begin{subfigure}[b]{0.53\textwidth}
        \centering
        \includegraphics[scale=0.90]{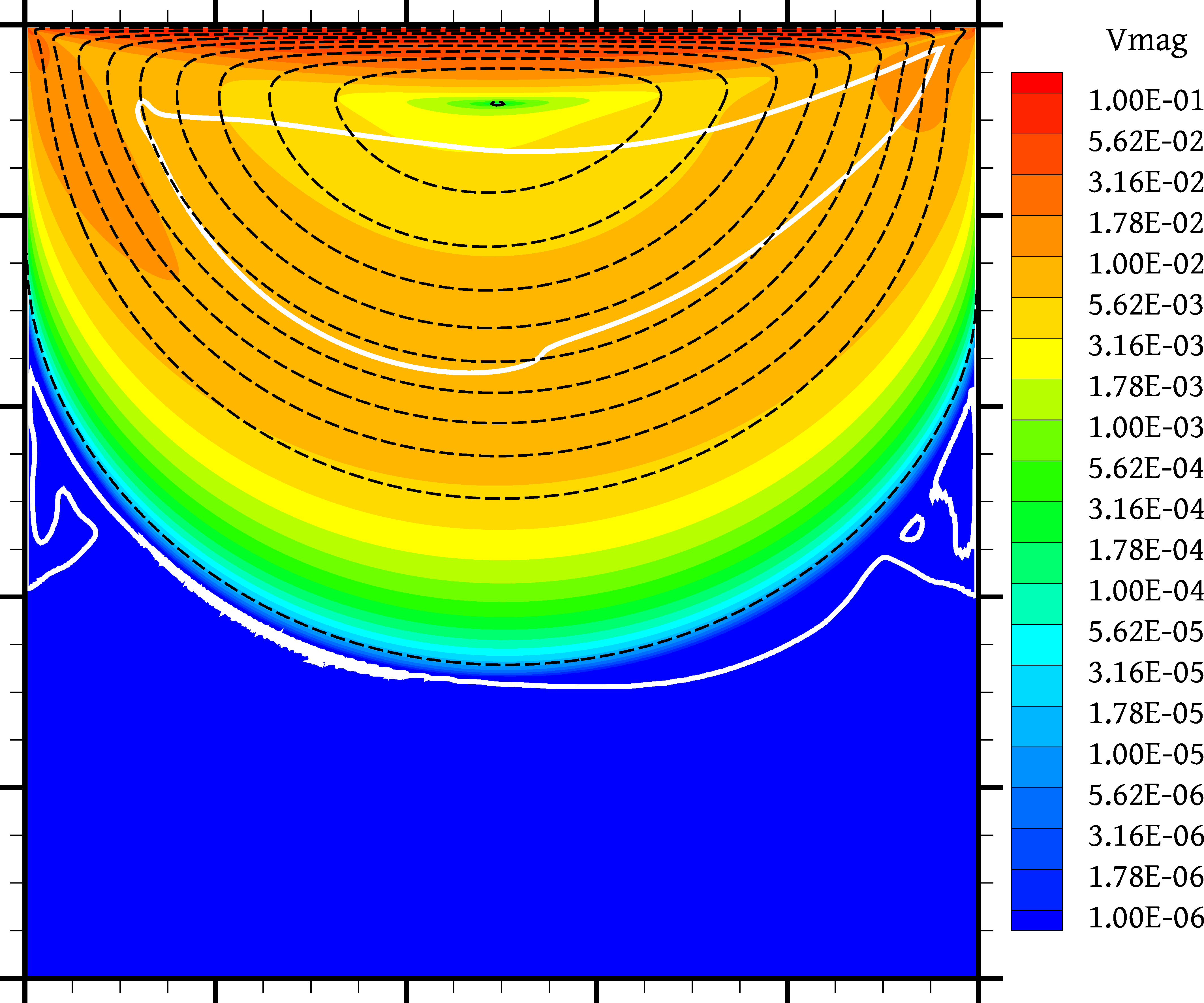}
        \caption{$\| \vf{u} \|$ [\si{m/s}]}
        \label{sfig: base case, Umag} 
    \end{subfigure}
    \caption{\subref{sfig: base case, yielded vs time} Yielded (white) and unyielded (colour) 
regions for the base case (Sec.\ \ref{ssec: results: base case}). Different colours denote the 
extent of the unyielded regions at $t$ = 30, 60, 90, 120, 150, 180 and 210 \si{s} (the lower 
unyielded region is expanding with time). \subref{sfig: base case, Umag} Colour contours of the 
magnitude of the velocity vector at $t$ = 210 \si{s}. The white lines are the yield lines ($\tau_d = 
\tau_y$) and the dashed black lines are streamlines, both at $t$ = 210 \si{s}. The streamlines are 
plotted at equal streamfunction intervals.} 
  \label{fig: base case flowfield}
\end{figure}

The flow field is visualised in Fig.\ \ref{fig: base case flowfield}. It resembles that for pure 
viscoplastic flow \cite{Syrakos_2013, Syrakos_2014}, in that there are two unyielded zones 
($\tau_d < \tau_y$), one at the bottom of the cavity touching the walls, containing stationary 
fluid, and one near the lid which is rotating with the flow and does not touch the walls, called a 
plug zone. Whether the material at a point is in a yielded or unyielded state is, of course, 
determined by whether $\tau_d$ is larger or smaller, respectively, than the yield stress $\tau_y$ 
there (Eq.\ \eqref{eq: constitutive}). Figure \ref{sfig: base case, Umag} shows that the streamlines 
do not cross into the lower unyielded zone, which therefore always consists of the same material, 
but they cross into and out of the plug zone. Thus, at every instant in time, liquefied particles 
are entering the plug zone, solidifying upon entry, while other, solidified particles, are exiting 
the zone, liquefying upon exit.

A striking feature of the flow evolution is the slowness with which the stationary unyielded zone 
tends to obtain its final shape. Figure \ref{sfig: base case, yielded vs time} shows that although 
the plug zone has reached its steady state already from $t$ = 30 \si{s}, the stationary zone 
continues to expand even at $t$ = 210 \si{s}. The shape of the yield line delimiting this zone is 
quite irregular. Furthermore, Fig.\ \ref{sfig: base case, Umag} shows that this zone is surrounded 
by an amount of fluid that is practically stationary and yet yielded. It is likely that this fluid 
will eventually become part of the stationary unyielded zone as $t \rightarrow \infty$, i.e.\ that 
the yield line will eventually lie somewhere close to the nearby streamline drawn in Fig.\ 
\ref{sfig: base case, Umag}, which separates the fluid with near-zero velocity from that whose 
velocity is larger. However, to ascertain this the simulation would have to be prolonged to 
prohibitively long times; already the expansion of the lower unyielded zone from $t$ = 180 \si{s} to 
$t$ = 210 \si{s}, marked in black colour in Fig.\ \ref{sfig: base case, yielded vs time}, is very 
small.

\begin{figure}[thb]
  \centering
  \includegraphics[scale=1.00]{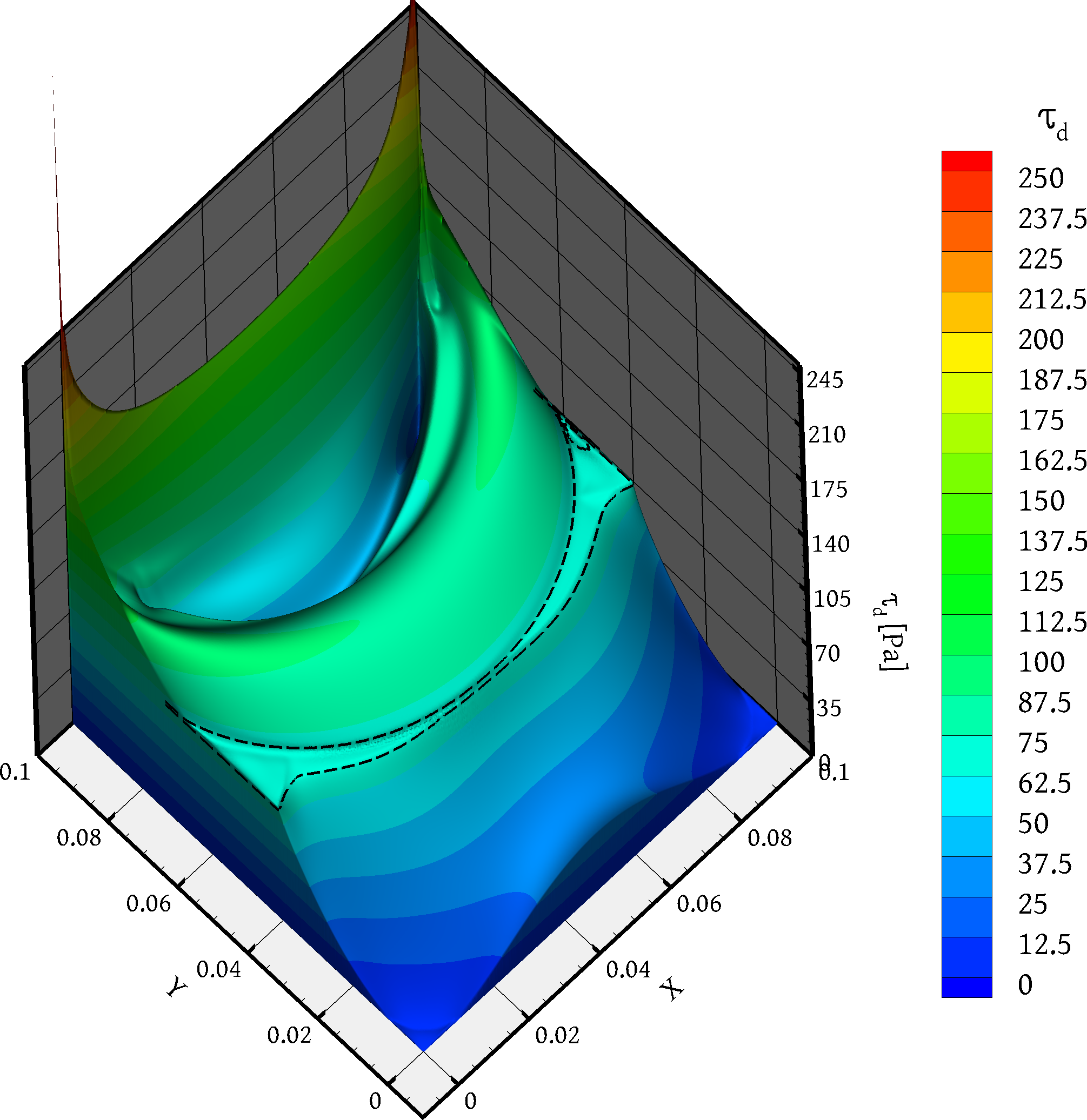}
  \caption{3D contour plot of the base case results (Sec.\ \ref{ssec: results: base case}) at $t$ = 
210 \si{s}. Both the colour contours and the plot height ($z$ axis) represent $\tau_d$. In addition, 
the dashed line encloses an area where $\tau_d \in [69,71]$ \si{Pa}.}
  \label{fig: base case 3D}
\end{figure}

A related feature is that the magnitude of the deviatoric stress tensor, $\tau_d$, is very close to 
the yield stress throughout the aforementioned near-zero velocity region into which the lower 
unyielded zone is expanding. Thus, this region appears as an almost completely flat surface in the 
three-dimensional plot of $\tau_d$ in Fig.\ \ref{fig: base case 3D} (the surface outlined in dashed 
line contains fluid where $\tau_d$ is within $\pm 1$ \si{Pa} of $\tau_y$). Due to its distinctive 
features, we will refer to this region as a ``transition zone''. In it, the fluid practically 
behaves as unyielded, although some of it can be formally yielded. The plug zone possesses no 
transition zone.

In order to explain the transition zone behaviour, we consider the constitutive equation \eqref{eq: 
constitutive} -- \eqref{eq: upper convected derivative} in the case that the fluid velocity is zero 
and $\tau_d$ is slightly above $\tau_y$. It becomes:
\begin{equation} \label{eq: constitutive stationary tensorial}
 \pd{\tf{\tau}}{t} \;=\; -\frac{G}{k^{\frac{1}{n}}} (\tau_d - \tau_y)^{\frac{1}{n}}
                         \frac{1}{\tau_d} \; \tf{\tau}
                   \;\equiv\; -r(\tau_d) \; \tf{\tau}
\end{equation}
Since the function $r(\tau_d)$ assumes only positive values, the minus sign on the right-hand side 
of Eq.\ \eqref{eq: constitutive stationary tensorial} drives the components of $\tf{\tau}$ towards 
zero as time passes, with a rate proportional to $r(\tau_d)$ (and to the magnitude of those 
components themselves). Hence $\tau_d$ decreases towards zero as well (Eq.\ \eqref{eq: deviatoric 
stress}). However, the rate at which they decrease, $r(\tau_d)$, diminishes as $\tau_d \rightarrow 
\tau_y$  and in fact $r(\tau_y) = 0$. Therefore, $\tau_d$ actually converges towards $\tau_y$ and 
not to zero. This is what we observe happening inside the transition zone.

It is useful to consider also a scalar version of Eq.\ \eqref{eq: constitutive stationary 
tensorial} ($\tau_d \equiv \tau$ in this case)\footnote{Eq.\ \eqref{eq: constitutive stationary 
scalar} can be viewed as describing the behaviour of the mechanical system of Fig.\ \ref{fig: model 
schematic} (with $\kappa = 0$) in the case that at some stress $\tau = \tau_0 > \tau_y$ we pin the 
right end of the spring to a fixed location, so that the total length $\gamma_v + \gamma_e$ remains 
constant henceforth, and we leave the system to relax. The spring will try to recover its 
equilibrium length by contracting or expanding $\gamma_e$, which will cause an opposite expansion 
or contraction of $\gamma_v$, resisted by the viscous and friction elements. Once the tension in 
the spring drops to $\tau_y$, the spring cannot overcome the friction $\tau_y$ of the friction 
element and motion stops, without the spring having attained its equilibrium length.}:
\begin{equation} \label{eq: constitutive stationary scalar}
 \frac{\mathrm{d}\tau}{\mathrm{d}t} \;=\;
   -\frac{G}{k^{\frac{1}{n}}} (\tau - \tau_y)^{\frac{1}{n}}
 \end{equation}
For $n = 1$, the solution to the above equation is
\begin{equation}
 \tau - \tau_y \;=\; (\tau_0 - \tau_y) \, e^{-t/\lambda'}
\end{equation}
where $\tau_0$ is the value of $\tau$ at $t = 0$ and $\lambda' \equiv k / G$ is a relaxation time. 
This behaviour is similar to that of a Maxwell viscoelastic fluid, only that now $\tau$ decays 
exponentially towards $\tau_y$ instead of towards zero. For $n \neq 1$, Eq.\ \eqref{eq: constitutive 
stationary scalar} can be written as ($\lambda' \equiv k^{1/n} / G$ now has units of 
\si{s/Pa^{1-1/n}}):
\begin{equation}
  \frac{\mathrm{d}(\tau - \tau_y)}{\mathrm{d}t} \;=\; 
   -\underbrace{\frac{1}{\lambda'} (\tau - \tau_y)}_{ \mathclap{\text{rate of decay for }n=1} }
   (\tau - \tau_y)^{\frac{1-n}{n}}
\end{equation}
so that the rate of decay of $\tau$ towards $\tau_y$ equals that for $n=1$ multiplied by a factor 
of $(\tau-\tau_y)^{(1-n)/n}$. If $n < 1$, as is the present choice but also the most common case, 
then $(1-n)/n > 0$ and as time progresses and $\tau \rightarrow \tau_y$, the extra factor 
$(\tau-\tau_y)^{(1-n)/n}$ tends to zero, and the rate of decay becomes progressively smaller than 
that of the $n = 1$ case, eventually becoming infinitesimal compared to it. This behaviour is 
explained physically by the fact that the relaxation time (Eq.\ \eqref{eq: lamda and eta}) is 
proportional to the fluid viscosity, which resists recovery from deformation (relaxation), and 
inversely proportional to the elastic modulus, which drives towards such recovery. For 
shear-thinning fluids ($n < 1$), the viscosity \textit{increases} as $\tau$ becomes smaller and, in 
the SHB and HB models, tends to infinity as $\tau \rightarrow \tau_y$\footnote{In the language of 
the mechanical analogue of Fig.\ \ref{fig: model schematic} this means that the damper component 
resisting the relaxation of the spring becomes stiffer as this relaxation proceeds.}. Thus, for 
these fluids the relaxation time tends to infinity as $\tau \rightarrow \tau_y$. The opposite 
happens in the less common case of $n > 1$.

By dividing both sides of Eq.\ \eqref{eq: momentum ND} by $\Rey$, because $Re$ is very small (Table 
\ref{table: dimensionless numbers}), it becomes apparent that the fluid particle accelerations (the 
left-hand side of that equation) are large and the velocity adjusts very quickly to force changes 
(the inertial time scales are very small). Thus, the velocity should vary at the same rate as these 
forces, i.e.\ the velocity and stress variations should go hand-in-hand (the boundary conditions 
are not a source of variation as the lid velocity remains constant for $t > T$ = 1 \si{s}). 
However, in Fig.\ \ref{fig: monitor} the velocity field appears to reach a steady state much faster 
than the stress field. This can be explained by observing that the stress actually also evolves 
quickly over most of the domain, the relaxation time $\lambda$ = 0.255 \si{s} (Table \ref{table: 
dimensionless numbers}) being quite small compared to the time scale of stress evolution in Fig.\ 
\ref{sfig: monitor trT}, except in the transition zone where the definition \eqref{eq: lamda and 
eta} is not representative and the actual relaxation time tends to infinity as time passes. There, 
the stress magnitude is of the order of $\tau_y$, so its local slow evolution has a noticeable 
impact on the average stress trace plotted in Fig.\ \ref{fig: monitor}. The velocity does follow 
the stress evolution, but since it is almost zero in this zone its local variations caused by the 
local stress evolution have a negligible impact on the overall kinetic energy plotted in Fig.\ 
\ref{fig: monitor}. We noticed that between $t$ = 30 \si{s} and $t$ = 210 \si{s} the change in the 
velocity components at any point in the domain is of the order of $10^{-6}$ \si{m/s}, except in the 
transition zone where it can be about five times larger.

\begin{figure}[!t]

    \centering
    \begin{subfigure}[t]{0.49\textwidth}
        \centering
        \vskip 0pt
        \includegraphics[scale=0.80]{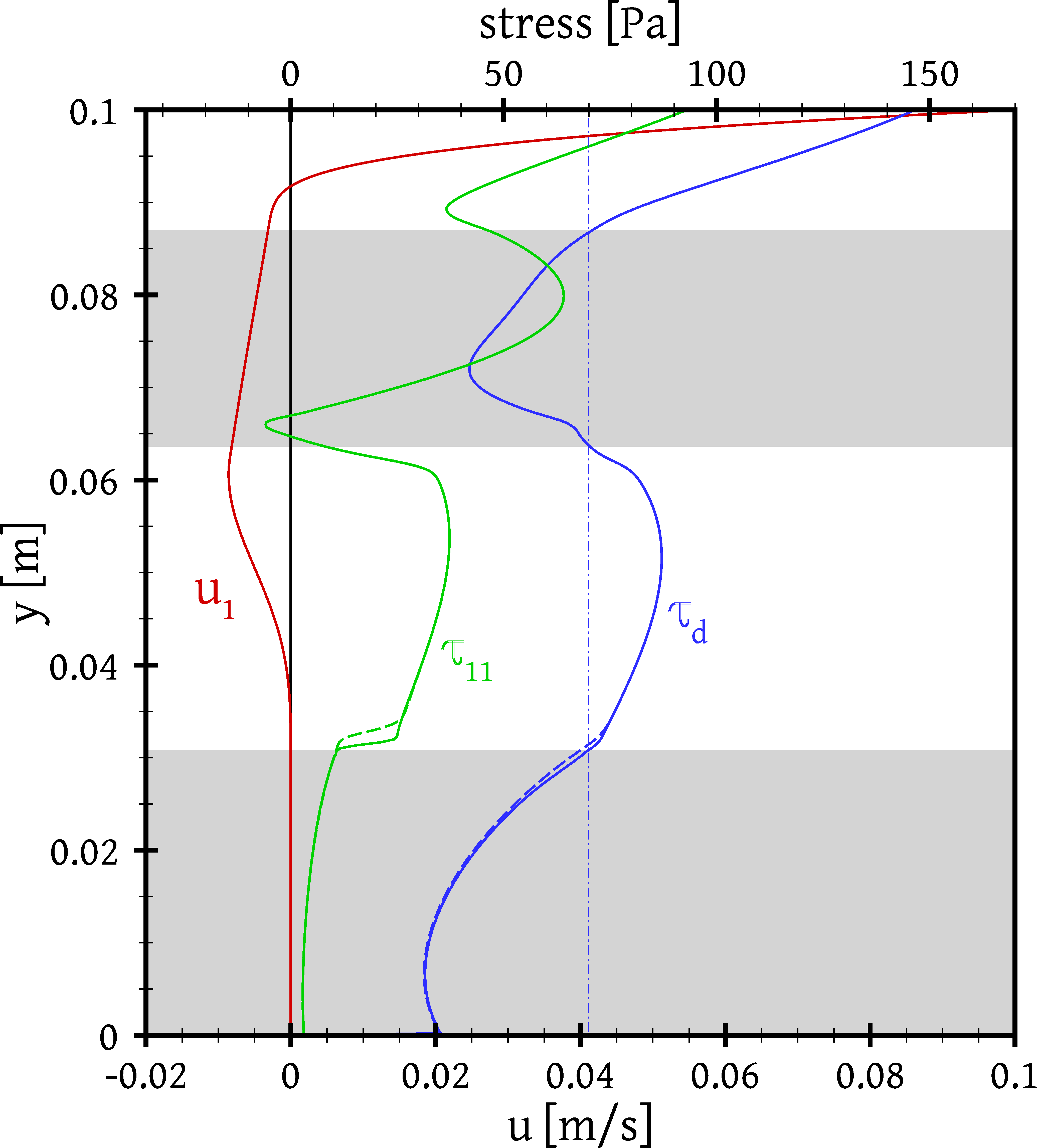}
        \caption{$u_1$, $\tau_{11}$ and $\tau_d$ at $x = L/2$.}
        \label{sfig: base case profiles C1}
    \end{subfigure}
    \begin{subfigure}[t]{0.49\textwidth}
        \centering
        \vskip 0pt
        \includegraphics[scale=0.80]{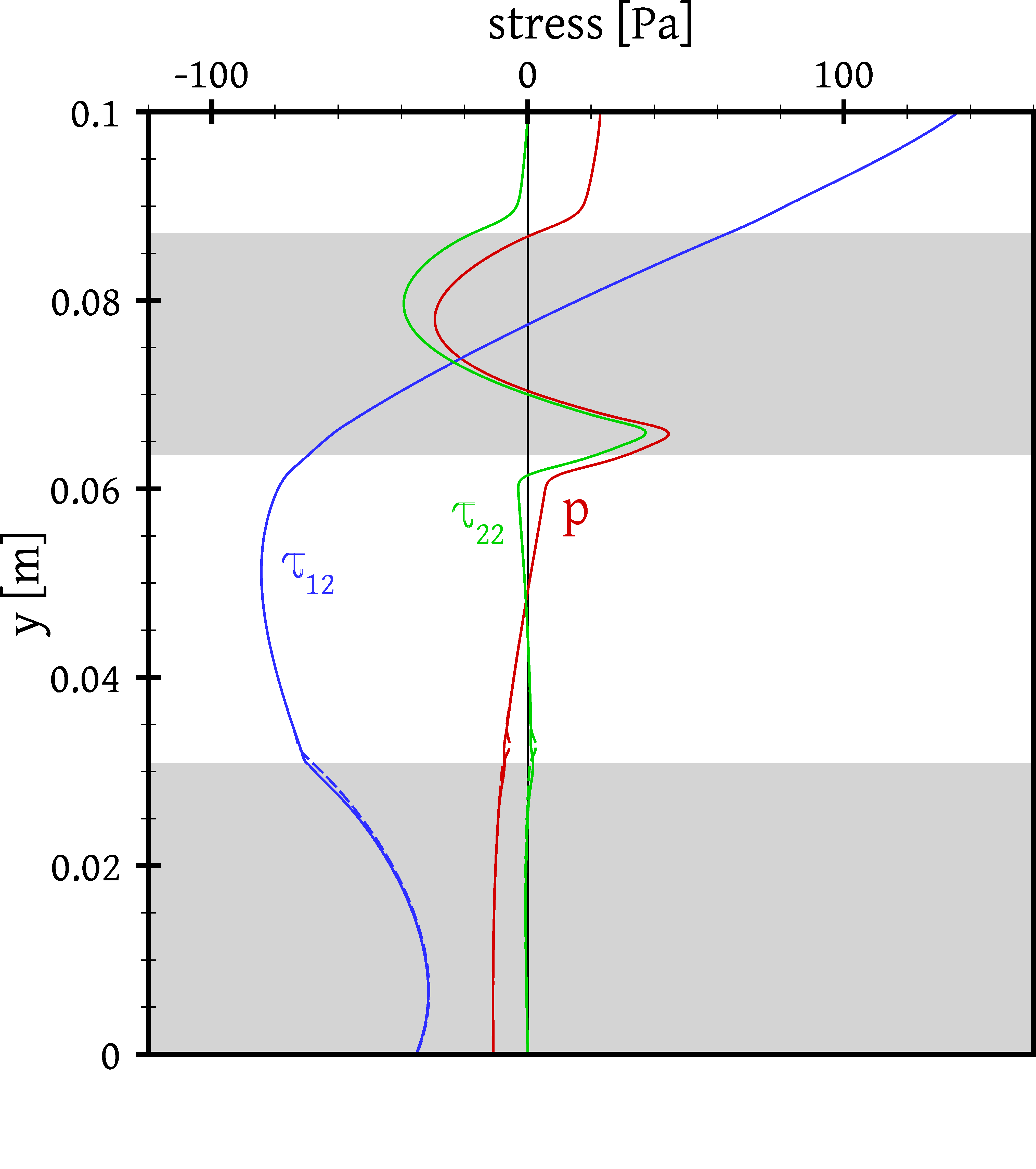}
        \caption{$p$, $\tau_{22}$ and $\tau_{12}$ at $x = L/2$.}
        \label{sfig: base case profiles C2} 
    \end{subfigure}
    
\vspace{0.5cm}

    \centering
    \begin{subfigure}[t]{0.49\textwidth}
        \centering
        \vskip 0pt
        \includegraphics[scale=0.80]{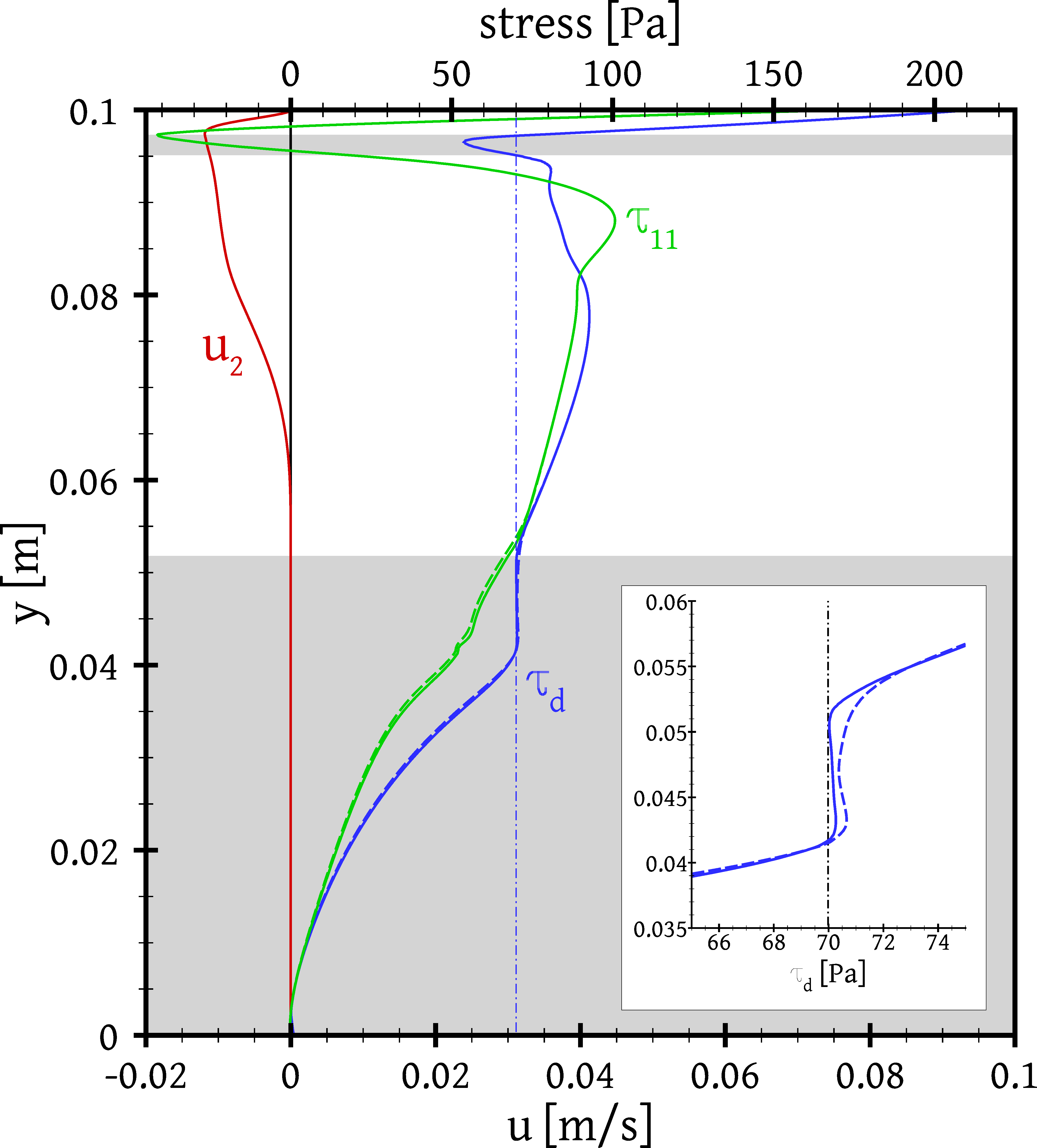}
        \caption{$u_2$, $\tau_{11}$ and $\tau_d$ at $x = 0.95L$.}
        \label{sfig: base case profiles R1}
    \end{subfigure}
    \begin{subfigure}[t]{0.49\textwidth}
        \centering
        \vskip 0pt
        \includegraphics[scale=0.80]{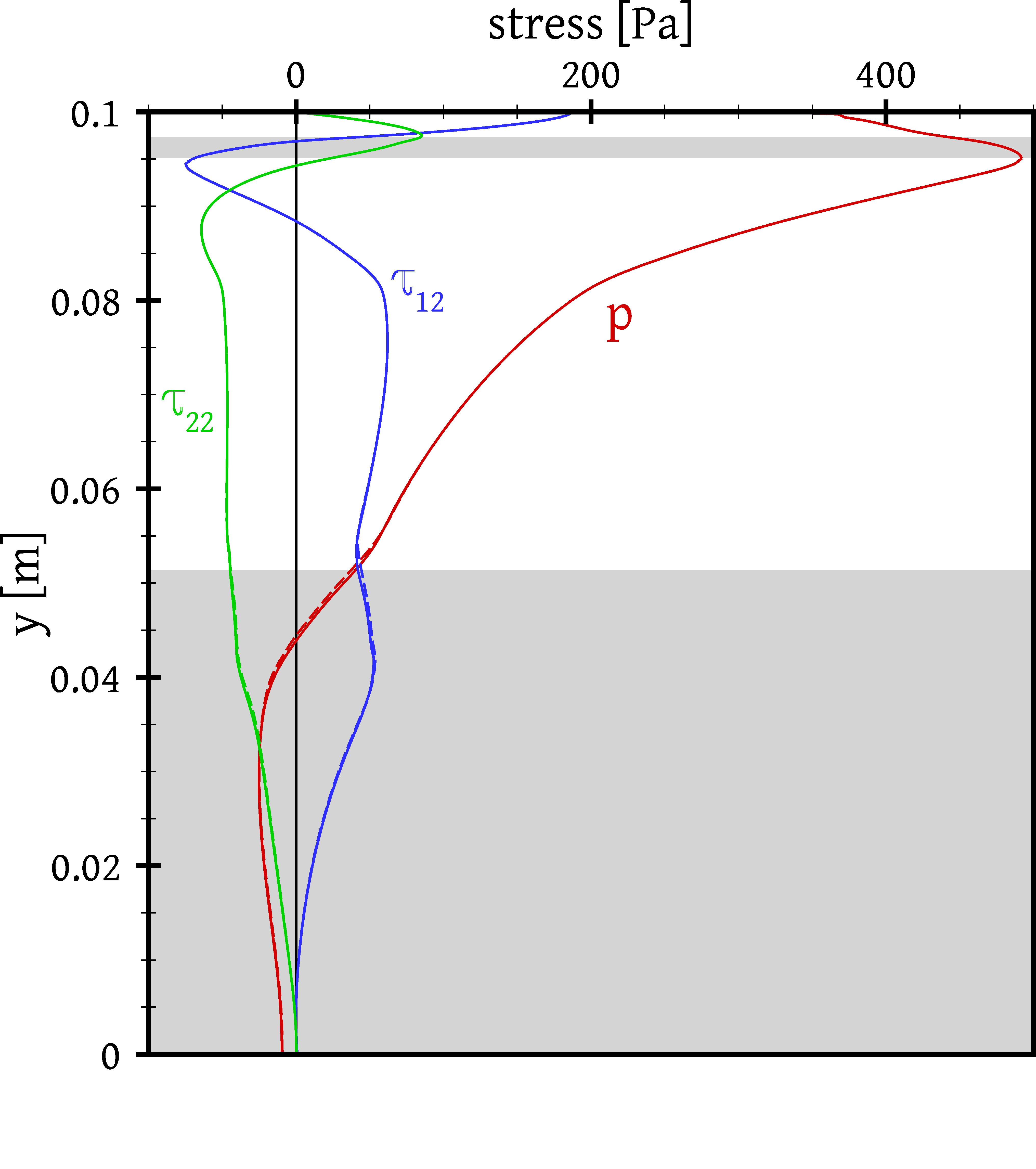}
        \caption{$p$, $\tau_{22}$ and $\tau_{12}$ at $x = 0.95L$.}
        \label{sfig: base case profiles R2} 
    \end{subfigure}
    
    \caption{Profiles of some dependent variables along vertical lines, for the base case (Sec.\ 
\ref{ssec: results: base case}): \subref{sfig: base case profiles C1}, \subref{sfig: base case 
profiles C2} at the vertical centreline ($x = 0.5L$); \subref{sfig: base case profiles R1}, 
\subref{sfig: base case profiles R2} close to the right wall ($x = 0.95L$). Dashed / continuous 
lines denote profiles at $t$ = 30 \si{s} / 210 \si{s}, respectively. The vertical dash-dot lines in 
\subref{sfig: base case profiles C1} and \subref{sfig: base case profiles R1} mark the yield stress 
value $\tau_y$. The unyielded and transition regions are shaded. The inset in \subref{sfig: base 
case profiles R1} shows a close-up view of the $\tau_d$ profile at $y \in [0.035,0.060]$ \si{m}.}
  \label{fig: base case profiles}
\end{figure}

Figure \ref{fig: base case profiles} shows vertical profiles of most flow variables along the 
centreline ($x = L/2$, Figs.\ \ref{sfig: base case profiles C1} and \ref{sfig: base case profiles 
C2}) and close to the right wall ($x = 0.95L$, Figs.\ \ref{sfig: base case profiles R1} and 
\ref{sfig: base case profiles R2}). Two profiles are drawn for each variable, one at time $t$ = 30 
\si{s} and one at $t$ = 210 \si{s}. These profiles are identical inside the yielded and plug zones, 
and only deviate very slightly inside the bottom unyielded zone. Thus the steady state has largely 
been reached already at $t$ = 30 \si{s}. The vertical line at $x = 0.95L$ cuts through a 
significant width of the transition zone, which can be recognised by the vertical line segment where 
$\tau_d \approx \tau_y$ in Fig.\ \ref{sfig: base case profiles R1} (approximately from $y$ = 0.041 
\si{m} to $y$ = 0.051 \si{m}). A close-up view of the variation of $\tau_d$ in the transition zone 
is shown in the inset of the same figure, where one can see that as time passes, $\tau_d$ decreases 
towards $\tau_y$ throughout the zone.

Interestingly, Fig.\ \ref{sfig: base case profiles C1} shows that the normal stress component 
$\tau_{11}$ tends to vary discontinuously across the boundary of the lower unyielded zone as time 
passes. The possibility of the SHB model producing solutions with discontinuous stress components 
and/or velocity gradients was noted and discussed in \cite{Cheddadi_2008, Cheddadi_2013}, where it 
was attributed to the combination of the upper convective derivative and  the viscoplastic ``max'' 
term. Nevertheless, as noted in \cite{Cheddadi_2008}, stress discontinuities must be such that the 
components of the force $\nabla \cdot \tf{\sigma} = \nabla \cdot (\tf{\tau} - p \tf{I})$ remain 
bounded (discontinuities in $\tf{\tau}$ that lead to infinite derivatives in $\nabla \cdot 
\tf{\tau}$ can exist if they are counterbalanced by opposite discontinuities in $- p \tf{I}$). 
Otherwise, at a point of discontinuity there will result a finite (i.e.\ non-zero) force acting on 
an infinitesimal mass, producing infinite acceleration and making the velocity field discontinuous, 
but this violates the continuity equation for an incompressible medium. In our case though, the 
variation of $\tau_{11}$ in the $x_2$ direction does not cause any force on the fluid (there is no 
$\partial \tau_{11} / \partial x_2$ derivative in $\nabla \cdot \tf{\tau}$) and is therefore allowed 
to be discontinuous.

\subsection{Varying the lid velocity}
\label{ssec: results: varying the lid speed}

Next we examine the flow driven by higher ($U$ = 0.4 \si{m/s}) and lower ($U$ = 0.025 \si{m/s}) lid 
velocities, which affects the dimensionless numbers as shown in Table \ref{table: dimensionless 
numbers}. Lowering the velocity increases the Bingham number and decreases the Weissenberg number, 
i.e.\ it accentuates the plastic character of the flow at the expense of its elastic character. The 
$\Rey$ values listed in Table \ref{table: dimensionless numbers} suggest that inertia effects should 
be negligible for $U$ = 0.025 and 0.100 \si{m/s}, but should be slightly noticeable for $U$ = 0.400 
\si{m/s}. The Table shows also that $\lambda$ increases as $U$ decreases; at lower velocities the 
apparent viscosity $\eta$, Eq.\ \eqref{eq: lamda and eta}, is higher and therefore the stresses 
relax more slowly. Thus, we expect the flow to evolve more slowly as $U$ is reduced.

\begin{figure}[!tb]
    \centering
    
    \begin{subfigure}[b]{0.32\textwidth}
        \centering
        \includegraphics[width=0.95\linewidth]{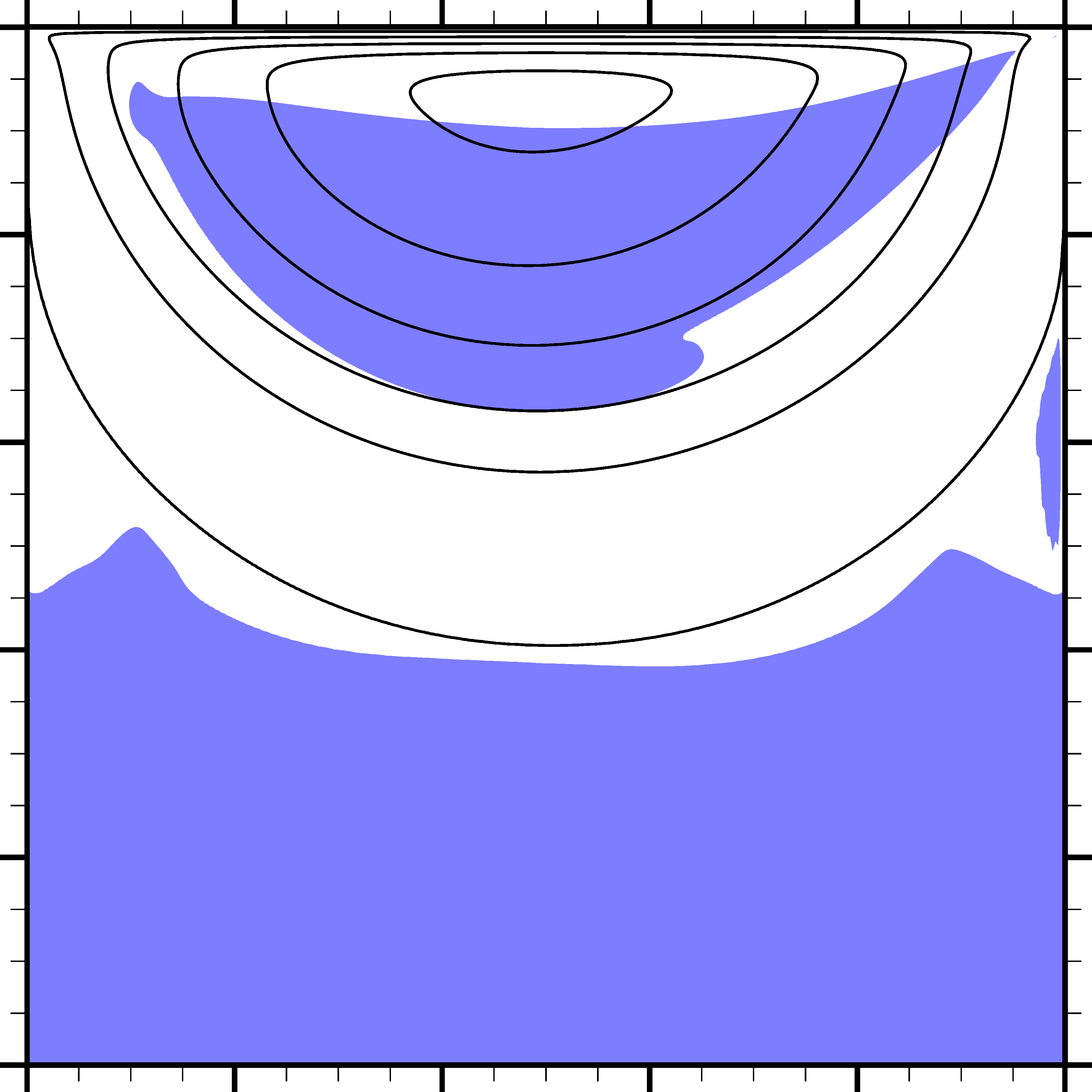}
        \caption{SHB, $U = 0.025$ \si{m/s}, $t$ = 90 \si{s}}
        \label{sfig: SHB flow U=0.025}
    \end{subfigure}
    \begin{subfigure}[b]{0.32\textwidth}
        \centering
        \includegraphics[width=0.95\linewidth]{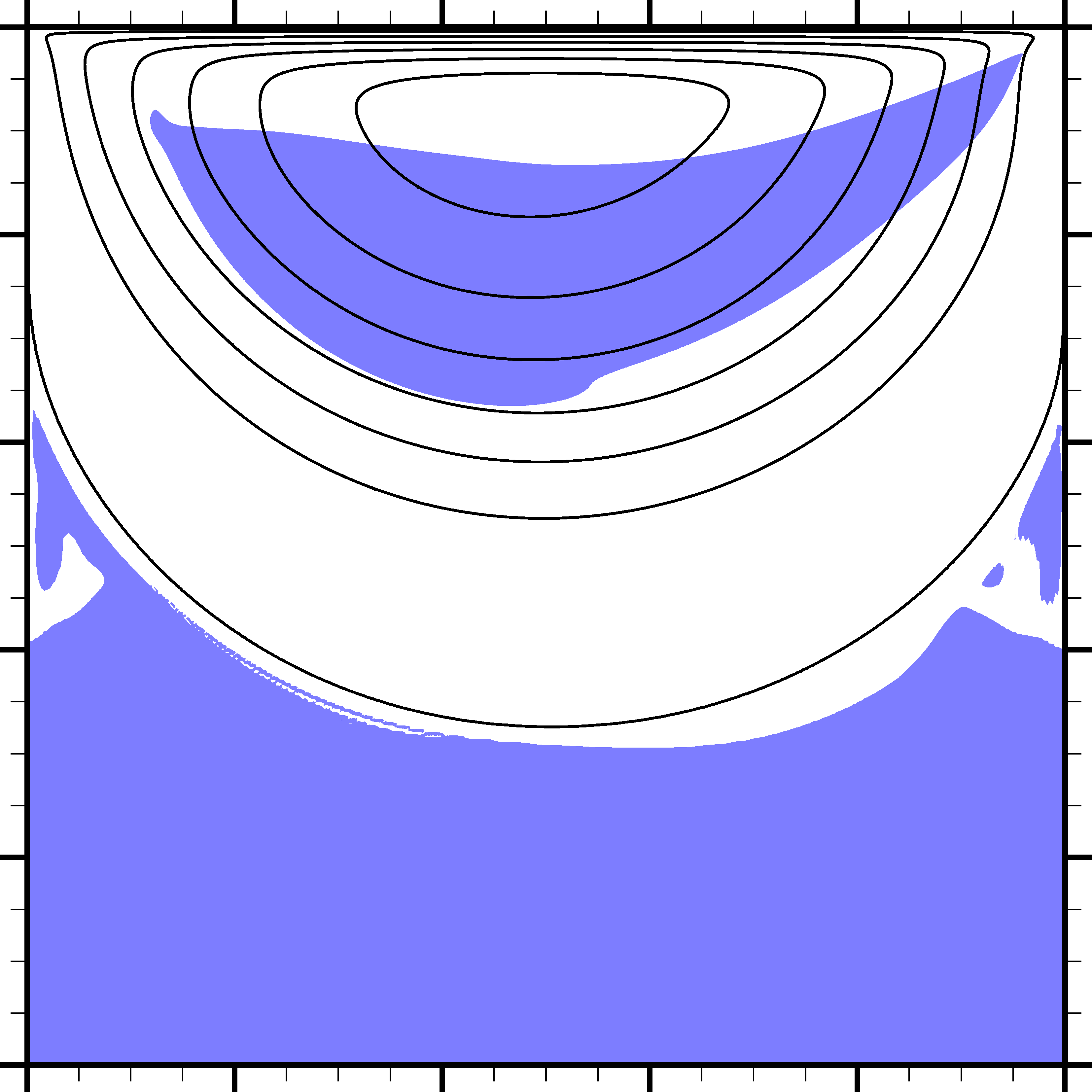}
        \caption{SHB, $U = 0.100$ \si{m/s}, $t$ = 210 \si{s}}
        \label{sfig: SHB flow U=0.100}
    \end{subfigure}
    \begin{subfigure}[b]{0.32\textwidth}
        \centering
        \includegraphics[width=0.95\linewidth]{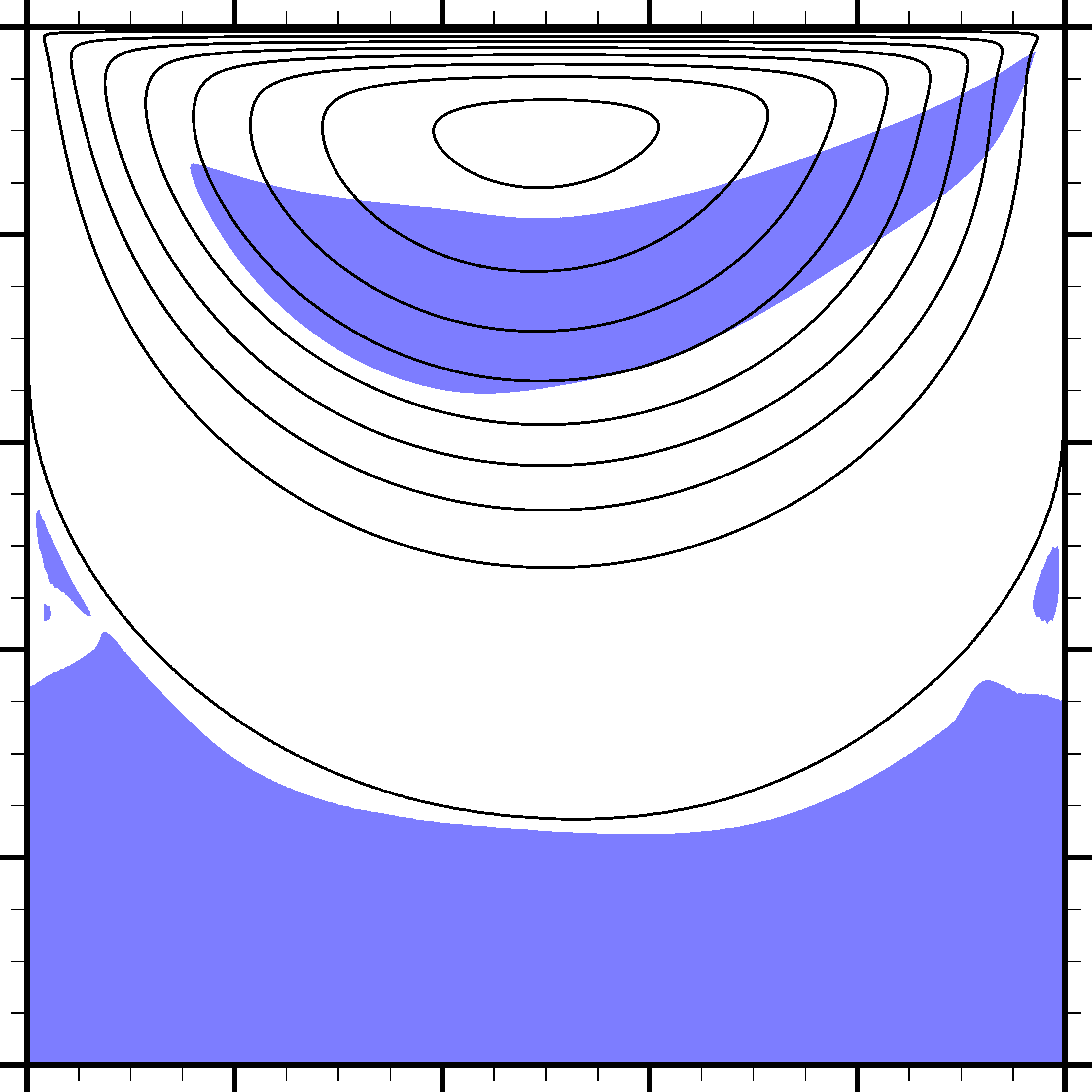}
        \caption{SHB, $U = 0.400$ \si{m/s}, $t$ = 60 \si{s}}
        \label{sfig: SHB flow U=0.400}
    \end{subfigure}

\vspace{0.25cm}    

    \begin{subfigure}[b]{0.32\textwidth}
        \centering
        \includegraphics[width=0.95\linewidth]{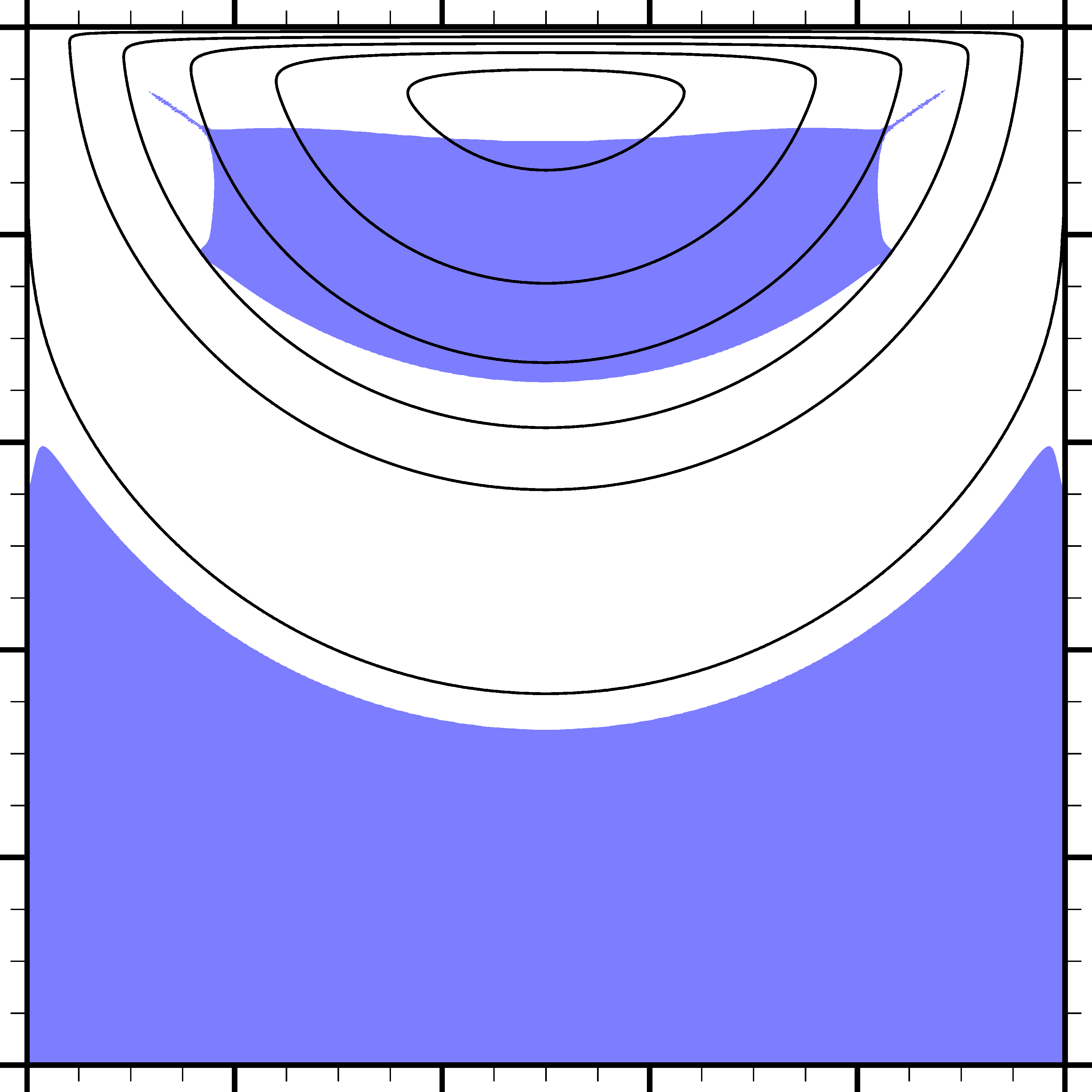}
        \caption{HB, $U = 0.025$ \si{m/s}}
        \label{sfig: HB flow U=0.025}
    \end{subfigure}
    \begin{subfigure}[b]{0.32\textwidth}
        \centering
        \includegraphics[width=0.95\linewidth]{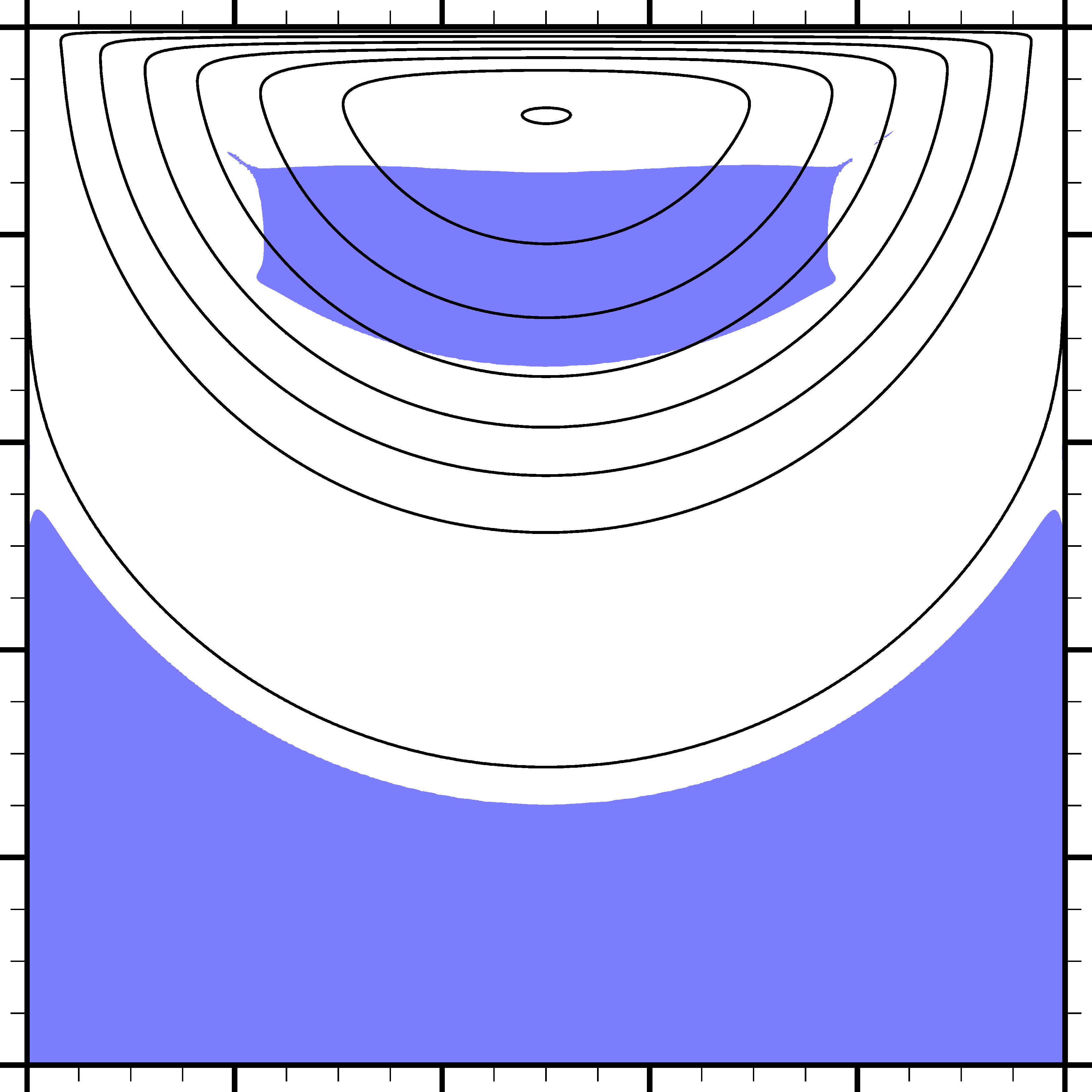}
        \caption{HB, $U = 0.100$ \si{m/s}}
        \label{sfig: HB flow U=0.100}
    \end{subfigure}
    \begin{subfigure}[b]{0.32\textwidth}
        \centering
        \includegraphics[width=0.95\linewidth]{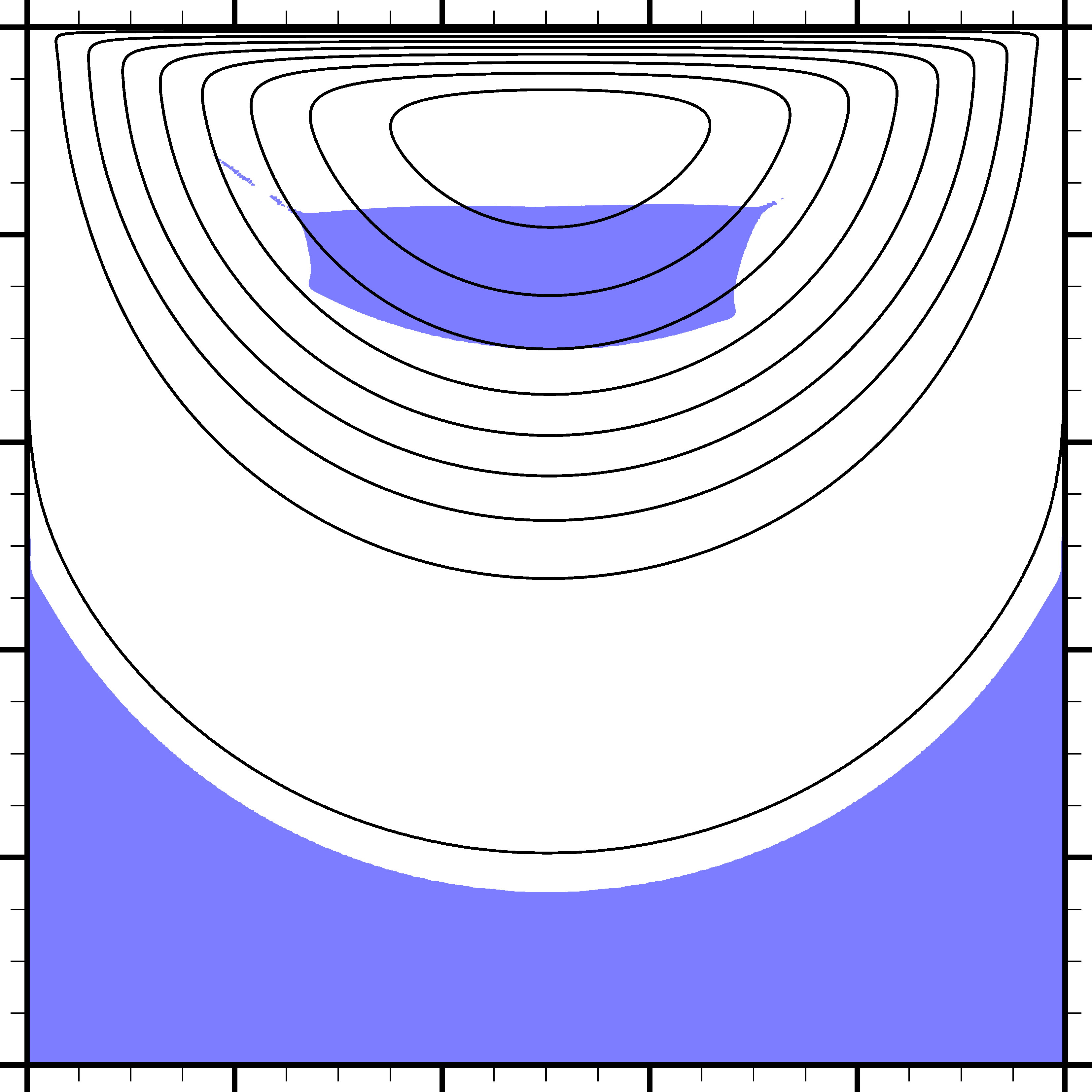}
        \caption{HB, $U = 0.400$ \si{m/s}}
        \label{sfig: HB flow U=0.400}
    \end{subfigure}

    \caption{Top row: SHB flow snapshots at the times indicated, for different lid velocities. 
Bottom row: corresponding HB steady-state flowfields. The unyielded regions are shown shaded; the 
lines are instantaneous streamlines, plotted at streamfunction values $\psi / (UL)$ = 0.04, 0.08, 
0.12 etc., plus a $\psi$ = \num{5e-6} \si{m^3/s} streamline just above the lower unyielded region.}
  \label{fig: flowfields}
\end{figure}

The lid velocity is again increased gradually according to Eq.\ \eqref{eq: lid velocity} with $T = 
L/U$. Figure \ref{sfig: monitor KE} shows that the KE of the fluid builds up in an oscillatory 
manner, usually with an overshoot, due to elastic effects. The larger $U$ is, the more prominent and 
persistent the oscillations. Figure \ref{sfig: monitor trT} confirms that the stress evolution is 
slower at lower $U$, as discussed above. The top row of Fig.\ \ref{fig: flowfields} shows the 
corresponding near-steady-state flow fields. Increasing the lid velocity leads to less unyielded 
material in the cavity, and the vortex has more free space to move away from the lid (see the $y_c$ 
coordinate in Table \ref{table: vortex metrics}). In each case there is a transition zone between 
the bottom unyielded zone and the yielded material, with the former not having yet expanded 
throughout the near-zero velocity region. The number and density of streamlines in Fig.\ \ref{fig: 
flowfields} shows that at higher $\Bin$ the flow is weaker and more confined to a thin layer below 
the lid while circulation is very weak in the rest of the domain. This is reflected also in the 
normalised KE diagram \ref{sfig: monitor KE}, and also in the normalised vortex strengths listed in 
Table \ref{table: vortex metrics}. That Table also shows that the vortex lies slightly to the left 
of the centreline, as is typical of viscoelastic lid-driven cavity flows, although this shift is not 
as pronounced as for the Oldroyd-B flows of Table \ref{table: validation}. Interestingly, as $U$ is 
increased the vortex centre moves towards the centreline despite $\Wei$ increasing; this could be 
due to inertia effects that will be discussed in Sec.\ \ref{ssec: results: comparison with HB}.

\begin{table}[!tb]
\caption{Dimensionless coordinates of the centre of the vortex $(\tilde{x}_c, \tilde{y}_c) \equiv 
(x_c/L, y_c/L)$ and value of the streamfunction there $\tilde{\psi}_c \equiv \psi_c / (LU)$, for 
various cases, at steady-state.}
\label{table: vortex metrics}
\begin{center}
\begin{small}   
\renewcommand\arraystretch{1.25}   
\begin{tabular}{ l | c c c | c c c }
\toprule
& \multicolumn{3}{c |}{ SHB } & \multicolumn{3}{c}{ HB }
\\
\midrule
                                &  $\tilde{x}_c$  &  $\tilde{y}_c$  &  $\tilde{\psi}_c$
                                &  $\tilde{x}_c$  &  $\tilde{y}_c$  &  $\tilde{\psi}_c$  \\
\midrule
 $U$ = 0.025 \si{m/s}           & $0.495$ & $0.935$ & $-0.0211$ & $0.500$ & $0.933$ & $-0.0214$ \\
 $U$ = 0.100 \si{m/s}           & $0.495$ & $0.917$ & $-0.0270$ & $0.500$ & $0.915$ & $-0.0281$ \\
 $U$ = 0.400 \si{m/s}           & $0.500$ & $0.899$ & $-0.0333$ & $0.505$ & $0.897$ & $-0.0352$ \\
 $U$ = 0.100 \si{m/s}, slip     & $0.482$ & $0.846$ & $-0.0125$ & $0.500$ & $0.853$ & $-0.0112$
\\
\bottomrule
\end{tabular}
\end{small}
\end{center}
\end{table}

\begin{figure}[tb]
    \centering
    \begin{subfigure}[b]{0.32\textwidth}
        \centering
        \includegraphics[width=0.95\linewidth]{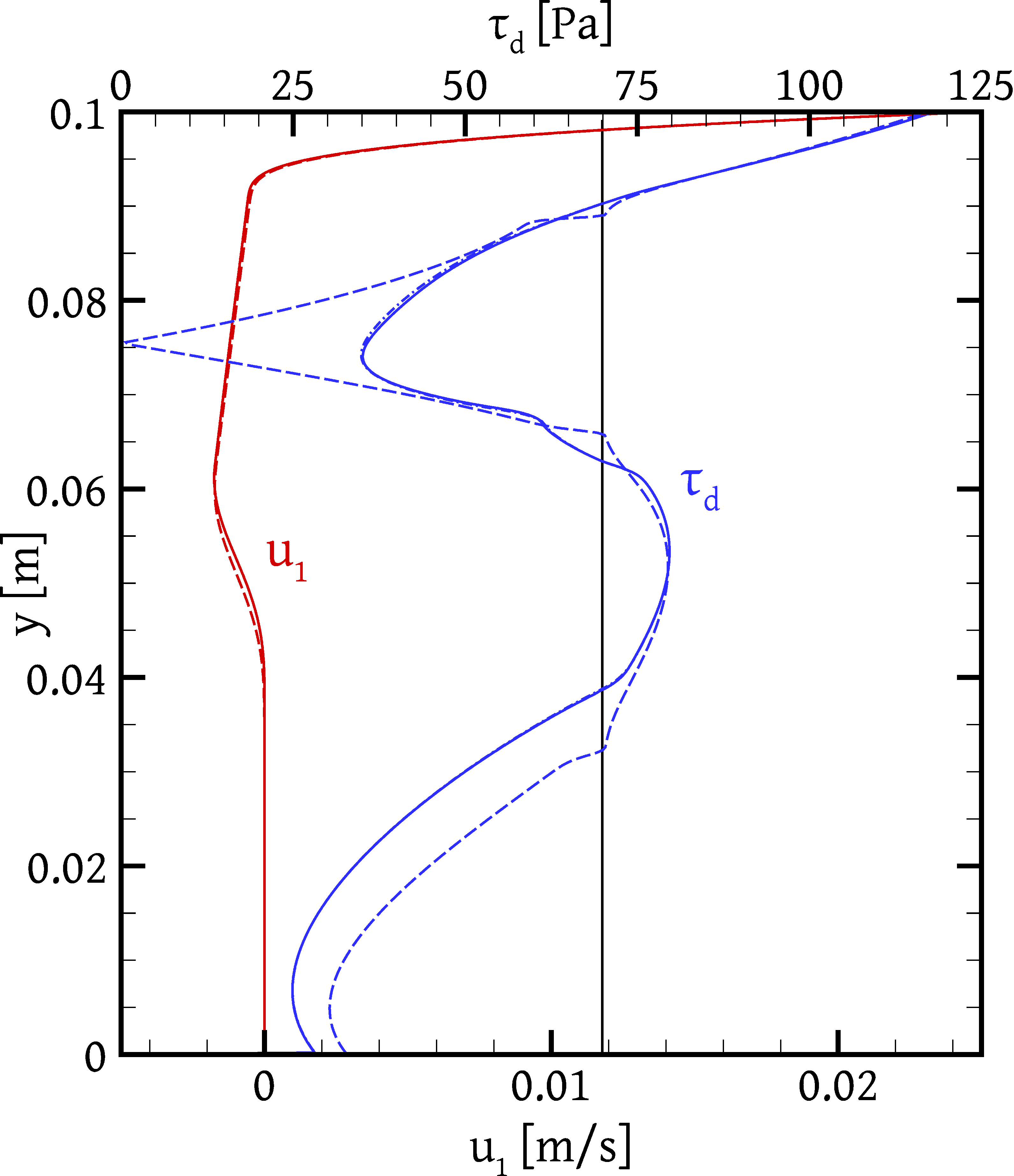}
        \caption{$U = 0.025$ \si{m/s}}
        \label{sfig: varU profiles U=0.025}
    \end{subfigure}
    \begin{subfigure}[b]{0.32\textwidth}
        \centering
        \includegraphics[width=0.95\linewidth]{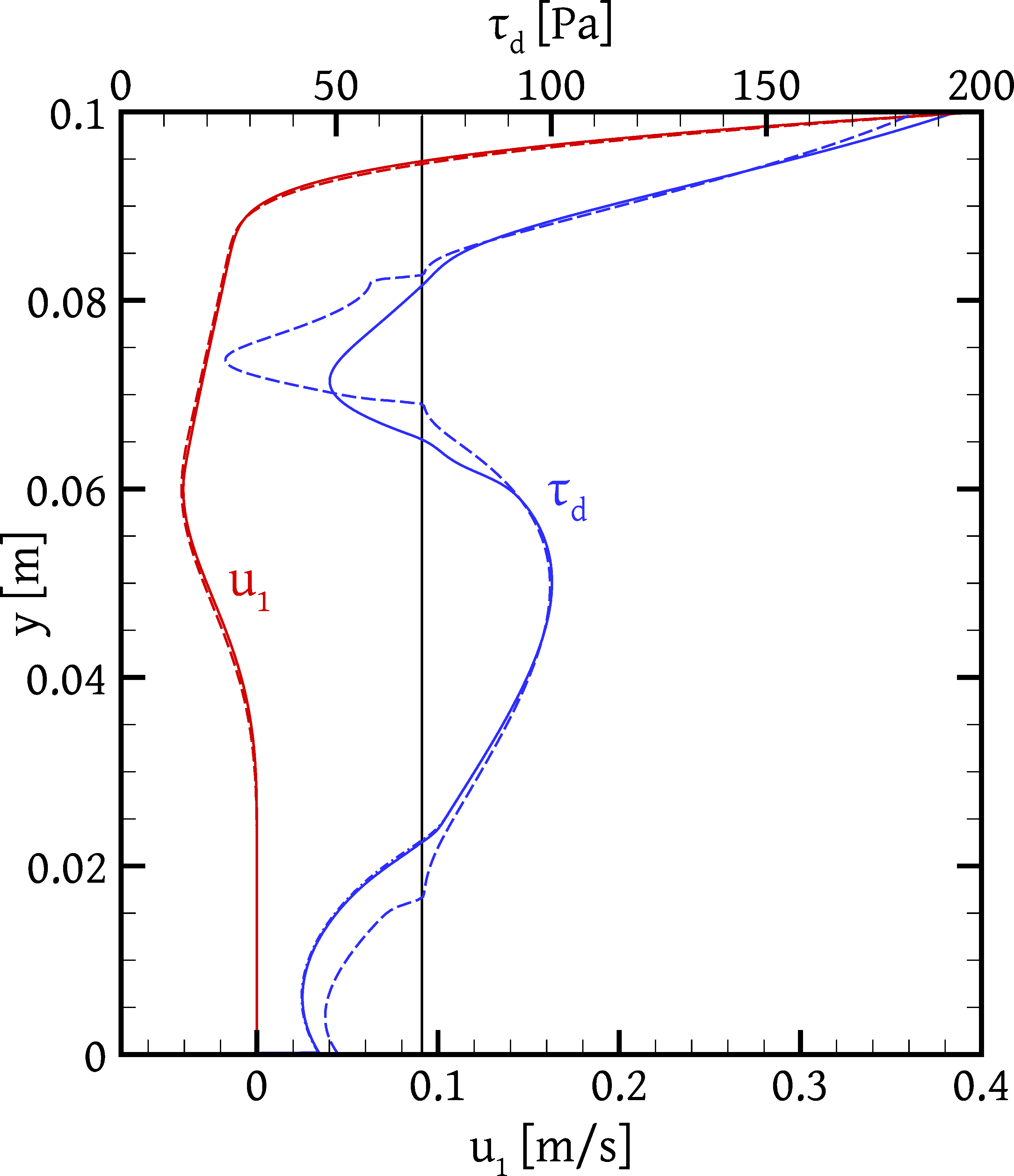}
        \caption{$U = 0.400$ \si{m/s}}
        \label{sfig: varU profiles U=0.400}
    \end{subfigure}
    \begin{subfigure}[b]{0.32\textwidth}
        \centering
        \includegraphics[width=0.95\linewidth]{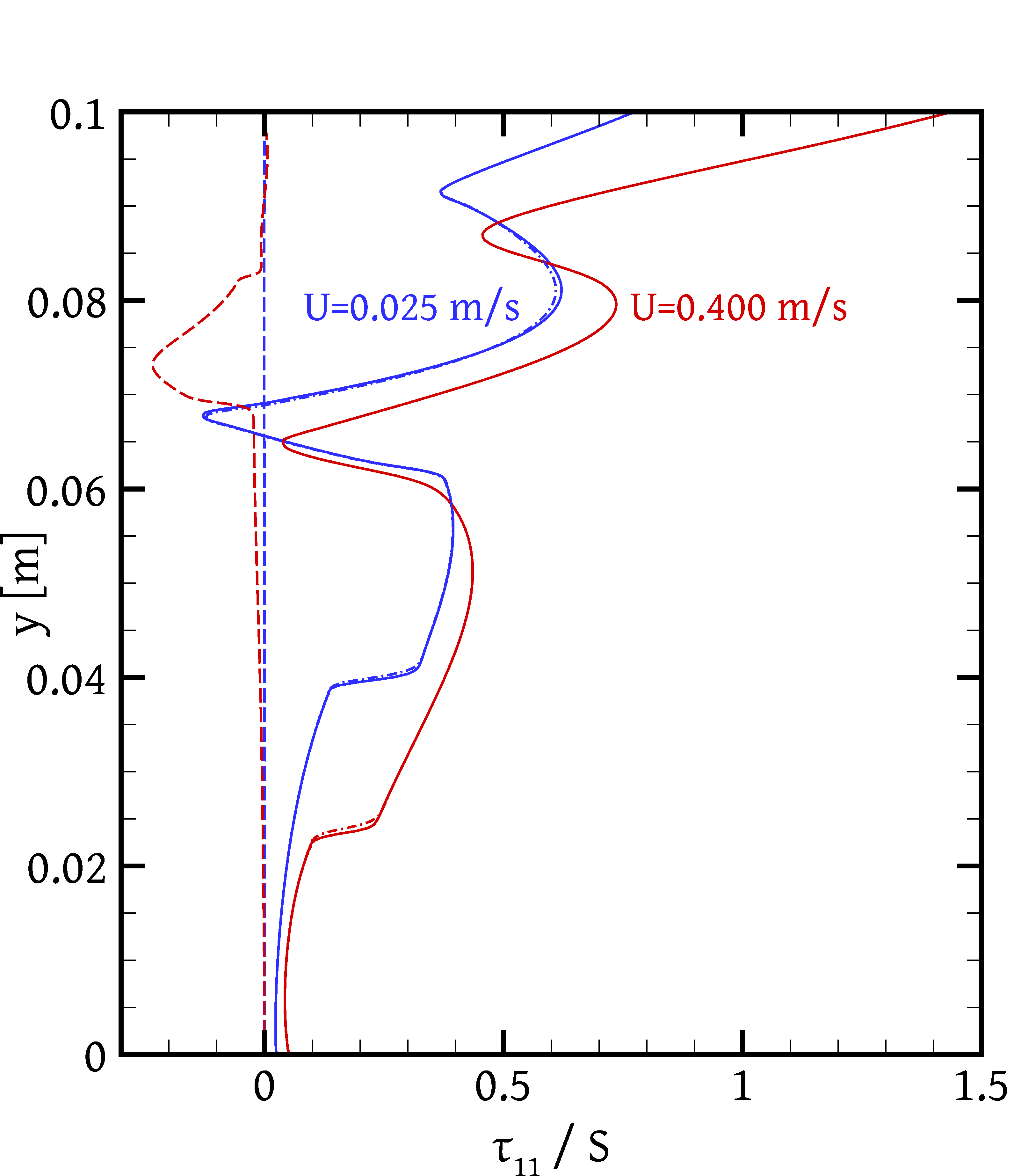}
        \caption{$\tau_{11} / S$}
        \label{sfig: varU Sxx profiles}
    \end{subfigure}
    \caption{Profiles of $u_1$ and $\tau_d$ for $U$ = 0.025 \si{m/s} \subref{sfig: varU profiles 
U=0.025} and $U$ = 0.400 \si{m/s} \subref{sfig: varU profiles U=0.400}, and profiles of $\tau_{11} 
/ S$ for both $U$'s \subref{sfig: varU Sxx profiles}. All profiles are along the vertical 
centreline ($x = L/2$). Continuous lines: SHB profiles at $t$ = 90 \si{s} ($U$ = 0.025 \si{m/s}) and 
$t$ = 60 \si{s} ($U$ = 0.400 \si{m/s}); dash-dot lines: SHB profiles at 30 \si{s} earlier; dashed 
lines: HB steady-state profiles. In \subref{sfig: varU profiles U=0.025} and \subref{sfig: varU 
profiles U=0.400} the vertical lines mark the yield stress.}
  \label{fig: varU profiles}
\end{figure}

\subsection{Comparison with the classic HB model}
\label{ssec: results: comparison with HB}

Since the HB equation is used extensively to model viscoplastic flows, we compare its predictions 
against those of the SHB equation in order to get a feel of the error involved in neglecting elastic 
effects. The HB simulations were carried out by performing Papanastasiou regularisation 
\cite{Papanastasiou_1987}, implemented as \cite{Sverdrup_2018}:
\begin{equation} \label{eq: HB Papanastasiou}
 \tf{\tau} \;=\; \eta \dot{\tf{\gamma}}
 \;,\qquad
 \eta \;=\; \frac{\tau}{\dot{\gamma}}
 \;\approx\;
 \frac{(\tau_y + k \dot{\gamma}^n)(1-e^{-\dot{\gamma}/\epsilon})}{\dot{\gamma}}
\end{equation}
where the regularisation parameter $\epsilon$ determines the accuracy of the approximation. This 
method was used for simulating lid-driven cavity Bingham flows in \cite{Syrakos_2013, Syrakos_2014}, 
where it was noticed that the equations became too stiff to solve for $\epsilon < 1/400$. However, 
with the present method, and using a continuation procedure where $\epsilon$ is progressively 
halved, solutions for $\epsilon$ = 1/128,000 were obtained. We performed steady-state simulations 
(since the steady-state of classic viscoplastic flows does not depend on the initial conditions) 
where the HB parameters $\tau_y$, $k$ and $n$ have the same values as for the SHB fluid (Table 
\ref{table: carbopol parameters}).

The flowfields, depicted in the second row of Fig.\ \ref{fig: flowfields}, are much more symmetric 
than their SHB counterparts. In HB lid-driven cavity flow, the only source of asymmetry is inertia. 
Inertial effects are imperceptible for $U$ = 0.025 and 0.100 \si{m/s} (Figs.\ \ref{sfig: HB flow 
U=0.025} and \ref{sfig: HB flow U=0.100}) and only slightly noticeable for $U$ = 0.400 \si{m/s} 
(Fig.\ \ref{sfig: HB flow U=0.025}), which is in accord with the corresponding $\Rey$ values listed 
in Table \ref{table: dimensionless numbers}. So, for $U$ = 0.4 \si{m/s} the vortex centre is 
shifted slightly towards the right (Table \ref{table: vortex metrics}), whereas at the lower $U$'s 
it is almost exactly on the centreline. Also, in Fig.\ \ref{sfig: HB flow U=0.400} one can see that 
the upper unyielded (plug) region is somewhat stretched towards the left. These features are 
opposite to those of the SHB case, where elasticity causes the vortex centre to move towards the 
left and the plug region to stretch towards the right. The opposite effects of inertia and 
elasticity have been noted also in \cite{Martins_2013, Hashemi_2017, Syrakos_2018}.

Figure \ref{fig: flowfields} and Table \ref{table: vortex metrics} show that the union of the lower 
unyielded and transition zones in SHB flow is slightly larger than the corresponding HB stationary 
regions, pushing the SHB vortex upwards and lowering its strength compared to the HB case. Figures 
\ref{sfig: varU profiles U=0.025} and \ref{sfig: varU profiles U=0.400} show that the SHB and HB 
velocity profiles along the vertical centreline are very similar; the $\tau_d$ profiles are somewhat 
more dissimilar but still not far apart, especially in the yielded regions. In the unyielded regions 
they cannot be expected to be similar due to the stress indeterminacy of the HB model (the currently 
predicted HB stress field in the unyielded regions is one of infinite possibilities, the one 
conforming to the Papanastasiou regularisation). Profiles of $\tau_{11}$ are plotted in Fig.\ 
\ref{sfig: varU Sxx profiles}; for HB flow, due to the symmetric flowfield and $\tau_{11}$ being 
proportional to $\partial u_1 / \partial x_1$, this stress component is nearly zero, except inside 
the plug region for $U$ = 0.400 \si{m/s}, which is somewhat asymmetric (Fig.\ \ref{sfig: HB flow 
U=0.400}). The SHB stresses, on the other hand, are significant due to elasticity, especially in 
the higher $U$ case which corresponds to higher $\Wei$. Finally, we note that in Fig.\ \ref{fig: 
varU profiles} two SHB profiles are plotted for each lid velocity: at $t$ = 60 and 90 \si{s} for $U$ 
= 0.025 \si{m/s}, and at $t$ = 30 and 60 \si{s} for $U$ = 0.400 \si{m/s}. These profiles are hardly 
distinguishable, indicating that the steady state for $U$ = 0.025 \si{m/s} has been practically 
reached at $t$ = 60 \si{s}, and for $U$ = 0.400 \si{m/s} at $t$ = 30 \si{s}.

\begin{figure}[tb]
    \centering
    \begin{subfigure}[b]{0.32\textwidth}
        \centering
        \includegraphics[width=0.95\linewidth]{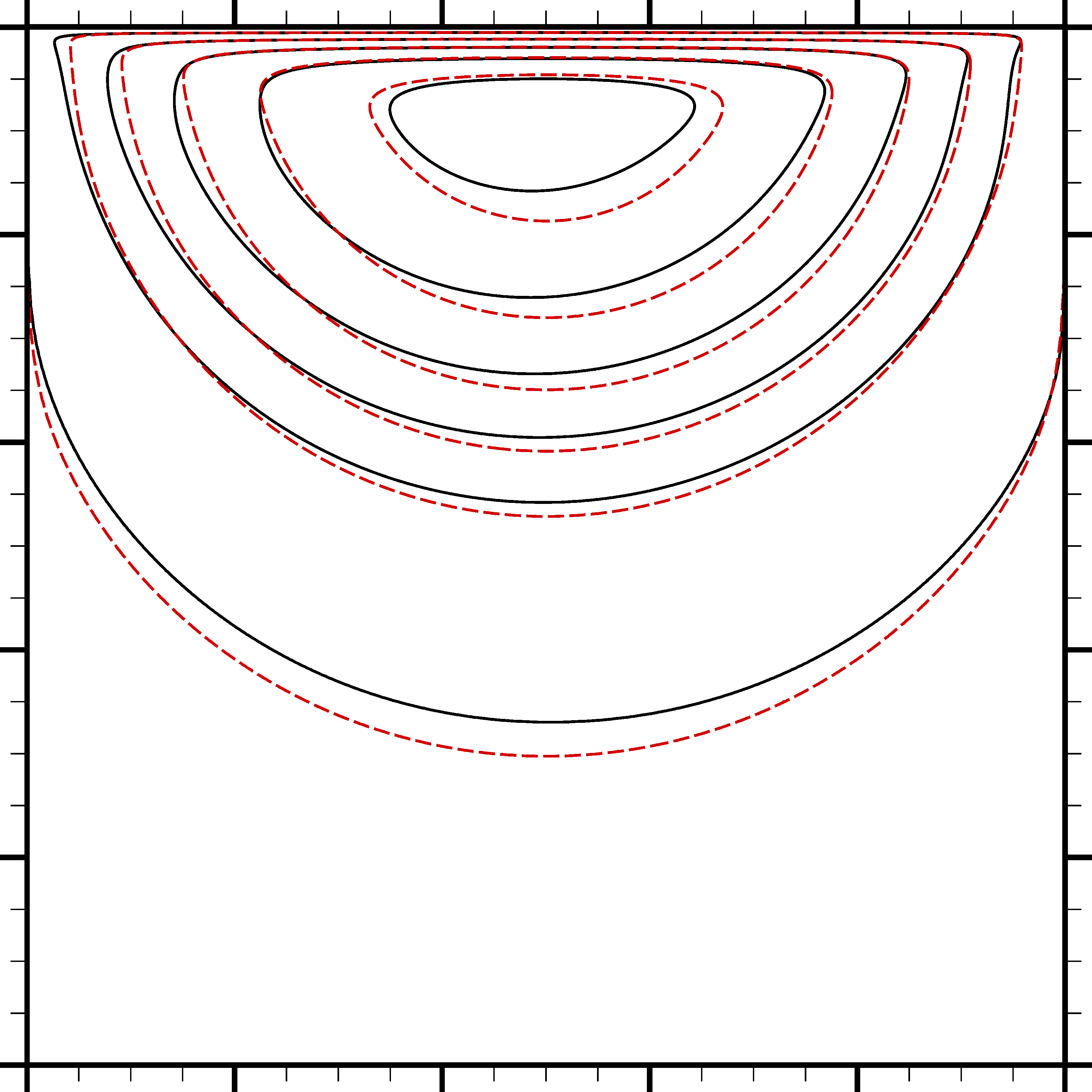}
        \caption{$G = 400$ \si{Pa}}
        \label{sfig: HB comparison: streamlines base}
    \end{subfigure}
    \begin{subfigure}[b]{0.32\textwidth}
        \centering
        \includegraphics[width=0.95\linewidth]{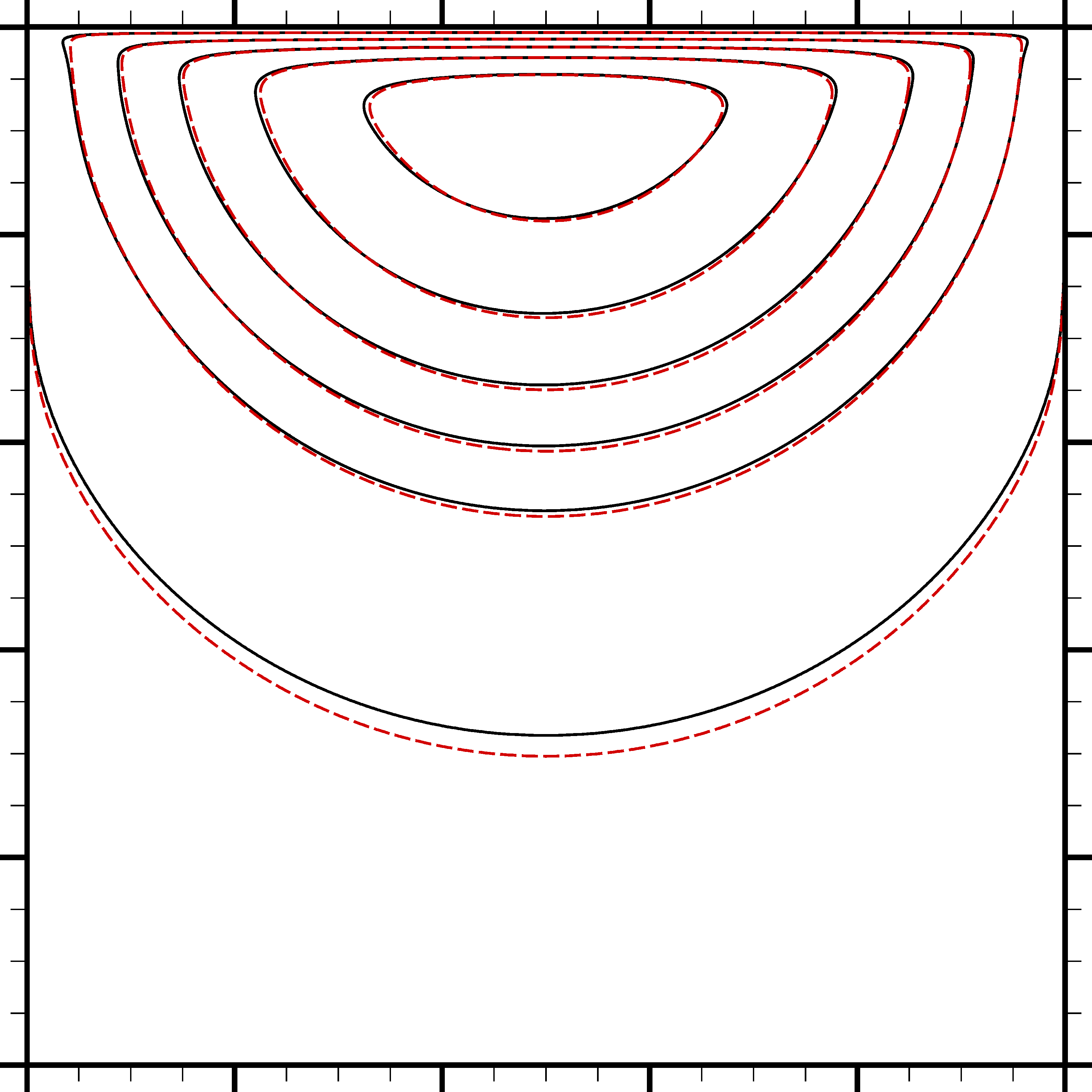}
        \caption{$G = 4,000$ \si{Pa}}
        \label{sfig: HB comparison: streamlines 10G}
    \end{subfigure}
    \begin{subfigure}[b]{0.32\textwidth}
        \centering
        \includegraphics[width=0.95\linewidth]{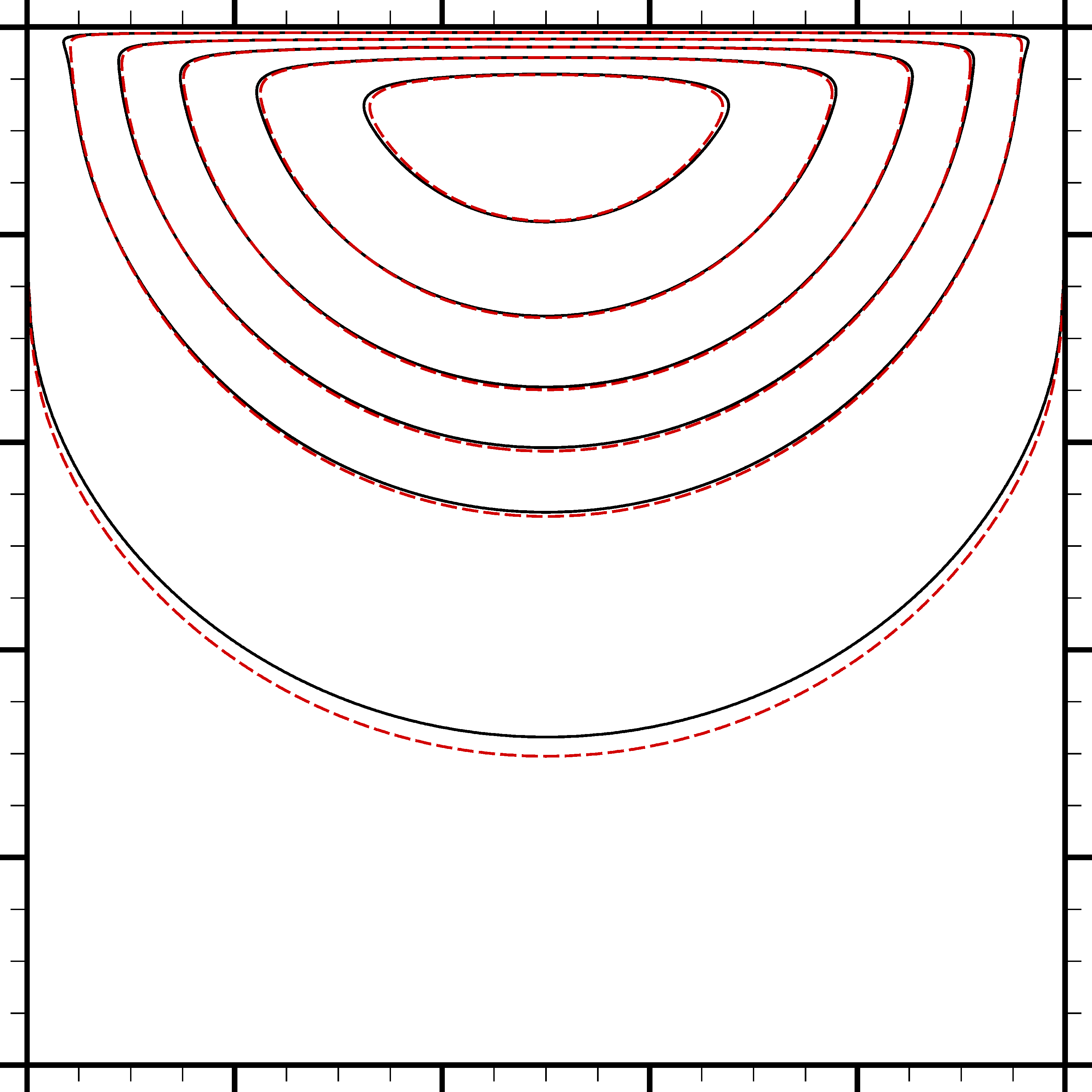}
        \caption{$G = 40,000$ \si{Pa}}
        \label{sfig: HB comparison: streamlines 100G}
    \end{subfigure}
    \caption{Continuous lines: SHB streamlines for various values of $G$, as indicated (the rest of 
the parameters are as in Table \ref{table: carbopol parameters}, and the lid velocity is $U$ = 0.1 
\si{m/s}). Dashed lines: corresponding HB streamlines, drawn at the same streamfunction values as 
for the SHB model. Since the HB predictions are independent of $G$, the dashed lines do not change
between plots. The streamlines are drawn at fixed streamfunction intervals.}
  \label{fig: HB comparison: streamlines}
\end{figure}

\begin{figure}[!t]
    \centering
    
    \begin{subfigure}[b]{0.32\textwidth}
        \centering
        \includegraphics[width=0.95\linewidth]{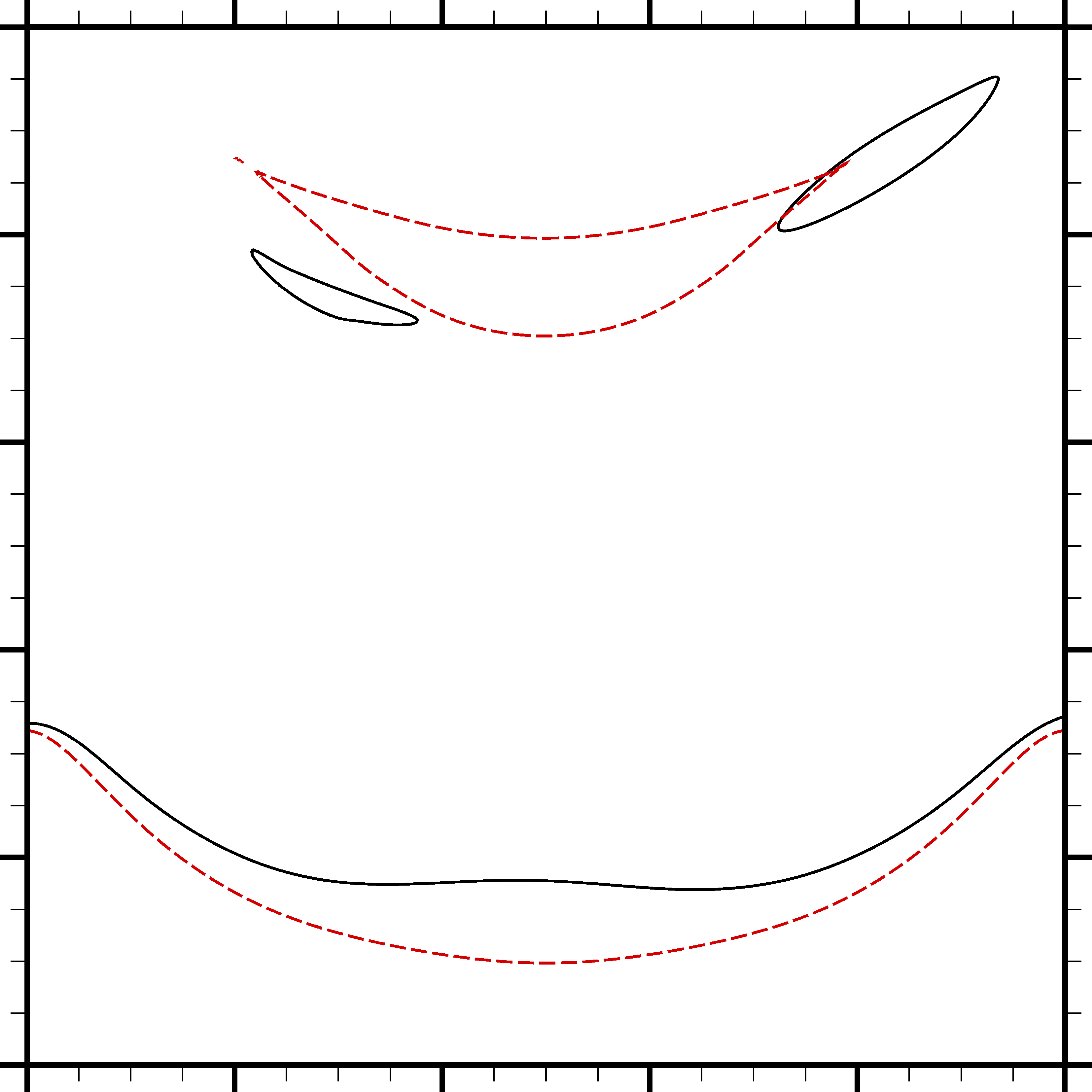}
        \caption{$\tau_d = 40$ \si{Pa}; $G = 400$ \si{Pa}}
        \label{sfig: HB comparison: tau 40 base}
    \end{subfigure}
    \begin{subfigure}[b]{0.32\textwidth}
        \centering
        \includegraphics[width=0.95\linewidth]{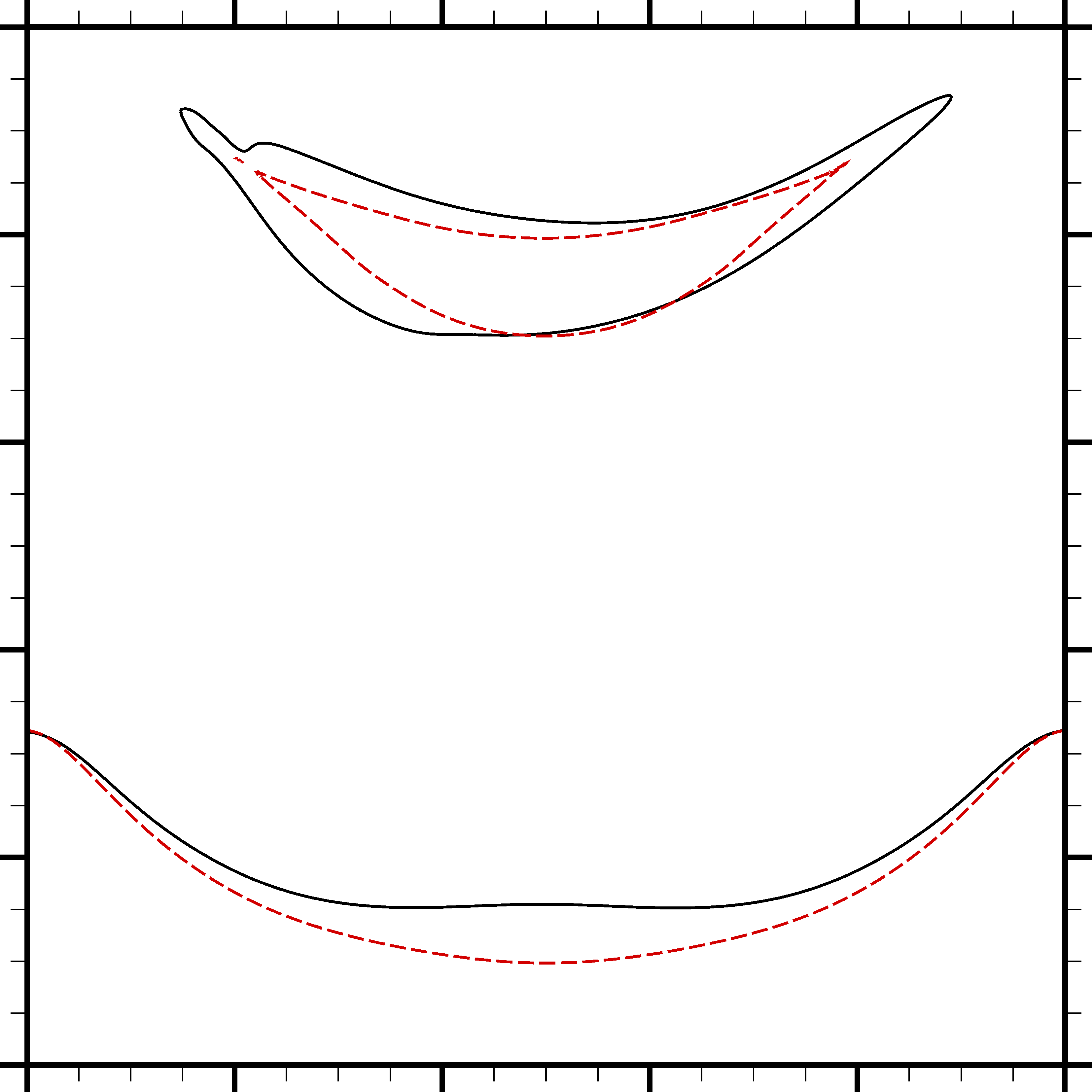}
        \caption{$\tau_d = 40$ \si{Pa}; $G = 4,000$ \si{Pa}}
        \label{sfig: HB comparison: tau 40 10G}
    \end{subfigure}
    \begin{subfigure}[b]{0.32\textwidth}
        \centering
        \includegraphics[width=0.95\linewidth]{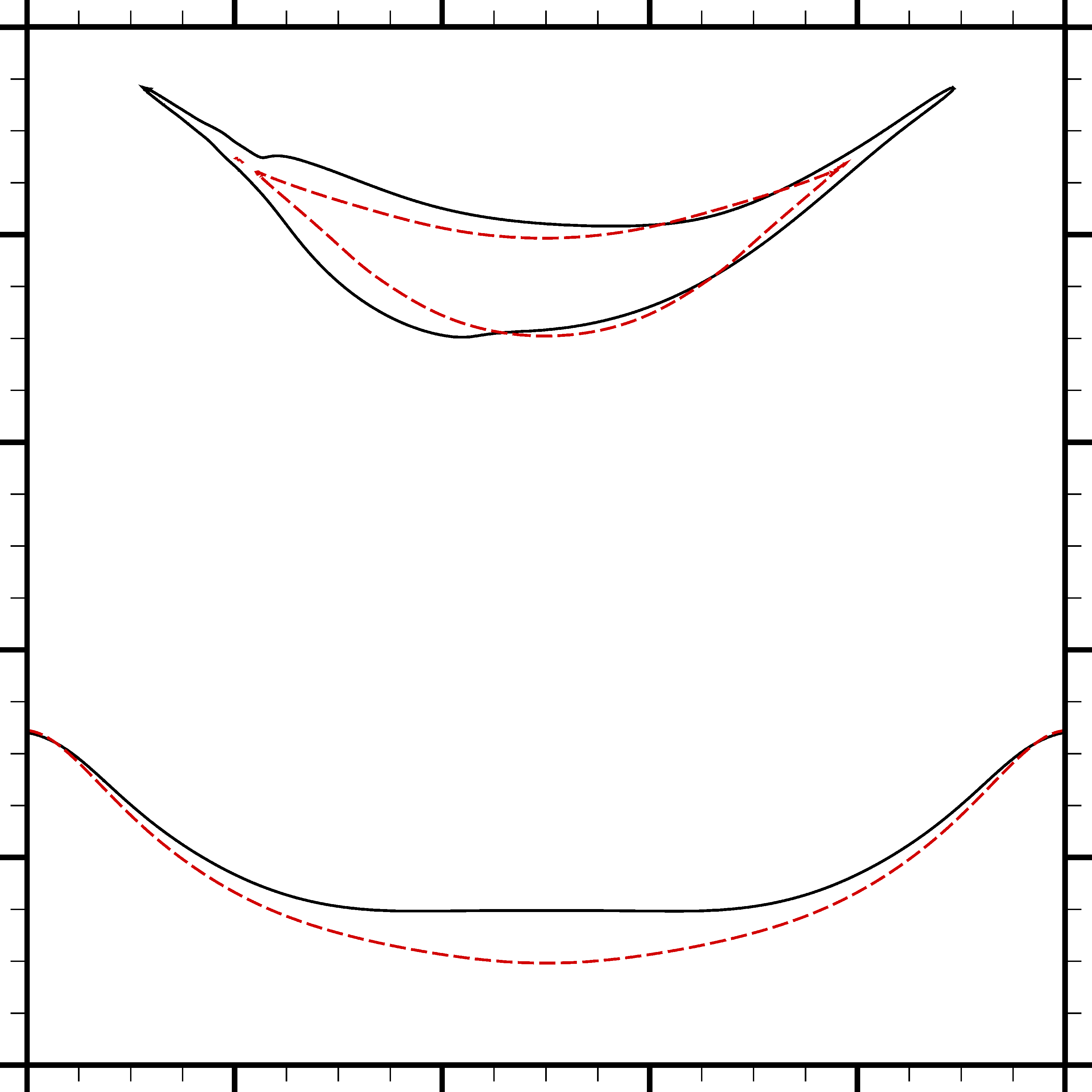}
        \caption{$\tau_d = 40$ \si{Pa}; $G = 40,000$ \si{Pa}}
        \label{sfig: HB comparison: tau 40 100G}
    \end{subfigure}
    
    \vspace{0.25cm}
    
    \begin{subfigure}[b]{0.32\textwidth}
        \centering
        \includegraphics[width=0.95\linewidth]{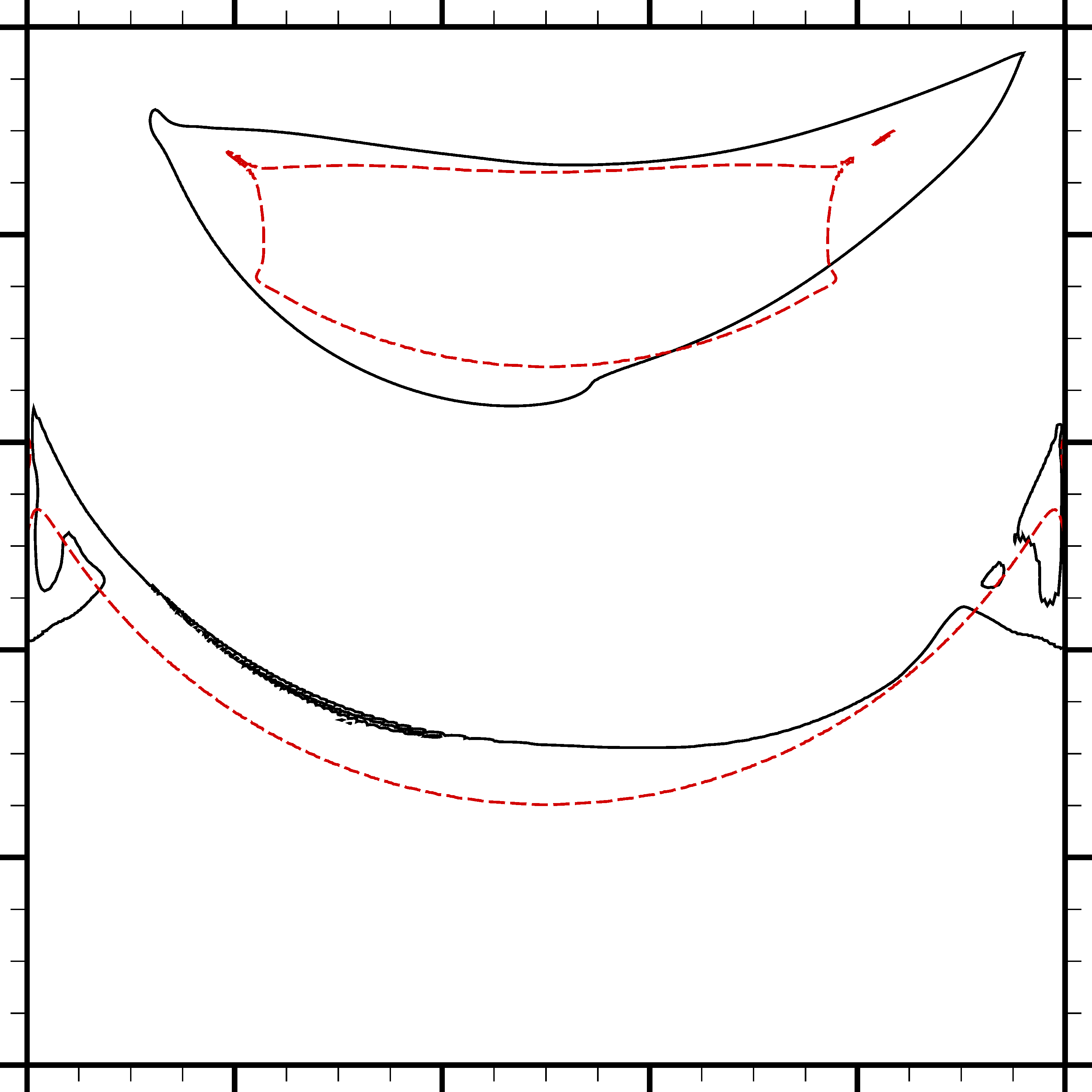}
        \caption{$\tau_d = 70$ \si{Pa}; $G = 400$ \si{Pa}}
        \label{sfig: HB comparison: tau 70 base}
    \end{subfigure}
    \begin{subfigure}[b]{0.32\textwidth}
        \centering
        \includegraphics[width=0.95\linewidth]{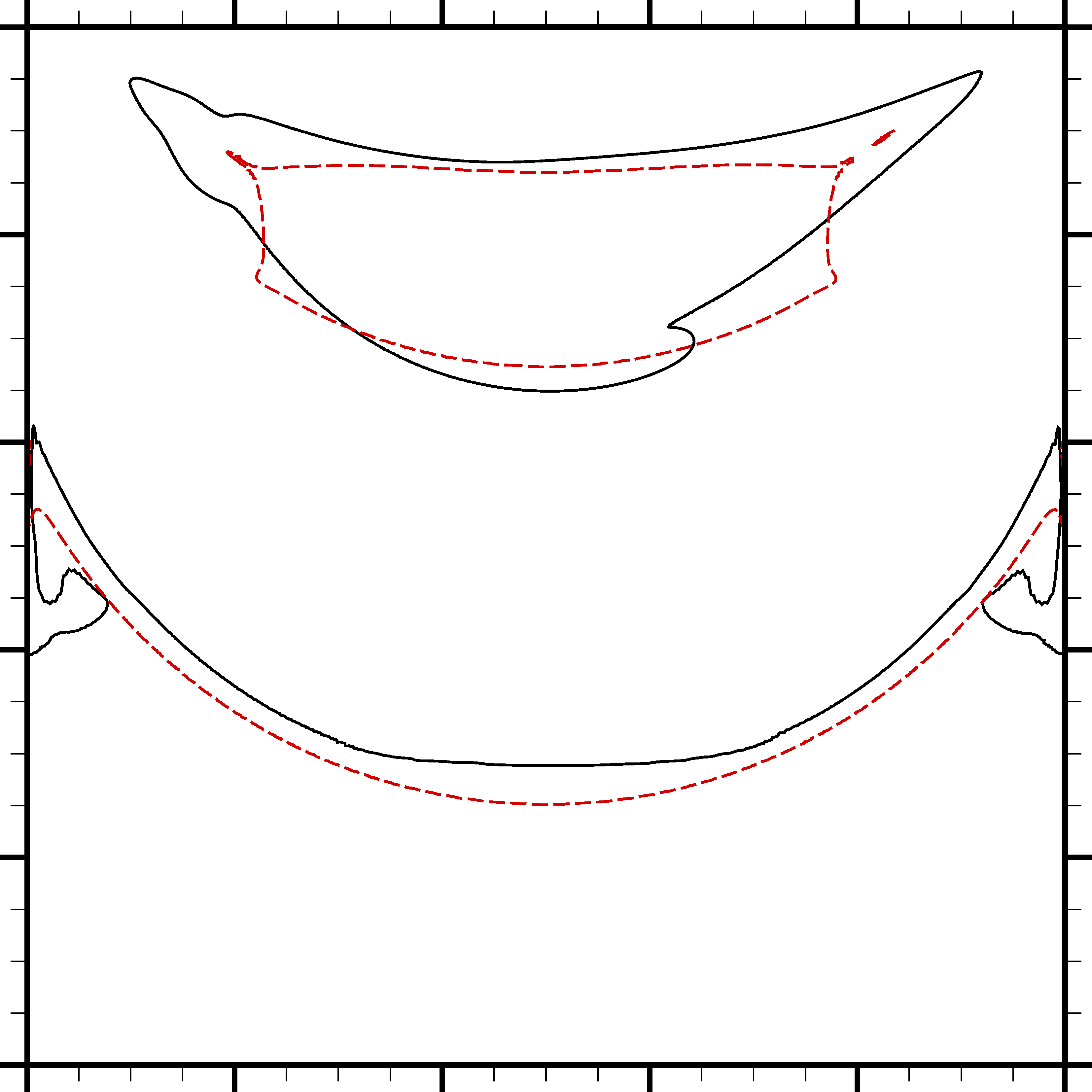}
        \caption{$\tau_d = 70$ \si{Pa}; $G = 4,000$ \si{Pa}}
        \label{sfig: HB comparison: tau 70 10G}
    \end{subfigure}
    \begin{subfigure}[b]{0.32\textwidth}
        \centering
        \includegraphics[width=0.95\linewidth]{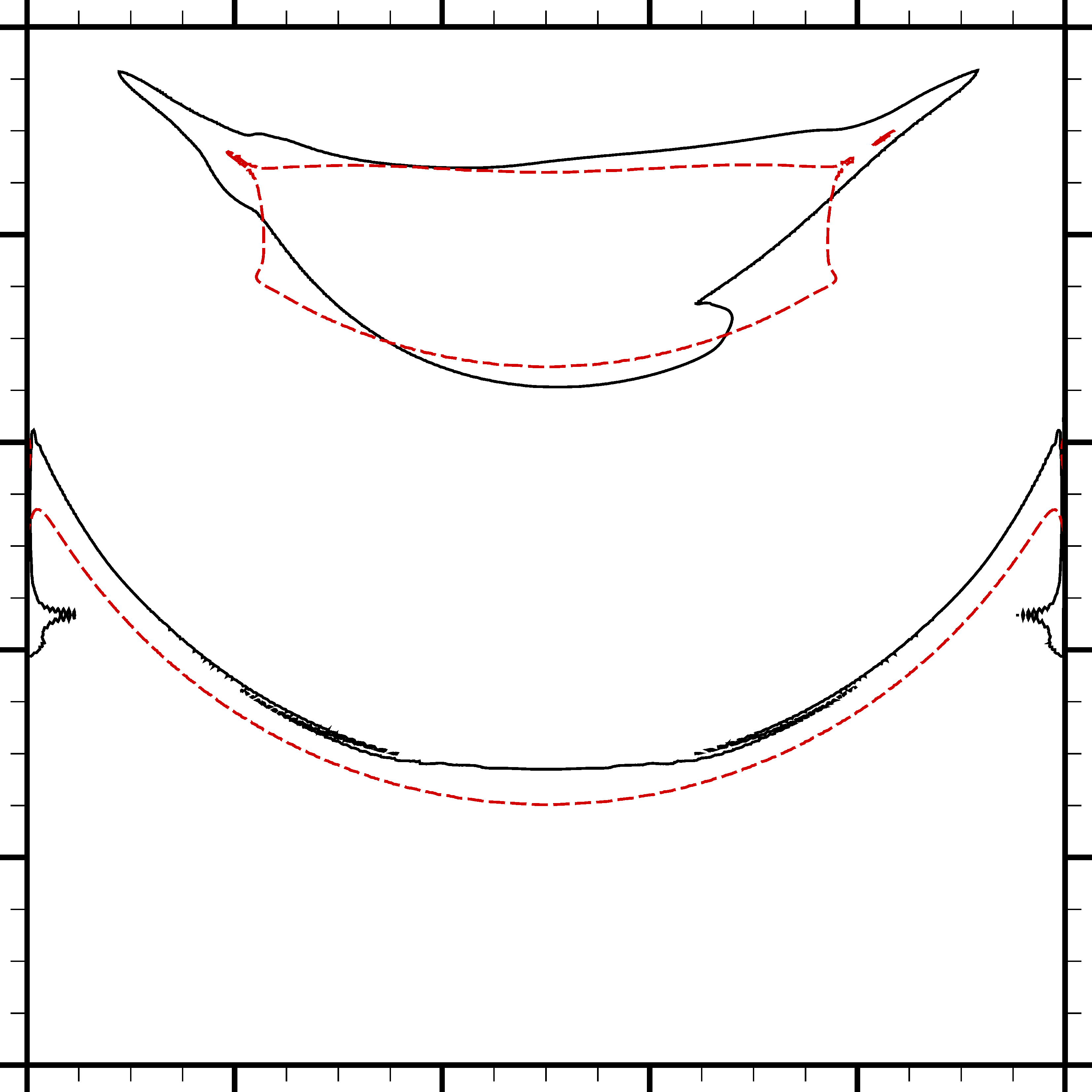}
        \caption{$\tau_d = 70$ \si{Pa}; $G = 40,000$ \si{Pa}}
        \label{sfig: HB comparison: tau 70 100G}
    \end{subfigure}

    \vspace{0.25cm}
    
    \begin{subfigure}[b]{0.32\textwidth}
        \centering
        \includegraphics[width=0.95\linewidth]{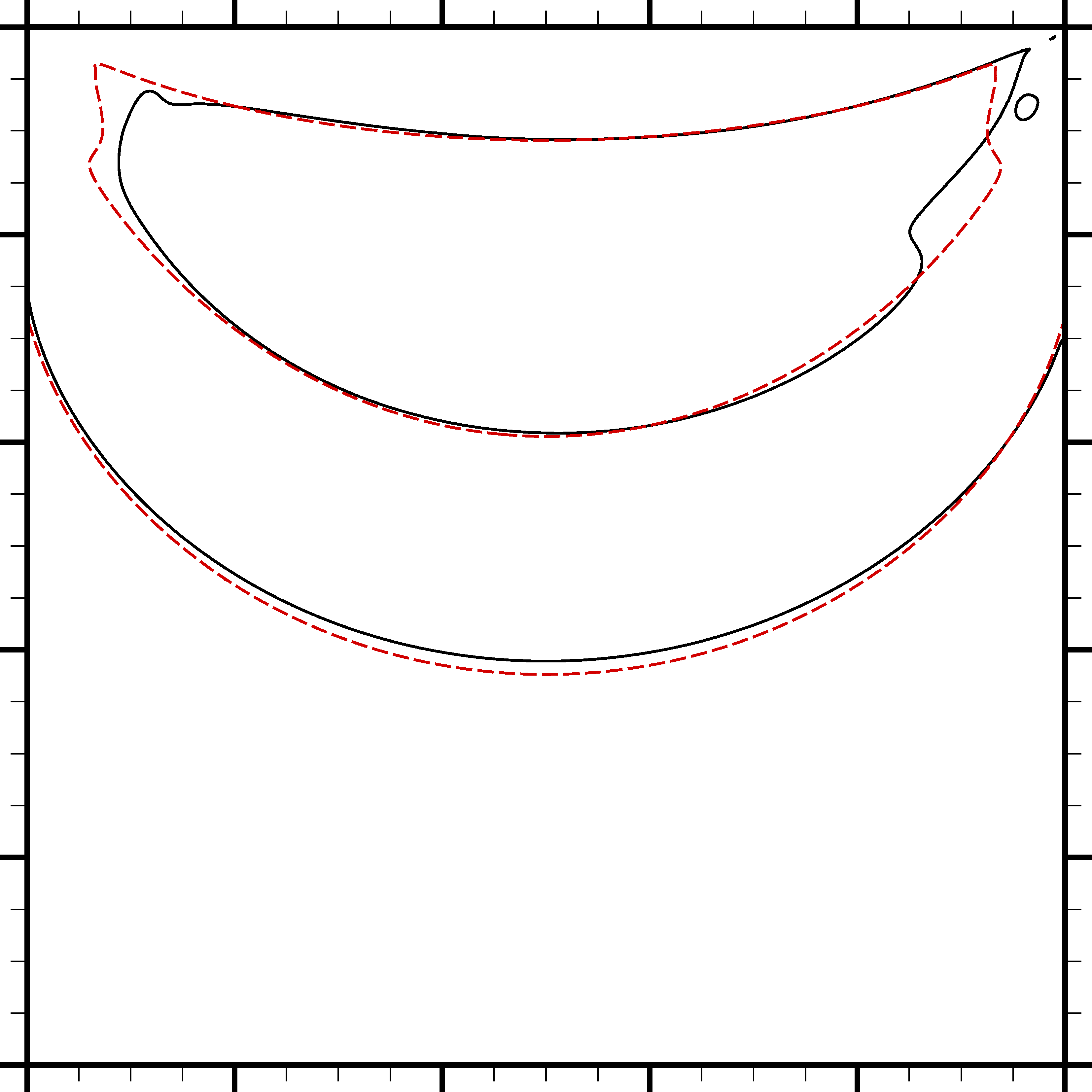}
        \caption{$\tau_d = 80$ \si{Pa}; $G = 400$ \si{Pa}}
        \label{sfig: HB comparison: tau 80 base}
    \end{subfigure}
    \begin{subfigure}[b]{0.32\textwidth}
        \centering
        \includegraphics[width=0.95\linewidth]{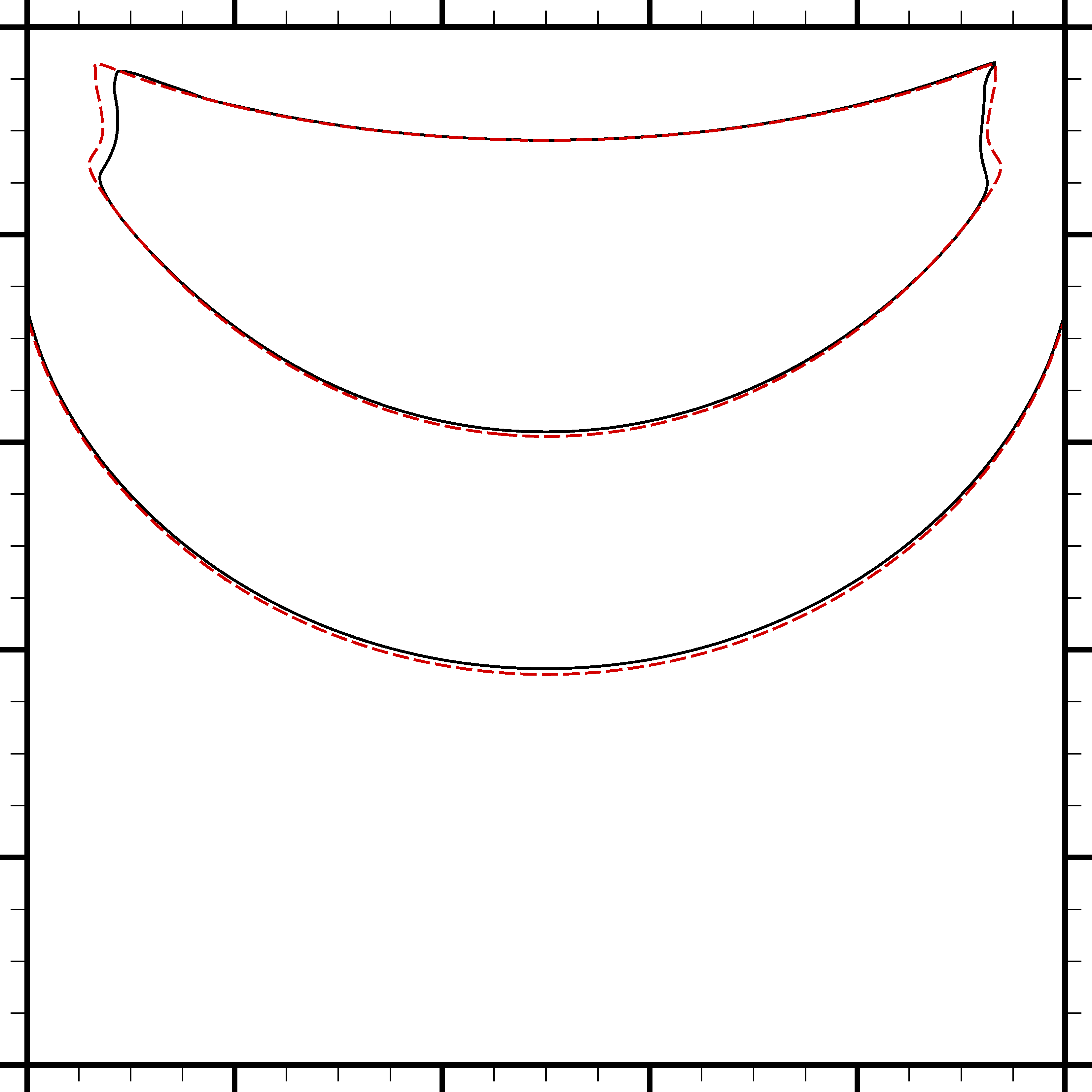}
        \caption{$\tau_d = 80$ \si{Pa}; $G = 4,000$ \si{Pa}}
        \label{sfig: HB comparison: tau 80 10G}
    \end{subfigure}
    \begin{subfigure}[b]{0.32\textwidth}
        \centering
        \includegraphics[width=0.95\linewidth]{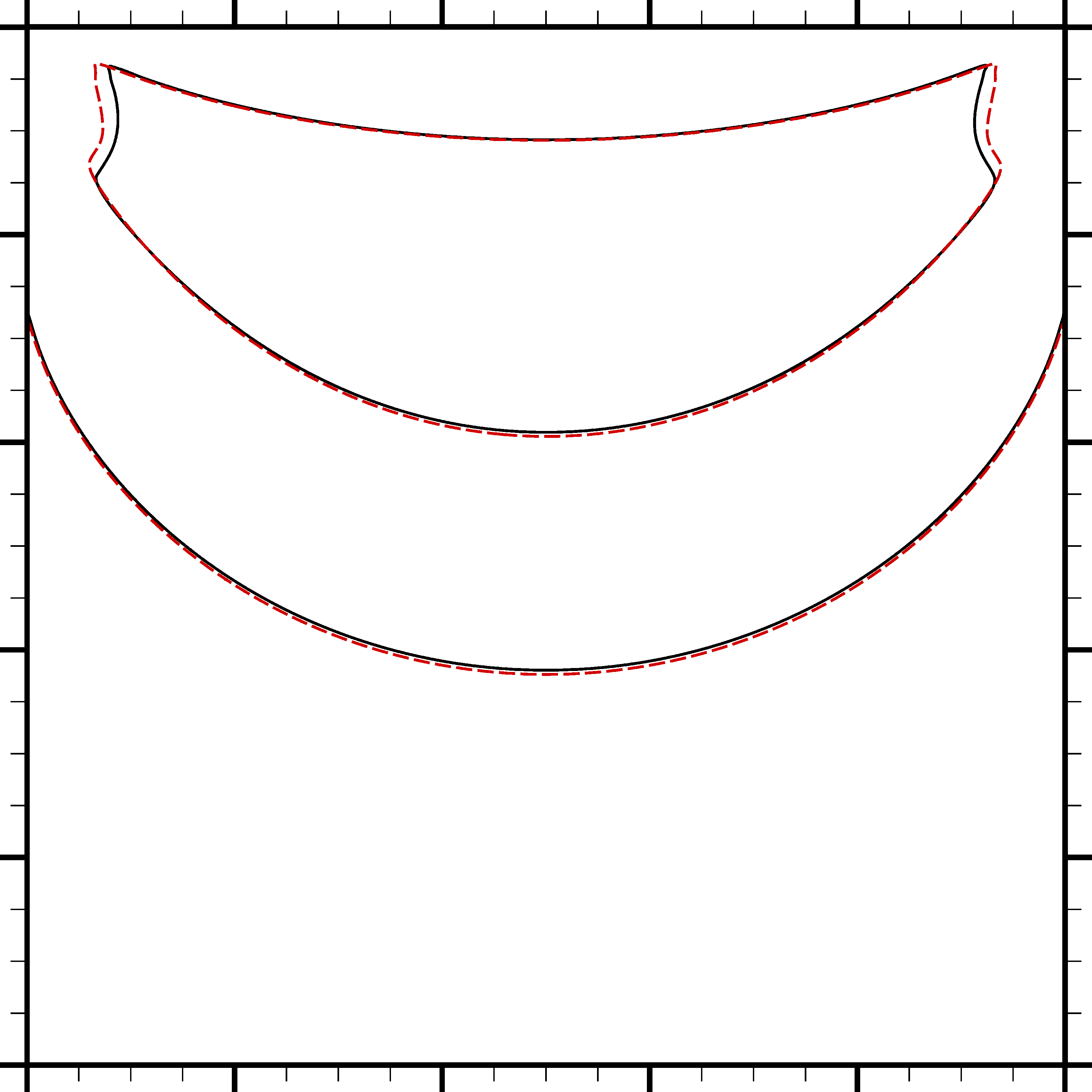}
        \caption{$\tau_d = 80$ \si{Pa}; $G = 40,000$ \si{Pa}}
        \label{sfig: HB comparison: tau 80 100G}
    \end{subfigure}

    \caption{Continuous lines: contours of $\tau_d$ predicted by the SHB model (top row: $\tau_d$ = 
40 \si{Pa}; middle row: $\tau_d = \tau_y$ = 70 \si{Pa}; bottom row: $\tau_d$ = 80 \si{Pa}) for 
various values of the elastic modulus $G$ (left column: $G$ = 400 \si{Pa}; middle column: $G$ = 
4,000 \si{Pa}; right column: $G$ = 40,000 \si{Pa}), the rest of the parameters having the values 
listed in Table \ref{table: carbopol parameters}, and the lid velocity being $U$ = 0.1 \si{m/s}.
Dashed lines: corresponding contours predicted by the HB model.}
    
  \label{fig: HB comparison: tau}
\end{figure}

Next, we performed a couple of simulations similar to the base case, only that we artificially 
increased the elastic modulus to 10 and 100 times the original value listed in Table \ref{table: 
carbopol parameters}. This reduces the relaxation times by factors of 10 and 100, respectively, so 
that the SHB material becomes more inelastic and its predictions should tend to match those of the 
HB model. Due to the smaller relaxation times, we also used smaller time steps: starting from an 
initial time step of \num{5e-5} \si{s} in both cases, we allowed the time step to increase up to 
$\Delta t_{\max}$ = \num{1.5e-3} \si{s} and \num{e-4} \si{s} in the $G$ = 4,000 \si{Pa} and $G$ = 
40,000 \si{Pa} cases, respectively. The simulations were stopped at $t$ = 50 \si{s} and 10 \si{s} 
respectively, and in the $G$ = 40,000 \si{Pa} case the lid acceleration period (Eq.\ \eqref{eq: lid 
velocity}) was reduced to 0.01 \si{s}. Figure \ref{fig: HB comparison: streamlines} shows that the 
SHB and HB streamlines converge rapidly with increasing $G$, while Fig.\ \ref{fig: HB comparison: 
tau} shows that convergence is not as rapid for the stresses. Of course, for $\tau_d < \tau_y$ (top 
row of Fig.\ \ref{fig: HB comparison: tau}) we cannot expect the SHB and HB stresses to ever match, 
due to the aforementioned indeterminacy of the HB stress tensor in the unyielded regions. The yield 
lines (second row of Fig.\ \ref{fig: HB comparison: tau}) appear to converge, but at a slow pace. 
Despite the shorter duration of the $G$ = 40,000 \si{Pa} simulation, Fig.\ \ref{sfig: HB comparison: 
tau 70 100G} shows that during this time the transition zone has evolved further than in the lower 
$G$ cases, due to the much shorter relaxation times. The same figure reveals some slight spurious 
stress oscillations on the lower yield line (can also be seen in Fig.\ \ref{sfig: HB comparison: tau 
70 base}). They could be due to the near-zero velocities there, which complicates the application of 
upwinding within CUBISTA. Finally, the last row of Fig.\ \ref{fig: HB comparison: tau} shows that 
within the yielded region the $\tau_d$ predictions of the two models are quite close, even for $G$ = 
400 \si{Pa}. A persistent discrepancy between the two models at the sides of the plug region seen in 
Fig.\ \ref{sfig: HB comparison: tau 80 100G} may be due to spatial or temporal discretisation 
errors, regularisation, etc.

\subsection{A case with slip}
\label{ssec: results: slip}

Carbopol gels are quite slippery \cite{Piau_2007, Perez_2012, Poumaere_2014}; the previous, no-slip 
results can be considered to correspond to roughened walls, like those used in rheological 
measurements to avoid slip \cite{Lacaze_2015}. However, if the walls are smooth then noticeable slip 
is expected. Viscoplastic and viscoelastic fluids can exhibit complex slip behaviour, e.g.\ 
power-law slip \cite{Kalyon_2005, Panaseti_2017}, pressure dependence \cite{Panaseti_2013}, or 
``slip yield stress'' \cite{Hatzikiriakos_2012, Philippou_2017}. However, experiments with Carbopol 
in \cite{Perez_2012} (0.2\% by weight) and \cite{Poumaere_2014} (0.08\% by weight) showed nearly 
Navier slip behaviour, $u_s = 5.151 \times 10^{-4} \, \tau_w^{0.876}$ and $u_s = 2.3 \times 10^{-4} 
\, \tau_w^{1.32}$, respectively, where $u_s$ is the slip velocity (the left-hand side of Eq.\ 
\eqref{eq: navier slip}) and $\tau_w$ is the wall tangential stress (the term $(\vf{n} \cdot 
\tf{\tau}) \cdot \vf{s}$ of the right-hand side of Eq.\ \eqref{eq: navier slip}). In 
\cite{Poumaere_2014} it is conjectured that this may be due to the formation of a thin layer of 
Newtonian fluid (solvent) that separates the wall surface from the gel micro-particles. In the 
present section we impose a Navier slip condition, Eq.\ \eqref{eq: navier slip}, on all walls, with 
a slip coefficient $\beta = 5 \times 10^{-4}$ \si{m / Pa.s}. For this case, concerning stress 
extrapolation to the walls, we could not get SIMPLE to converge with the $D$ coefficients given by 
\eqref{eq: D+ boundary} -- \eqref{eq: D- boundary} and we set them equal to zero.

\begin{figure}[tb]
    \centering
    \begin{subfigure}[b]{0.32\textwidth}
        \centering
        \includegraphics[width=0.95\linewidth]{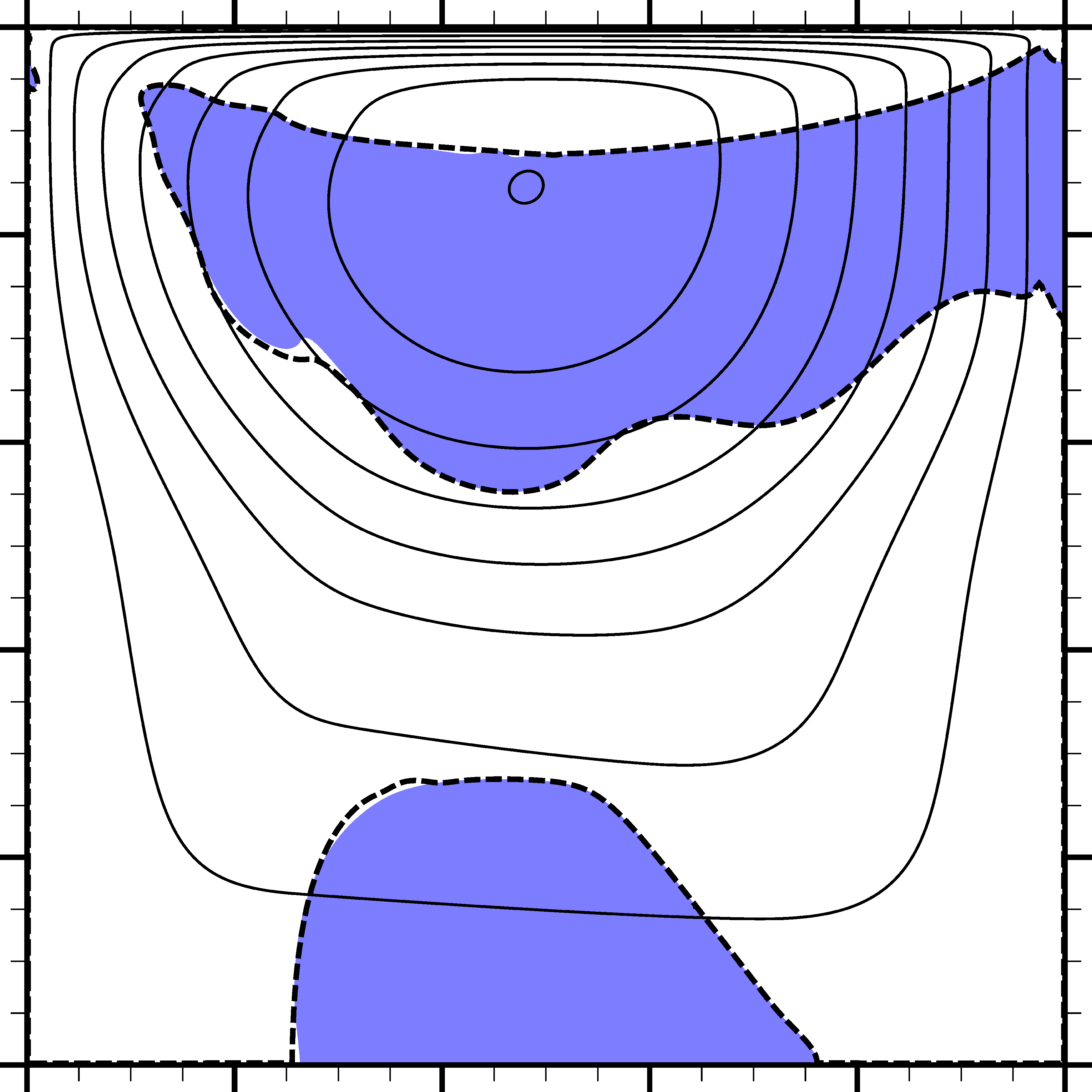}
        \caption{SHB flowfield}
        \label{sfig: slip flowfield}
    \end{subfigure}
    \begin{subfigure}[b]{0.32\textwidth}
        \centering
        \includegraphics[width=0.95\linewidth]{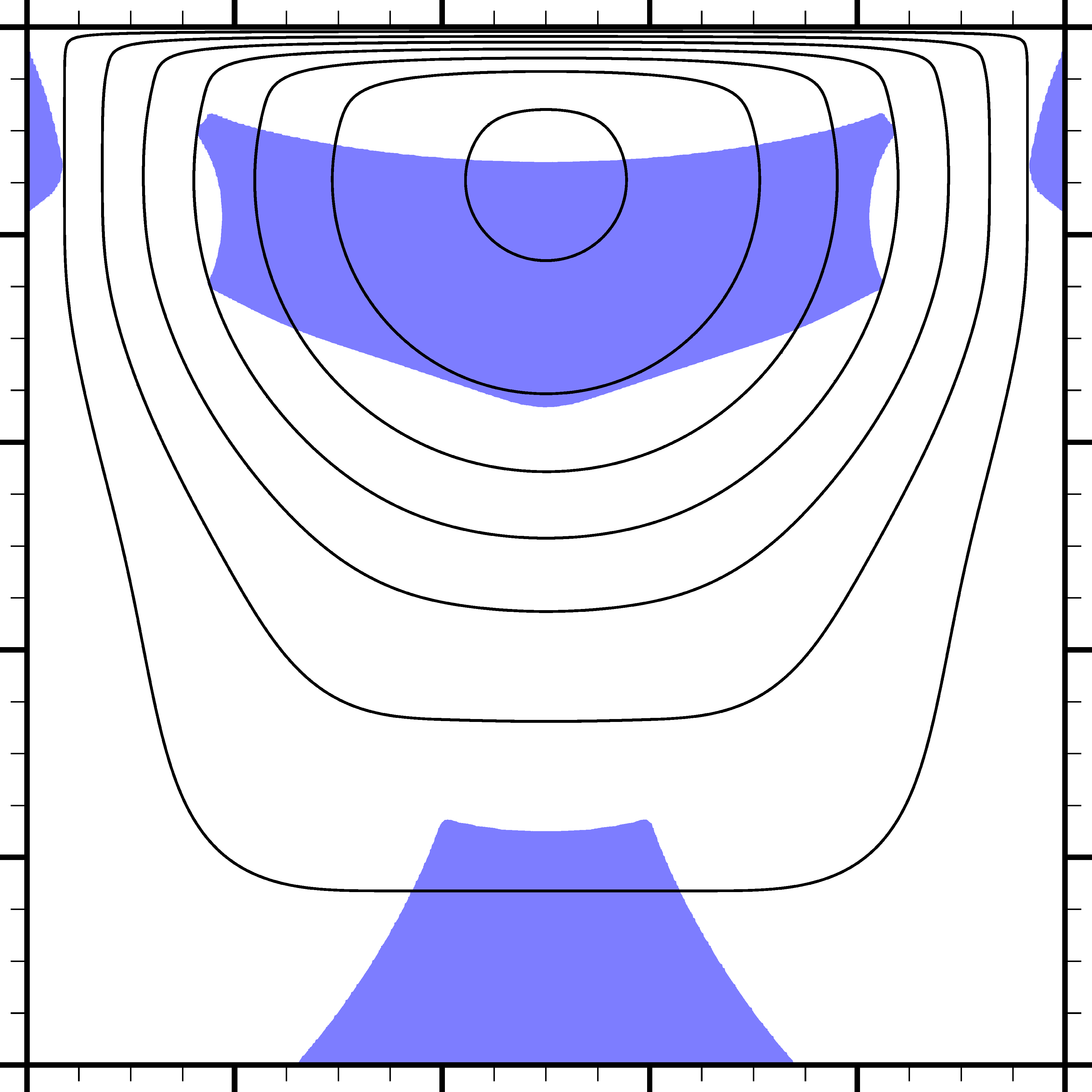}
        \caption{HB flowfield}
        \label{sfig: slip flowfield HB}
    \end{subfigure}
    \begin{subfigure}[b]{0.32\textwidth}
        \centering
        \includegraphics[width=0.95\linewidth]{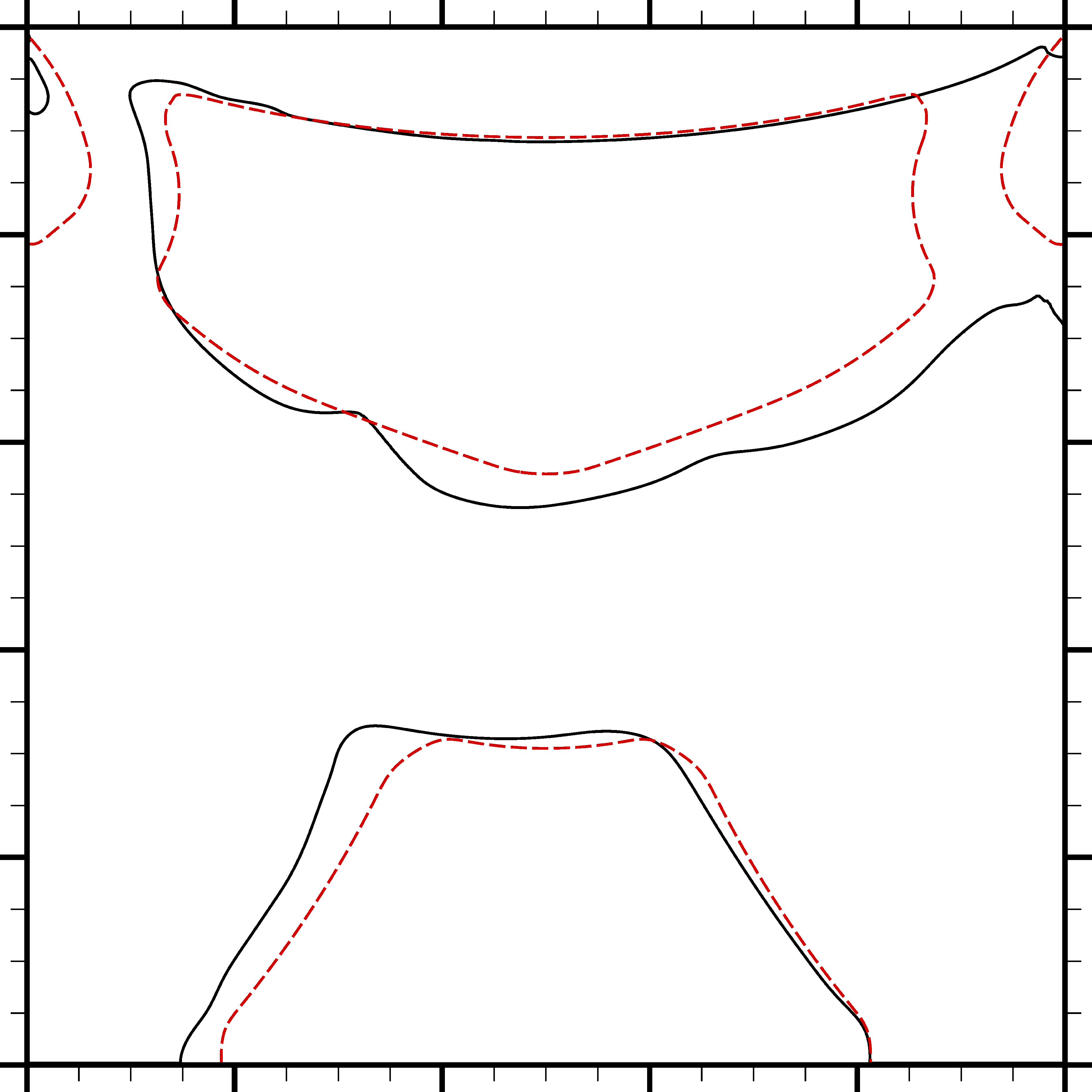}
        \caption{$\tau_d$ = 75 \si{Pa}}
        \label{sfig: slip tau_d 75}
    \end{subfigure}
    \caption{\subref{sfig: slip flowfield} SHB flowfield for the slip case; shaded region: 
unyielded material at $t$ = 60 \si{s}; dashed lines: yield lines at $t$ = 30 \si{s}; continuous 
lines: streamlines (they are drawn at equal streamfunction intervals). \subref{sfig: slip 
flowfield HB} Corresponding steady-state HB flowfield (the streamlines are drawn at the same values 
of streamfunction as in \subref{sfig: slip flowfield}). \subref{sfig: slip tau_d 75} The $\tau_d$ 
= 75 \si{Pa} contour of the SHB (continuous line) and HB (dashed line) flowfields.}
  \label{fig: slip flowfields}
\end{figure}

\begin{figure}[tb]
    \centering
    \begin{subfigure}[t]{0.43\textwidth}
        \centering
        \includegraphics[scale=0.82]{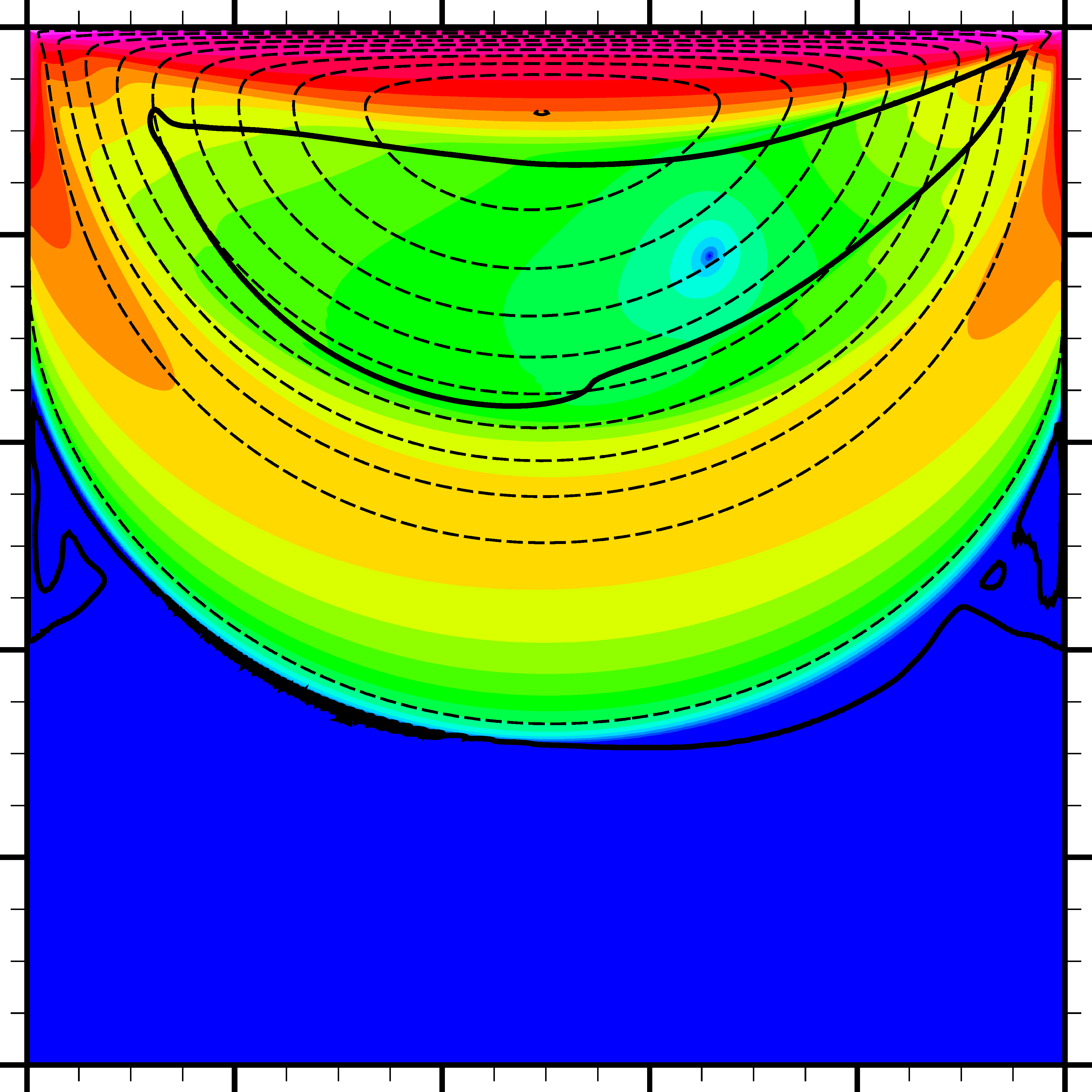}
        \caption{No slip}
        \label{sfig: base case, gamma}
    \end{subfigure}
    \begin{subfigure}[t]{0.12\textwidth}
        \centering
        \includegraphics[scale=0.85]{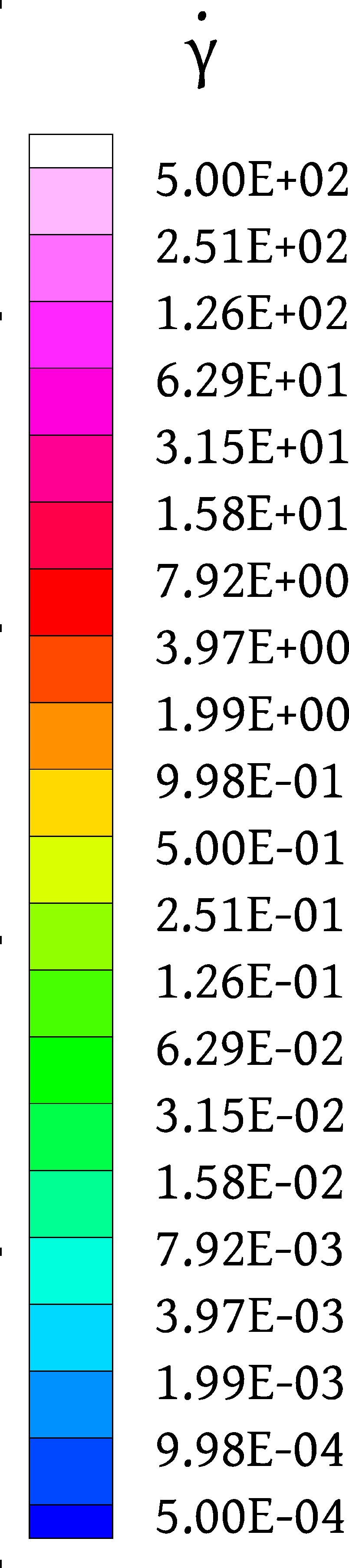}
    \end{subfigure}
    \begin{subfigure}[t]{0.43\textwidth}
        \centering
        \includegraphics[scale=0.82]{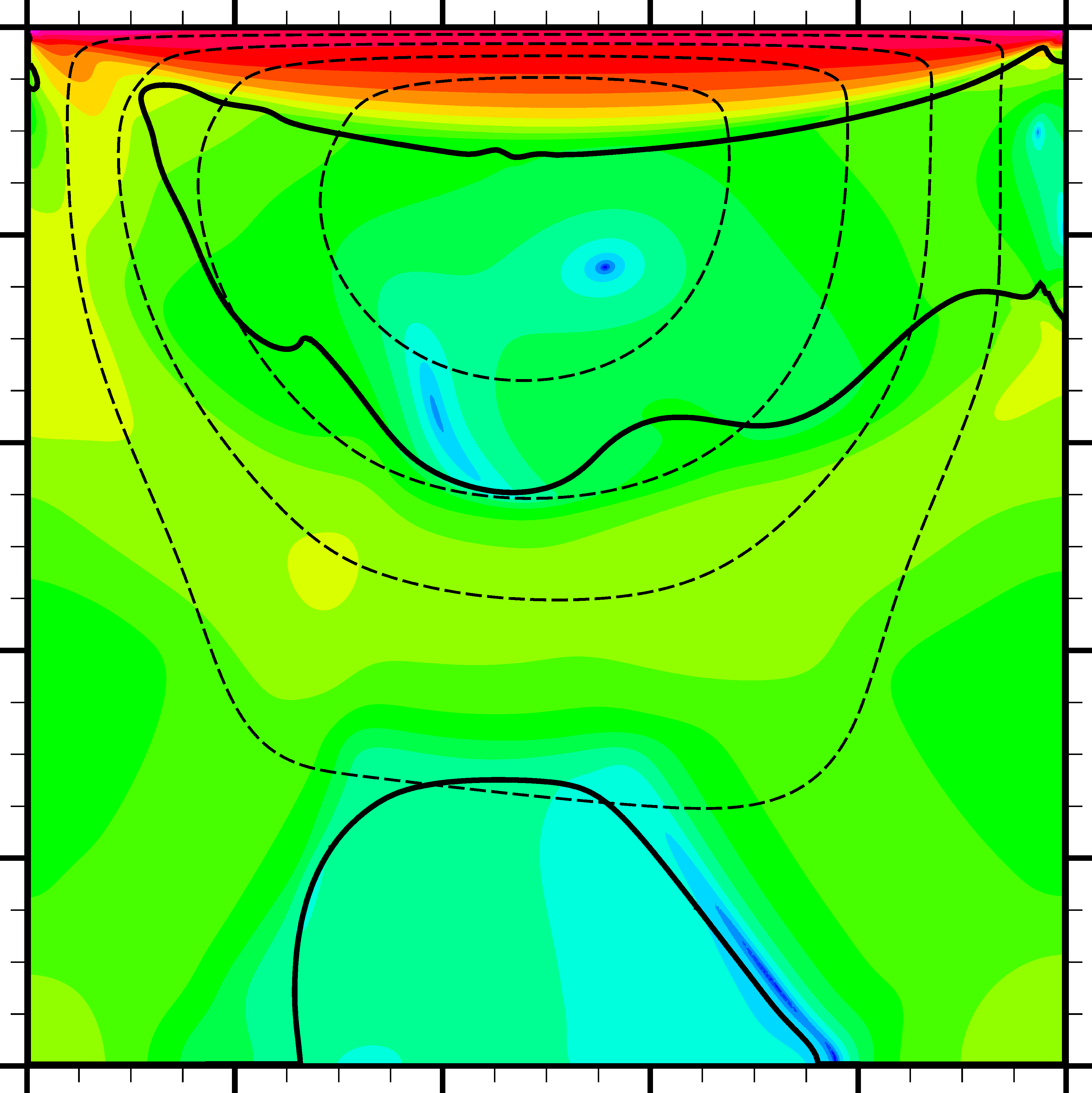}
        \caption{Slip}
        \label{sfig: slip, gamma}
    \end{subfigure}
    \caption{Contours of $\dot{\gamma}$ [\si{s^{-1}}] for $U$ = 0.100 \si{m/s} without wall slip
\subref{sfig: base case, gamma} (base case) or with wall slip \subref{sfig: slip, gamma}. The thick 
continuous lines are yield lines and the dashed lines are streamlines. The latter are plotted at 
equal streamfunction intervals (the same intervals in both figures).}
  \label{fig: base and slip gamma}
\end{figure}

\begin{figure}[tb]
    \centering
    \begin{subfigure}[b]{0.49\textwidth}
        \centering
        \vskip 0pt
        \includegraphics[width=0.88\linewidth]{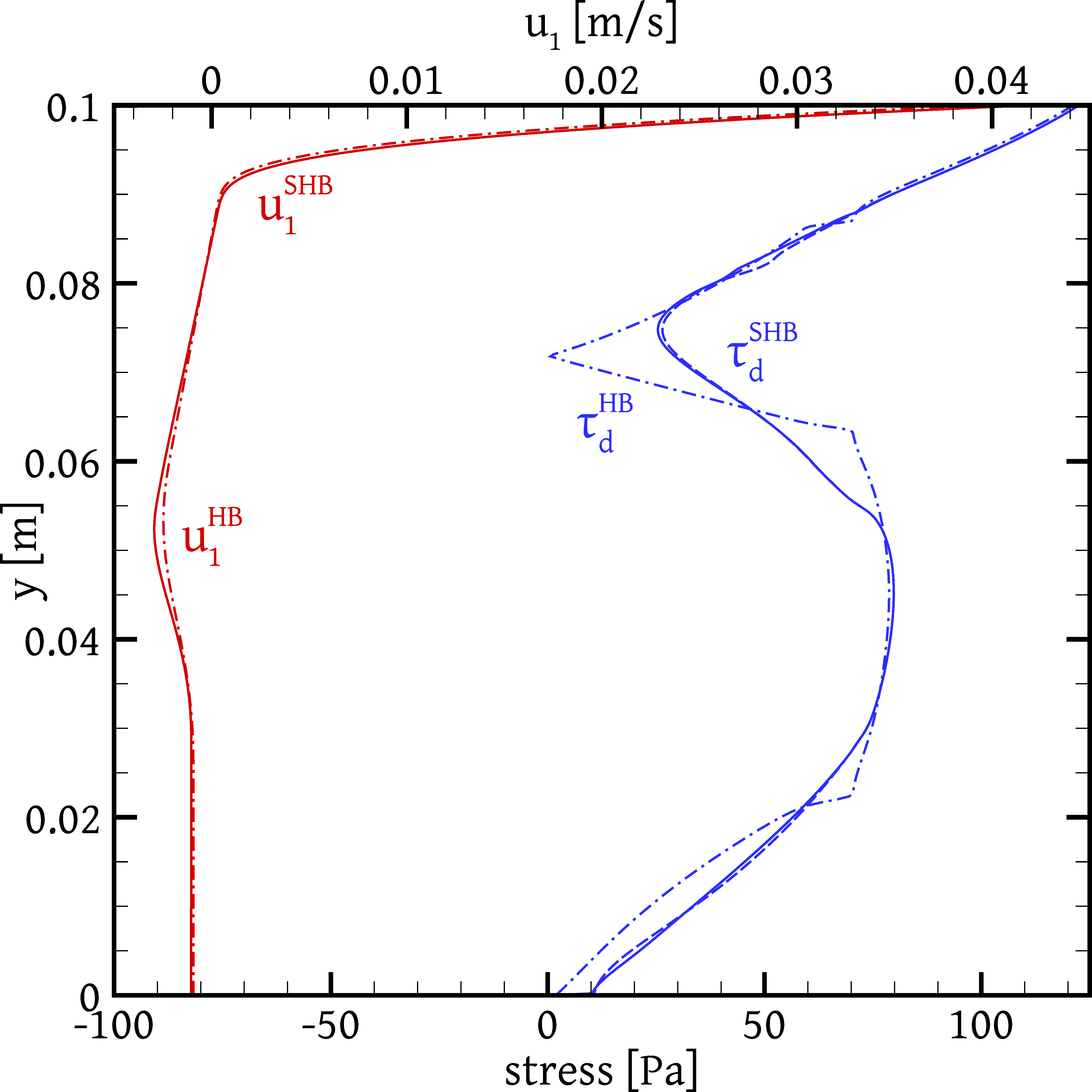}
        \caption{$u_1$ and $\tau_d$ at $x=L/2$}
        \label{sfig: slip profiles u}
    \end{subfigure}
    \begin{subfigure}[b]{0.49\textwidth}
        \centering
        \includegraphics[width=0.88\linewidth]{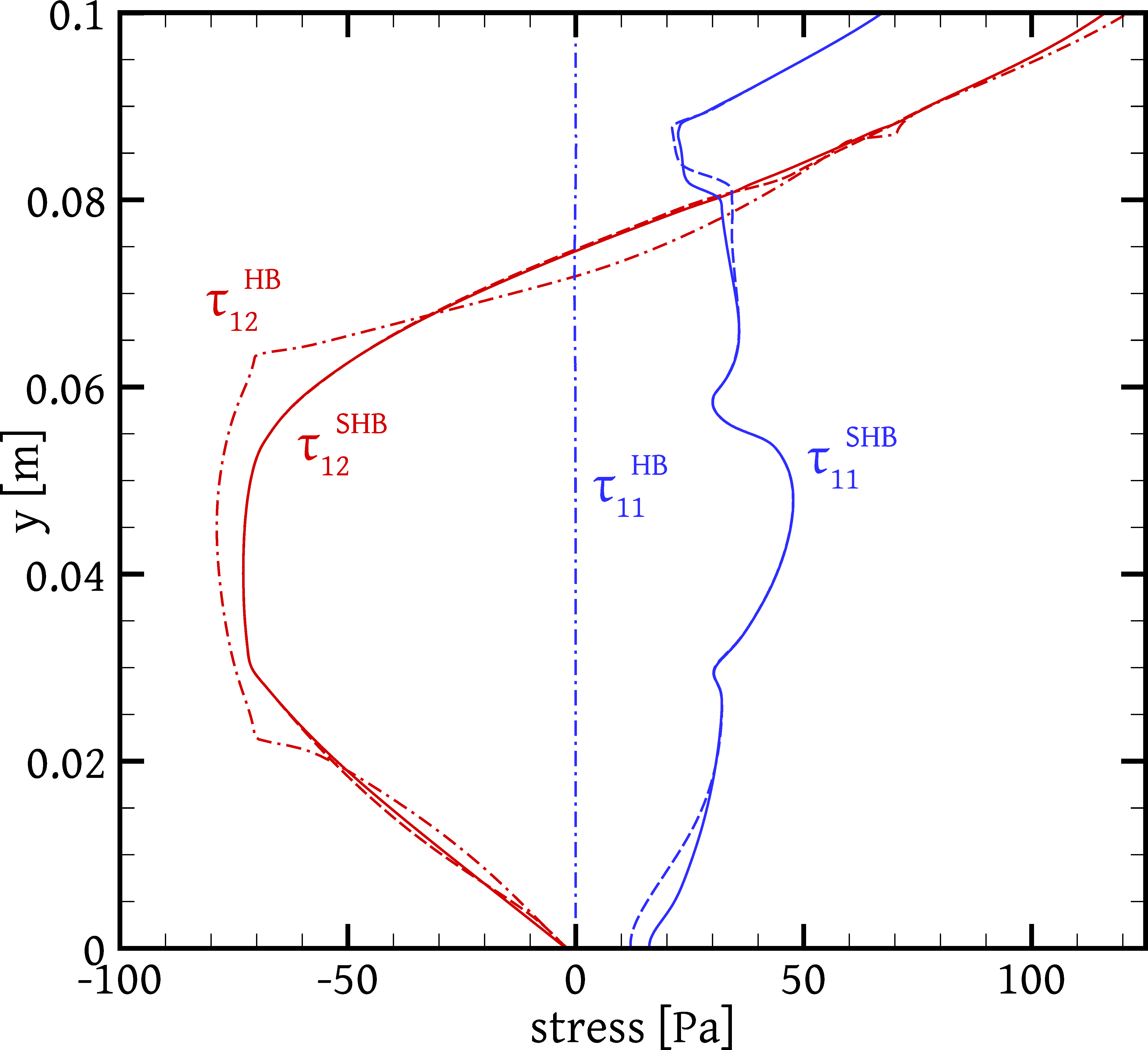}
        \caption{$\tau_{11}$ and $\tau_{12}$ at $x=L/2$}
        \label{sfig: slip profiles stress}
    \end{subfigure}
    \caption{Profiles of some dependent variables along the vertical centreline, for the slip case. 
SHB results at $t$ = 60 and 30 \si{s} are plotted in continuous and dashes lines, respectively, and 
steady state HB results in dash-dot line.}
  \label{fig: slip profiles}
\end{figure}

Figure \ref{sfig: slip flowfield} shows the flowfield. A distinguishing feature is that the two 
unyielded zones touch the cavity walls and yet contain moving material which is sliding over the 
wall. This feature is common also to the HB flowfield, shown in Fig.\ \ref{sfig: slip flowfield 
HB}, which is, again, much more symmetric. Figure \ref{sfig: slip tau_d 75} shows that SHB and HB 
contours of $\tau_d$ are more similar for $\tau_d > \tau_y$. Due to slip, the flow induced by the 
lid is much weaker than in the no-slip case, as can be seen from the lower kinetic energy in Fig.\ 
\ref{fig: monitor}, the smaller vortex strength in Table \ref{table: vortex metrics}, and the wider 
streamline spacing in Fig.\ \ref{sfig: slip, gamma} than in Fig.\ \ref{sfig: base case, gamma}. On 
the other hand, the streamlines in Fig.\ \ref{sfig: slip, gamma} are more evenly spaced near the 
bottom of the cavity compared to Fig.\ \ref{sfig: base case, gamma} as slip allows the circulation 
to extend all the way down to the bottom wall, where now there is no \textit{stationary} unyielded 
zone. Consequently, there is also no transition zone. Indeed, comparing in Fig.\ \ref{sfig: slip 
flowfield} the yield lines at $t$ = 30 \si{s} (dashed lines) and $t$ = 60 \si{s} (the boundary of 
the shaded regions) one sees that there is very little change, and the bottom unyielded zone even 
slightly contracts with time. The absence of a transition zone can be explained by examining the 
$\dot{\gamma}$ distributions of Fig.\ \ref{fig: base and slip gamma}; whereas in the no-slip case 
(Fig.\ \ref{sfig: base case, gamma}) the velocity gradients are practically zero inside the bottom 
unyielded zone, in the slip case (Fig.\ \ref{sfig: slip, gamma}) they are non-zero throughout the 
domain, thus excluding the transition zone situation where the constitutive equation reduces to Eq.\ 
\eqref{eq: constitutive stationary tensorial}. Figure \ref{fig: base and slip gamma} also shows that 
in the upper part of the cavity $\dot{\gamma}$ is much lower in the presence of slip. Due to the 
shear-thinning nature of the fluid, this results in higher viscosities and associated relaxation 
times which may explain why the stress evolution is slower in the slip case than in the no-slip one 
as seen in Fig.\ \ref{sfig: monitor trT} (and also in Figs.\ \ref{sfig: slip flowfield} and 
\ref{fig: slip profiles}, where there are slight changes between $t$ = 30 and 60 \si{s}). 
Surprisingly, Fig.\ \ref{sfig: monitor trT} shows that $\mathrm{tr}(\tf{\tau})_{\mathrm{avg}}$ is 
actually higher in the slip case, probably due to the absence of a bottom stationary zone. For 
benchmarking purposes we plot profiles of some dependent variables along the vertical centreline in 
Fig.\ \ref{fig: slip profiles}.

\subsection{Multiplicity of solutions}
\label{ssec: results: multiplicity of solutions}

It was mentioned in Sec.\ \ref{sec: equations} that Cheddadi et al.\ \cite{Cheddadi_2012} found 
that the steady state of a SHB flow (including the velocity field) can depend in a continuous 
manner on the initial conditions. That was observed in simple cylindrical Couette flow, where the 
shear stress is determined by the geometry and kinematics but the circumferential normal stress 
depends also on its initial value, which affects the final extent of the yield zone and therefore 
also the final velocity field. We performed analogous experiments to see if this can be observed 
also in the present, more complex flow. So, we solved two more cases that are identical to the 
``base case'' of Sec.\ \ref{ssec: results: base case}, except for the initial stress conditions: 
$\tau_{11} = \tau_{22} = +\sqrt{3} \tau_y$ (first case) or $-\sqrt{3} \tau_y$ (second case) 
throughout, the remaining stress components being zero. These initial conditions correspond, 
respectively, to tension and compression states that are isotropic in the $xy$ plane; from Eq.\ 
\eqref{eq: deviatoric stress} it follows that $\tau_d(t=0) = \tau_y$ everywhere, i.e.\ the material 
is on the verge of yielding. The material is initially at rest ($\vf{u} = 0$). The simulations were 
carried on until time $t$ = 60 \si{s}.

\begin{figure}[b]
    \centering
    \begin{subfigure}[b]{0.48\textwidth}
        \centering
        \includegraphics[width=0.95\linewidth]{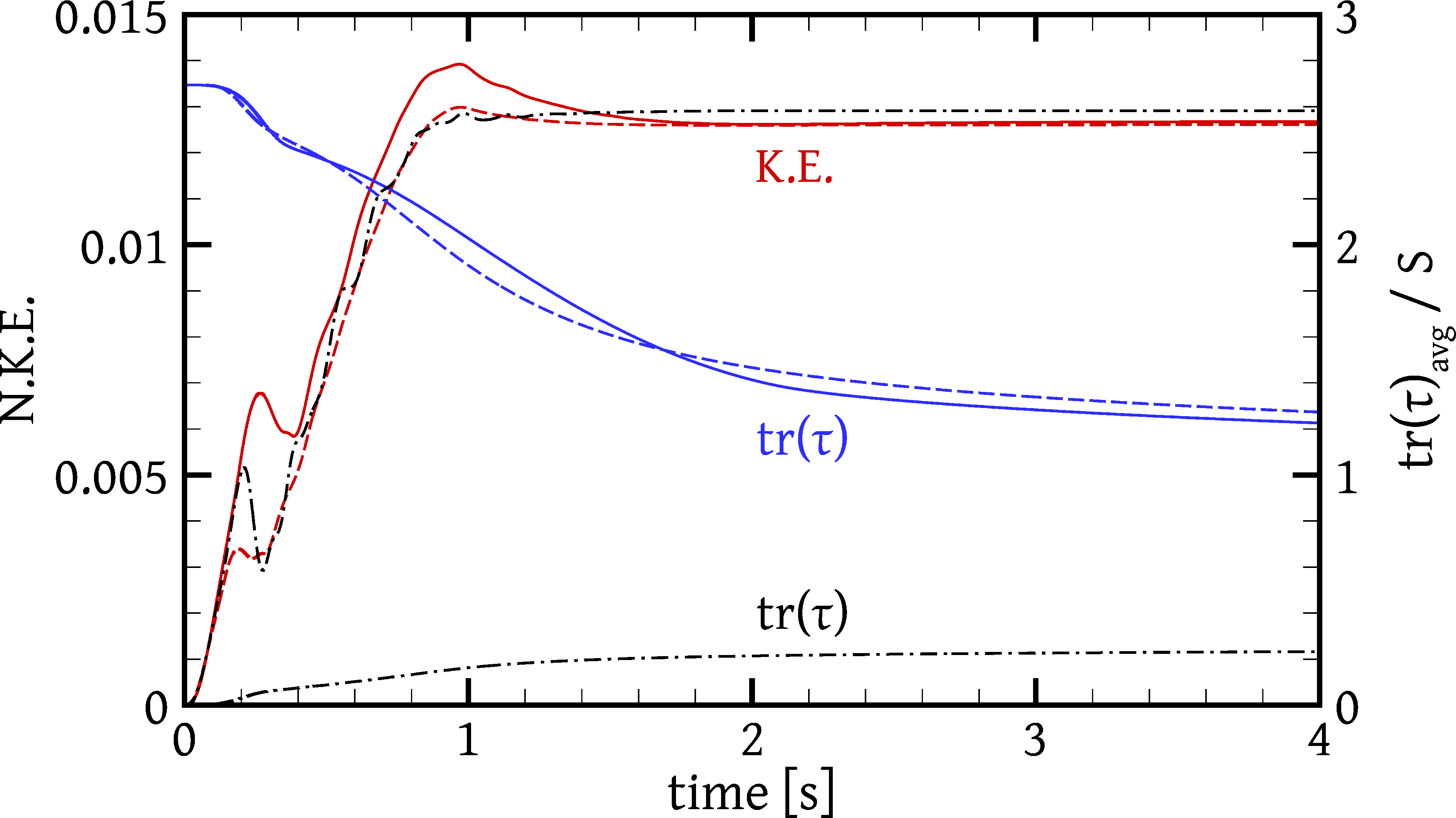}
        \caption{$t \in [0,4]$ \si{s}}
        \label{sfig: multiplicity monitor zoom}
    \end{subfigure}
    \begin{subfigure}[b]{0.48\textwidth}
        \centering
        \includegraphics[width=0.95\linewidth]{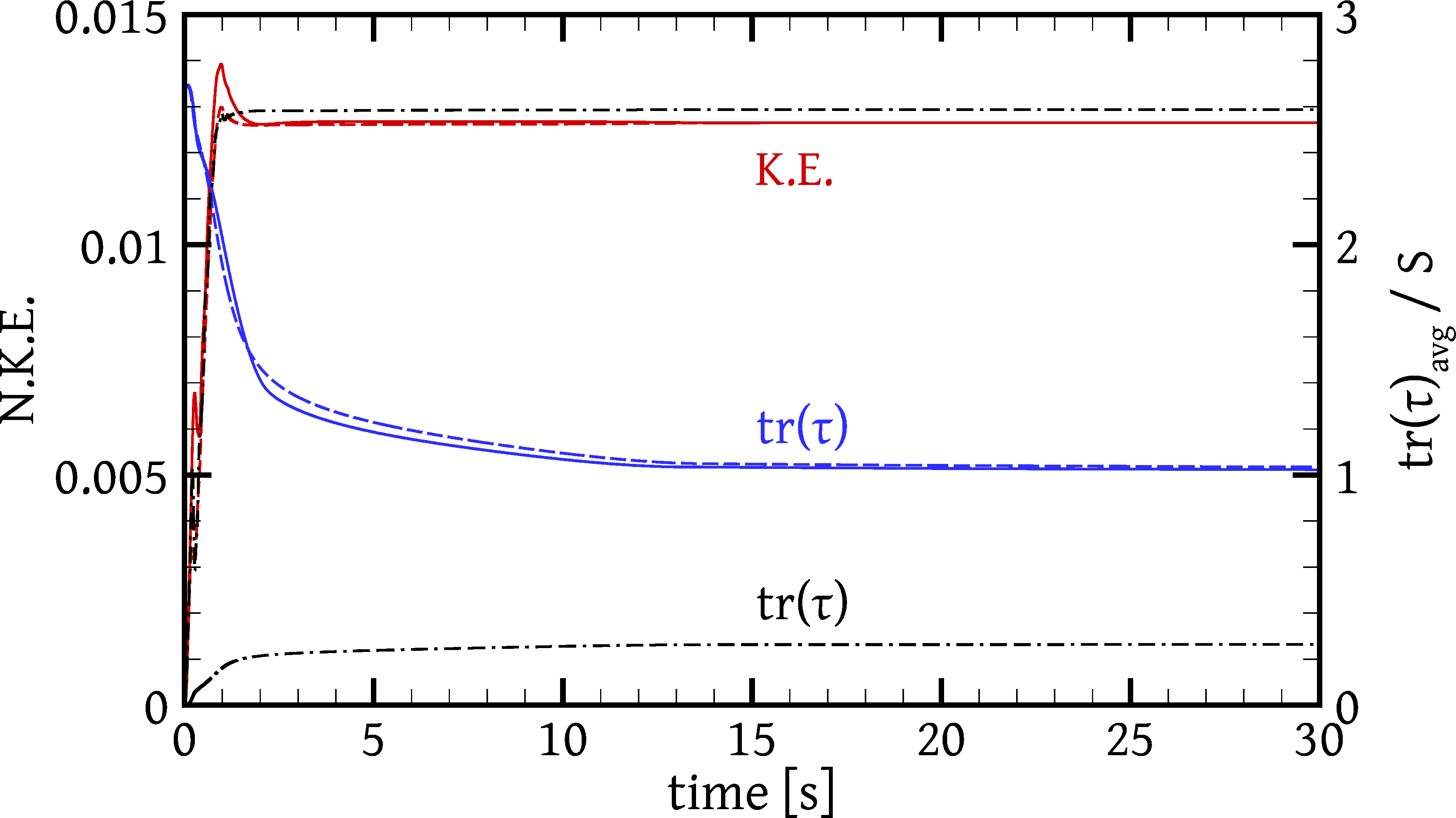}
        \caption{$t \in [0,30]$ \si{s}}
        \label{sfig: multiplicity monitor}
    \end{subfigure}
    \caption{Histories of N.K.E.\ (Eq.\ \eqref{eq: kinetic energy}) and 
$\mathrm{tr}(\tf{\tilde{\tau}})_{\mathrm{avg}}$ (Eq.\ \eqref{eq: average trace}) for 
the cases with initial conditions $\tau_{11} = \tau_{22} = -\sqrt{3} \tau_y$ (continuous lines),
$\tau_{11} = \tau_{22} = +\sqrt{3} \tau_y$ (dashed lines) and $\tau_{11} = \tau_{22} = 0$ 
(dash-dot lines, the base case).}
  \label{fig: multiplicity monitor}
\end{figure}

\begin{figure}[tb]
    \centering
    \begin{subfigure}[b]{0.32\textwidth}
        \centering
        \includegraphics[width=0.95\linewidth]{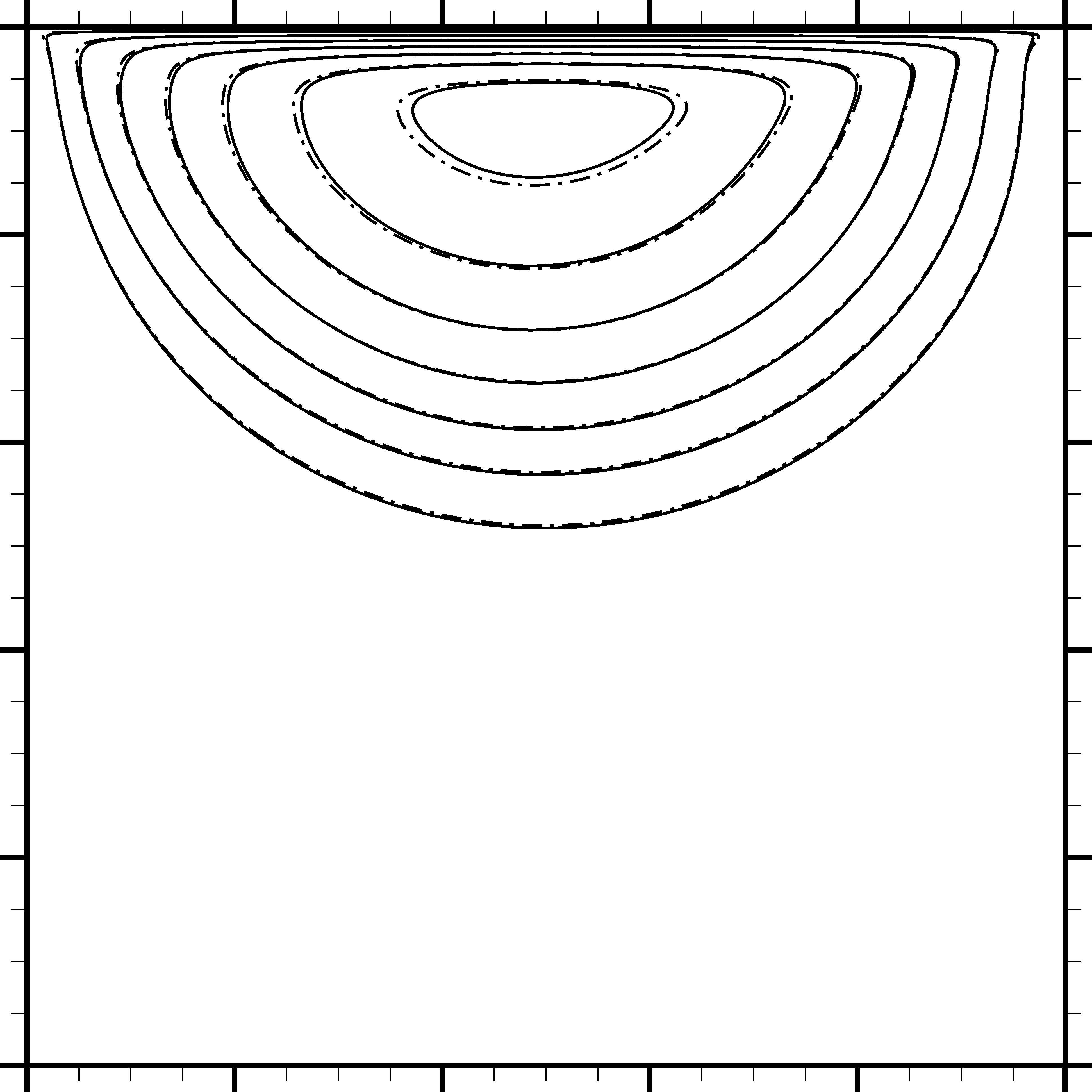}
        \caption{streamlines}
        \label{sfig: multiplicity streamlines}
    \end{subfigure}
    \begin{subfigure}[b]{0.32\textwidth}
        \centering
        \includegraphics[width=0.95\linewidth]{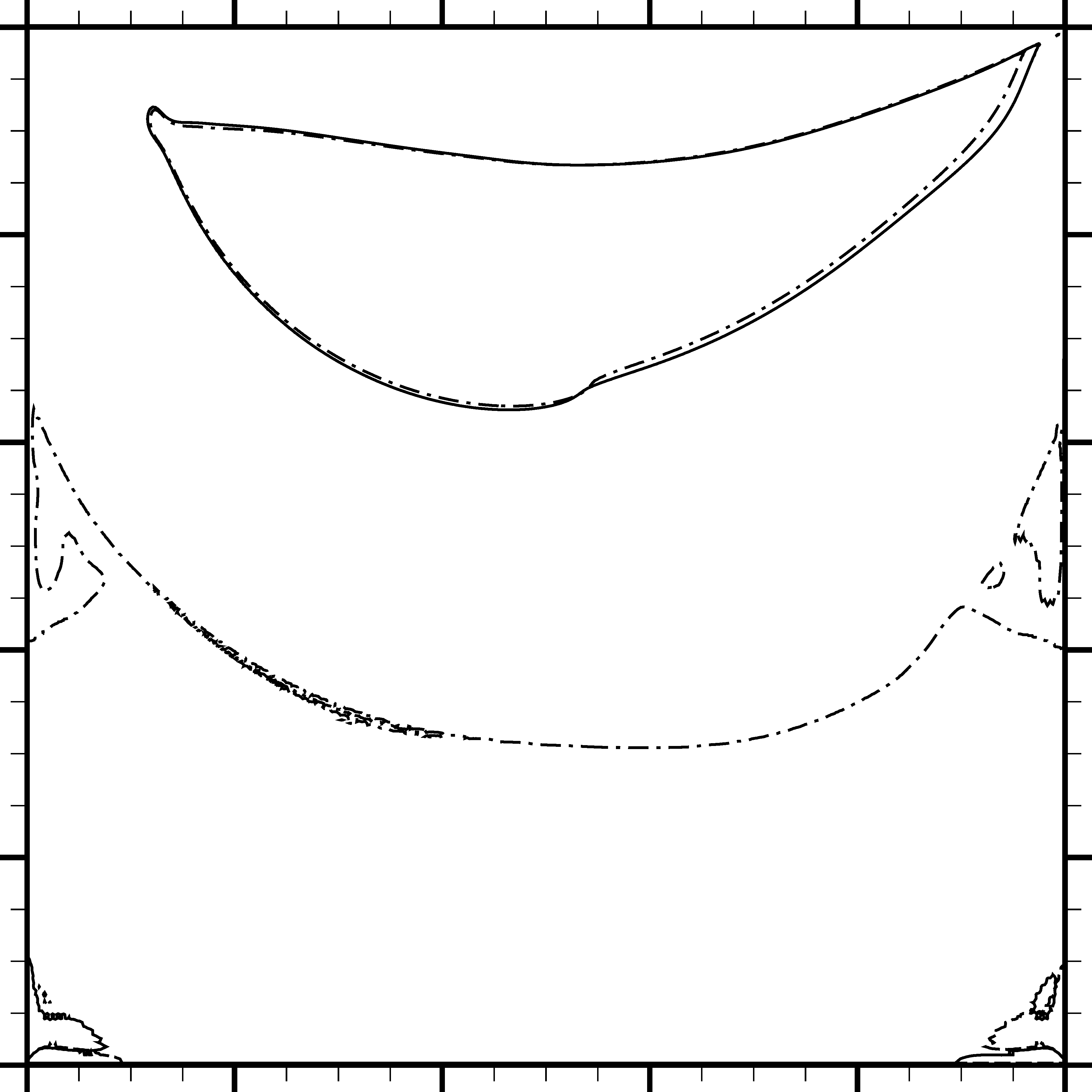}
        \caption{yield lines}
        \label{sfig: multiplicity yield lines}
    \end{subfigure}
    \begin{subfigure}[b]{0.32\textwidth}
        \centering
        \includegraphics[width=0.95\linewidth]{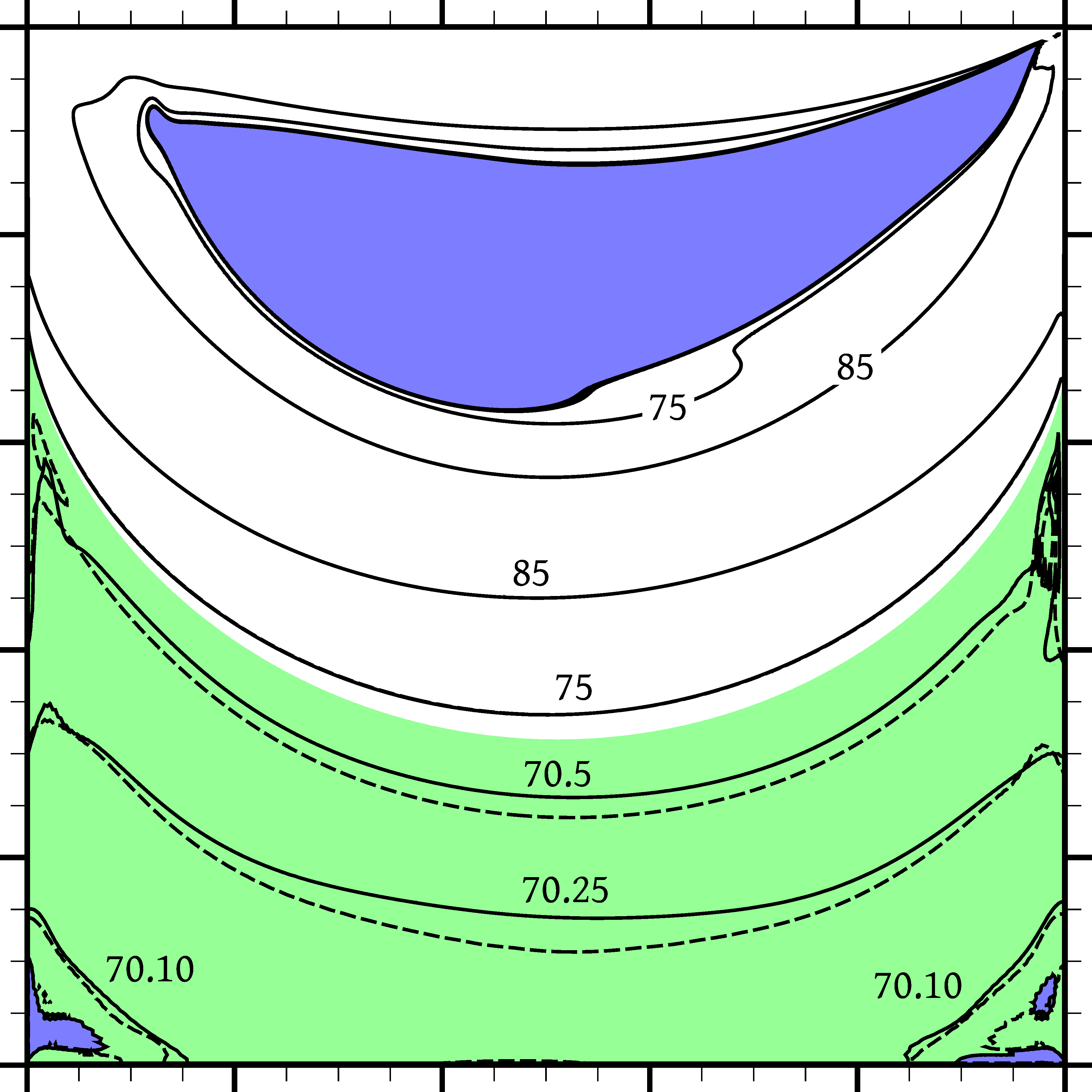}
        \caption{$\tau_d$}
        \label{sfig: multiplicity tau_d}
    \end{subfigure}
    \caption{Steady-state streamlines \subref{sfig: multiplicity streamlines} (streamfunction 
interval $\delta \phi / (\rho U)$ = \num{3.6e-4}), yield lines \subref{sfig: multiplicity yield 
lines}, and $\tau_d$ contours \subref{sfig: multiplicity tau_d} for the cases with initial 
conditions $\tau_{11} = \tau_{22} = -\sqrt{3} \tau_y$ (continuous lines), $+\sqrt{3} \tau_y$ 
(dashed lines), and zero (dash-dot lines, only in \subref{sfig: multiplicity streamlines} and 
\subref{sfig: multiplicity tau_d}). In \subref{sfig: multiplicity tau_d} the values of $\tau_d$, in 
\si{Pa}, are indicated next to each contour. The unyielded region of the $\tau_{11} = \tau_{22} = 
-\sqrt{3} \tau_y$ case is shaded blue. The region where $\|\vf{u}\| < 5\times 10^{-5}U$ is shaded 
green.} \label{fig: multiplicity flow fields}
\end{figure}

Interestingly, the velocity fields of both of these new cases arrive at practically the same steady 
state; this can be seen in Fig.\ \ref{fig: multiplicity monitor}, where the kinetic energies of both 
cases converge, and more clearly in Fig.\ \ref{sfig: multiplicity streamlines} where the respective 
streamlines can be seen to be identical. However, this steady-state is not the same as that arrived 
at in the base case ($\tau_{11} = \tau_{22} = 0$); Fig.\ \ref{fig: multiplicity monitor} shows that 
unlike in the two new cases, the kinetic energy of the base case does not exhibit an overshoot and 
eventually increases to a value which is about 2.25\% larger than that of the other two cases. In 
Fig.\ \ref{sfig: multiplicity streamlines} the base case streamlines are also not identical to 
those of the other two cases. As far as the main vortex is concerned, the new cases have it located 
at $(\tilde{x}_c, \tilde{y}_c) = (0.497, 0.917)$ with a strength of $\tilde{\psi}_c = -0.0266$, 
which is located slightly to the right and is slightly weaker than that of the base case (Table 
\ref{table: vortex metrics}). Of course, all these differences are small, but nevertheless they 
confirm the observation of Cheddadi et al.\ \cite{Cheddadi_2012}.

\begin{figure}[tb]
    \centering
    \begin{subfigure}[b]{0.32\textwidth}
        \centering
        \includegraphics[width=0.95\linewidth]{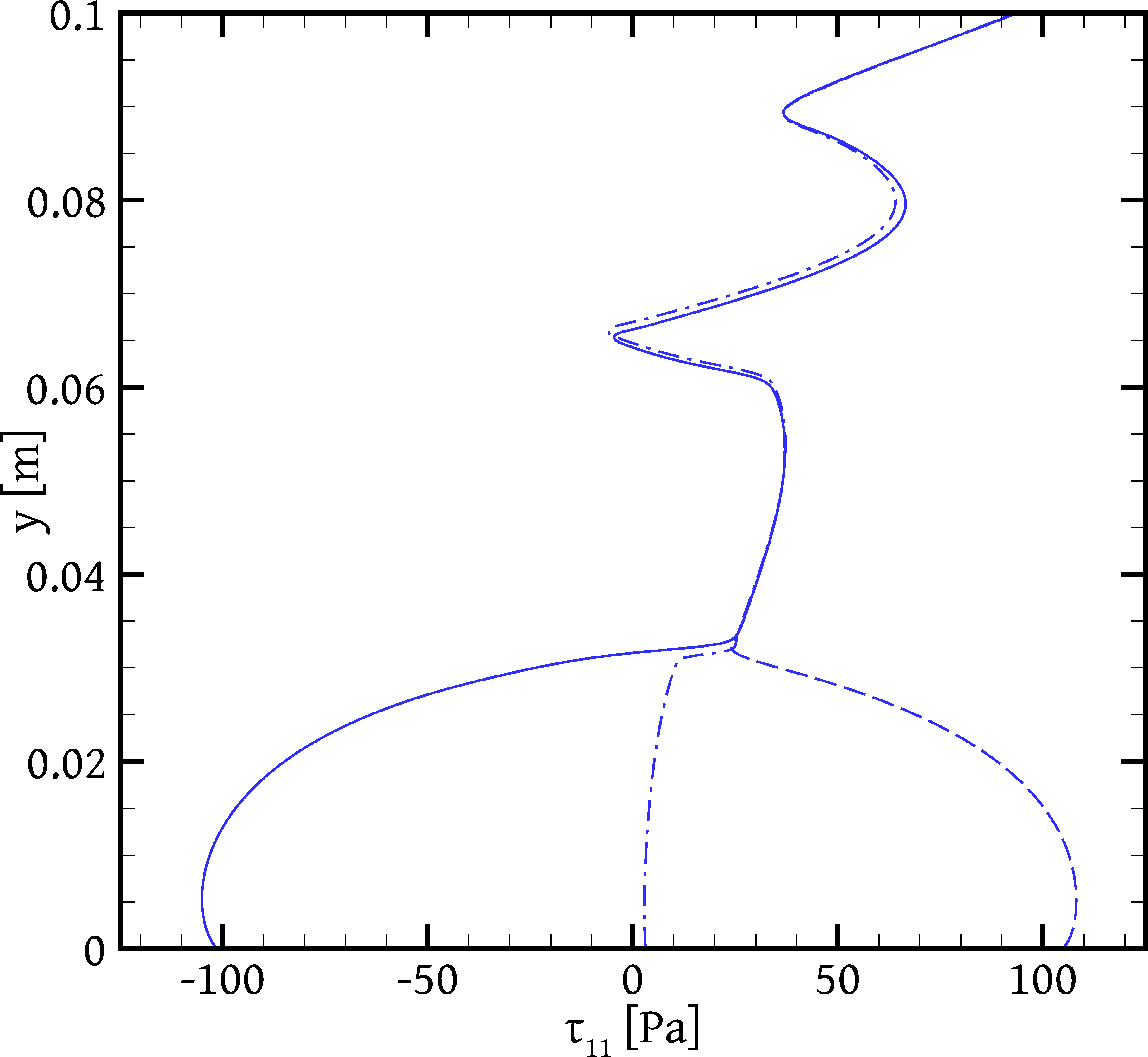}
        \caption{$\tau_{11}$}
        \label{sfig: multiplicity profiles s11}
    \end{subfigure}
    \begin{subfigure}[b]{0.32\textwidth}
        \centering
        \includegraphics[width=0.95\linewidth]{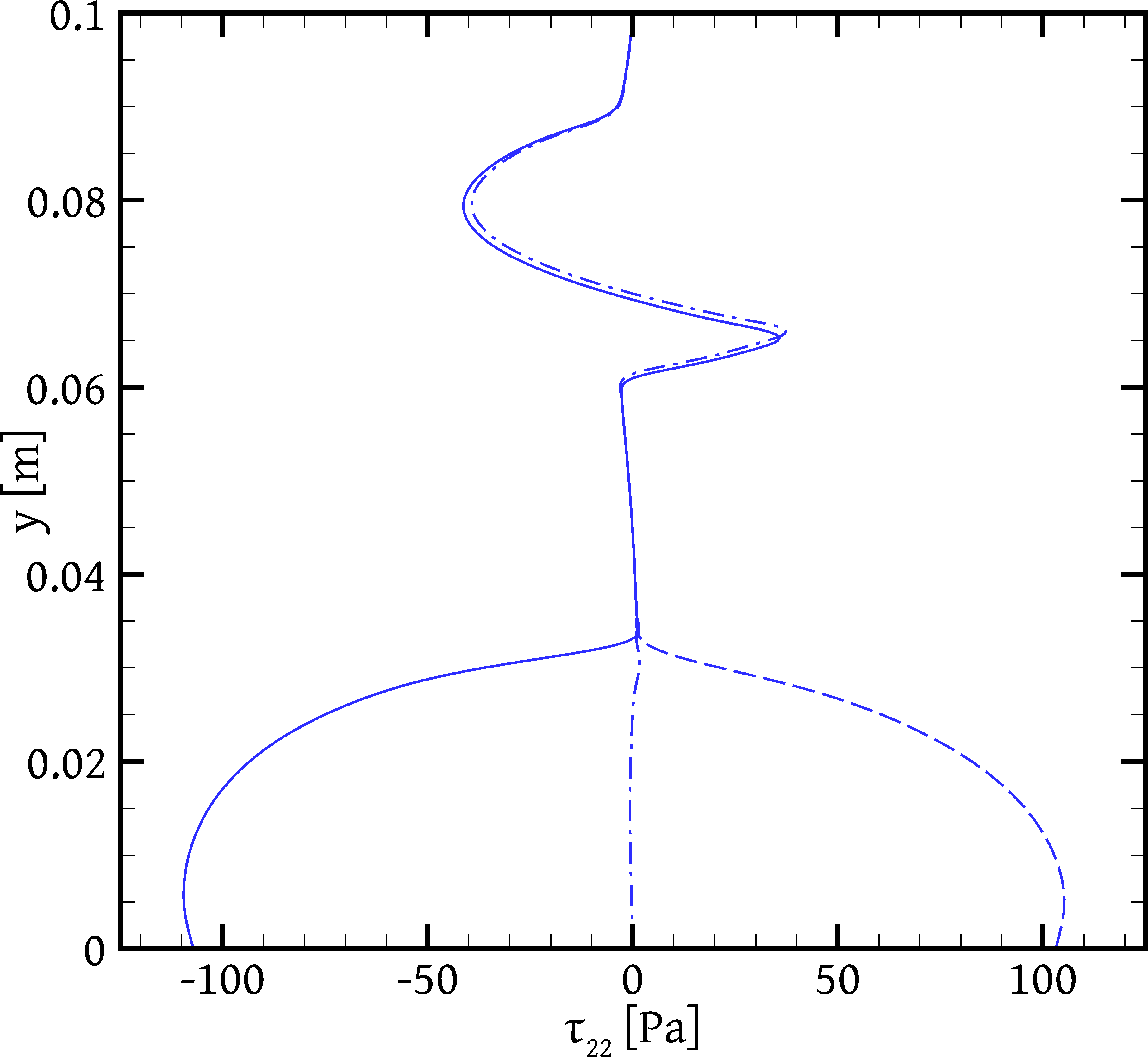}
        \caption{$\tau_{22}$}
        \label{sfig: multiplicity profiles s22}
    \end{subfigure}
    \begin{subfigure}[b]{0.32\textwidth}
        \centering
        \includegraphics[width=0.95\linewidth]{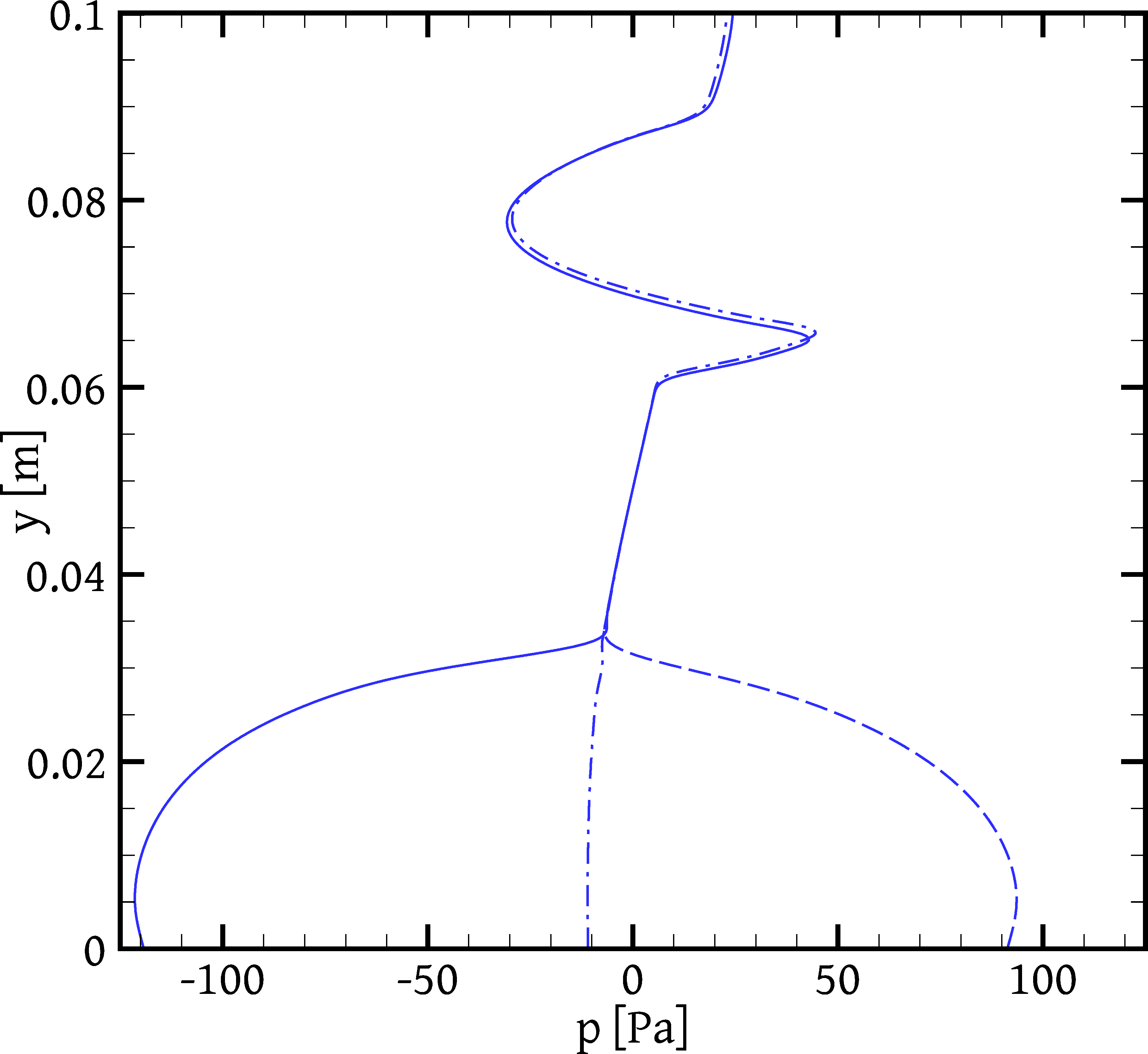}
        \caption{pressure}
        \label{sfig: multiplicity profiles p}
    \end{subfigure}
    \caption{Normal stress components \subref{sfig: multiplicity profiles s11}-\subref{sfig: 
multiplicity profiles s22} and pressure \subref{sfig: multiplicity profiles p} along the vertical 
centreline ($x = L/2$) for the cases with initial conditions $\tau_{11} = \tau_{22} = -\sqrt{3} 
\tau_y$ (continuous lines), $+\sqrt{3} \tau_y$ (dashed lines), and zero (dash-dot lines).}
  \label{fig: multiplicity profiles}
\end{figure}

Concerning the stress fields, we can note the following. Firstly, a striking difference between the 
new cases and the base case is that in the former the bottom unyielded zone is largely absent and is 
replaced by a large transition zone; this is shown in Fig.\ \ref{sfig: multiplicity tau_d} where the 
transition zone is approximately the green region, throughout which $\tau_d$ is slightly above 
$\tau_y$. Inside the transition region the $\tau_d$ fields of the two new cases are similar (Fig.\ 
\ref{sfig: multiplicity tau_d}), but the individual $\tau_{11}$ and $\tau_{22}$ stress components 
have opposite signs which they have inherited from the initial conditions, as seen in Fig.\ 
\ref{fig: multiplicity profiles}; in fact, near the bottom wall these stress components retain 
values close to the initial ones, $\pm \sqrt{3} \tau_y \approx 121$ \si{Pa}. On the other hand, the 
base case stresses are close to zero in that region, again an inheritance of the initial conditions. 
This reflects on the much lower stress trace values seen in Fig.\ \ref{fig: multiplicity monitor}. 
Outside of the transition region, where the rates of deformation are non-zero (including the plug 
zone), the two new cases have identical steady-state stress fields, as seen in Figs.\ \ref{sfig: 
multiplicity tau_d} and \ref{fig: multiplicity profiles}, whereas those of the base case deviate 
slightly. This includes the yield line that forms the boundary of the plug zone: Fig.\ \ref{sfig: 
multiplicity yield lines} shows that the plug zones of the two new cases are identical, but that of 
the base case is slightly smaller, which is likely the cause of the slightly different velocity 
field.

\subsection{Cessation}
\label{ssec: results: cessation}

Our final investigation concerns the sudden stopping of the lid and the study of the subsequent flow 
decay. Classic viscoplastic models are known to predict flow cessation in finite time after the 
driving agent is removed \cite{Frigaard_2019}. This can be proved rigorously using variational 
inequalities (e.g.\ upper bounds for the cessation times of some one-dimensional flows are derived 
in \cite{Glowinski_1984, Huilgol_2002, Muravleva_2010b}), but roughly speaking it is due to the 
effect of the yield stress on the rate of energy dissipation: it keeps it high enough during the 
flow decay such that all of the fluid's kinetic energy (KE) is dissipated in finite time -- a rough 
explanation is sketched in Appendix \ref{appendix: energy dissipation}. The cessation of 
viscoplastic (Bingham) lid-driven cavity flow was studied in \cite{Syrakos_2016a}, according to the 
findings of which we would expect HB flow to cease completely at some time $t_c \ll T_c \equiv \rho 
U L / S \approx$ 0.11 \si{s} ($T_c$ is the time needed for a force of magnitude $SL^2$ to bring a 
mass of momentum $\rho U L^3$ to rest).

\begin{figure}[tb]
    \centering
    \begin{subfigure}[b]{0.49\textwidth}
        \centering
        \includegraphics[scale=0.95]{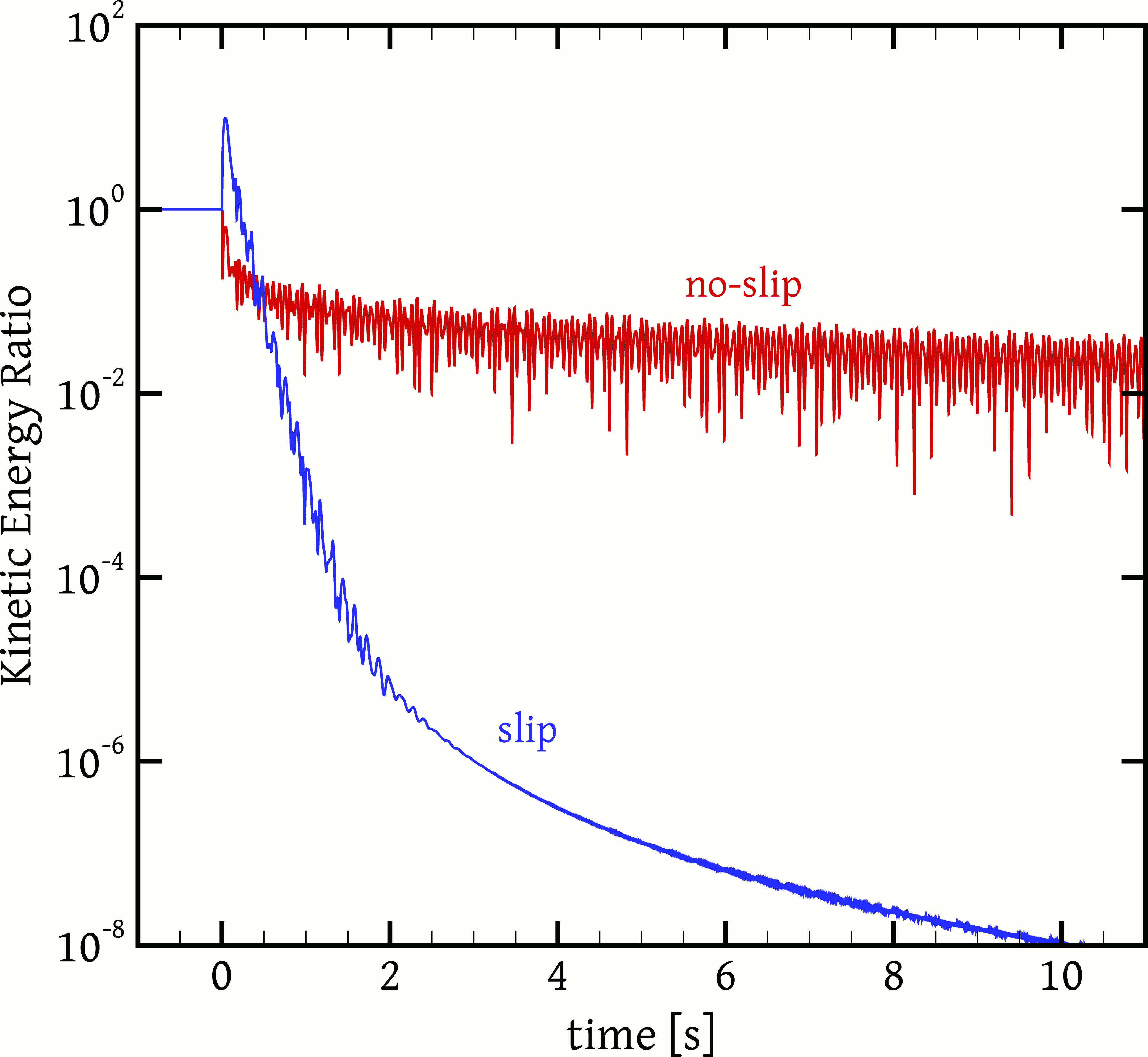}
        \caption{}
        \label{sfig: cessation monitor wide}
    \end{subfigure}
    \begin{subfigure}[b]{0.49\textwidth}
        \centering
        \includegraphics[scale=0.95]{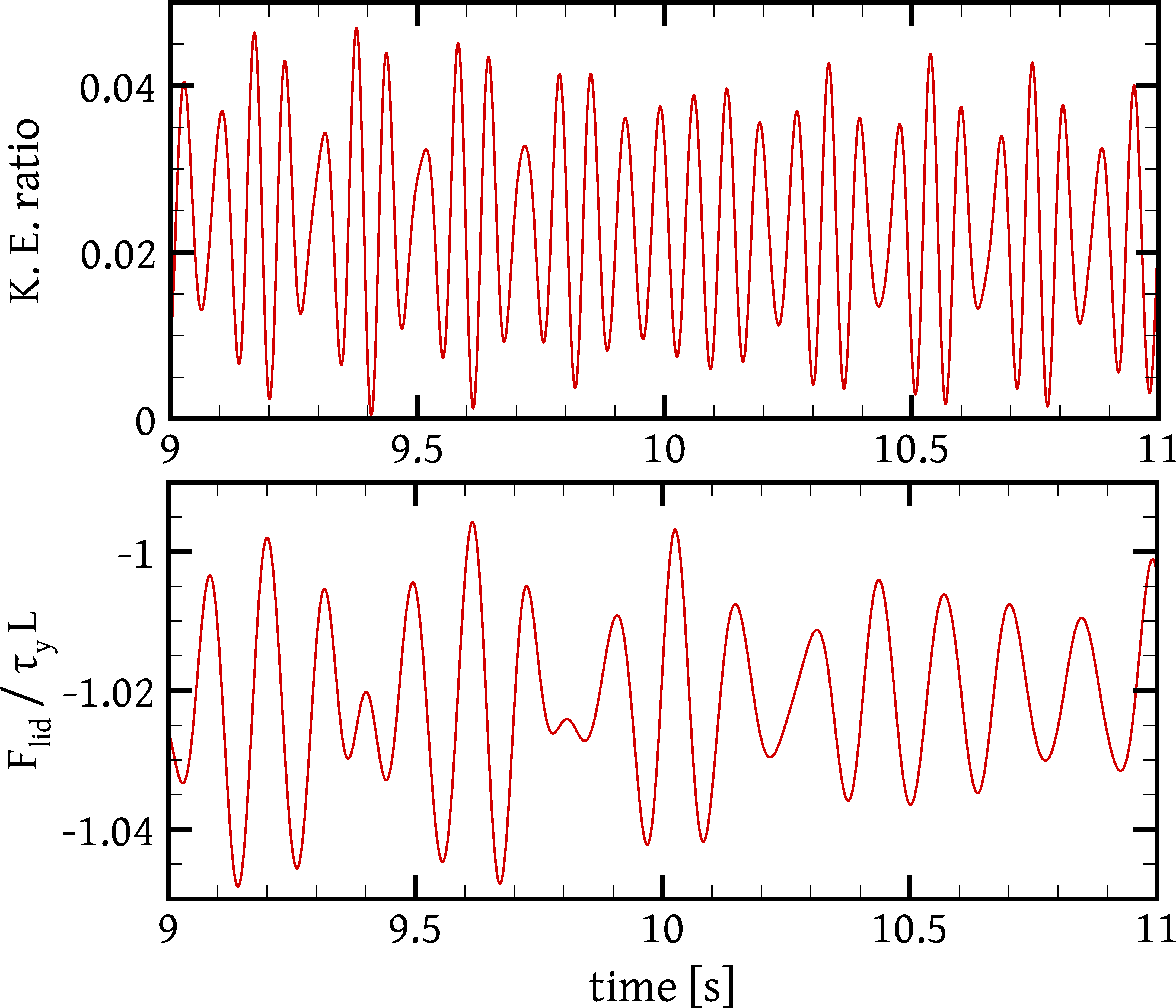}
        \caption{}
        \label{sfig: cessation monitor zoom} 
    \end{subfigure}
    \caption{\subref{sfig: cessation monitor wide}: Fluid kinetic energy (Eq.\ \eqref{eq: kinetic 
energy}), normalised by its value at the instant the lid is halted, versus time elapsed (the lid is 
halted at time $t$ = 0) for the no-slip and slip cases. \subref{sfig: cessation monitor zoom}: The 
top diagram is a close-up view of a portion of the no-slip curve in \subref{sfig: cessation monitor 
wide}. The bottom diagram plots the corresponding values of the force exerted by the fluid on the 
lid, normalised by $\tau_y L$.}
  \label{fig: cessation monitor}
\end{figure}

The situation concerning SHB flow is expected to be somewhat different. Firstly, even though the 
energy conversion rate $\int_{\Omega} \tf{\tau}\!:\!\nabla \vf{u} \: \mathrm{d}\Omega$ is still 
expected to be large enough to convert all the KE of the material in finite time, this rate now 
includes not only energy dissipation, but also energy storage in the form of elastic (potential) 
energy, which can later be converted back into KE \cite{Winter_1987}. In fact, once all of the 
material becomes unyielded there is no mechanism for energy dissipation and it is expected that the 
remaining energy will perpetually change form from elastic to kinetic and vice versa, resulting in 
oscillatory motions. Furthermore, the formation of transition regions in previous simulations makes 
it uncertain whether all of the material will become unyielded in finite time. To investigate these 
issues, using the ``steady-state'' base flow (Sec.\ \ref{ssec: results: base case}) as initial 
condition, we suddenly stopped the lid motion and carried out a simulation to see how the flow 
evolves. We repeated the procedure also with the slip case of Sec.\ \ref{ssec: results: slip}. 
Figure \ref{sfig: cessation monitor wide} shows the evolution of the KE with time ($t$ = 0 is the 
instant the lid is halted). In neither case does the flow cease in finite time.

\begin{figure}[!t]
    \centering
    
    \begin{subfigure}[b]{0.245\textwidth}
        \centering
        \includegraphics[width=0.95\linewidth]{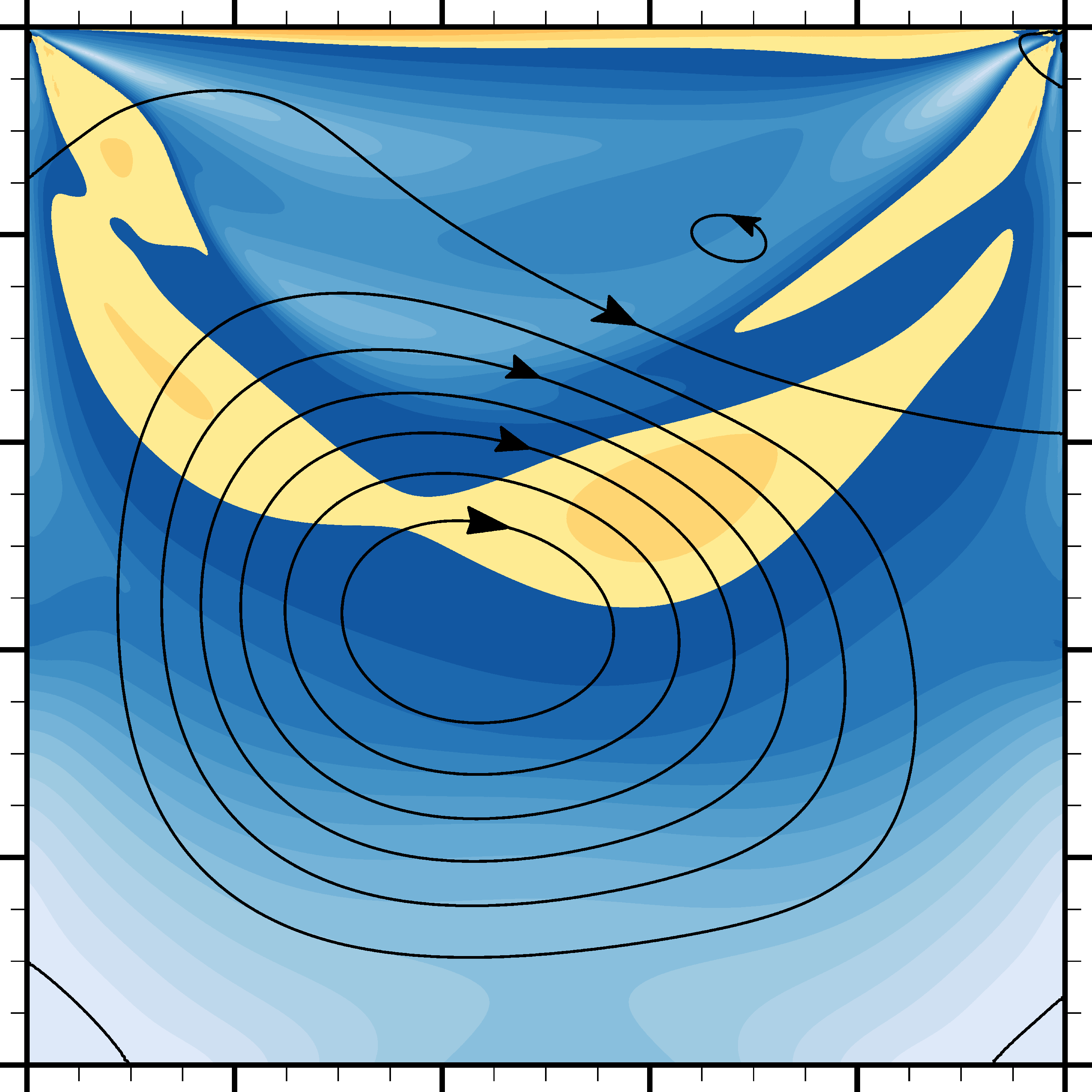}
        \caption{$t$ = 3 \si{s}}
        \label{sfig: cessation flowfield t=3}
    \end{subfigure}
    \begin{subfigure}[b]{0.245\textwidth}
        \centering
        \includegraphics[width=0.95\linewidth]{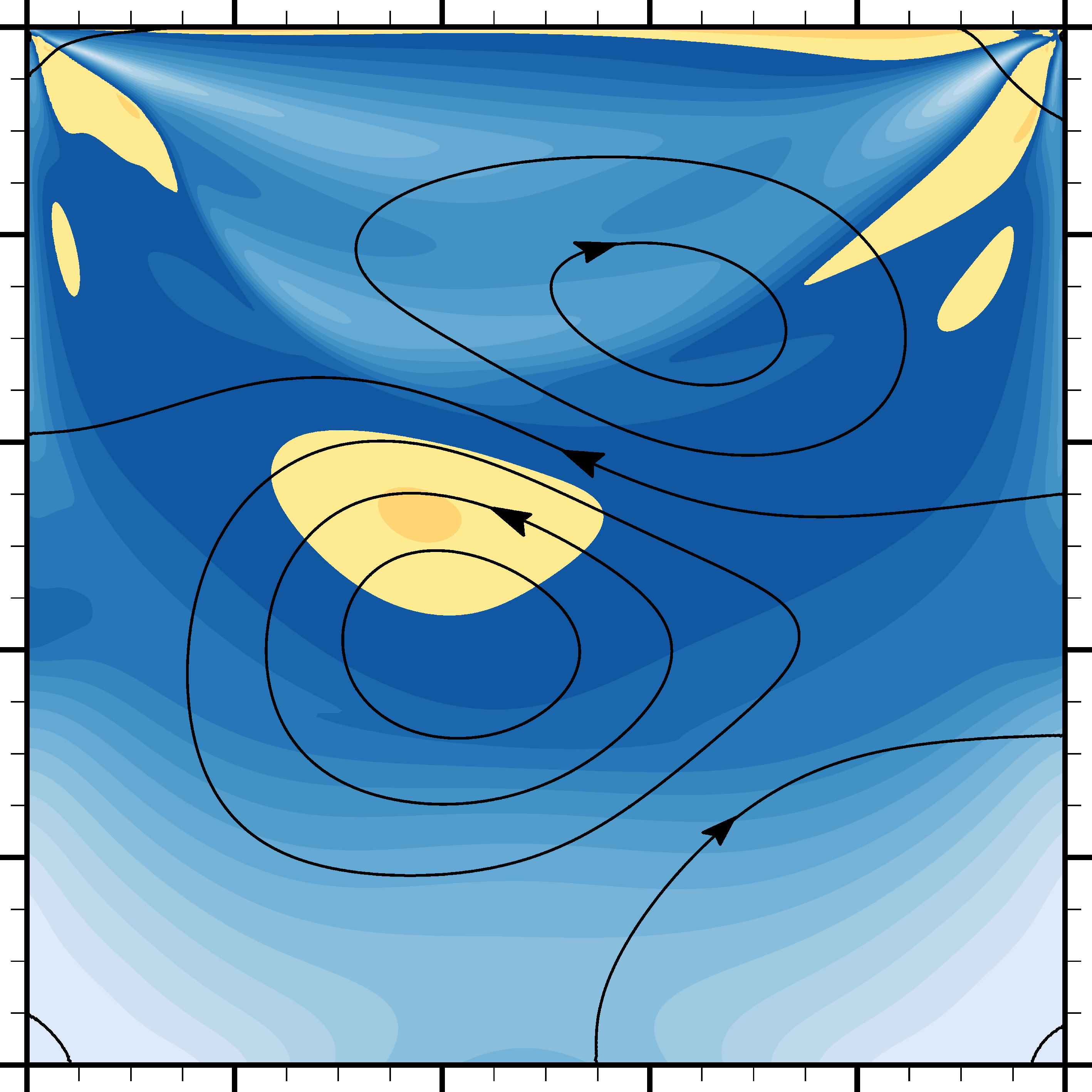}
        \caption{$t$ = 3.05 \si{s}}
        \label{sfig: cessation flowfield t=3.05}
    \end{subfigure}
    \begin{subfigure}[b]{0.245\textwidth}
        \centering
        \includegraphics[width=0.95\linewidth]{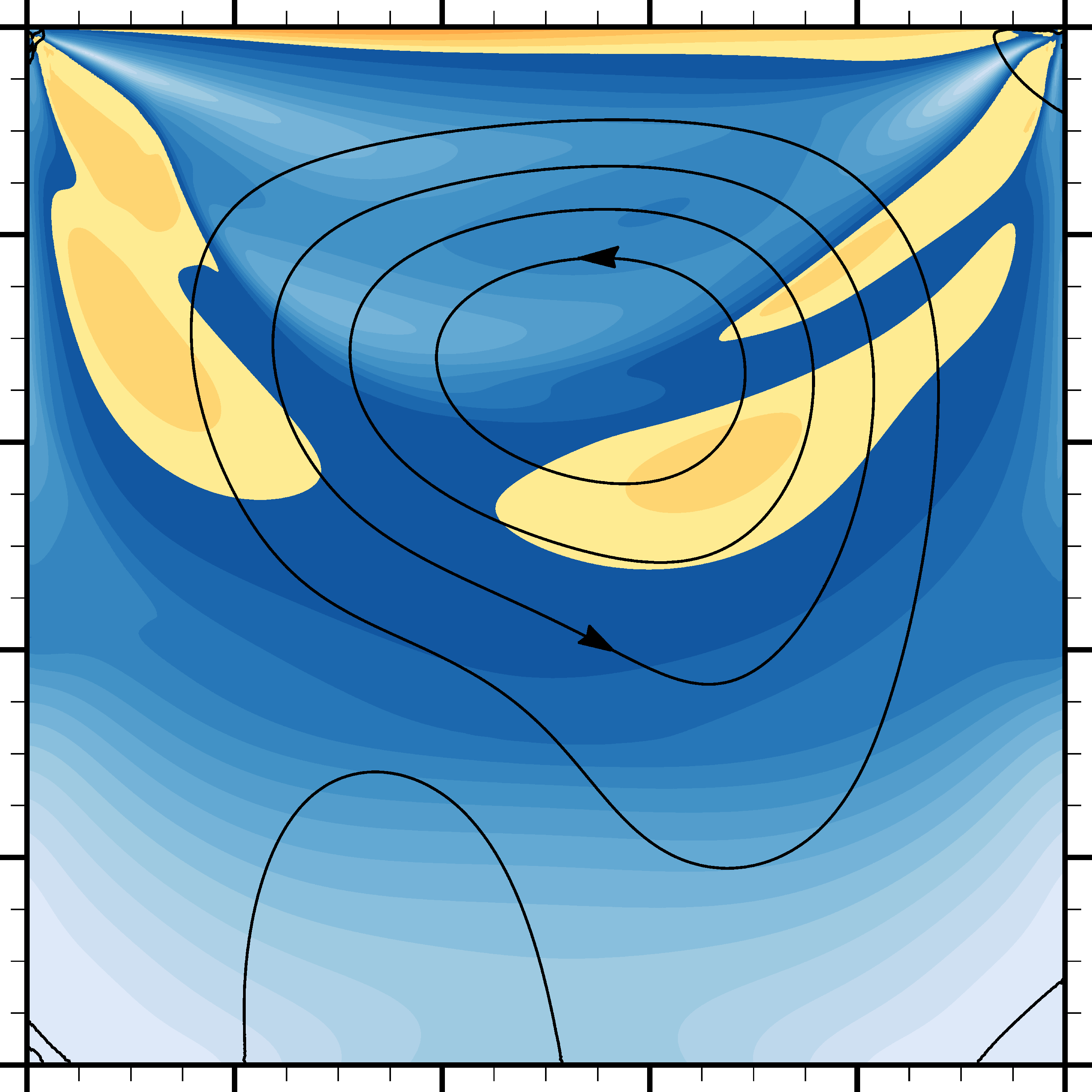}
        \caption{$t$ = 3.10 \si{s}}
        \label{sfig: cessation flowfield t=3.10}
    \end{subfigure}
    \begin{subfigure}[b]{0.245\textwidth}
        \centering
        \includegraphics[width=0.95\linewidth]{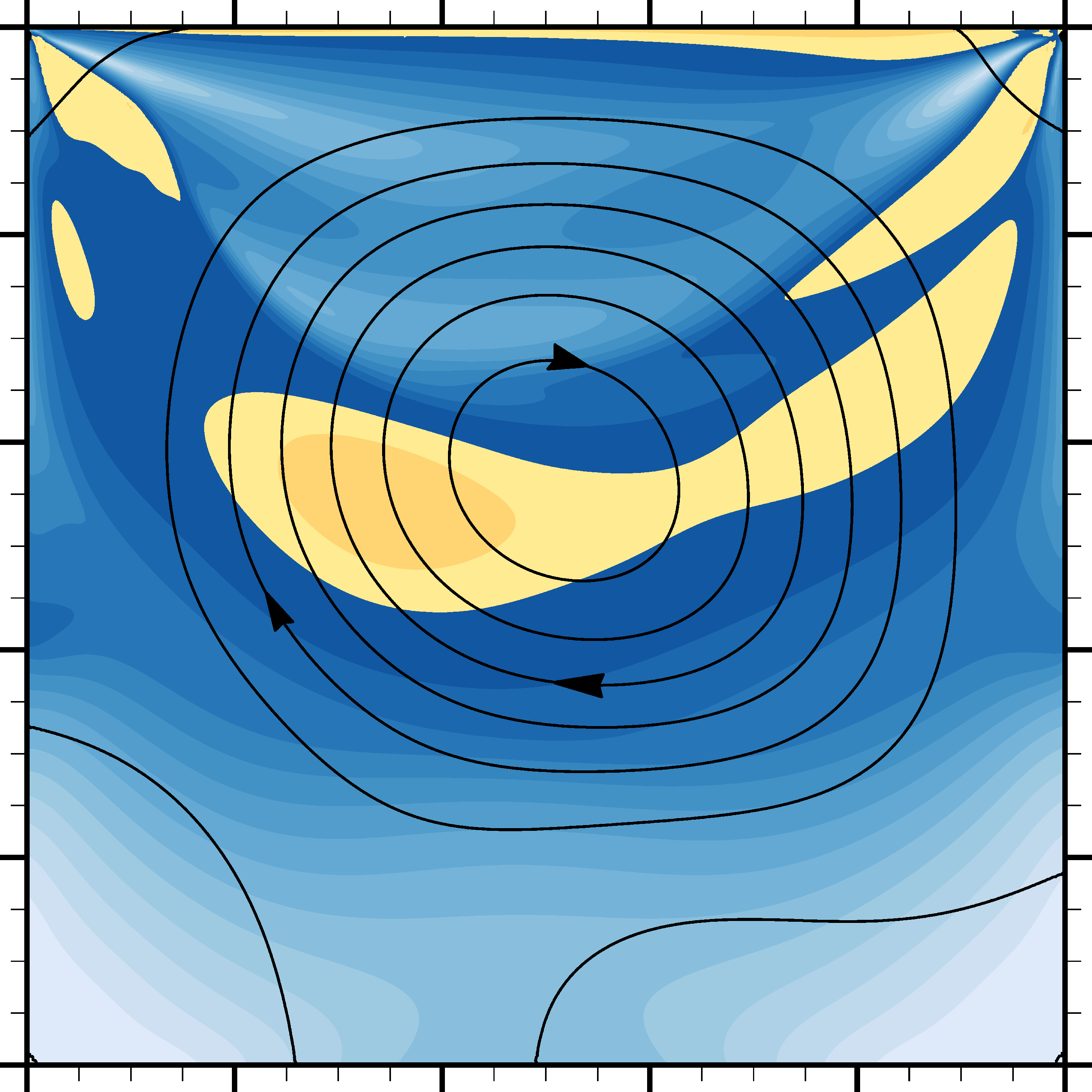}
        \caption{$t$ = 3.15 \si{s}}
        \label{sfig: cessation flowfield t=3.15}
    \end{subfigure}
    
    \vspace{0.25cm}
    
    \begin{subfigure}[b]{0.245\textwidth}
        \centering
        \includegraphics[width=0.95\linewidth]{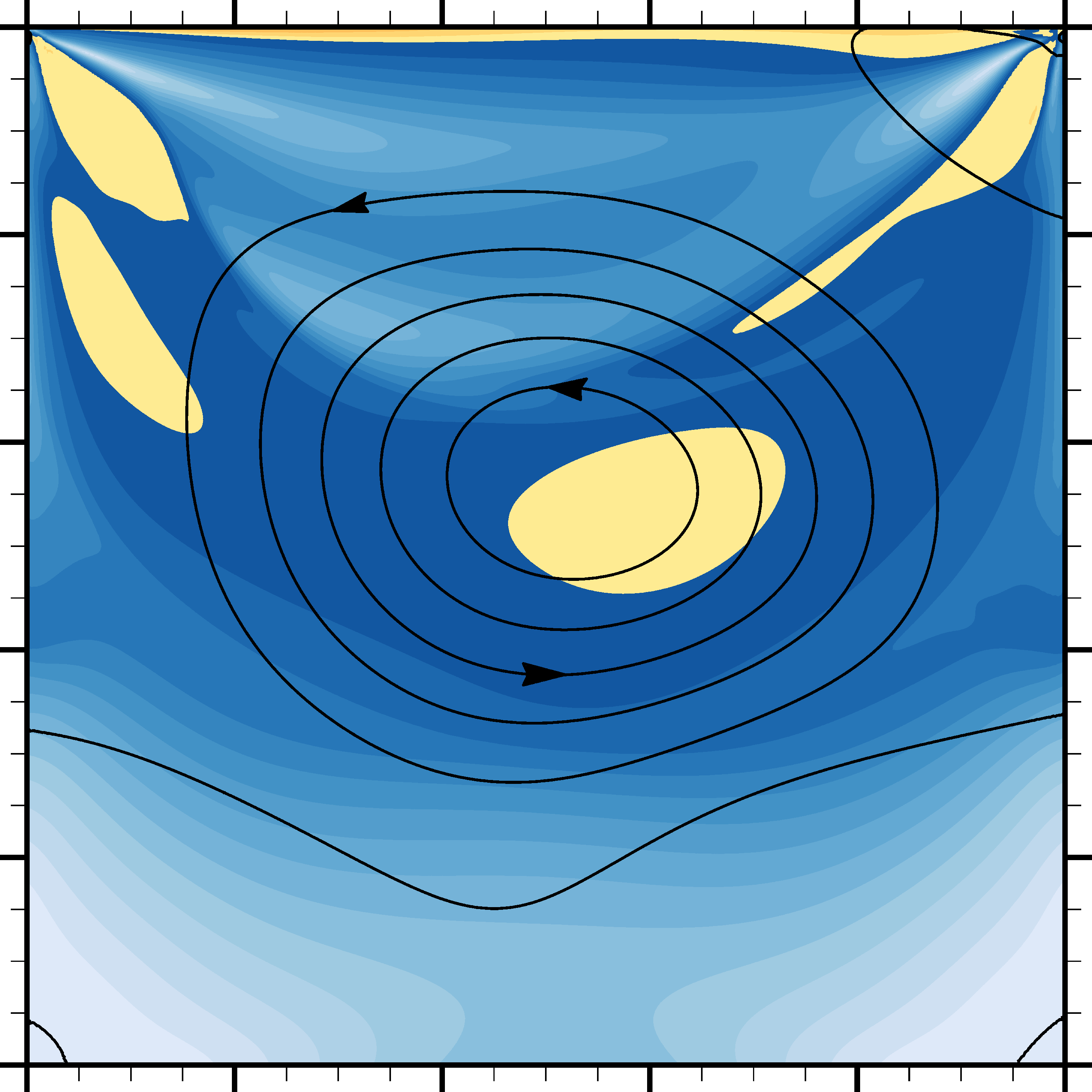}
        \caption{$t$ = 3.20 \si{s}}
        \label{sfig: cessation flowfield t=3.20}
    \end{subfigure}
    \begin{subfigure}[b]{0.245\textwidth}
        \centering
        \includegraphics[width=0.95\linewidth]{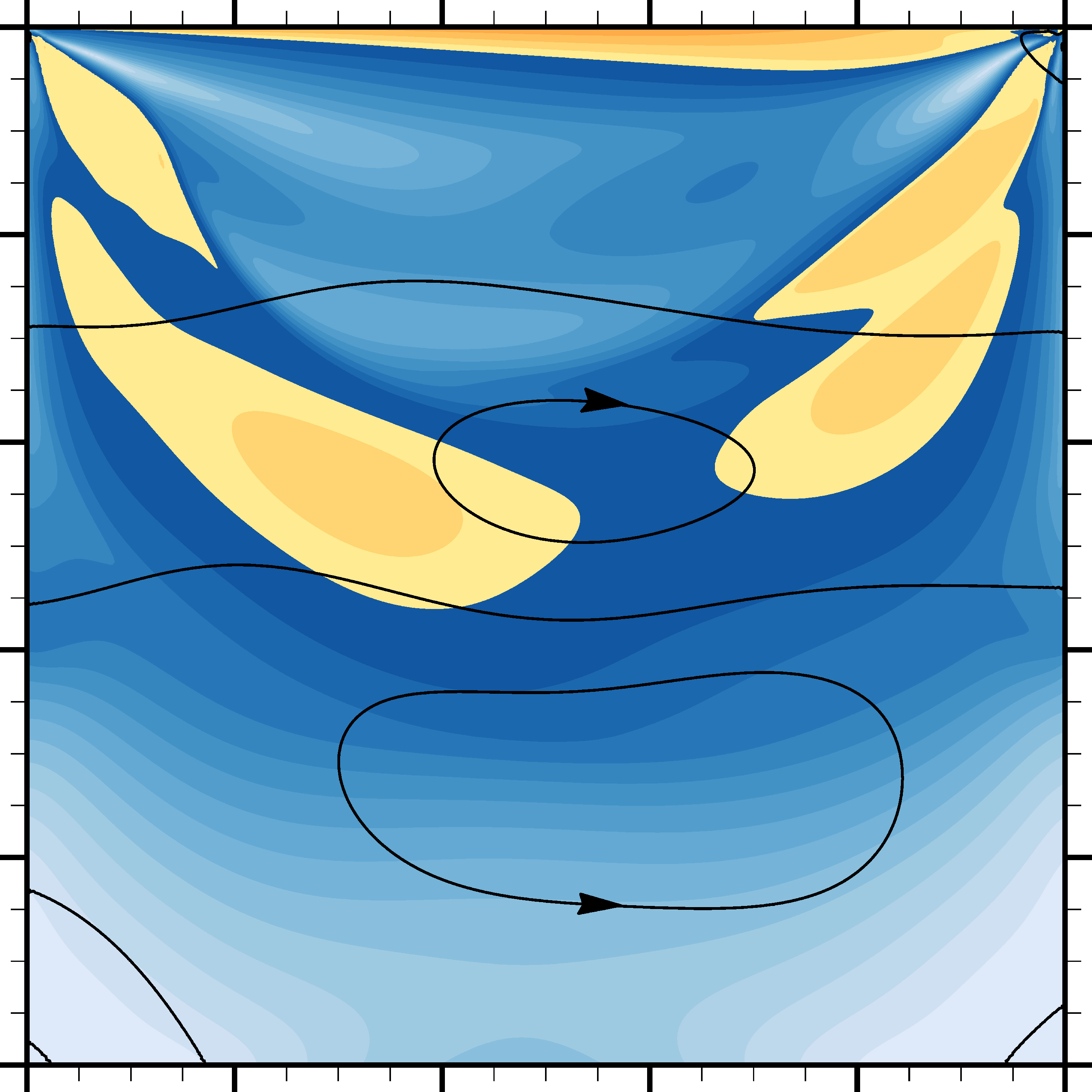}
        \caption{$t$ = 3.25 \si{s}}
        \label{sfig: cessation flowfield t=3.25}
    \end{subfigure}
    \begin{subfigure}[b]{0.245\textwidth}
        \centering
        \includegraphics[width=0.95\linewidth]{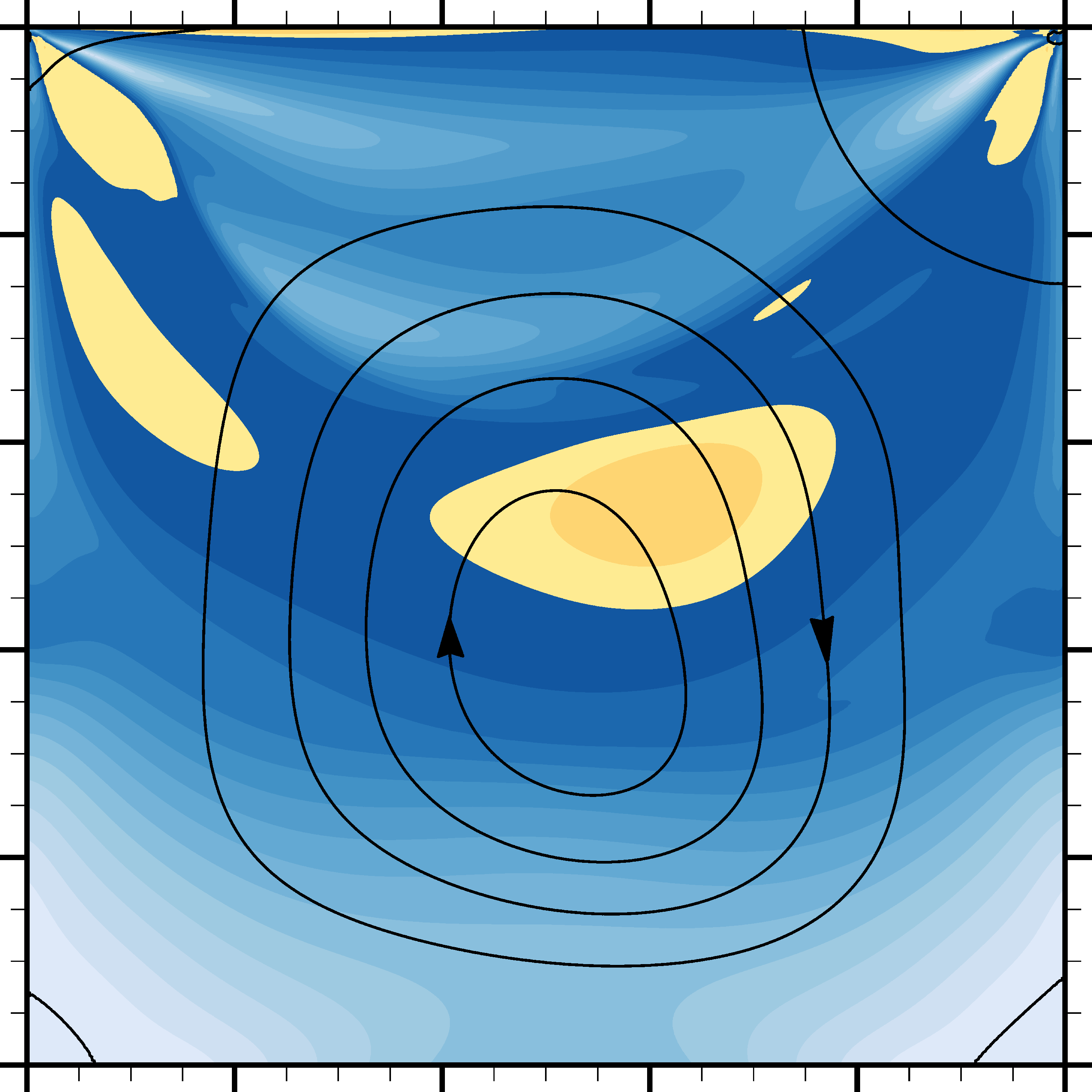}
        \caption{$t$ = 3.30 \si{s}}
        \label{sfig: cessation flowfield t=3.30}
    \end{subfigure}
    \begin{subfigure}[b]{0.245\textwidth}
        \centering
        \includegraphics[width=0.43\linewidth]{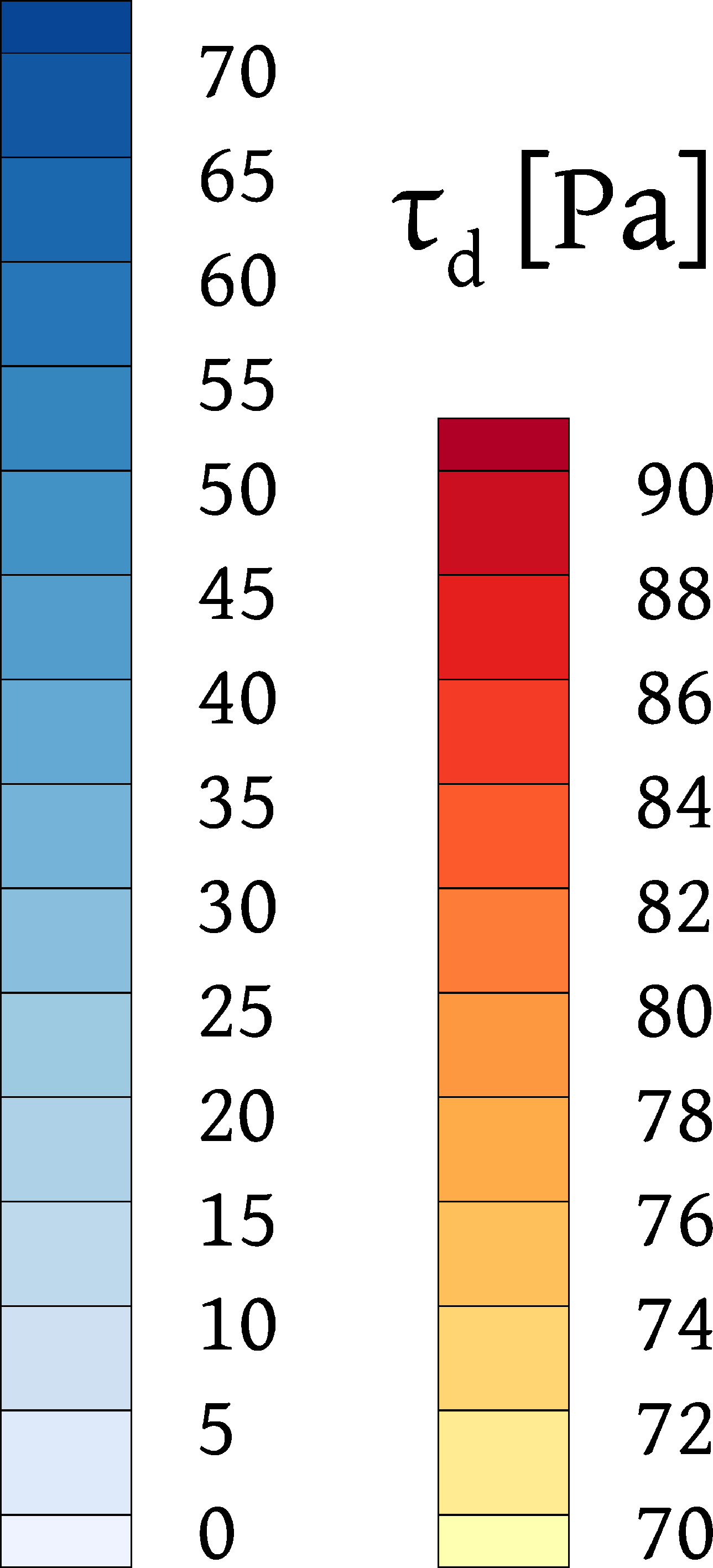}
        \caption{colour map}
        \label{sfig: cessation flowfields legend}
    \end{subfigure}

    \vspace{0.25cm}
    
    \begin{subfigure}[b]{0.49\textwidth}
        \centering
        \includegraphics[width=0.95\linewidth]
            {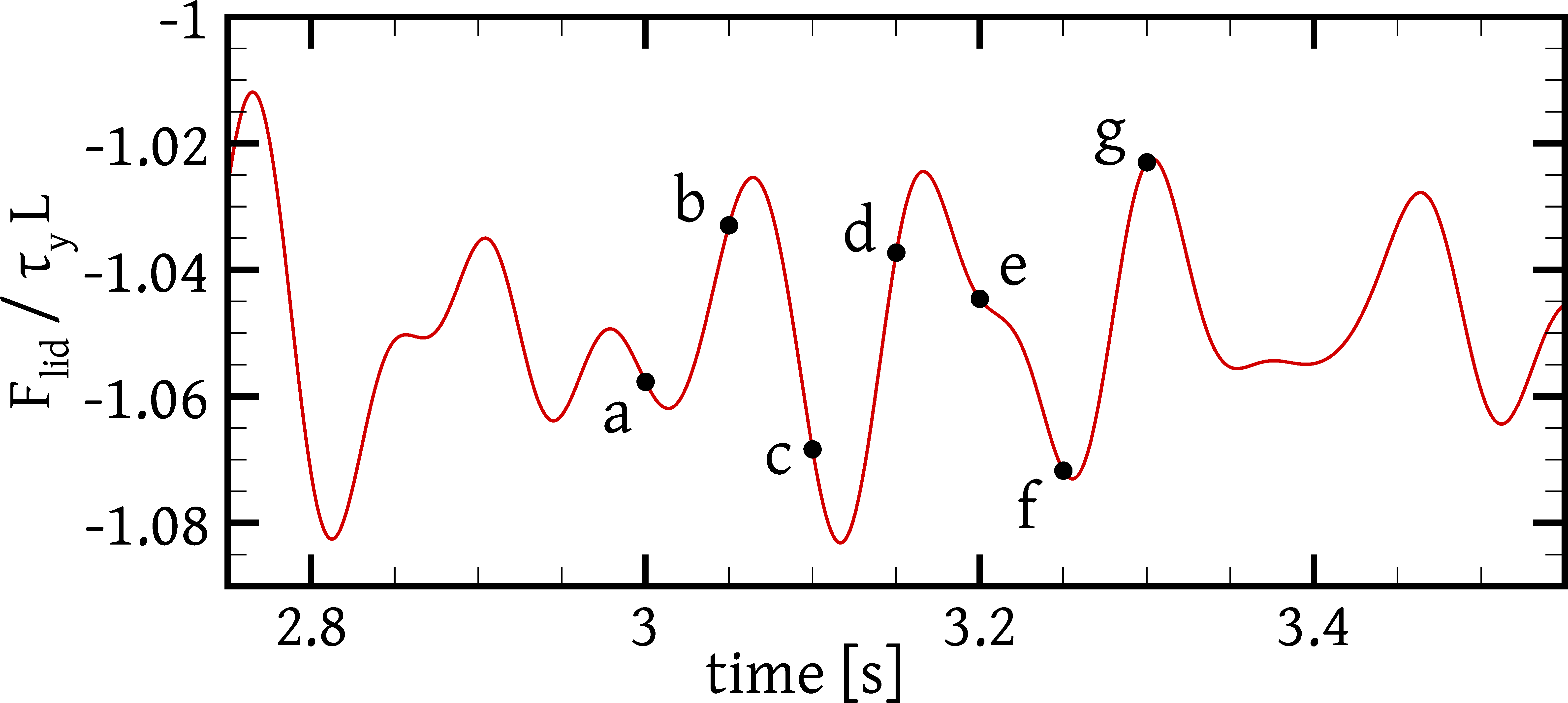}
        \caption{History of $F_{\text{lid}}$}
        \label{sfig: cessation flowfields Flid}
    \end{subfigure}
    \begin{subfigure}[b]{0.245\textwidth}
        \centering
        \includegraphics[width=0.95\linewidth]{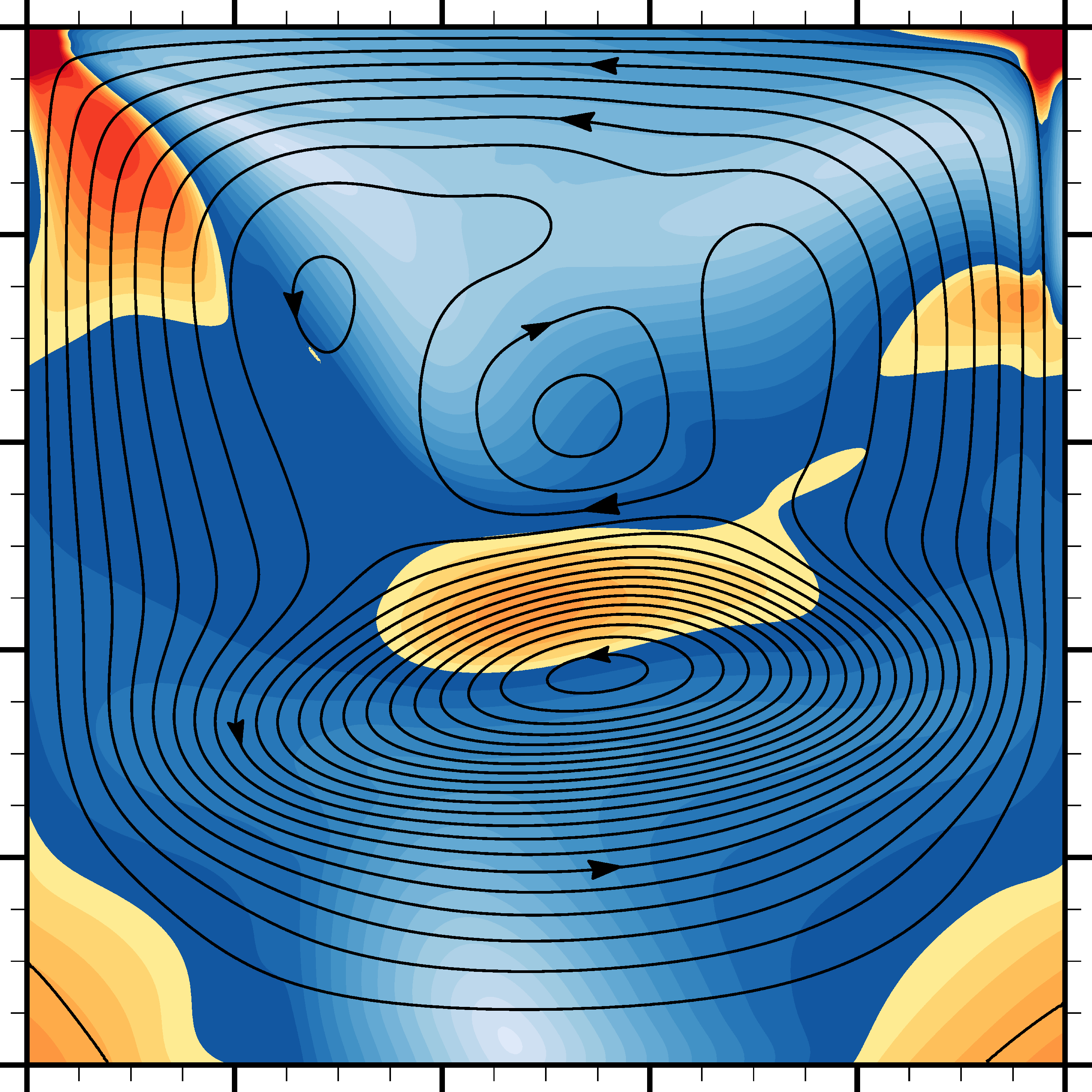}
        \caption{(slip) $t$ = 0.10 \si{s}}
        \label{sfig: cessation flowfield slip start}
    \end{subfigure}
    \begin{subfigure}[b]{0.245\textwidth}
        \centering
        \includegraphics[width=0.95\linewidth]{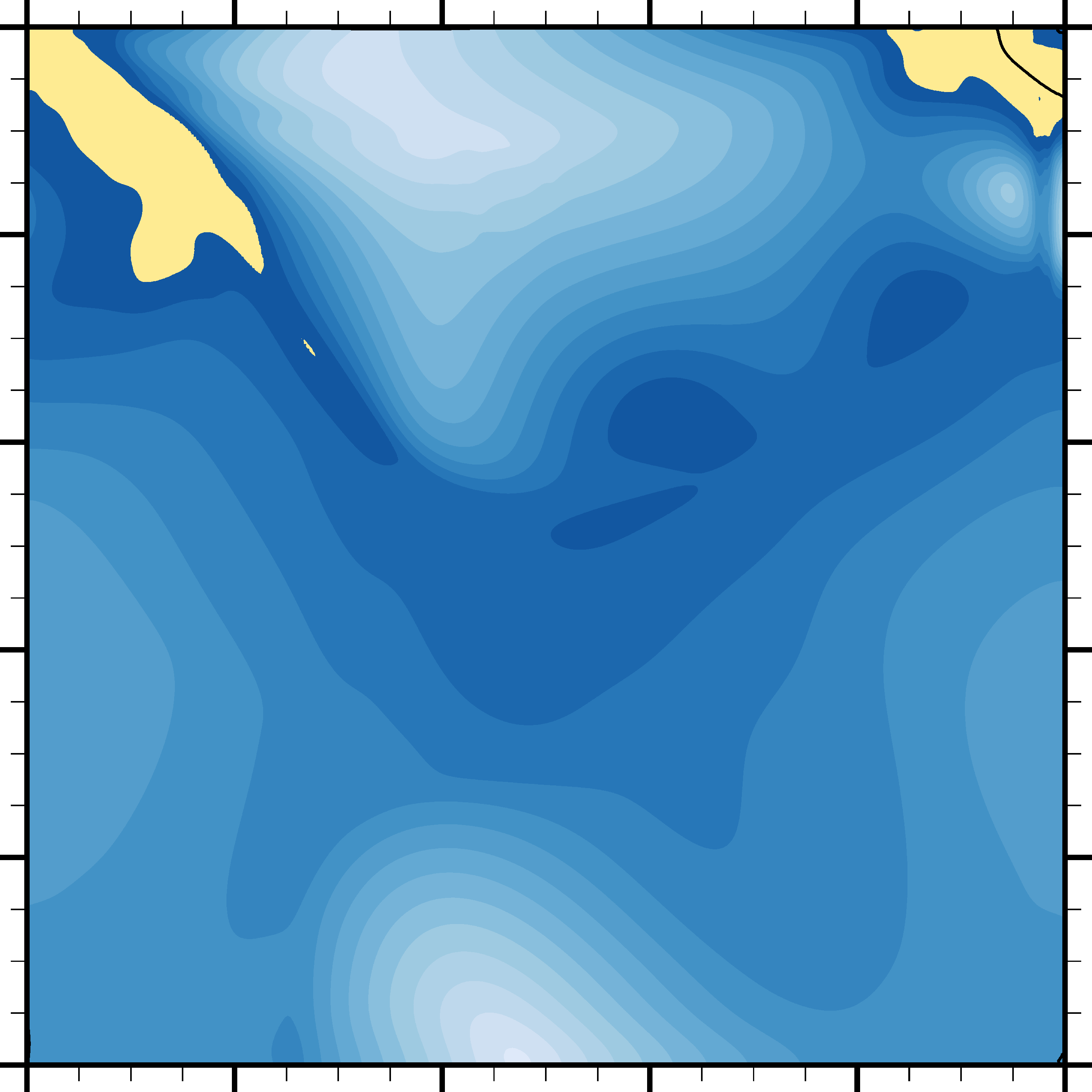}
        \caption{(slip) $t$ = 60 \si{s}}
        \label{sfig: cessation flowfield slip end}
    \end{subfigure}
    
    \caption{\subref{sfig: cessation flowfield t=3}-\subref{sfig: cessation flowfield t=3.30}: 
Snapshots of the no-slip cessation flow field, with colour contours of $\tau_d$, of different base 
colour for yielded (yellow) and unyielded (blue) regions (see the colour map \subref{sfig: cessation 
flowfields legend}), and instantaneous streamlines plotted at streamfunction intervals of $\delta 
\psi / LU = 0.002$, with $\psi = 0$ as one of the contours. \subref{sfig: cessation flowfields 
Flid}: Part of the history of the force exerted by the fluid on the lid, normalised by $\tau_y L$, 
with the instants corresponding to snapshots \subref{sfig: cessation flowfield t=3}-\subref{sfig: 
cessation flowfield t=3.30} marked on the plot. \subref{sfig: cessation flowfield slip 
start}-\subref{sfig: cessation flowfield slip end}: Snapshots of the slip cessation flow field.}
  \label{fig: cessation flowfields}
\end{figure}

In the no-slip case, the KE history is highly oscillatory, confirming the anticipated perpetual 
back-and-forth conversion between kinetic and elastic energies. The top diagram of Fig.\ \ref{sfig: 
cessation monitor zoom} shows a clearer picture over a narrower time window. The KE peaks 29 times 
within that 2 \si{s} window, i.e.\ a single oscillation lasts about 69 \si{ms}; the time step is 
kept to about $\Delta t$ = \num{2e-4} \si{s} by the adjustable time step scheme, so each KE 
oscillation period is resolved into about 350 time steps. The diagram shows that the KE variation 
does not consist of a single harmonic component, and Fig.\ \ref{fig: cessation flowfields} shows 
that the flow is indeed quite complex. Usually at any given instant there appears one main vortex 
rotating in either the clockwise or anticlockwise sense, but its location and shape are not fixed, 
while smaller vortices may also appear. The material is not completely unyielded, and the yielded 
zones are transported along as the material oscillates while their size varies with time. The 
existence of these yielded zones allows some energy dissipation, and hence the mean KE in Fig.\ 
\ref{fig: cessation monitor} very slowly drops. The bottom diagram of Fig.\ \ref{sfig: cessation 
monitor zoom} shows the force $F_{\mathrm{lid}}$ exerted by the fluid on the lid. This force is 
negative, pushing the lid towards the left, i.e. in the opposite direction than it was moving prior 
to its halting. The magnitude of the force oscillates slightly above the value $\tau_y L$, and an 
inspection shows that this is because the magnitude of $\tau_{12}$ on the lid slowly drops towards 
$\tau_y$ (the lid is in touch with a transition zone). Figure \ref{sfig: cessation flowfields Flid} 
shows that the magnitude of $F_{\mathrm{lid}}$ drops when the lid touches a clockwise vortex (Figs.\ 
\ref{sfig: cessation flowfield t=3.05}, \ref{sfig: cessation flowfield t=3.15}, \ref{sfig: cessation 
flowfield t=3.30}) and it increases when it touches a counter-clockwise vortex (Figs.\ \ref{sfig: 
cessation flowfield t=3.10}, \ref{sfig: cessation flowfield t=3.20}). A comparison between the KE 
and $F_{\mathrm{lid}}$ diagrams in Fig.\ \ref{sfig: cessation monitor zoom} shows that 
$F_{\mathrm{lid}}$ oscillates at a frequency that is roughly half of that of the KE, which is 
expected since during a single $F_{\mathrm{lid}}$ period both the clockwise and anticlockwise vortex 
velocities are maximised, which leads to two KE peaks.

In the slip case, the KE peaks immediately after the lid halt (Fig.\ \ref{sfig: cessation monitor 
wide}) but then the oscillations die out relatively quickly and the KE diminishes. The KE peak is 
due to the material recoiling right after the lid halts (Fig.\ \ref{sfig: cessation flowfield slip 
start}), which is possible as the walls provide limited resistance, due to slip. Then, the KE is 
dissipated through a mechanism that is absent in the no-slip case: as the material oscillates inside 
the cavity, it slides past the walls due to slip, and the friction at the wall / material interface 
dissipates the KE. Even in this case, however, transition zones exist long after the lid has halted 
(Fig.\ \ref{sfig: cessation flowfield slip end}).

\section{Conclusions}
\label{sec: conclusions}

This work presented a FVM for elastoviscoplastic flows described by the SHB model. The method is 
applicable to collocated structured and unstructured meshes; it incorporates a new variant of 
momentum interpolation, a variant of ``two sides diffusion'', and the CUBISTA scheme, in order to 
suppress spurious pressure, velocity, and stress oscillations, respectively. The method was shown 
to achieve this goal, and also to be consistent on both smooth and irregular meshes. Similar 
stabilisation techniques have recently been applied in a Finite Element context to allow use of 
equal-order polynomial basis functions for all variables \cite{Varchanis_2019}. An implicit 
temporal discretisation with adaptive time step is employed.

The method was applied to the simulation of EVP flow in a lid-driven cavity, with the SHB parameters 
chosen so as to represent Carbopol. The results can serve as benchmark solutions, but furthermore 
the simulations were designed so as to allow investigation of several aspects of the SHB model 
behaviour. Complementary simulations with the classic HB model were also performed for comparison. 
It was noticed that the SHB model can predict ``transition zones'', regions where the velocity is 
near-zero and it takes a very long, or infinite, time to transition from a formally yielded to an 
unyielded state. The differences between the SHB and HB velocity fields are rather small for the lid 
velocities tested, although some elastic effects are noticeable in the SHB case, such as a slight 
upstream displacement of the vortex, a downstream stretching of the plug zone, and a small swelling 
of the bottom unyielded zone, compared to the HB results. These differences diminish by increasing 
the elastic modulus $G$, although the convergence of the SHB and HB yield lines is rather slow and 
requires very large values of $G$. The velocity fields converge much faster. We also applied slip at 
the walls, as EVP materials are slippery, in which case no transition zones developed.

Motivated by the observation of Cheddadi et al.\ \cite{Cheddadi_2012} that different initial 
conditions of stress can lead to different steady states even as concerns the velocity, we 
performed two additional simulations where the initial state of the material was stationary but on 
the verge of yielding, with either compressive or tensile residual stresses, respectively. Indeed, 
although both of these simulations converged to the same steady-state velocity field, that field was 
slightly but noticeably different than that obtained when the initial stresses were zero. Due to 
this property of the SHB model, an accurate calculation of a steady state requires not only 
sufficient spatial resolution (grid spacing) but also sufficient temporal resolution (time step 
size), which is facilitated by the use of an adaptive time discretisation scheme such as the one 
proposed herein.

Finally, simulations of the cessation of the flow after the lid is suddenly halted were performed, 
which showed that, unlike what is predicted by the HB model, the flow does not cease in finite time 
but there is a perpetual back-and-forth conversion between kinetic and elastic energies. In the 
no-slip case, a very slow net energy dissipation was observed, which is possible due to the 
persisting existence of yielded regions. The energy dissipation is much faster under wall slip, due 
to friction between the material and the walls.

Incorporation of additional rheological phenomena such as thixotropy is planned for the future.


\section*{Acknowledgements}
This research was funded by the LIMMAT Foundation under the Project ``MuSiComPS''.



\begin{appendices}
\renewcommand\theequation{\thesection.\arabic{equation}}
\setcounter{equation}{0}

\section{The SHB extra stress tensor in the limit of infinite elastic modulus}
\label{appendix: SHB tau in limit of large G}

The SHB constitutive equation \eqref{eq: constitutive} distinguishes between $\tf{\tau}$ and 
$\tf{\tau}{}_d$, unlike the HB equation \eqref{eq: HB rate-of-strain}. For the two models to become 
equivalent as $G \rightarrow \infty$ it is necessary that $\tf{\tau} \rightarrow \tf{\tau}{}_d$ or, 
equivalently, $\mathrm{tr}(\tf{\tau}) \rightarrow 0$.

In the limiting case $G \rightarrow \infty$ the first term on the left-hand side of Eq.\ \eqref{eq: 
constitutive} diminishes. This makes yielded material (where the ``max'' term is non-zero) tend to 
behave as a generalised Newtonian fluid so that $\mathrm{tr}(\tf{\tau}) \rightarrow 0 \Rightarrow 
\tf{\tau} \rightarrow \tf{\tau}{}_d$ and the two models do become equivalent in yielded regions. On 
the other hand, in unyielded regions (where the ``max'' term is zero) Eq.\ \eqref{eq: constitutive} 
tends to reduce to $\dot{\tf{\gamma}} = 0$, which does not, at first glance, imply that $\tf{\tau} 
\rightarrow \tf{\tau}{}_d$. However, the following observation can be made. In an unyielded region 
the SHB model can be written as (substituting Eq.\ \eqref{eq: upper convected derivative} into Eq.\ 
\eqref{eq: constitutive})
\begin{equation} \label{eq: SHB in unyielded region}
 \frac{D\tf{\tau}}{Dt} \;\equiv\; \pd{\tf{\tau}}{t} \;+\; \vf{u}\cdot \nabla \tf{\tau}
 \;=\; G \dot{\tf{\gamma}}
 \;+\; G \left( 
       (\nabla \vf{u})^{\mathrm{T}} \cdot \tf{\tau} \;+\;
       \tf{\tau} \cdot \nabla \vf{u}
       \right)
\end{equation}
The material derivative $D\tf{\tau}/Dt$ is the rate of change of the stress tensor with respect to 
time in a fluid particle moving with the flow. This derivative is applied component-wise to the 
stress tensor, and therefore the rate of change of $\mathrm{tr}(\tf{\tau})$ is
\begin{equation} \label{eq: rate of change of trace}
 \frac{D(\mathrm{tr}(\tf{\tau}))}{Dt} \;=\; \mathrm{tr}\left( \frac{D\tf{\tau}}{Dt} \right)
\end{equation}
or, using Eq.\ \eqref{eq: SHB in unyielded region},
\begin{equation} \label{eq: rate of change of trace 2}
 \frac{D(\mathrm{tr}(\tf{\tau}))}{Dt}
 \;=\;
 G \: \mathrm{tr}(\dot{\tf{\gamma}})
 \;+\;
 G \: \mathrm{tr} \left( 
       (\nabla \vf{u})^{\mathrm{T}} \cdot \tf{\tau} \;+\;
       \tf{\tau} \cdot \nabla \vf{u}
       \right)
\end{equation}
In the right-hand side, $\mathrm{tr}(\dot{\tf{\gamma}}) \rightarrow 0$ because in the unyielded 
regions $G \rightarrow \infty$ forces $\dot{\tf{\gamma}} \rightarrow 0$. For the second term in 
the right-hand side, in index notation, we have
\begin{equation}
 (\nabla \vf{u})^{\mathrm{T}} \cdot \tf{\tau} \;+\; \tf{\tau} \cdot \nabla \vf{u}
 \;=\;
 \sum_{i=1}^3 \sum_{j=1}^3 \left(
   \sum_{k=1}^3 \pd{u_i}{x_k}\tau_{kj}
   \;+\;
   \sum_{k=1}^3  \pd{u_j}{x_k}\tau_{ik}
 \right) \vf{e}_i \vf{e}_j
\end{equation}
where $\vf{e}_i$ is the unit vector in the direction $i$. Taking the trace of the above 
expression we obtain
\begin{equation} \label{eq: trace of upper convective terms}
 \mathrm{tr}\left( 
   (\nabla \vf{u})^{\mathrm{T}} \cdot \tf{\tau} \;+\; \tf{\tau} \cdot \nabla \vf{u} 
 \right)
 \;=\;
 \sum_{i=1}^3 \sum_{j=1}^3 \left( \pd{u_j}{x_i} \,+\, \pd{u_i}{x_j} \right) \tau_{ij}
\end{equation}
The terms in parentheses in the sum \eqref{eq: trace of upper convective terms} are just the 
components of $\dot{\tf{\gamma}}$, and so they tend to zero as $G \rightarrow \infty$. Thus the 
whole right-hand side of Eq.\ \eqref{eq: rate of change of trace 2} tends to zero in the unyielded 
regions as $G \rightarrow \infty$. Therefore, under these conditions, $D(\mathrm{tr}(\tf{\tau}))/Dt 
\rightarrow 0$, i.e.\ $\mathrm{tr}(\tf{\tau})$ remains constant in any particle moving in an 
unyielded region. Given that $\mathrm{tr}(\tf{\tau}) = 0$ for yielded particles, as mentioned above, 
there appears to be no mechanism for $\mathrm{tr}(\tf{\tau})$ to acquire any value different than 
zero. Thus we expect that $\tf{\tau} \rightarrow \tf{\tau}{}_d$ also in the unyielded regions. But 
even if $\mathrm{tr}(\tf{\tau}) \neq 0 \Rightarrow \tf{\tau} \neq \tf{\tau}{}_d$ in such a region, 
for practical purposes the stress field would still be equivalent to a HB stress field where the 
isotropic part of $\tf{\tau}$ is incorporated into the pressure.

\section{Effect of grid refinement on the pressure stabilisation scheme}
\label{appendix: pressure stabilisation}
\setcounter{equation}{0}

\begin{figure}[tb]
  \centering
  \includegraphics[scale=1.]{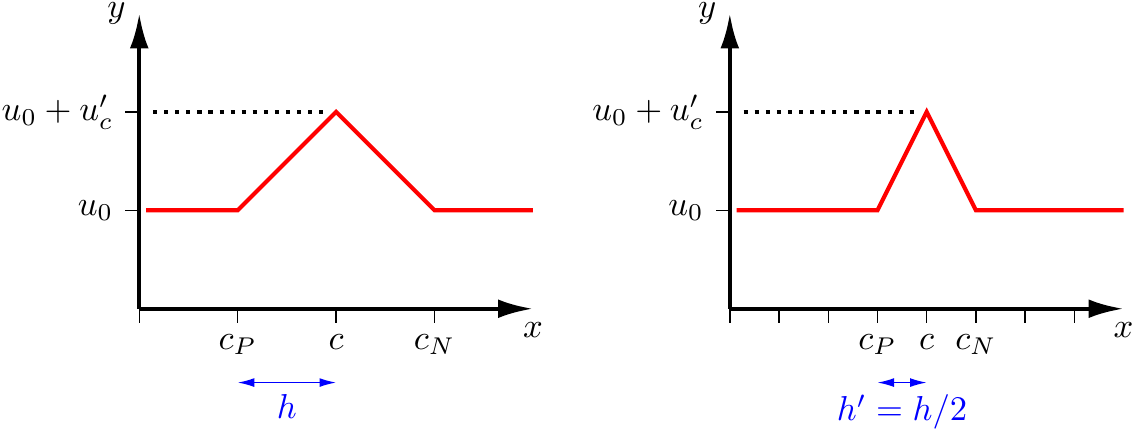}
  \caption{A local velocity perturbation $u'_c$ with wavelength equal to the grid spacing $h$ 
causes local perturbations $\mathrm{d}u'/\mathrm{d}x$ to the velocity gradient that are 
proportional to $u'_c/h = O(h^{-1})$; it follows that the corresponding perturbations to the second 
derivative of velocity $\mathrm{d}^2u'/\mathrm{d}x^2$ are proportional to $O(h^{-1})/h = 
O(h^{-2})$.}
  \label{fig: velocity perturbations}
\end{figure}

It follows from the definition of $a_f^{mi}$ \eqref{eq: ami} that as the grid is refined ($h_f 
\rightarrow 0$) the viscous term in the denominator dominates over the inertial one: at fine grids 
$a_f^{mi} \approx h_f / (2\kappa + 2\eta_a)$. This can be explained with the aid of the simple 
one-dimensional example used for deriving the scheme. From Eq.\ \eqref{eq: ami 1D}, if we want to 
locally perturb a velocity field at point $c$ by $u'_c$ (Fig.\ \ref{fig: velocity perturbations}) by 
locally perturbing the pressure gradient by $-\mathrm{d}p'/\mathrm{d}x|_c \approx (p'_P-p'_N)/h$ 
then these perturbations are related by
\begin{equation} \label{eq: ami 1D analysis}
 u'_c \;=\; \frac{1}{
 \rho \left( \left. \dfrac{\mathrm{d}u}{\mathrm{d}x} \right|_c \:+\; \dfrac{1}{\Delta t} \right)
 \;+\; \dfrac{2\eta_c}{h^2}}
 \; \cdot \; \frac{p'_P - p'_N}{h}
 \;=\; O(h^2) \cdot \left. \frac{\mathrm{d}p'}{\mathrm{d}x} \right|_c
 \;\;\Rightarrow\;\; \left. \frac{\mathrm{d}p'}{\mathrm{d}x} \right|_c \;=\; O(h^{-2}) \cdot u'_c
\end{equation}
Thus, in order to effect a local velocity perturbation of wavelength equal to the grid spacing, as 
shown in Fig.\ \ref{fig: velocity perturbations}, the pressure gradient must be adjusted locally by 
an amount that scales as $(h^{-2})$, i.e.\ it must become larger and larger as the grid is refined. 
This is because it has to overcome the viscous resistance which scales as $\mathrm{d}^2u' / 
\mathrm{d}x^2|_c = O(h^{-2})$: grid refinement not only increases the velocity slopes (and 
associated viscous stresses) but also changes these slopes over a shorter distance (Fig.\ 
\ref{fig: velocity perturbations}). On the contrary, the part of the pressure gradient that is 
needed to balance the inertial contribution of $u'_c$ to the momentum balance is independent of the 
grid spacing. Conversely, Eq.\ \eqref{eq: ami 1D analysis} shows that fixed localised perturbations 
of the pressure gradient of wavelength equal to the grid spacing cause local velocity perturbations 
of magnitude $O(h^2)$, which become smaller and smaller with grid refinement due to increased 
viscous resistance.

The pseudo-velocities $u_f^{p+}$ and $u_f^{p-}$ defined by Eqs.\ \eqref{eq: up+} and \eqref{eq: 
up-}, have magnitudes of $O(h^2)$ (because $a_f^{mi} = O(h)$ and, in smooth pressure fields, both 
$p_P - p_N$ and $\nabla p \cdot (\vf{P} - \vf{N})$ are $O(h)$), whereas the actual velocity 
$\bar{\vf{u}}{}_{c_f}$ in \eqref{eq: mass flux} does not diminish with refinement but tends to a 
finite value, the exact velocity at the given point. Therefore, as the grid is refined the mass flux 
becomes dominated by the interpolated velocity $\bar{\vf{u}}{}_{c_f}$ while the stabilisation 
pseudo-velocities $u_f^{p+}$ and $u_f^{p-}$ diminish. This may raise concerns that the stabilising 
effect of the scheme may also diminish. However, this is not the case. Suppose that spurious 
pressure oscillations do arise, of amplitude $\Delta p^o$ (in the absence of preventive measures, 
this amplitude is unaffected by grid refinement \cite{Syrakos_06a}). The total pressure can be 
decomposed into a smooth part, which is close to the exact solution, and a spurious oscillatory 
part: $p = p^s + p^o$. Then the sum of pseudo-velocities at point $c$ of the grid shown in Fig.\ 
\ref{fig: 1D grid} is equal to
\begin{align}
\nonumber
 u_c^{p+} \;-\; u_c^{p-} \;&=\; a_c^{mi} \left[ 
 (p^s_P + p^o_P) - (p^s_N + p^o_N) \;-\; \bar{\nabla}^q_h (p^s+p^o)_c \cdot (\vf{P} - \vf{N}{}_f)
 \right]
\\[0.2cm]
\nonumber
 &=\; a_c^{mi} \Big[
 \Delta p^o \;+\;
 \underbrace{ (p^s_P - p^s_N) \;-\; 
              \bar{\nabla}^q_h p^s_c \cdot (\vf{P} - \vf{N}{}_f) }_{O(h^2) \;\mathrm{or}\; O(h^3)}
 \Big]
\\
\label{eq: pseudovelocity sum magnitude c}
 &\approx\; a_c^{mi} \, \Delta p^o
\end{align}
because $p^o_P - p^o_N = \Delta p^o$ and $\nabla^q_h p^o_P = \nabla^q_h p^o_N = 0$ (the 
$\nabla^q_h$ operator is insensitive to oscillations). Also, according to the discussion of Eq.\ 
\eqref{eq: momentum inteprolation truncation error}, the underlined terms add up to $O(h^2)$ (or 
$O(h^3)$ on smooth structured grids), which is small compared to $\Delta p^o$. In other words, 
$u_f^{p+}$ cancels out with the smooth part of $u_f^{p-}$ but leaves out the oscillatory part. A 
similar consideration for point $c_P$ (Fig.\ \ref{fig: 1D grid}) leads to
\begin{equation} \label{eq: pseudovelocity sum magnitude c_P}
 u_{c_P}^{p+} \;-\; u_{c_P}^{p-} \;\approx\; - a_{c_P}^{mi} \, \Delta p^o
\end{equation}
where the minus sign comes from the fact that, according to the nature of the spurious oscillation, 
if the oscillatory pressure component decreases by $\Delta p^o$ from $P$ to $N$, then it 
\textit{increases} by $\Delta p^o$ from $PP$ to $P$. Then the combined contribution of faces $c$ 
and $c_P$ to the image of the continuity operator for cell $P$ (the left-hand side of Eq.\ 
\eqref{eq: continuity integral discrete}) divided by the cell volume is
\begin{equation*}
 \frac{1}{\Omega_P} \left[ \dot{M}_c + \dot{M}_{c_P} \right] \;=\;
 \frac{1}{h^2} \rho \, h \left[
   ( \bar{u}_c \:+\: u_c^{p+} \:-\: u_c^{p-} )  \;-\;
   ( \bar{u}_{c_P} \:+\: u_{c_P}^{p+} \:-\: u_{c_P}^{p-} )
 \right]
\end{equation*}
which, using Eqs.\ \eqref{eq: pseudovelocity sum magnitude c} and \eqref{eq: pseudovelocity sum 
magnitude c_P}, and also assuming that $a_c^{mi} \approx a_{c_P}^{mi} \approx Ch$ for some constant 
$C$ ($C = 2(\kappa + \eta_a)$ if the grid is fine enough, from Eq.\ \eqref{eq: ami}), becomes
\begin{equation} \label{eq: mass flux sum: c + c_P}
 \frac{1}{\Omega_P} \left[ \dot{M}_c + \dot{M}_{c_P} \right] \;=\;
 \rho \frac{\bar{u}_c - \bar{u}_{c_P}}{h} \;+\; 2 \rho C \Delta p^o
\end{equation}
The first term on the right-hand side is the contribution of the actual velocities to the 
continuity image, and can be seen to tend to the $h$-independent value $\rho \partial u / \partial 
x|_P$ with grid refinement. The second term on the right-hand side is the contribution of the 
pseudo-velocities, \textit{which can also be seen to be $h$-independent}, i.e.\ it does not 
diminish with grid refinement, although the pseudo-velocities themselves do diminish compared to 
the actual velocities, as mentioned above. If we take into account also the other two faces of cell 
$P$ (the horizontal ones) then the total contribution of the pseudo-velocities to the continuity 
image is $4 \rho C \Delta p^o$. If we repeat this analysis for the neighbouring cells $N$ and $PP$, 
then it will turn out that the contributions of the pseudo-velocities to the continuity images of 
those cells are $-4 \rho C \Delta p^o$, i.e.\ they have opposite sign to that of cell $P$. Overall, 
in the presence of pressure oscillations the contributions of the pseudo-velocities to the 
continuity image over the entire grid have a checkerboard (oscillatory) pattern like the one shown 
in Fig.\ \ref{fig: oscillations schematic}, with amplitude proportional to the amplitude of the 
pressure oscillations, $\Delta p^o$. As discussed in relation to Eq.\ \eqref{eq: FVM system}, when 
solving the system of all continuity equations \eqref{eq: continuity integral discrete}, because 
the right-hand side is smooth (it is zero), spurious pressure oscillations, which would produce an 
oscillatory right-hand side, are excluded.

Thus, the effectiveness of momentum interpolation in suppressing spurious pressure oscillations 
does not degrade with grid refinement. The amplitude of the oscillations produced by momentum 
interpolation to the continuity image is proportional to the amplitude of the spurious pressure 
oscillations themselves, and independent of the grid spacing $h$.

\section{Energy dissipation and flow cessation}
\label{appendix: energy dissipation}
\setcounter{equation}{0}

This appendix sketches an explanation for the finite cessation times of HB fluids, but for rigorous 
proofs the reader is referred to the literature cited in Sec.\ \ref{ssec: results: cessation}. For 
inelastic fluids, once the lid motion ceases, the rate of energy dissipation equals the rate of 
decrease of the kinetic energy of the fluid \cite{Winter_1987}.
\begin{equation} \label{eq: cessation energy balance}
 \frac{\mathrm{d}}{\mathrm{d}t} \int_{\Omega} \tfrac{1}{2} \rho \vf{u} \cdot \vf{u} \, 
\mathrm{d}\Omega
 \;=\;
 -\int_{\Omega} \tf{\tau} : \nabla \vf{u} \, \mathrm{d}\Omega
\end{equation}
For generalised Newtonian fluids ($\tf{\tau} \;=\; \eta(\dot{\gamma}) \, \dot{\tf{\gamma}}$) we have
\begin{equation} \label{eq: energy dissipation GN}
 \tf{\tau} : \nabla \vf{u}
 \;=\;
 \eta(\dot{\gamma}) \, \dot{\tf{\gamma}} : \nabla \vf{u}
 \;=\;
 \eta(\dot{\gamma}) \, \dot{\gamma}^2
\end{equation}
For a HB fluid, $\eta(\dot{\gamma}) = (\tau_y + k\dot{\gamma}^n) / \dot{\gamma}$ and the energy 
dissipation rate \eqref{eq: energy dissipation GN} becomes $(\tau_y + k\dot{\gamma}^n) \, 
\dot{\gamma}$; in fact, this expression holds even in unyielded regions as it predicts zero energy 
dissipation there\footnote{In HB unyielded regions the energy dissipation is zero due to 
$\dot{\tf{\gamma}} = 0$ since, by symmetry of the stress tensor, we have $\tf{\tau} : \nabla \vf{u}$ 
= $\frac{1}{2} \, \tf{\tau} : \nabla \vf{u} \,+\, \frac{1}{2} \, \tf{\tau} : \nabla \vf{u}$ = 
$\frac{1}{2} \, \tf{\tau} : \nabla \vf{u} \,+\, \frac{1}{2} \, \tf{\tau}^{\mathrm{T}} : \nabla 
\vf{u}^{\mathrm{T}}$ = $\tf{\tau} : ( \frac{1}{2} \nabla \vf{u} + \frac{1}{2} \nabla 
\vf{u}^{\mathrm{T}})$ = $\tf{\tau} : \dot{\tf{\gamma}}$ = 0.} due to $\dot{\gamma} = 0$. So, for HB 
flow the energy balance \eqref{eq: cessation energy balance} becomes
\begin{equation} \label{eq: cessation energy balance HB}
 \frac{\mathrm{d}}{\mathrm{d}t} \int_{\Omega} \tfrac{1}{2} \rho \vf{u} \cdot \vf{u} \, 
\mathrm{d}\Omega
 \;=\;
 -\int_{\Omega} \left( \tau_y \,+\, k \dot{\gamma}^n \right) \dot{\gamma} \, \mathrm{d}\Omega
\end{equation}

Next we will assume that the velocity at any point $\vf{x}$ at any instant $t$ can be expressed as 
the product of a function of time, $\chi$, and a function of space, $u_0$, as
\begin{equation} \label{eq: cessation velocity assumption}
 \vf{u}(\vf{x},t) \;=\; \chi(t-t_0) \, u_0(\vf{x})
\end{equation}
where $u_0(\vf{x})$ is the velocity at time $t_0$, if we set $\chi(0) = 1$. We therefore assume that 
the velocity field retains its shape in time, but just downscales by a factor of $\chi(t-t_0)$ at 
time $t$ relative to time $t_0$. This assumption is correct for the cessation of Newtonian flow if 
$t_0$ is large enough \cite{Syrakos_2016a}, but obviously involves an error in the viscoplastic 
case, where the unyielded regions expand in time; nevertheless, we will assume this error to be 
acceptable, not invalidating the final conclusion. Then, let $V(t)$ be a measure of the velocity in 
the domain, e.g.\ the mean or maximum velocity magnitude. Whatever the precise choice of $V$, it 
follows from \eqref{eq: cessation velocity assumption} that $V(t) = \chi(t-t_0) \, V(0)$. We can 
then normalise the velocity as
\begin{equation} \label{eq: cessation velocity normalisation}
 \frac{\vf{u}(\vf{x},t)}{V(t)} \;=\;
 \frac{\chi(t-t_0) \, u_0(\vf{x})}{\chi(t-t_0) \, V(0)}
 \;=\;
 \frac{u_0(\vf{x})}{V(0)}
 \;\equiv\; \tilde{\vf{u}}{}_0(\vf{x})
 \;\Rightarrow\;
 \vf{u}(\vf{x},t) \;=\; \tilde{\vf{u}}{}_0(\vf{x}) \, V(t)
\end{equation}
Substituting for $\vf{u}$ from \eqref{eq: cessation velocity normalisation} into the left and right 
side of Eq.\ \eqref{eq: cessation energy balance HB} we get, respectively,
\begin{equation} \label{eq: cessation lhs}
 \int_{\Omega} \tfrac{1}{2} \rho \vf{u} \cdot \vf{u} \, \mathrm{d}\Omega 
 \;=\;
 V^2 \, \int_{\Omega} \tfrac{1}{2} \rho \tilde{\vf{u}}{}_0 \cdot \tilde{\vf{u}}{}_0 \, 
   \mathrm{d}\Omega
 \;=\;
 C_k \, V^2
\end{equation}
\begin{equation} \label{eq: cessation rhs}
  \int_{\Omega} \left( \tau_y \,+\, k \dot{\gamma}^n \right) \dot{\gamma} \, \mathrm{d}\Omega
  \;=\;
  \tau_y \, V \int_{\Omega} \tilde{\dot{\gamma}}_0 \, \mathrm{d}\Omega
  \;+\;
  k \, V^{n+1} \int_{\Omega} \tilde{\dot{\gamma}}_0^{n+1} \, \mathrm{d}\Omega
  \;=\;
  C_p \, \tau_y \, V
  \;+\;
  C_v \, k \, V^{n+1}
\end{equation}
where $\tilde{\dot{\gamma}}{}_0$ is the magnitude of $\tilde{\dot{\tf{\gamma}}}{}_0 \equiv \nabla 
\tilde{\vf{u}}{}_0 + (\nabla \tilde{\vf{u}}{}_0)^{\mathrm{T}}$ (it is dimensional). Since 
$\tilde{\vf{u}}{}_0$ is not a function of $t$, neither are the constants $C_k$, $C_p$ and $C_v$, 
and substituting \eqref{eq: cessation lhs} and \eqref{eq: cessation rhs} into \eqref{eq: cessation 
energy balance HB} we are left with the following ordinary differential equation:
\begin{equation} \label{eq: cessation ODE}
 \frac{\mathrm{d}(V^2)}{\mathrm{d}t}
 \;=\;
 -C'_p \tau_y V \;-\; C'_v k V^{n+1} \;\leq\; -C'_p \tau_y V
\end{equation}
where $C'_p = C_p / C_k$ and $C'_v = C_v / C_k$ are positive constants. The above inequality means 
that the flow will decay at least as fast as if the plastic term ($-C'_p \tau_y V$) was the only one 
present. In that case, solving the equation would give
\begin{equation} \label{eq: cessation V plastic}
 V(t) \;=\; V(t_0) \;-\; \frac{C'_p \tau_y}{2} (t-t_0)
\end{equation}
so that the flow reaches complete cessation in finite time: $V(t_c) = 0 \Rightarrow t_c = t_0 + 
2V(t_0)/(C'_p \tau_y)$. Thus, the plastic dissipation (due to $\tau_y$) alone suffices to bring the 
whole material to rest in finite time $t_c$; including also the viscous dissipation term, $-C'_v k 
V^{n+1}$, will only make this process faster.

\begin{figure}[thb]
  \centering
  \includegraphics[scale=0.80]{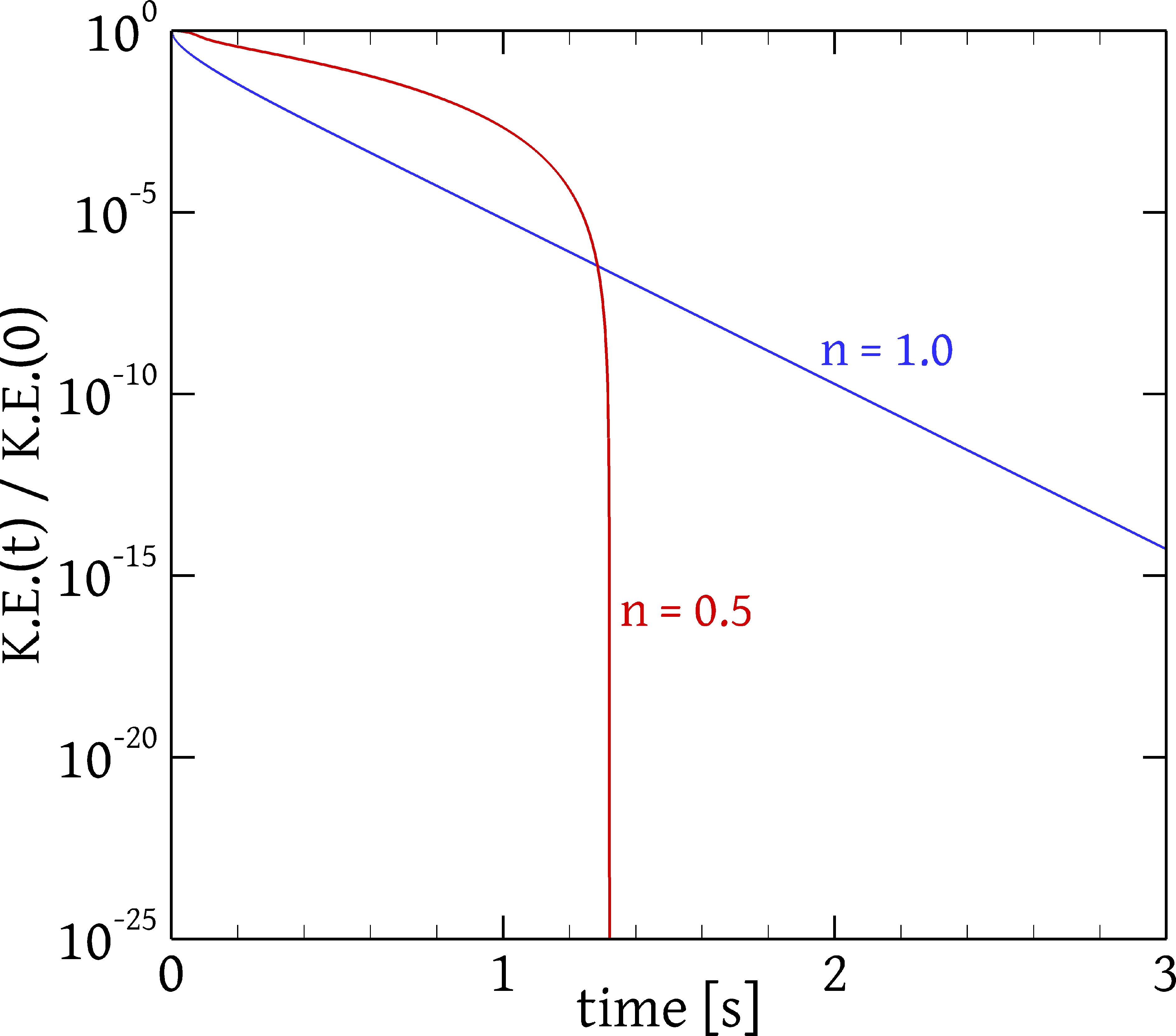}
  \caption{Decay of kinetic energy with time after lid motion cessation of two power law fluids 
(Eq.\ \eqref{eq: HB stress} with $\tau_y$ = 0, $k$ = 1 \si{Pa.s^n}, and $\rho$ = 1000 \si{kg/m^3}) 
of exponents $n$ = 1.0 and 0.5, respectively. The domain and grid are the same as in Sec.\ 
\ref{sec: results}, and the initial condition ($t$ = 0) is the steady state flow for a lid velocity 
of 1 \si{m/s}. The kinetic energy is normalised by its value at $t$ = 0. The lid is brought to a 
halt suddenly for $n$ = 1, but gradually over a time interval of 0.1 \si{s} for $n$ = 0.5 (to avoid 
iterative solver convergence difficulties).}
  \label{fig: PL cessation}
\end{figure}

It is interesting to briefly look at the case $\tau_y = 0$, when the plastic term is absent from 
\eqref{eq: cessation ODE}. If $n = 1$, the flow is Newtonian and the solution of the equation is
\begin{equation} \label{eq: cessation V Newtonian}
 V(t) \;=\; V(t_0) \, e^{-\frac{C'_v k}{2}(t-t_0)}
\end{equation}
This result is identical with what was found in \cite{Syrakos_2016a}, and means that the flow 
continuously decays but never completely ceases. If $n \neq 1$ (power-law fluid) the solution is
\begin{equation} \label{eq: cessation V power-law}
 V(t) \;=\; \left[ 
 V(t_0)^{1-n} \;-\; (1-n) \frac{C'_v k}{2} (t-t_0)
 \right]^{\frac{1}{1-n}}
\end{equation}
For $n>1$, the exponent $1/(1-n)$ is negative, the flow decays more slowly than in the Newtonian 
case as the viscosity drops with flow decay, and complete cessation is never reached. However, 
in the $n < 1$ case the viscosity rises to infinity as the flow decays towards cessation; the 
exponent $1/(1-n)$ is positive, and cessation is reached in finite time, $V(t_c) = 0 \Rightarrow 
t_c = t_0 + 2 V(t_0)^{1-n} / ((1-n) C'_v k)$. Thus, the existence of a yield stress is not a 
prerequisite for finite cessation time (Fig.\ \ref{fig: PL cessation}).

\section{Second order accurate reconstruction of cell centre value from face values}
\label{appendix: derivation of reconstruction scheme}
\setcounter{equation}{0}

Suppose that the exact (or approximated to at least second-order accuracy) values $\phi_f$ of a 
quantity $\phi$ are known at the face centres of a cell $P$ (Fig.\ \ref{fig: grid nomenclature}), 
and we want to approximate from these values the value of $\phi$ at the cell centre, 
$\overline{\phi_P}$. One way to proceed is the following. The fact that the known values are 
located at the cell boundary provides an incentive to try to derive a scheme based on the 
divergence theorem. Let $\vf{r}(\vf{x}) = \vf{x} - \vf{P}$ be the vector function that returns the 
relative position of location $\vf{x}$ relative to the centroid $\vf{P}$. Then $\nabla \cdot \vf{r} 
= \nabla \cdot \vf{x} = D$ where $D = 2$ or 3 in two- and three-dimensional space, respectively. We 
can then use the product rule to the divergence of the product $\phi \vf{r}$:
\begin{equation*}
 \nabla \cdot (\phi \vf{r}) \;=\; \nabla \phi \cdot \vf{r} \;+\; \phi \, \nabla \cdot \vf{r}
                            \;=\; \nabla \phi \cdot \vf{r} \;+\; D \, \phi
\end{equation*}
Integrating both sides over the whole cell $P$ and applying the divergence theorem on the left-hand 
side we get
\begin{equation} \label{eq: integrated product rule}
 \sum_{f=1}^F \int_{s_f} \phi \, \vf{r} \cdot \vf{n}_f \, \mathrm{d}s
 \;=\;
 \int_{\Omega_P} \nabla \phi \cdot \vf{r} \, \mathrm{d}\Omega
 \;+\;
 D \int_{\Omega_P} \phi \, \mathrm{d} \Omega
\end{equation}
This is an exact equation, but now we will approximate each of the integrals by the midpoint rule:
\begin{align}
\label{eq: midpoint rule term 1}
  \int_{s_f} \phi \, \vf{r} \cdot \vf{n}_f \, \mathrm{d}s
  \;&=\;
  \phi(\vf{c}_f) \, (\vf{r}(\vf{c}_f) \cdot \vf{n}_f) \, s_f \;+\; O(h^2) \, s_f
\\
\label{eq: midpoint rule term 2}
  \int_{\Omega_P} \nabla \phi \cdot \vf{r} \, \mathrm{d} \Omega
  \;&=\;
  \nabla \phi(\vf{P}) \cdot \vf{r}(\vf{P}) \, \Omega_P \;+\; O(h^2) \, \Omega_P
  \;=\;
  0 \;+\; O(h^2) \, \Omega_P
\\
\label{eq: midpoint rule term 3}
  \int_{\Omega_P} \phi \, \mathrm{d} \Omega
  \;&=\;
  \phi(\vf{P}) \, \Omega_P \;+\; O(h^2) \, \Omega_P
\end{align}
where in \eqref{eq: midpoint rule term 2} we have used that $\vf{r}(\vf{P}) = \vf{P} - \vf{P} = 0$. 
Substituting these into \eqref{eq: integrated product rule} we get
\begin{equation} \label{eq: reconstruction error exaggerated}
 \phi(\vf{P}) \;=\; \underbrace{\frac{1}{D \, \Omega_P} \sum_{f=1}^F \left[ 
                    \phi(\vf{c}_f) \, s_f \, \vf{r}(\vf{c}_f) \cdot \vf{n}_f
                    \right]}_{\overline{\phi_P}}
              \;+\; \underbrace{\sum_{f=1}^F O(h^2) \frac{s_f}{\Omega_P}}_{= O(h)}
\end{equation}
The first term on the right hand side is the approximation $\overline{\phi_P}$, as can be seen by 
comparing to Eq.\ \eqref{eq: stress reconstruction}. The second term on the right hand side is the 
difference between this approximation and the exact value $\phi(\vf{P})$, i.e.\ the error, which is 
$O(h)$ because $s_f/\Omega_P = O(h^{-1})$. Therefore, the approximation \eqref{eq: stress 
reconstruction} appears to be only first-order accurate, of the same order as simply setting 
$\overline{\phi_P} = \phi(\vf{c}_f)$ for any arbitrary face $f$!

Is it possible that this error estimation is too pessimistic and does not account for some error 
cancellation that occurs when adding the truncation errors of the midpoint rule approximations 
\eqref{eq: midpoint rule term 1} -- \eqref{eq: midpoint rule term 3}? In fact this happens to be 
the case\footnote{A calculation involving Taylor expansions along the faces shows that, in the sum 
of the truncation errors of Eq.\ \eqref{eq: midpoint rule term 1} over all cell faces, their 
leading order terms cancel out.} and it turns out that the approximation \eqref{eq: stress 
reconstruction} is second-order accurate, which is easiest to prove by showing that it is exact for 
linear functions. Indeed, if it is exact for linear functions then expanding $\phi$ in a Taylor 
series about $\vf{P}$ we find that the error is
\begin{align*}
 \phi(\vf{x}) \;&=\; \phi(\vf{P}) \;+\; \nabla \phi (\vf{P}) \cdot \vf{r}(\vf{x}) \;+\; O(h^2)
 \;\Rightarrow
\\
 \overline{\phi_P} \;&=\; 
 \overline{[\phi(\vf{P}) \;+\; \nabla \phi (\vf{P}) \cdot \vf{r}(\vf{x})]_P}
 \;+\; \overline{O(h^2)_P}
\\
 &=\; \phi(\vf{P}) \;+\; \nabla \phi(\vf{P}) \cdot \vf{r}(\vf{P}) \;+\; \overline{O(h^2)_P}
\\
 &=\; \phi(\vf{P}) \;+\; O(h^2)
\end{align*}
where we have used the facts that our interpolation operator \eqref{eq: stress reconstruction} is 
linear ($\overline{[\phi + \psi]_P} = \overline{\phi_P} + \overline{\psi_P}$), that it is exact for 
linear functions (and $\phi(\vf{P}) + \nabla \phi(\vf{P}) \cdot \vf{r}(\vf{x})$ is a linear 
function), that $\vf{r}(\vf{P}) = 0$, and finally that $\overline{O(h^2)_P} = O(h^2)$ (which can be 
seen by replacing $\tau_{ij,f}$ with $O(h^2)$ in the formula \eqref{eq: stress reconstruction} and 
noting that both $s_f (\vf{c}_f-\vf{P})\cdot \vf{n}_f$ and $\Omega_P$ are of the same order, 
$O(h^2)$ in 2D and $O(h^3)$ in 3D).

\begin{figure}[thb]
  \centering
  \includegraphics[scale=1.00]{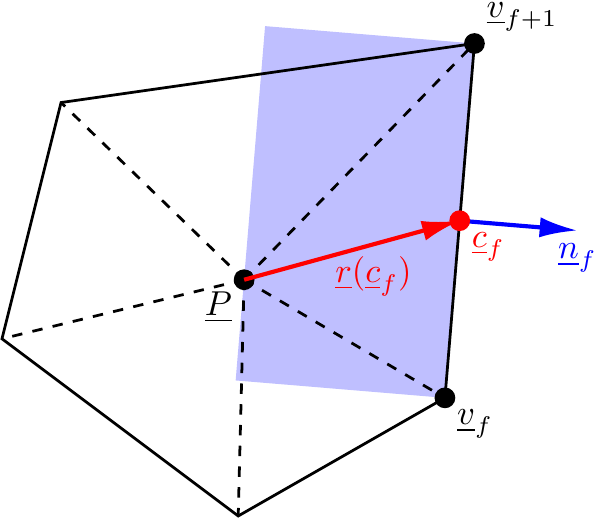}
  \caption{Notation concerning the geometry of cell $P$.}
  \label{fig: geometry theorems}
\end{figure}

It remains to show that the interpolation operator is exact for linear functions. Suppose a linear 
function $\phi(\vf{x}) = \phi(\vf{P}) + \vf{g} \cdot \vf{r}(\vf{x})$ where $\vf{g}$ is a constant 
vector (it is the gradient of $\phi$). We want to show that $\overline{\phi_P} = \phi(\vf{P})$. 
Applying the approximation \eqref{eq: stress reconstruction} to $\phi$ we get
\begin{align}
\nonumber
 \overline{\phi_P} \;&=\; \frac{1}{D\,\Omega_P} \sum_{f=1}^F \left[
 \left( \phi(\vf{P}) + \vf{g} \cdot \vf{r}(\vf{c}_f) \right) (\vf{r}(\vf{c}_f) \cdot \vf{n}_f)\,s_f
 \right]
\\
\label{eq: interpolate linear function}
 &=\;  \phi(\vf{P}) \, \frac{1}{D\,\Omega_P} \sum_{f=1}^F \left[ (\vf{r}(\vf{c}_f) \cdot \vf{n}_f) 
       \, s_f \right]
 \;+\; \frac{1}{D\,\Omega_P} \, \vf{g} \cdot \sum_{f=1}^F \left[ \vf{r}(\vf{c}_f) (\vf{r}(\vf{c}_f) 
       \cdot  \vf{n}_f) \, s_f \right]
\end{align}
To proceed further we need a couple of geometrical results. Figure \ref{fig: geometry theorems} 
shows cell $P$ divided into $F$ triangles by the dashed lines which connect the centroid $\vf{P}$ 
to the vertices $\vf{v}_f$. Consider the triangle $(\vf{P},\vf{v}_f,\vf{v}_{f+1})$, which has 
face $f$ as one of its sides. The product $\vf{r}(\vf{c}_f) \cdot \vf{n}_f$ is the perpendicular 
distance from $\vf{P}$ to face $f$, and therefore the product $(\vf{r}(\vf{c}_f) \cdot \vf{n}_f) \, 
s_f$ is equal to the shaded area of Fig.\ \ref{fig: geometry theorems}, which is enclosed in a 
rectangle with one side coinciding with face $f$, another parallel side through $\vf{P}$, and two 
perpendicular sides passing through the vertices $\vf{v}_f$ and $\vf{v}_{f+1}$, respectively. The 
area of the triangle $(\vf{P},\vf{v}_f,\vf{v}_{f+1})$ is half that of this rectangle (in three 
dimensions, the volume of the cone-like shape obtained by joining $\vf{P}$ with the edges of face 
$f$ is one third of the volume of the shaded rectangular box). By adding the areas of all triangles 
we get the total area of cell $P$\footnote{Result \eqref{eq: volume of cell} is also obtainable by 
setting $\phi = 1$ in Eq.\ \eqref{eq: integrated product rule}.}:
\begin{equation} \label{eq: volume of cell}
 \frac{1}{D} \sum_{f=1}^F \left( \vf{r}(\vf{c}_f) \cdot \vf{n}_f \right) s_f \;=\; \Omega_P
\end{equation}

A second geometric result that will be needed is the following. The centroid of the triangle
$(\vf{P},\vf{v}_f,\vf{v}_{f+1})$ is $(\vf{P} + \vf{v}_f + \vf{v}_{f+1}) / 3 = (\vf{P} + 2 \vf{c}_f) 
/ 3$. The centroid of the whole cell, $\vf{P}$, is equal to the sum of the individual centroids of 
the triangles, $\vf{P}_f$ say, weighted by their areas, $\Omega_f = (1/2)(\vf{r}(\vf{c}_f) \cdot 
\vf{n}_f)s_f$:
\begin{equation*}
\nonumber
 \vf{P} \;=\; \frac{1}{\Omega_P} \sum_{f=1}^F \vf{P}_f \, \Omega_f
 \;\; \Rightarrow \;\;
 \vf{P} \Omega_P \;=\; \sum_{f=1}^F \frac{1}{3}\left( \vf{P} + 2\vf{c}_f \right) 
                                 \frac{1}{2} \left(\vf{r}(\vf{c}_f) \cdot \vf{n}_f \right) s_f
\end{equation*}
Substituting for $\Omega_P$ from \eqref{eq: volume of cell} (with $D=2$), moving everything to the 
left hand side, and merging the two sums, we arrive at
\begin{equation} \label{eq: geometric result 2}
 \sum_{f=1}^F \vf{r}(\vf{c}_f) \left( \vf{r}(\vf{c}_f) \cdot \vf{n}_f \right) s_f \;=\; 0
\end{equation}
where we have used also that $\vf{c}_f - \vf{P} = \vf{r}(\vf{c}_f)$. In three dimensions, the exact 
same equation \eqref{eq: geometric result 2} holds, but is derived by noting that the volume of the 
cone- or pyramid-like shape whose base is face $f$ and its apex is at $\vf{P}$ is $\Omega_f = (1/3) 
(\vf{r}(\vf{c}_f) \cdot \vf{n}_f)s_f$ ($D = 3$ in Eq.\ \eqref{eq: volume of cell}) while its 
centroid is $\vf{P}_f = (\vf{P} + 3\vf{c}_f)/4$.

Finally, substituting Eqs.\ \eqref{eq: volume of cell} and \eqref{eq: geometric result 2} in the 
first and second terms, respectively, of the right-hand side of Eq.\ \eqref{eq: interpolate linear 
function}, we arrive at $\overline{\phi_P} = \phi(\vf{P})$: the interpolation scheme \eqref{eq: 
stress reconstruction} is exact for linear functions, and is therefore second-order accurate.

\end{appendices}




\bibliographystyle{ieeetr}
\bibliography{elastoviscoplastic_FVM}









\end{document}